# A terminal-group and fluorination matrix for designing polar nematic liquid crystals

Calum J. Gibb*[1], Jordan Hobbs[2], William C. Ogle[1,2,3], Richard. J. Mandle[1,2]

[1] School of Chemistry, University of Leeds, Leeds, UK, LS2 9JT
[2] School of Physics and Astronomy, University of Leeds, Leeds, UK, LS2 9JT
[3] School of Applied Mathematics, University of Leeds, Leeds, UK, LS2 9JT

*Author for correspondence e-mail: c.j.gibb@leeds.ac.uk

## Abstract

Lateral fluorination has emerged as a powerful design variable in the design of polar liquid crystals, but its interplay with terminal-group identity remains poorly understood. Here we report a 40-member matrix of biphenyl-ester-phenyl mesogens that independently varies lateral core fluorination and terminal-group identity. Increasing lateral fluorination generally suppresses apolar mesophases and clearing points, however this same change can enable polar nematic phase behaviour in conjunction with suitable terminal units. Graph Fourier analysis of conformer averaged atomic charges shows polar materials occupy a high-frequency region of charge-spectral space, consistent with a bond-scale electrostatic motif. Different terminal units can bias this towards higher or lower frequencies, and in doing so can enable polar order to emerge.

## Introduction

Spontaneous ferroelectric order in nematic liquid crystals is now a firmly established phenomenon. The discovery of ferroelectric nematic ($N_F$) phase demonstrated that macroscopic polarisation could exist in a three-dimensional fluid, liquid-crystalline state [1-6]. Spontaneous polar ordering has also been found in established phases with lamellar ordering including orthogonal and tilted smectic phases, $SmA_F$[7] and $SmC_P$[8-10] (respectively), as well as additional phases with more complex structures[11-13]. In parallel, antiferroelectric nematic and smectic variants of these phases also exist, such as the splay nematic phase ($N_S$ [14], also referred to as $SmZ_A$ [15, 16], $N_X$, and $N_{AF}$ [17]), the anti-ferroelectric smectic A ($SmA_{AF}$ [11, 18]), and anti-ferroelectric smectic C ($SmC_{AF}$ [19]) phases. The combination of fluidity, orientational order, and macroscopic polarisation make these materials attractive for electro-optic, nonlinear-optical, sensing, and responsive soft-matter applications [20-23]. However, the development of new materials, and of transferable design principles that govern polar phase formation, remains a major stumbling block.

Fluorination has long been central to the design of liquid-crystalline materials [24-27]. In the case of polar liquid crystals, the fluorination pattern can be a decisive design variable which controls polar vs apolar organisation [28, 29] and the balance between nematic and lamellar organisation [30]. These effects are striking, but they arise primarily from lateral substitution and indirect structural/electronic effects: fluorine atoms modify the electrostatic surface[28], steric and conformational preferences of the molecule [31], while contributing only indirectly to the electric dipole projection along the molecular long axis. Indeed, when employed as a terminal group, fluorine is generally less effective at promoting polar phases than the analogous nitrile or nitro compounds [32-39].

By contrast, the role of the terminal groups as a systematic design variable in polar liquid crystals has not been explored to the same depth. Terminal groups, by definition, act along the principal molecular axis and are therefore directly implicated in longitudinal polarity and end-to-end interactions. Terminal group choice is central to conventional *apolar* liquid-crystal design, where nitrile, nitro, fluoro, trifluoromethyl, isothiocyanato, and pentafluorosulfanyl groups have all been used to tune mesophase stability and physical properties. However, it is not yet clear how these terminal groups (or other less commonly encountered, such as $OCF_3$, $SCF_3$, $NFCF_3$ etc) interact with lateral fluorination patterns to promote or suppress *polar* organisation.

Here, we examine this relationship using a 40-member matrix of biphenyl-ester-phenyl materials in which lateral core fluorination and terminal group identity are varied independently. This design allows matched-pair compositions across fluorination pattern and terminal motif, revealing terminal-group identity to be a key design variable that gates polar order.

## Results And Discussion

The transitional properties of the 40 materials are summarised **Fig. 1** and **Tables S1-4**. For simplicity, we refer to each material alphanumerically, where the number refers to the number of fluorine atoms on the biphenyl unit (i.e. **1-4**) and the letter refers arbitrarily to the terminal-group (**a-j**). For example, the material with one F atom on the biphenyl unit and the nitrile head moiety with no appended F atoms is denoted as **1a**. The chemical synthesis and structural characterisation of these materials can be found in in the ESI to this article. We generally found that increasing the degree of fluorination of materials correlated directly to a decrease in the yield of the product, from an average of 76% for those with 1 appended F atom to 32% with 4 appended F atoms. Notably, **4h** could not be isolated presumably due to the combination of both a highly electron poor phenol and carboxylic acid.

Initial assignments for the materials were made based on polarised optical microscopy (POM) by the observation of characteristic textures (**Fig. 2a)**, and differential scanning calorimetry (DSC, **Fig. 2b** and **2c**) before the identity of thermally accessible lamellar phases were confirmed by small angle X-ray scattering (SAXS) (**Fig. 2d** and **e**). Current response measurements were used to confirm the polar nature of the $N_S$ and $N_F$ phases (**Fig. 2f** and **g**). POM observations of nematic phases showed either schlieren or marble textures while SmA phases formed focal conics. For SmA phases formed directly from the isotropic the texture grew from bâtonnets. The $N_F$ phase formed from droplets upon direct transition from the isotropic and then formed the characteristic banded texture upon sufficient cooling. The SmA phases exhibited by these materials generally exhibited unusually high enthalpies which is explained by rather defined layers, as evidenced by the number of higher order harmonics of the smectic layer scattering peak observed using X-ray scattering. Normally for a SmA phase the layers are diffuse density modulations while for the materials here the layers will have rather sharp boundaries between regions of high and low electron density. A similar effect has been observed for $SmA_F$ and $SmC_P$ exhibiting materials [10]. Ferroelectric order in the nematic phases was confirmed using a current response measurement where a single peak indicative of polarisation reversal was observed for all $N_F$ exhibiting materials. For **4b** and **4d** we observed a transition into a mesophase whose optical textures were judged consistent with the $N_F$ on deep supercooling, however crystallisation of the samples precludes confirmation of this assignment by current response measurement. The indicated transition temperature values in **Fig. 2** are therefore considered tentative.

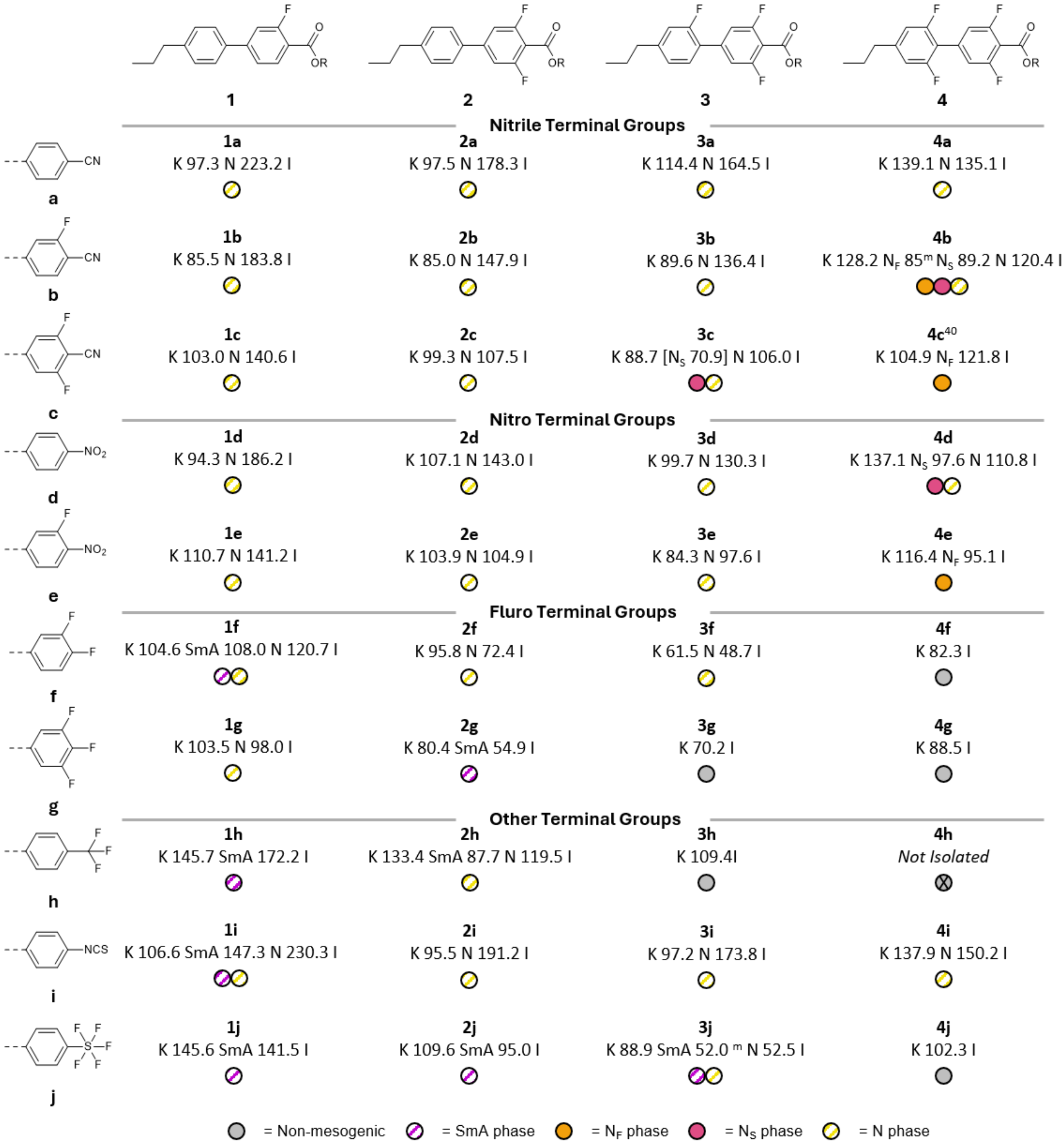

**Fig. 1:** Molecular design matrix for the 40 materials in the terminal-group and lateral-fluorination screening. The associated transitional properties and transition temperatures (given in °C) are given below each material which are grouped by fluorination pattern on the precursor biphenyl acid (columns) and the identity of the terminal group in the precursor phenol (rows). $^m$ indicates a transition

temperature as observed by POM NB. material **4c** has been reported previously [40].

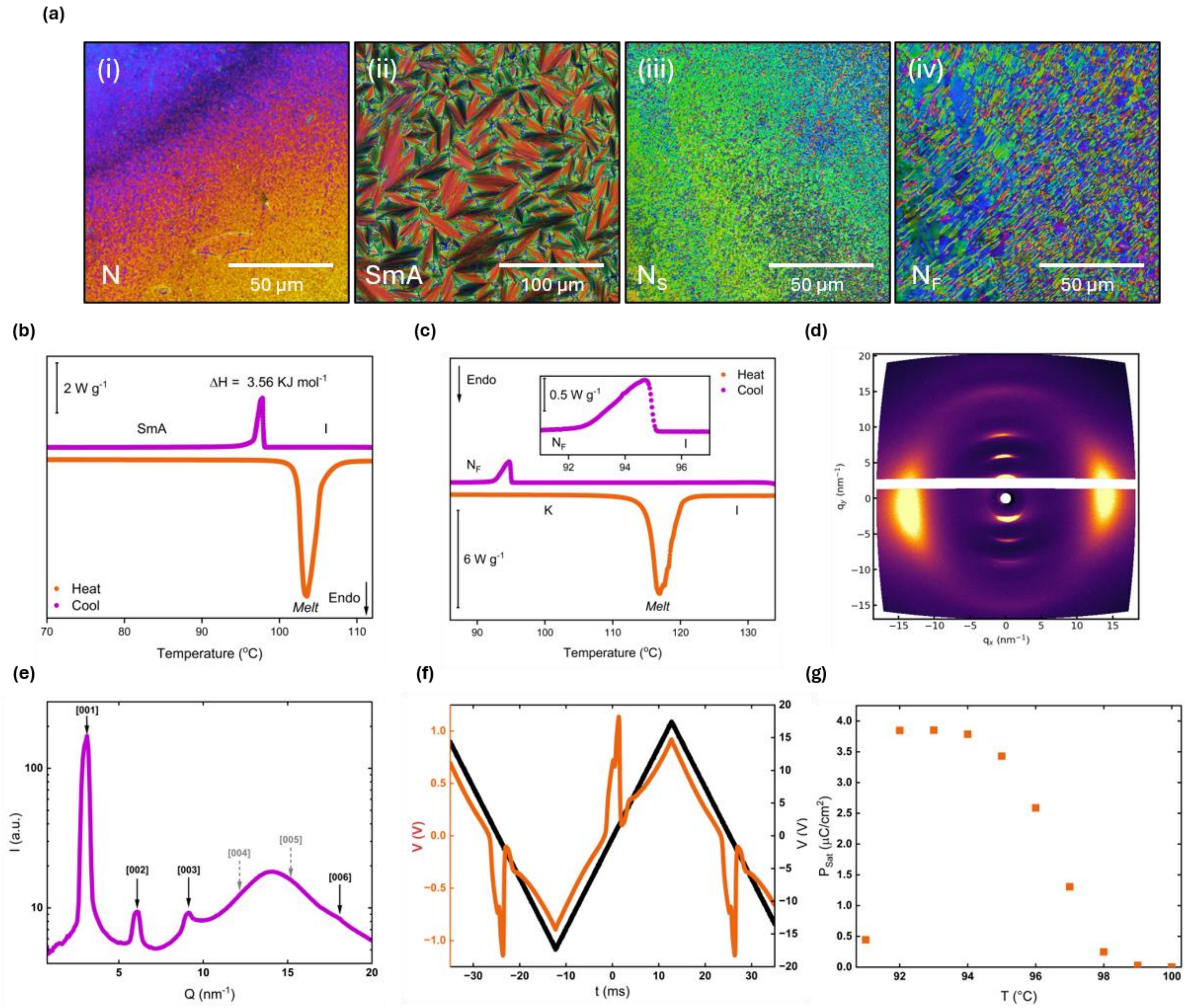

**Fig 2:** Characterisation of materials in this work: **(a)** polarised optical microscopy images of the **(i)** N (**1a**, 180 °C), **(ii)** SmA (**4h**, 160 °C), **(iii)** $N_S$ (**3c**, 68 °C), and **(iv)** $N_F$ (**4c**, 110 °C) phases. Micrographs were taken of thin samples sandwiched between untreated glass coverslips; differential scanning calorimetry traces of **(b) 1g** and **(c) 4e**; **(d)** 2D X-ray scattering pattern for for **1h**; **(e)** 1D scattering pattern obtained by radially integrating fig. 2d. Marked are the peaks and their indexing. Dashed peaks are not observed in the 1D integration but can be observed in the 2D image. **(f)** current response measurement for **4e,** measured at 20 Hz and 10 $V_{RMS}$ in a 5 µm thick cell with no alignment layer; **(g)** temperature dependence of the saturated polarization ($P_S$) for **4e.**

A general trend is apparent from **Fig. 1**: increasing the degree of lateral fluorination suppresses the apolar nematic clearing point ($T_{N-I}$). Across matched pairs, defined here as molecules with the same terminal motif and differing by a single fluorine atom, each additional fluorine lowers $T_{N\text{-}I}$ by an mean of 26 ± 13°C. The destabilisation effect is sufficiently strong that several terminal-group families become non-mesogenic at the highest lateral fluorine count, however, the same increase in fluorination can stabilise, or even induce, polar nematic-type ordering. In the present series this behaviour is only seen for nitrile- and nitro- terminated materials, and not for pentafluorosulphanyl-, isothiocyanato-, trifluoromethyl-, or fluoro-; however, we note that examples of polar organisation in fluoro- [41], and trifluoromethyl- [36, 42] terminated materials are widespread for other core-systems. Nevertheless, this selectivity for nitro and nitrile in this materials family is surprising – the unsuccessful terminal groups are strongly electron withdrawing, polarisable, and used to varying degrees in conventional nematic liquid crystal design [43, 44]. The emergence of polar order is not a consequence of increased electron withdrawing character, polarity, or any simple descriptor. Instead, polar ordering appears to require a specific compatibility between the terminal group and the laterally fluorinated core, such that the molecule combines sufficient local electrostatic structure with a shape and packing motif capable of supporting polar organisation.

The nitro group appears to be slightly more effective than the nitrile in supporting polar order. Notably, the nitrile-terminated **4a** exhibits only an apolar nematic, where the analogous nitro-terminated material (**4d**) exhibits an $N_S$ phase. Similarly, **4b** exhibits the phase sequence I-N-$N_S$-$N_F$ whereas the equivalent nitro-terminated compound (**4e**) has a direct I-$N_F$ transition. On average, the nitro group lowers the conventional nematic clearing point relative to a nitrile by 35 ± 7 °C alongside this modest increase in the thermal stability of polar order. These effects are not accompanied by a systematic change in melting point, with $T(K)_{NO_2}$- $T(K)_{CN}$ = -0.9 ± 13 °C. Although inadept at generating polar organisation, we note the isothiocyanate group (NCS) gives reliably higher clearing points for apolar phases than nitrile or nitro terminated materials with equivalent fluorination pattern.

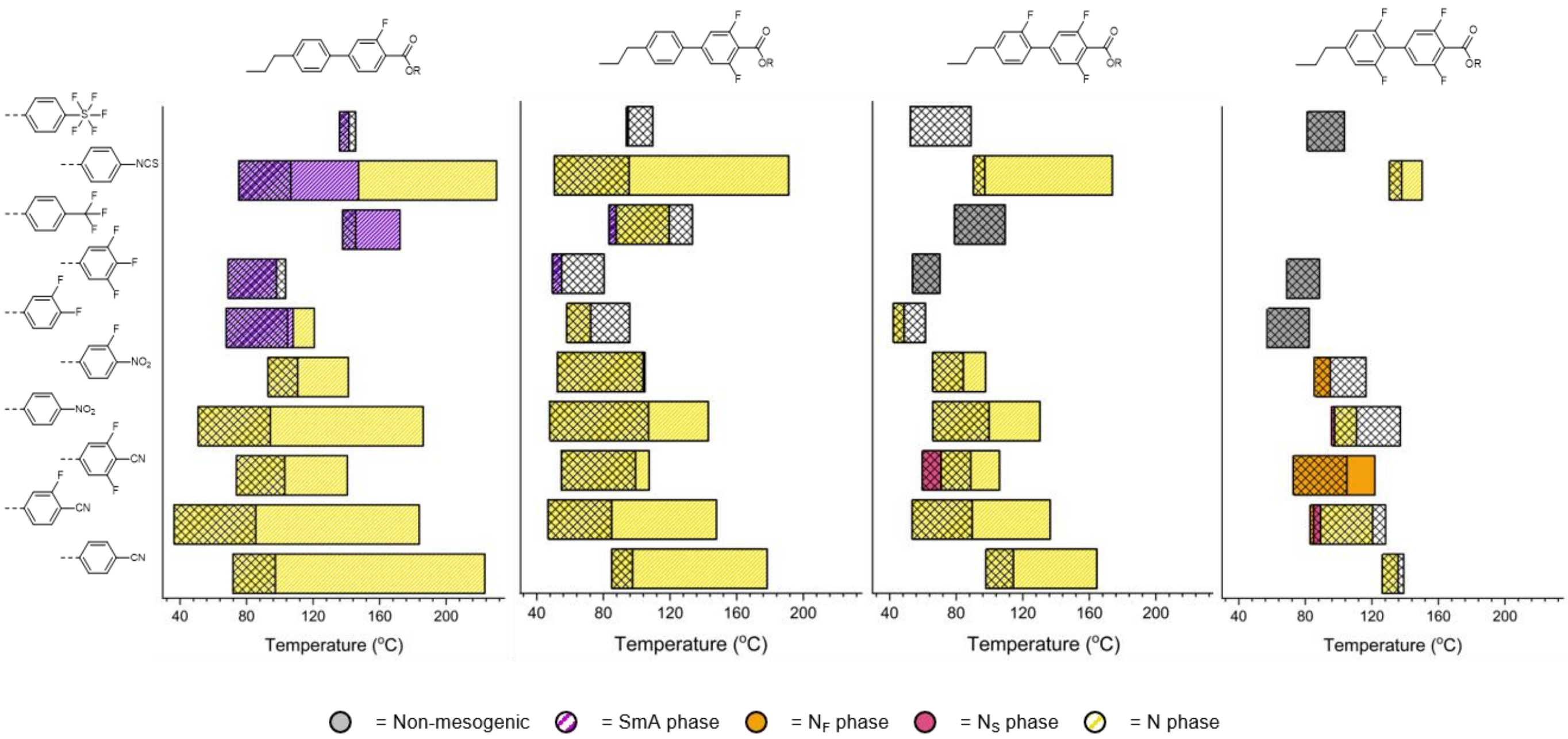

**Fig. 3.** The dependence of the transition temperatures in the molecular design matrix for the 40 materials in the terminal-group and lateral-fluorination screening. The hash bar indicates the meting point of the material; any transition within the hashed region is supercooled below the melting point.

We next consider the relationship between the 40 present materials and the fluorinated biphenyl benzoates reported previously by Gibb et al. (**Fig. 4**) [28]. Several of the materials reported here can be regarded as close constitutional analogues of these materials, differing in the placement of the carboxylate ester within the molecular framework. This subtle change preserves the overall polarity of the molecule yet leads to a systematic reduction in the onset of the $N_F$ (or $N_S$) phase, from ~ 12 °C in the case of **(1)** to ~ 60 °C in the case of **(2)**. Conversely, extending the present scaffold by incorporating an additional phenyl ring produces a dramatic increase in the onset temperature of either $N_F$ or $N_S$ phases (**NC2221**, **NC1222**, **NC2222**) [30]. Augmenting the structure of **4a** – which exhibits non-polar nematic phase only – with an additional ester linkage leads to the material **3CN²** [45] which is reported to exhibit both $N_F$ and $N_S$ phases. These comparisons emphasise that lateral fluorination alone does not determine polar nematic behaviour; rather, the emergence and stability of polar organisation instead depend upon the combination of fluorination, terminal-group choice, and the broader mesogenic core structure.

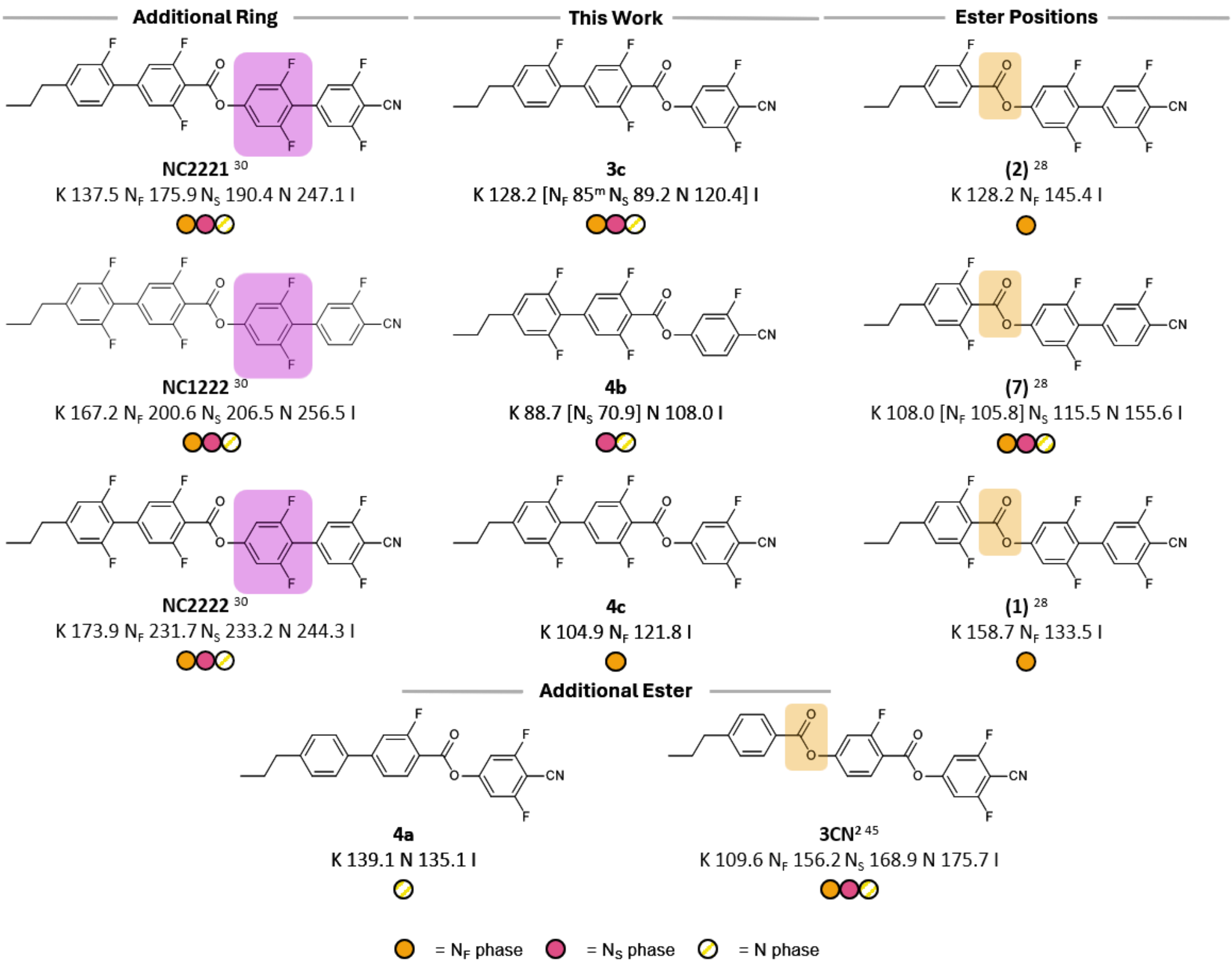

**Fig. 4:** Comparison of phase behaviour across homologous material families. Rows represent the same hydrogen/fluorine (H/F) substitution pattern; columns represent different families of related materials with an additional ring (left: **NC2221**, **NC1222**, **NC2222**; ref [30]); this work (centre); and with varied ester

position (right: **2**, **7**, **1**; ref [28]). The inclusion of an additional ester group to **4a** to give **3CN²**, reported in ref [45], induces polar order. [m] indicates a transition temperature as observed by POM.

Comparison of these constitutional and structurally comparable analogous (**Fig. 4**) illustrates gaps remain in our understanding. Polar organisation may not be described by considering terminal group or fluorination pattern alone, behaviours in one family do not necessarily translate directly to related sets of materials, and apparently plausible terminal-groups fail to generate polar order even when complemented by favourable fluorination patterns [46]. This suggests polar ordering is not enabled or described by a simple unimolecular feature, but the more nuanced way in which electrostatic structure maps across the molecular framework, and how this is then implicated in the interactions between pairs (or groups) of molecules. Models such as the commonly cited Madhusudana model [47] are useful in providing rational for why certain chemical modifications result in polar phases or not but still fall short in allowing for the transfer of structure-property understanding in one set of materials to another.

To probe this, we analysed the conformer-averaged partial atomic charges using a graph Fourier transform (GFT), treating the molecular graph as the basis on which the charge distribution is decomposed into graph-frequency components as per Palacio-Betancur and Jackson [48]. In this representation, low graph frequencies correspond to charge distributions that vary smoothly over the molecular connectivity. For the normalised graph, Laplacian eigenvalues lie in the interval $0 \leq \lambda \leq 2$. At the limiting cases, the $\lambda = 0$ mode corresponds to a constant mode over the entire molecule, whereas $\lambda = 2$ corresponds to the highest frequency oscillation supported by the molecular graph, in which neighbouring atoms have eigenvector amplitudes of opposite sign. Accordingly, charge-spectral power at $\lambda \approx 2$ reflects the projection of the atomic charge distribution onto a bond-alternating electrostatic mode. This should not be misunderstood as molecular polarity in the conventional dipolar sense; rather, it describes a highly localised pattern of charge alternation across the molecular framework.

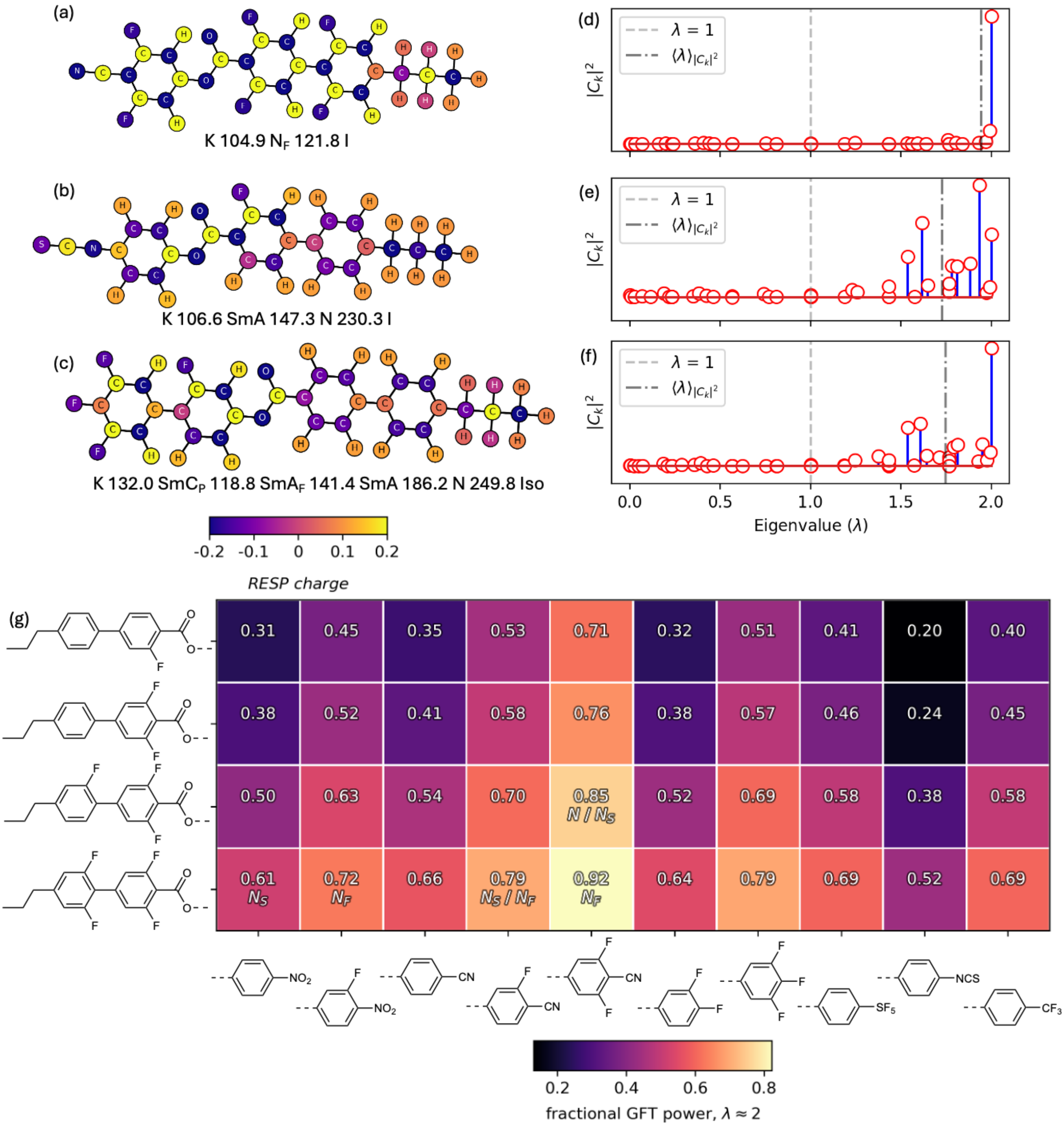

**Fig. 5:** Molecular graph representations of (a) compound **1** (polar nematic; ref [28]), (b) **4i** (apolar; this work), and (c) **F2100** (polar smectic; ref [30]), with mapped partial atomic charges *via* the RESP scheme. Normalised graph Fourier transform (GFT) power spectra (d-f) for compound **1** (d), **4i** (e), and **F2100** (f) obtained using RESP atomic partial charges. (g) Heatmap of fractional non-zero GFT power at $\lambda \approx 2$ as a function of terminal group identity and fluorination pattern; polar phase types ($N_F$, $N_S$) are indicated for materials that exhibit these experimentally.

We first use three representative materials to establish the chemical meaning of the charge-spectral descriptors: compound **1** exhibits an $N_F$ phase (**Fig. 5a**); compound **4i** exhibits only apolar mesophases (**Fig. 5b**); compound **F2100** exhibits polar smectic phases (**Fig. 5c**; $SmA_F$ and $SmC_P$ [30]). The $N_F$ material (**1**) is dominated by charge-spectral power close to λ= 2, indicating that its charge distribution is projected onto the highest-frequency, bond-alternating graph mode (**Fig. 5a**). The apolar material (**4i**) also contains substantial high-frequency power, but this is distributed over many modes at λ ≥ 1.5 rather than being distilled into a single feature as with **1**. The polar smectic material (**F2100**) retains a strong λ ≈ 2 mode but with additional discrete spectral weight around λ ≈ 1.5. This intermediate-frequency component is consistent with charge modulation over larger molecular fragments, superimposed on the bond-scale alternation associated with λ ≈ 2. We interpret the polar smectic spectrum as containing an $N_F$ like electrostatic motif with an additional segment-scale electrostatic contrast that may contribute to lamellar organisation.

We interrogate the full 40-member matrix using the workflow above, representing the materials with a single high-frequency GFT descriptor corresponding to the fraction of spectral power at λ ≈ 2. We then grouped materials by the identity of the terminal group and fluorination pattern (**Fig. 5g**). Without exception, increasing lateral fluorination increases the fractional spectral power at λ ≈ 2, in line with expectations considering the C-F bond is strongly polarised. Materials exhibiting polar phases (both $N_F$ and $N_S$) are located towards the higher-frequency end of this distribution, consistent with the representative spectra (discussed above).

That said, exhibiting a high population of this spectral power does not guarantee that polar order is observed. Several materials which form only apolar phases (i.e. **4a**) or are non-mesogenic (i.e. **4j** and **4g**) occupy similar regions of GFT space to those that exhibit polar phases (i.e. **4d** and **4e**). Evidently, a single GFT descriptor is not a complete phase classifier, but rather it identifies an electrostatic regime in which polar organisation may be accessible; whether this tendency is expressed experimentally depends on, among other things, the melting point and supercooling range. Together these act as an additional experimental filter on the observability of latent polar behaviour. For example, 3,4,5-trifluorobenzene confers comparable λ ≈ 2 fraction to the 3-fluorobenzonitrile motif but matched pair analysis for materials in **Fig. 1** shows this leads to a vast reduction in clearing point, and polar order is not observed for these materials. We also note that, the near maximal fluorination patterns conferred by conjunction of several 1,3-difluorobenzene motifs – an arrangement commonly found in polar LC materials – promotes alternating atomic charge patterns which give rise to a strong GFT spectral feature at $\lambda \approx 2$, irrespective of whether the material exhibits an $N_F$ phase.

Similarly, the pentafluorosulphanyl ($SF_5$) terminated materials occupy a higher frequency region of GFT space than the analogous nitro and nitrile materials despite being non-mesogenic. Additional fluorine atoms adjacent to the $SF_5$ unit might further enhance the relevant charge-spectral features, and in doing so might generate polar order. While examples of such systems are known, their synthesis presents a high barrier due to the required use of elemental fluorine [49] or $SF_5Cl$ gas [50]. More broadly, these results motivate the study of alternate terminal groups beyond the relatively small set employed in polar nematic materials. We believe that future studies would benefit from constructing the terminal group library around a molecular core system which is more performant in terms of its ability to generate polar phase types, such as that used in **NC2222** and **F2100** [30].

The GFT workflow offers a practical means to triage this chemical space prior to synthesis.

## Conclusions

We report a molecular matrix comprised of 40 materials which independently varies terminal group and lateral core fluorination pattern for a family of biphenyl-ester-phenyl mesogens. Across the series, increasing lateral fluorination generally suppresses conventional apolar mesomorphism and lowers clearing points, but this same structural change can enable polar nematic-type behaviour when combined with suitable terminal units. Polar phases are observed only for nitrile- and nitro-terminated materials, demonstrating that lateral fluorination is enabling but alone not sufficient for this particular core system.

By comparing terminal-groups, we find that conventional nematic stability and polar-phase propensity are distinct design targets. Isothiocyanate terminated materials give robust apolar nematic phases and high clearing points, but do not exhibit polar order. Conversely, a nitro terminus lowers the conventional nematic clearing points relative to nitrile but appears to favour the expression of polar phases in the most highly fluorinated cores. Polar order cannot be rationalised simply, instead requiring compatibility between the terminal group, lateral fluorination pattern, and broader mesogenic framework.

Graph Fourier analysis of conformer averaged atomic charges provides an electrostatic rationale for these trends. Materials exhibiting polar phases occupy the high-frequency region of charge-spectral space, consistent with projection of the charge distribution onto bond-scale electrostatic motifs. However, this descriptor is not a complete phase classifier; it identifies an electrostatic regime in which polar organisation becomes accessible, while melting point, crystallisation tendency, packing efficiency and supercooling determine whether that latent polar tendency can be realised experimentally. These results establish terminal-group identity as a primary design variable in fluorinated polar liquid crystals, rather than a secondary substituent effect.

## Methods

Fluorinated 4-(4-propylphenyl)benzoic acids were synthesised in house [30]. Phenols and other chemicals were purchased from commercial suppliers (Merck, Apollo Scientific, Fluorochem, Manchester Organics) and used as received. Reactions were performed in standard laboratory glassware under ambient pressure, with purification made by flash chromatography over silica gel with a hexane/DCM gradient using a CombiFlash NextGen300+ system (Teledyne Isco). Chromatographed materials were filtered as solutions through a PTFE frit (~500 nm) to remove particulate matter, then recrystalised from an appropriate solvent system. See the ESI for full details and chemical characterisation.

Electronic structure calculations were performed using the Orca 6.1 software package [51-53]. We generated an ensemble of conformers with the GOAT algorithm [54] with GFN2-xTB [55, 56]. Conformers within 5 kcal mol$^{-1}$ of the global minimum were retained; for each, a single point calculation with the wB97x-3c composite DFT method [57], yields atomic partial charges using the MBIS [58] and RESP [59] charge models for each conformer. The population weighted atomic partial charges were mapped onto the molecular graph, with atoms treated as nodes and covalent bonds as edges. Graph Fourier transforms were then performed following the approach of Palacio-Betancur and Jackson [48], using the eigenvectors of the normalised graph Laplacian as the graph-frequency basis.

## Data availability

The data associated with this paper are openly available from the University of Leeds Data Repository at https://doi.org/10.5518/1945**.**

Python code for calculation of the graph Fourier transform power spectra is available via GitHub at https://github.com/RichardMandle/gft_descriptors.

## Acknowledgements

RJM thanks UKRI for funding via a Future Leaders Fellowship, grant number R/W006391/1, and the University of Leeds for funding via a University Academic Fellowship. RJM and JH thank the Royal Society for funding via Research Grant RGS\R2\242503. Computational work was performed on the AIRE HPC, part of the high-performance computing facilities at the University of Leeds.

# A terminal-group and fluorination matrix for designing polar nematic liquid crystals

# ESI

Calum J. Gibb*[1], Jordan Hobbs[2], William C. Ogle[1,2,3], Richard. J. Mandle[1,2]

[1] School of Chemistry, University of Leeds, Leeds, UK, LS2 9JT
[2] School of Physics and Astronomy, University of Leeds, Leeds, UK, LS2 9JT
[3] School of Applied Mathematics, University of Leeds, Leeds, UK, LS2 9JT

*Author for correspondence e-mail: c.j.gibb@leeds.ac.uk

**Contents**

# 1. Supplementary Methods

## 1.1. Chemical Synthesis

Chemicals were purchased from commercial suppliers (Fluorochem, Merck, ChemScene, Ambeed) and used as received. Solvents were purchased from Merck and used without further purification. Reactions were performed in standard laboratory glassware at ambient temperature and atmosphere and were monitored by TLC with an appropriate eluent and visualised with 254 nm light. Chromatographic purification was performed using a Combiflash NextGen 300+ System (Teledyne Isco) with a silica gel stationary phase and a hexanes/ethyl acetate (EtOAc) or hexanes: dichloromethane (DCM) gradient as the mobile phase, with detection made in the 200-800 nm range. Chromatographed materials were filtered through 200 nm PTFE frits and then subjected to re-crystallisation from an appropriate solvent system where appropriate.

## 1.2. Chemical Characterisation

Chemical materials were characterised by NMR spectroscopy using a Bruker Avance III HDNMR spectrometer operating at 400 MHz, 100.5 MHz or 376.4 MHz ($^{1}$H, $^{13}$C{$^{1}$H} and $^{19}$F, respectively) or a Bruker AV4 NEO 11.75T spectrometer operating at 500 MHz, 125.5 MHz ($^{1}$H, and $^{13}$C{$^{1}$H}, respectively).

## 1.3. Thermal Analysis

Differential scanning calorimetry (DSC) measurements were performed using a TA Instruments Q2000 DSC instrument (TA Instruments, Wilmslow UK), equipped with a RCS90 Refrigerated cooling system (TA Instruments, Wilmslow UK). The instrument was calibrated against an Indium standard, and data were processed using TA Instruments Universal Analysis Software. Samples were analysed under a nitrogen atmosphere, in hermetically sealed aluminium TZero crucibles (TA Instruments, Wilmslow, UK) and subjected to three analysis cycles. In all cases, samples were subject to heating and cooling at a rate of 10 K min$^{-1}$. Phase transition temperatures were measured as onset values on cooling cycles for consistency between monotropic and enantiotropic phase transitions, while crystal melts were obtained as onset values on heating.

## 1.4. Polarized Optical Microscopy (POM)

Polarised light optical microscopy (POM) was performed using a Leica DM2700P polarised light microscope (Leica Microsystems (UK) Ltd., Milton Keynes, UK), equipped with 10x and 50x magnification, and a rotatable stage. A Linkam TMS 92 heating stage (Linkam Scientific Instruments Ltd., Redhill, UK) was used for temperature control, and samples were studied sandwiched between two untreated glass coverslips. Images were recorded using a Nikon D3500 Digital Camera (Nikon UK Ltd., Surbiton, UK), using DigiCamControl software.

## 1.5. X-ray Scattering

X-ray scattering measurements, both small angle (SAXS) and wide angle (WAXS) were recorded using an Anton Paar SAXSpoint 5.0. This was equipped with a primux 100 Cu X-ray source with a 2D EIGER2 R detector with a variable sample-detector distance. The X-rays had a wavelength of 0.154 nm. Samples were filled into either thin-walled quartz capillaries or held between Kapton tape. Temperature was controlled using an Anton Paar heated sampler with a range of 20 °C to 300 °C and the samples held in a chamber with an atmospheric pressure of >1 mbar.

Background scattering patterns were recorded, scaled according to the sample transmission and then subtracted from the samples' obtained 2D SAXS pattern. 1D patterns were obtained by radially integrating the 2D SAXS patterns. Peak positions and FWHM was recorded and then converted into d spacing following Bragg's law.

### 1.6 Current Response Measurements

Spontaneous polarisation measurements are undertaken using the current reversal technique. Triangular waveform AC voltages are applied to the sample cells with an Agilent 33220A signal generator (Keysight Technologies), and the resulting current outflow is passed through a current-to-voltage amplifier and recorded on a RIGOL DHO4204 high-resolution oscilloscope (Telonic Instruments Ltd, UK). Heating and cooling of the samples during these measurements is achieved with an Instec HCS402 hot stage controlled to 10 mK stability by an Instec mK1000 temperature controller. The LC samples are held in 4μm thick cells with no alignment layer, supplied by Instec. The measurements consist of cooling the sample at a rate of 1 Kmin$^{-1}$ and applying a set voltage at a frequency of 20 Hz to the sample every 1 K. The voltage was set such that it would saturate the measured $P_S$ and was determined before final data collection.

There are three contributions to the measured current trace: accumulation of charge in the cell ($I_c$), ion flow ($I_i$), and the current flow due to polarisation reversal ($I_p$). To obtain a $P_S$ value, we extract the latter, which manifests as one or multiple peaks in the current flow, and integrate as:

$$P_S = \int \frac{I_p}{2A} dt \qquad \textbf{(2)}$$

where A is the active electrode area of the sample cell.

## 2 Supplementary Results

### 2.1 Tabulated transitional properties of the 40-membered materials matrix.

**Table S1.** Tabulated transitional properties, including transition temperatures (°C) and associated enthalpies of transition (KJ mol$^{-1}$) for **1a-1j**.

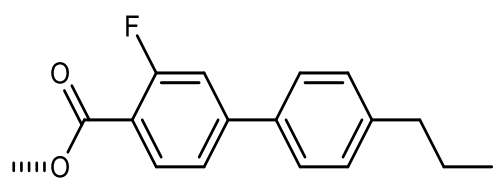

| Comp. 1 | Head-group | Phase Sequence | |
|---|---|---|---|
| | | Transitional properties (°C) | Enthalpy (KJ mol$^{-1}$) |
| a | 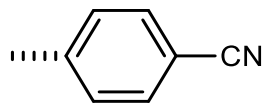 | K 97.3 N 223.2 I | K 21.9 N 0.89 I |
| b | 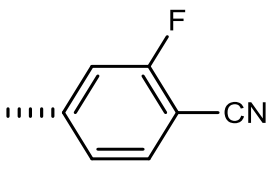 | K 85.5 N 183.8 I | K 26.0 N 0.40 I |
| c | 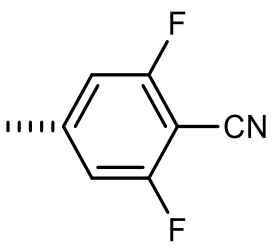 | K 103.0 N 140.6 | K 29.9 N 0.21 I |
| d | 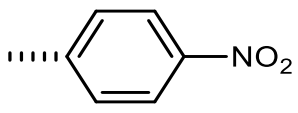 | K 94.3 N 186.2 I | K 23.9 N 0.19 I |
| e | 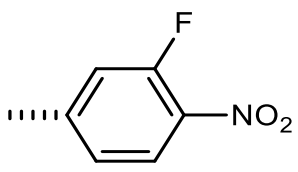 | K 110.7 N 141.2 I | K 31.5 N 0.12 I |
| f | 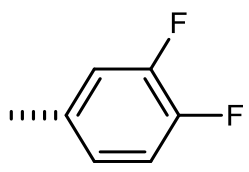 | K 104.6 SmA 108.0 N 120.7 I | K 22.6 SmA 2.24 N 0.23 I |
| g | 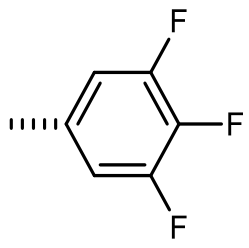 | K 103.5 SmA 98.0 I | K 26.5 SmA 3.56 I |
| h | 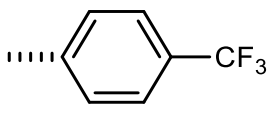 | K 145.7 SmA 172.2 I | K 20.8 SmA 7.00 I |
| i | 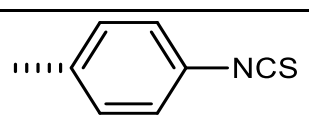 | K 106.6 SmA 147.3 N 230.3 I | K 62.7 SmA 0.23 N 0.61 I |
| j | 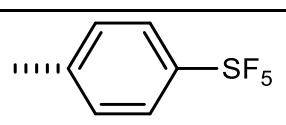 | K 145.6 SmA 141.5 I | K 34.3 SmA 4.49 I |

**Table S2.** Tabulated transitional properties, including transition temperatures (°C) and associated enthalpies of transition (KJ $mol^{-1}$) for **2a-2j**. $^{m}$ indicates a transition temperature as observed by POM.

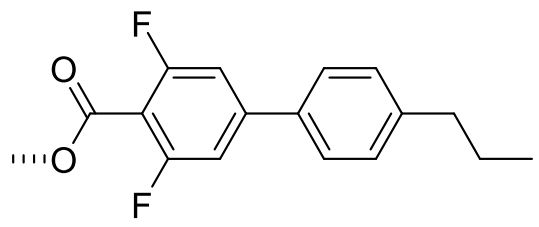

| Comp. 2 | Head-group | Phase Sequence | |
|---|---|---|---|
| | | Transitional properties (°C) | Enthalpy (KJ $mol^{-1}$) |
| a | 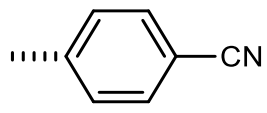  | K 97.5 N 178.3 I | K 21.9 N 0.59 . I |
| b | 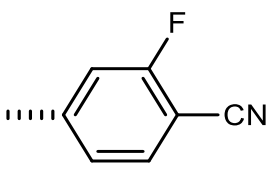  | K 85.0 N 147.9 I | K 27.0 N 0.22 I |
| c | 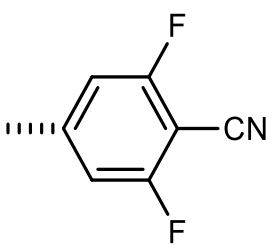  | K 99.3 N 107.5 I | K 27.8. N 0.21. I |
| d | 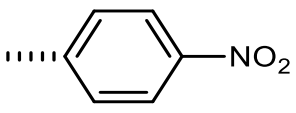  | K 133.4 SmA 119.5 I | K 37.8 SmA 6.56 I |
| e | 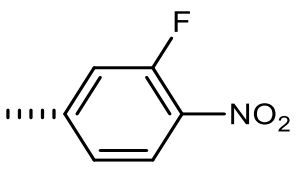  | K 80.4 SmA 54.9 I | K 27.2 SmA 3.50 I |
| f | 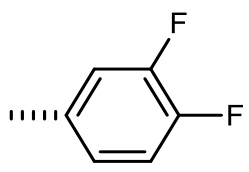  | K 95.8 N 72.4 I | K 29.3 N 0.14 I |
| g | 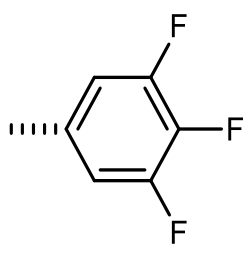  | K 107.1 N 143.0 I | K 31.7 N 0.17 I |
| h | 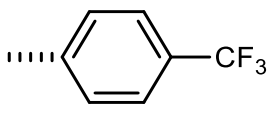  | K 103.9 N 104.9 I | K 32.1 N 0.08 I |
| i | 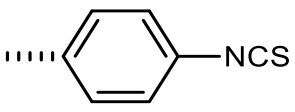  | K 109.8 SmA 95.0$^{m}$ I | K 28.1 [SmA] N/A I |
| j | 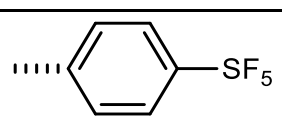  | K 95.5 N 191.2 I | K 13.7 N 0.27 I |

**Table S3.** Tabulated transitional properties, including transition temperatures (°C) and associated enthalpies of transition (KJ $mol^{-1}$) for **3a-3j**. $^{m}$ indicates a transition temperature as observed by POM.

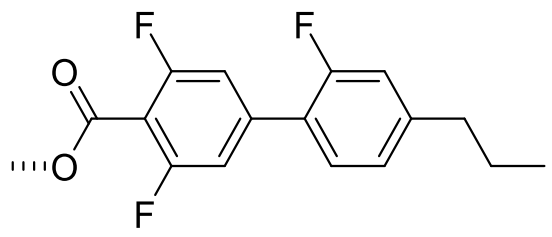

| Comp. 3 | Head-group | Phase Sequence | |
|---|---|---|---|
| | | Transitional properties (°C) | Enthalpy (KJ $mol^{-1}$) |
| a | 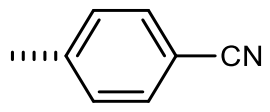 | K 114.4 N 164.5 I | K 27.2 N 0.45 I |
| b | 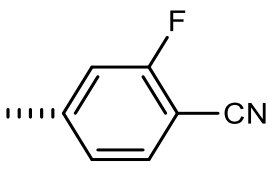 | K 89.6 N 136.4 I | K 22.2 N 0.32 I |
| c | 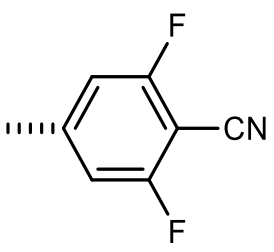 | K 88.7 $N_S$ 70.9 N 108.0 | K 21.5 $N_S$ 0.003 N 0.22 |
| d | 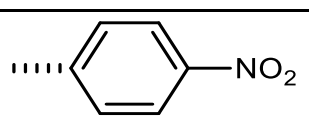 | K 109.4 I | K 30.3 I |
| e | 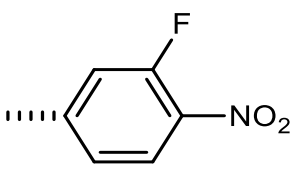 | K 70.2 I | K 23.7 I |
| f | 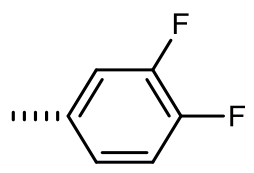 | K 61.5 N 48.7 I | K 26.4 N 0.14 I |
| g | 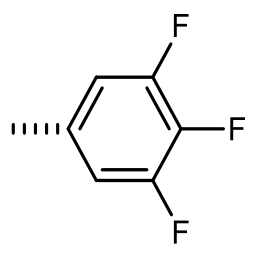 | K 99.7 N 130.3 I | K 27.2 N 0.19 I |
| h | 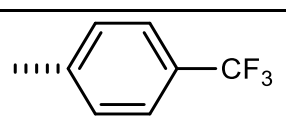 | K 84.3 N 97.6 I | K 25.1 N 0.13 I |
| i | 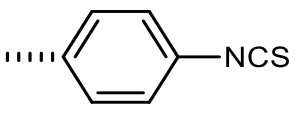 | K 88.9 SmA $52.0^{m}$ N 52.5 I | K 25.1 SmA N/A N 0.07 I |
| j | 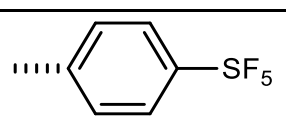 | K 109.8 SmA $95.0^{m}$ I | K 28.1 [SmA] N/A I |

**Table S4.** Tabulated transitional properties, including transition temperatures (°C) and associated enthalpies of transition (KJ $mol^{-1}$) for **4a-4j**. $^m$ indicates a transition temperature as observed by POM

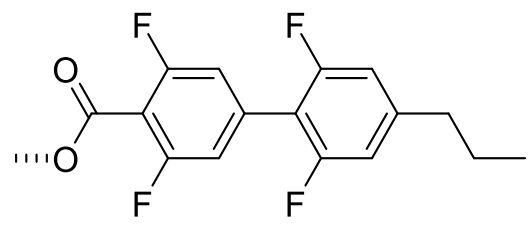

| Comp. 4 | Head-group | Phase Sequence | |
|---|---|---|---|
| | | Transitional properties (°C) | Enthalpy (KJ $mol^{-1}$) |
| a | 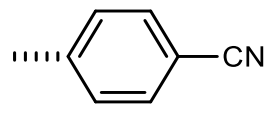 | K 139.1 N 135.1 I | K 30.1 N 0.53 I |
| b | 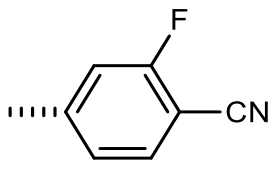 | K 128.2 $N_F$ $85^m$ $N_S$ 89.2 N 120.4 I | K 35.8 $N_F$ N/A $N_S$ 0.003 N 0.85 I |
| c | 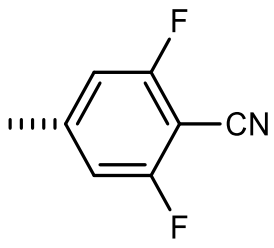 | K 104.9 $N_F$ 121.8 I | K 29.0 $N_F$ 3.55 I |
| d | 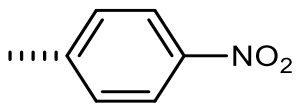 | K 137.1 $N_S$ 97.8 N 110.8 I | K 40.7 $N_S$ 0.047 N 0.54 I |
| e | 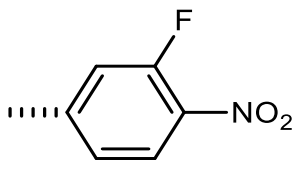 | K 116.4 $N_F$ 95.1 I | K 32.4 $N_F$ 2.77 I |
| f | 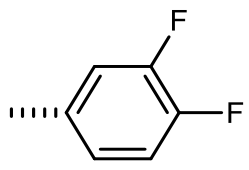 | K 82.3 I | K 32.0 I |
| g | 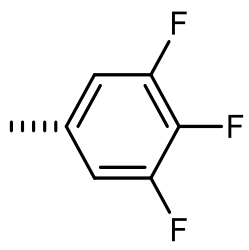 | K 88.5 I | K 22.3 I |
| h | 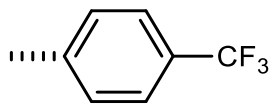 | N/A | N/A |
| i | 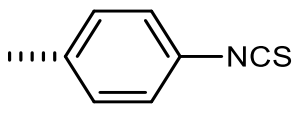 | K 137.9 N 150.2 I | K 31.8 N 0.64 I |
| j | 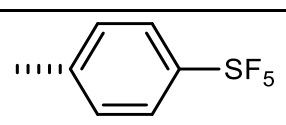 | K 102.3 I | K 33.1 I |

### 2.2 Supplementary phase characterisation

#### 2.2.1 Polarized optical Microscopy (POM)

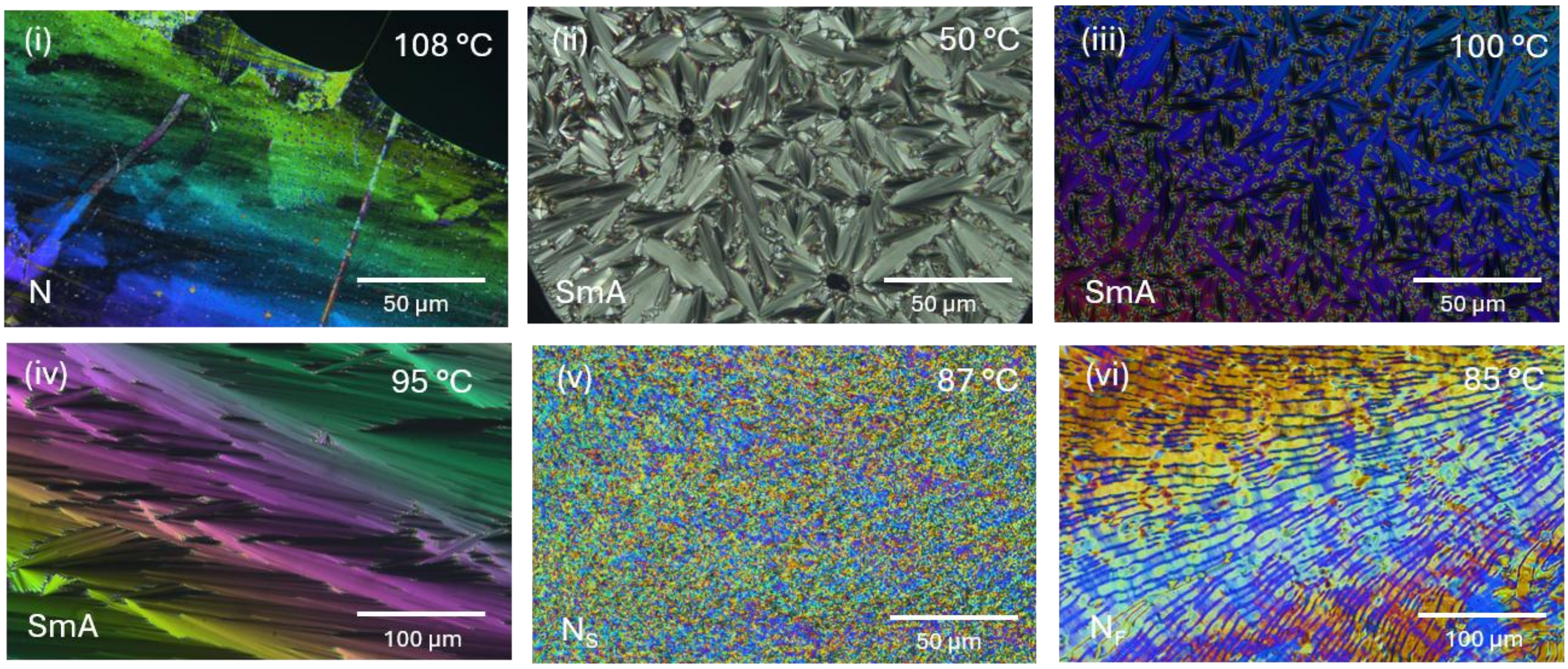

**Fig. S1.** Representative example POM micrographs taken between untreated glass slides for **(i)** the N phase of **1b**, **(ii)** the SmA phase of **2g**, **(iii)** the SmA phase of 1i, **(iv)** the SmA phase of **2j**, **(v)** the NS phase of **4b**, and **(vi)** the N$_F$ phase of **4b**.

#### 2.2.2 Differential scanning calorimetry (DSC)

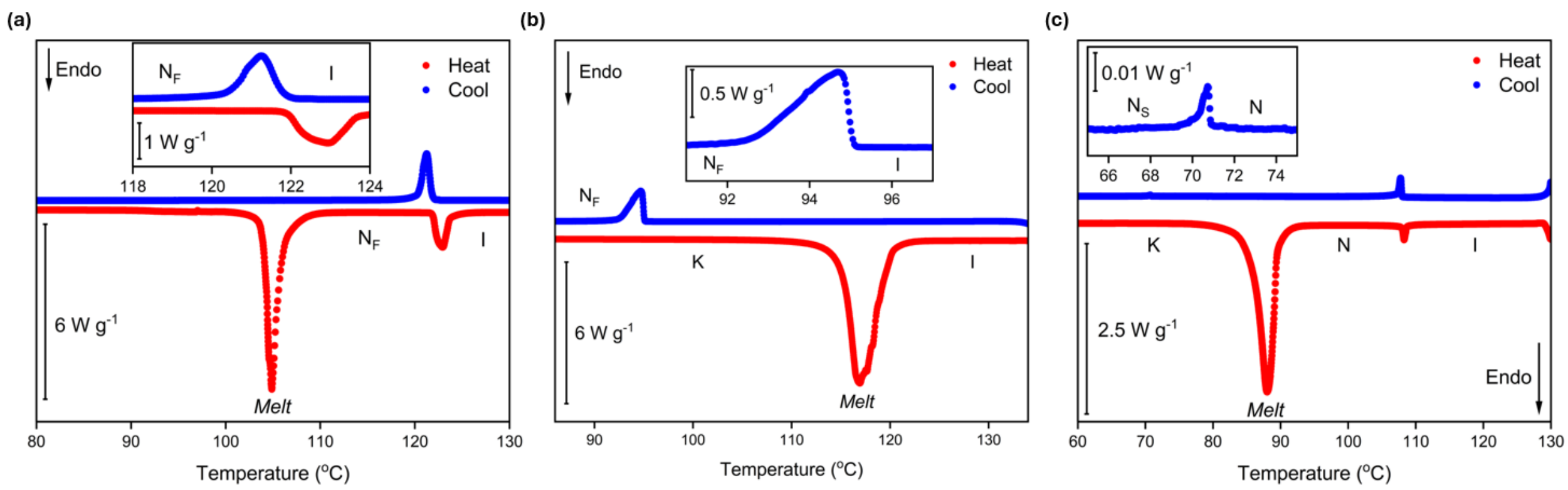

**Fig. S2.** Example DSC thermograms for **(a) 4c**, **(b) 4e**, and **(c) 3c** recorded at a heat/cool rate of 10 K min$^{-1}$.

### 2.2.3 Spontaneous Polarization ($P_S$)

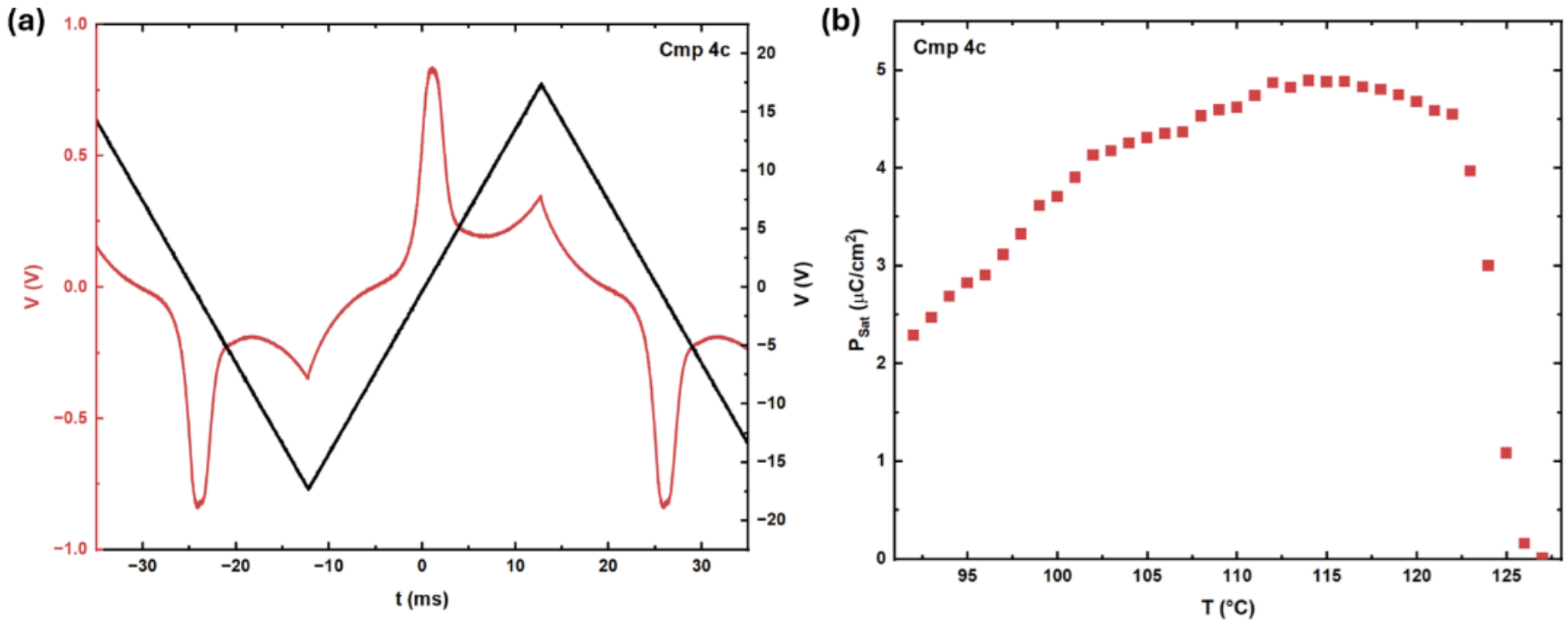

**Fig. S3.** **(a)** Example current response for **4c** in the $N_F$ phase recorded in a 5 µm cell with no alignment layer, **(b)** temperature dependence of the saturated polarisation for **4c**

### 2.2.4 X-ray Scattering

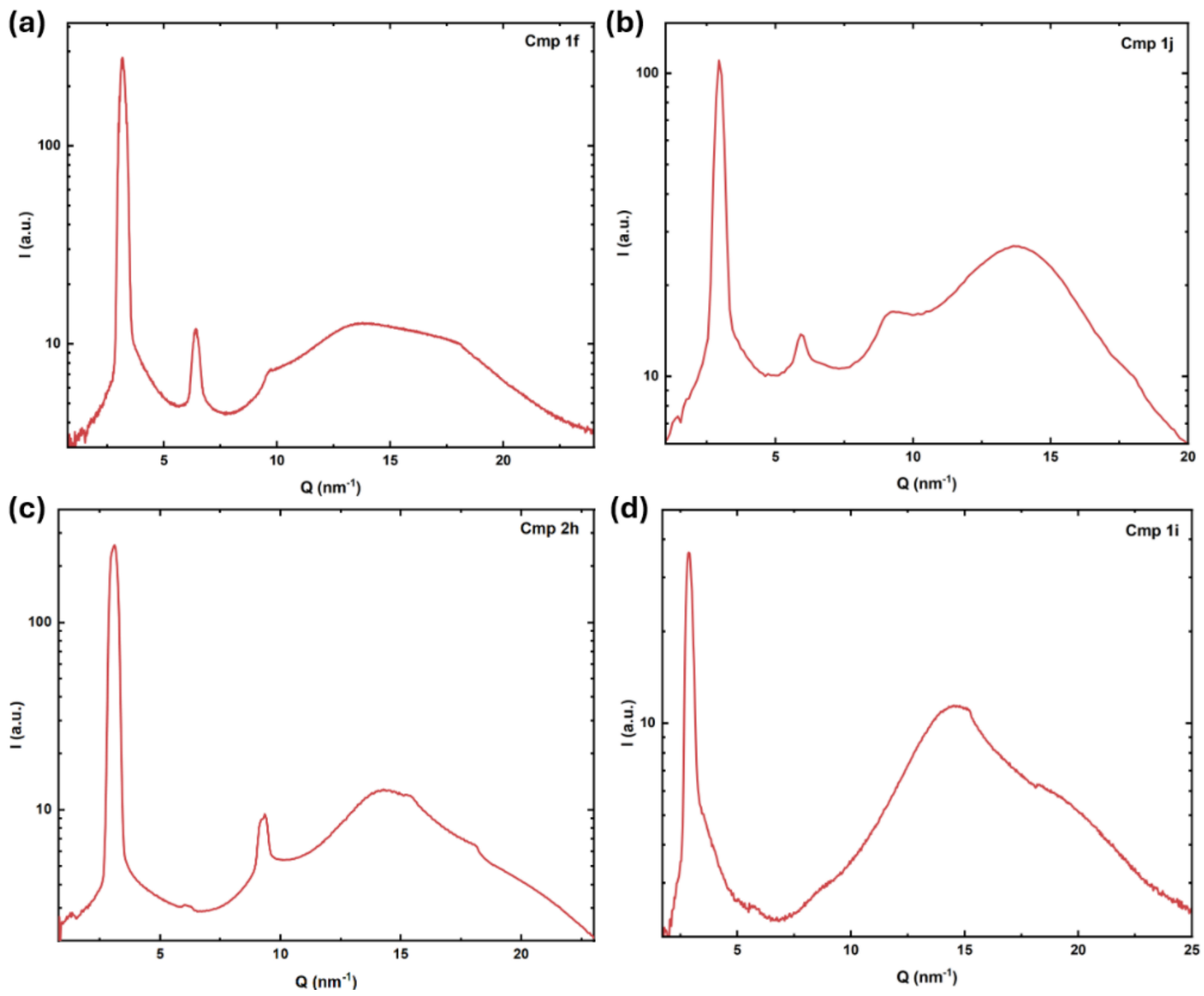

**Fig. S4.** 1D X-ray scattering patterns for **(a) 1f**, **(b) 1j**, **(c) 2h**, **(d) 1i**. All patterns were recorded in the SmA phase of each material and were obtained by 360° radial integration of 2D data.

## 2.3 Supplementary GFT

We performed the same GFT analysis in the main manuscript but using population weighted MBIS atomic partial charges. Doing so gives somewhat different values for the fractional GFT power of each individual molecule within the plot (**Fig. SI-Å**), but the conclusions based on RESP charges are respected; namely that, irrespective of terminal group choice, increasing the lateral fluorination shifts the molecules towards a larger fractional high-frequency spectral power. The materials that exhibit polar phases ($N_S$, $N_F$) are located towards the higher-frequency end of this distribution but overlap in fractional GFT power with non-polar materials.

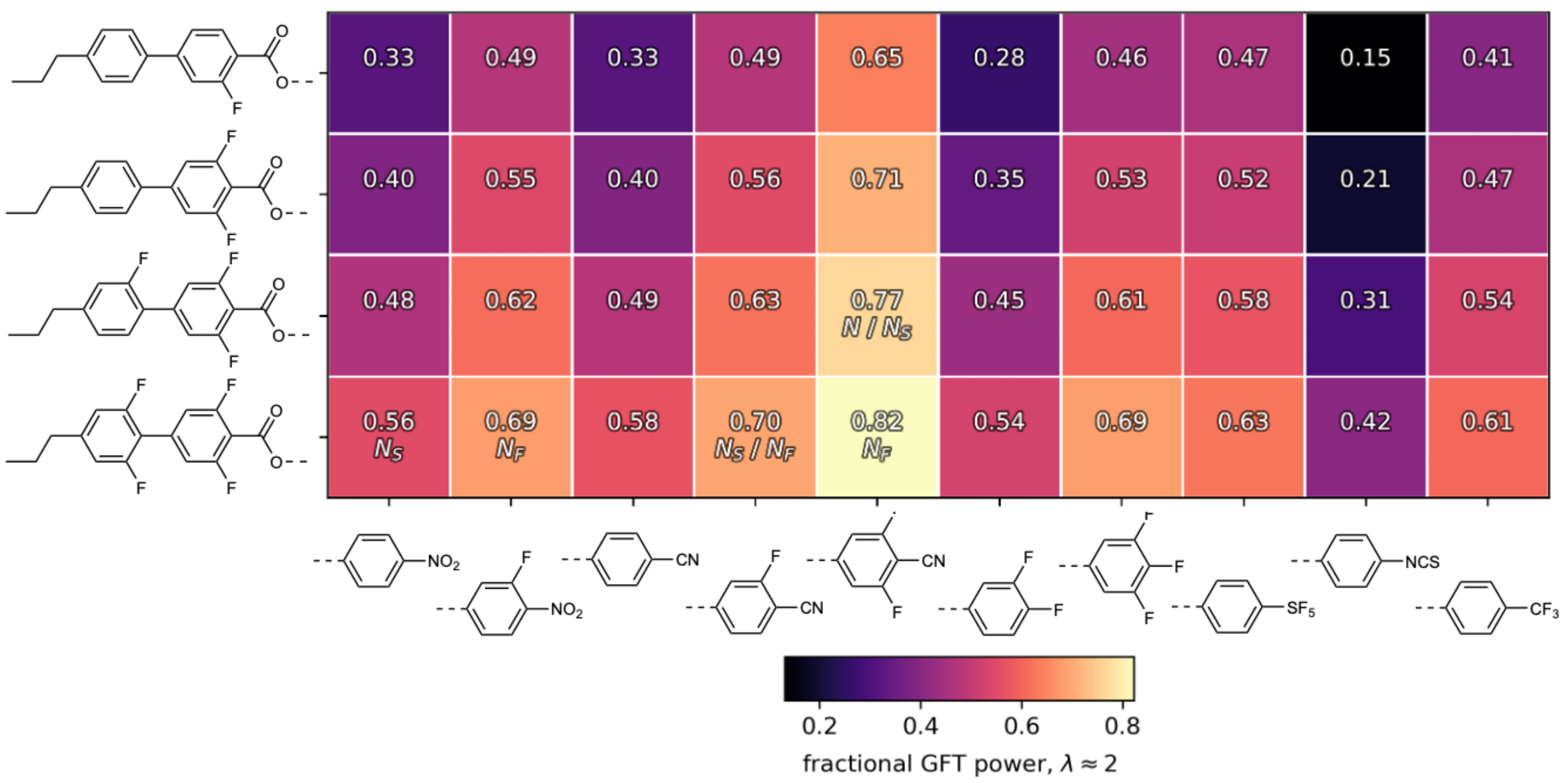

**Fig. S5:** Terminal-group and fluorination-pattern dependence of high-frequency charge-spectral power from population weighted MBIS charges computed with the wB97x-3c composite DFT method.

## 3 Organic Synthesis

### 3.1 General esterification procedure used in the s*ynthesis of materials 1-40*

The general synthetic procedure is described in **Scheme S1**. The phenolic reagents are all items of commerce, and the synthesis of the fluorinated biphenyl acids have been reported previously [30]. A round bottomed flask was charged with the appropriate phenol (0.5 mmol, 1.0 eq), benzoic acid (0.6 mmol, 1.1 eq.), EDC.HCl (0.75 mmol, 1.5 eqv.) and DMAP (~ 2 mol%). Dichloromethane was added (conc. ~ 0.1 M) and the suspension stirred until complete consumption of the phenol as judged by TLC. Once complete, the reaction solution was concentrated and purified by flash chromatography over silica gel with a gradient of hexane/DCM using a Combiflash NextGen300+ system. To remove ionic impurities, a pre-column containing a small amount of neutral alumina was used before samples entered the silica gel column. The chromatographed material was dissolved into the minimum quantity of DCM, filtered through a 0.2 micron PTFE filter, concentrated to dryness and finally recrystalised from MeOH.

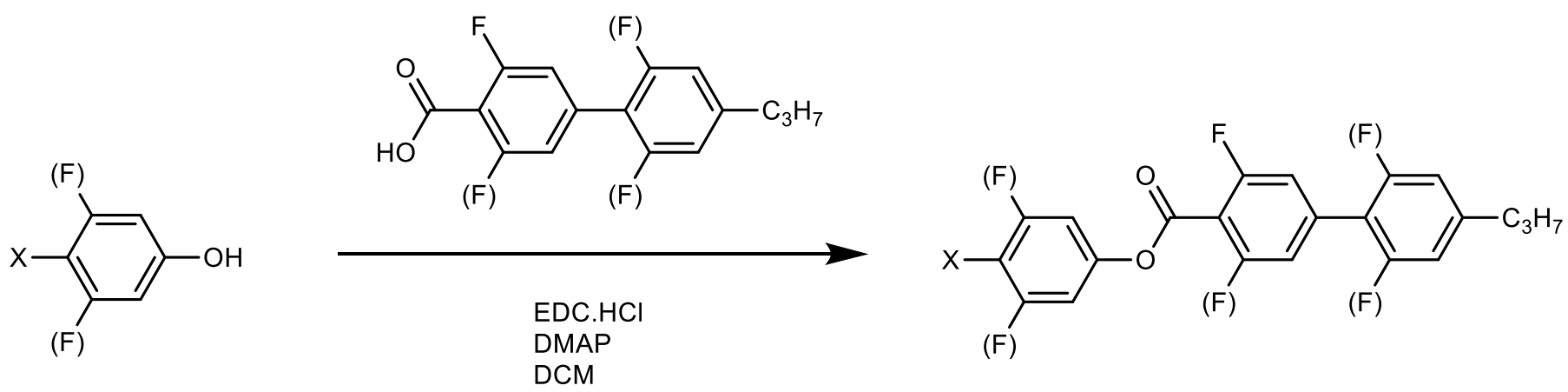

**Scheme S1** General esterification procedure used in the preparation of **1-40**; where X = CN, F, $CF_3$, $NO_2$, $SF_5$, NCS.

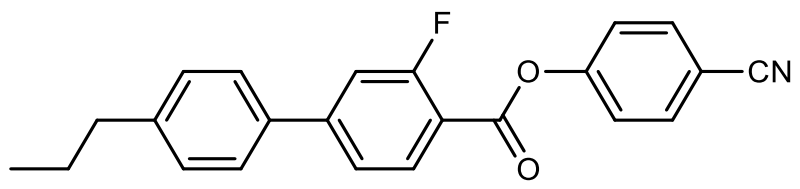

**1a**

*4-cyanophenyl 3-fluoro-4'-propyl-[1,1'-biphenyl]-4-carboxylate*

| | |
|---|---|
| Yield: | (White solid) 146 mg, 81 % |
| $^1$H NMR (400 MHz): | 8.14 (t, *J* = 7.5 Hz, 1H, Ar-**H**), 7.76 (ddd, *J* = 8.8, 2.4, 2.1 Hz, 2H, Ar-**H**), 7.56 (ddd, *J* = 8.2, 1.9, 1.8 Hz, 2H, Ar-**H**), 7.52 (dd, *J* = 8.2, 1.7 Hz, 1H, Ar-**H**), 7.44 (dd, *J* = 12.2, 1.7 Hz, 1H, Ar-**H**), 7.41 (ddd, *J* = 8.7, 2.4, 2.1 Hz, 2H, Ar-**H**), 7.31 (ddd, *J* = 8.3, 2.1, 1.9 Hz, 2H, Ar-**H**), 2.66 (t, *J* = 7.5 Hz, 2H, Ar-C$\mathbf{H_2}$-$CH_2$), 1.69 (h, *J* = 7.4 Hz, 2H, $CH_2$-C$\mathbf{H_2}$-$CH_3$), 0.98 (t, *J* = 7.3 Hz, 3H, $CH_2$-C$\mathbf{H_3}$). |
| $^{13}$C{$^1$H} NMR (101 MHz): | 164.12 (d, *J* = 261.9 Hz), 161.86 (d, *J* = 4.4 Hz), 153.97, 149.26 (d, *J* = 9.2 Hz), 144.15, 135.60, 135.59, 133.76, 132.99, 129.33, 127.06, 122.95, 122.56 (d, *J* = 3.3 Hz), 118.28, 115.21 (d, *J* = 9.8 Hz), 109.97, 37.73, 24.47, 13.83. |
| $^{19}$F NMR (376 MHz): | -107.17 (dd, $J_{F\text{-}H}$ = 12.1, 7.5 Hz, 1F, Ar-**F**). |

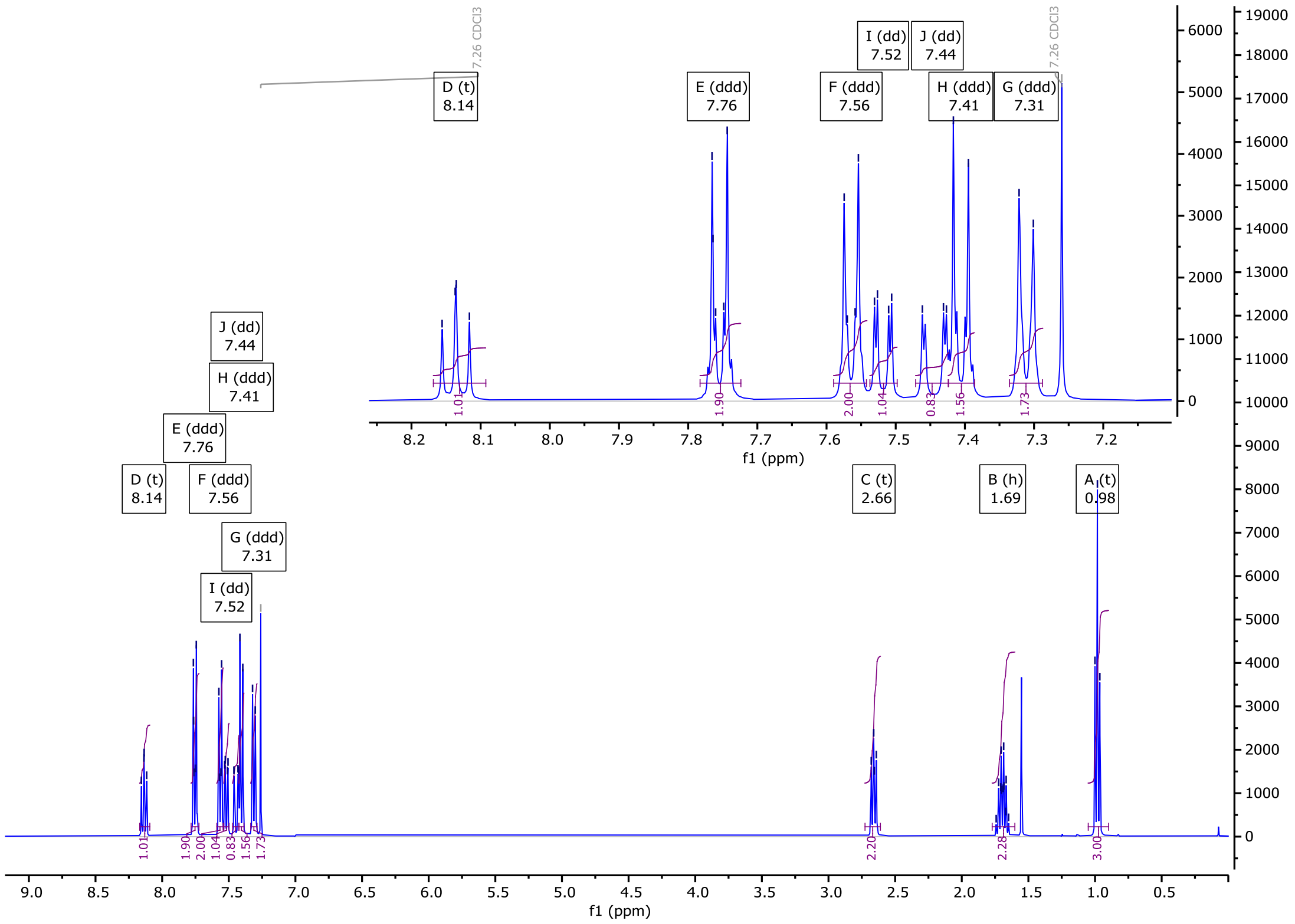

**Fig. S6.** $^{1}H$ NMR of **1a** in $CDCl_3$.

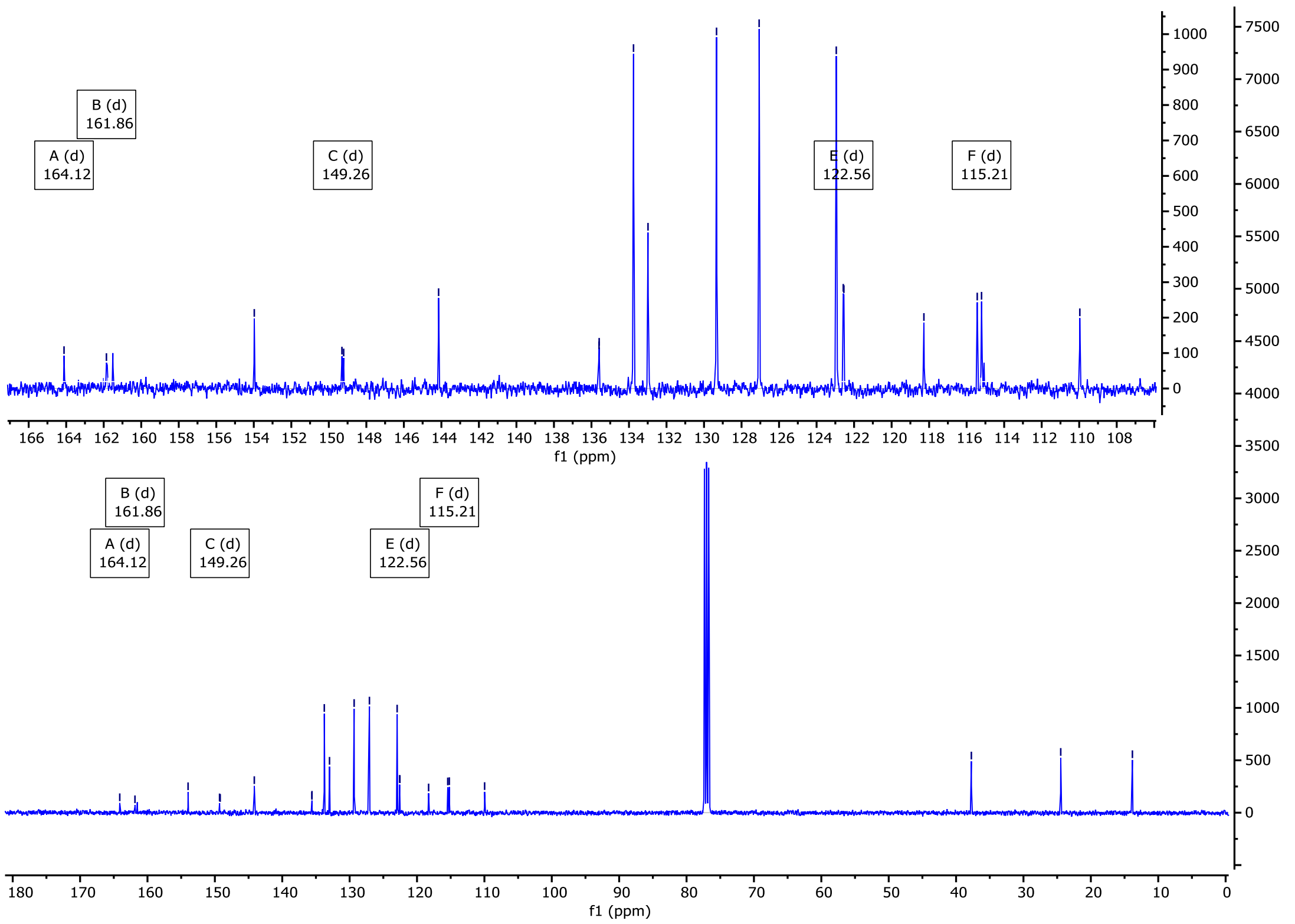

**Fig. S7.** $^{13}C\{^{1}H\}$ NMR of **1a** in $CDCl_3$.

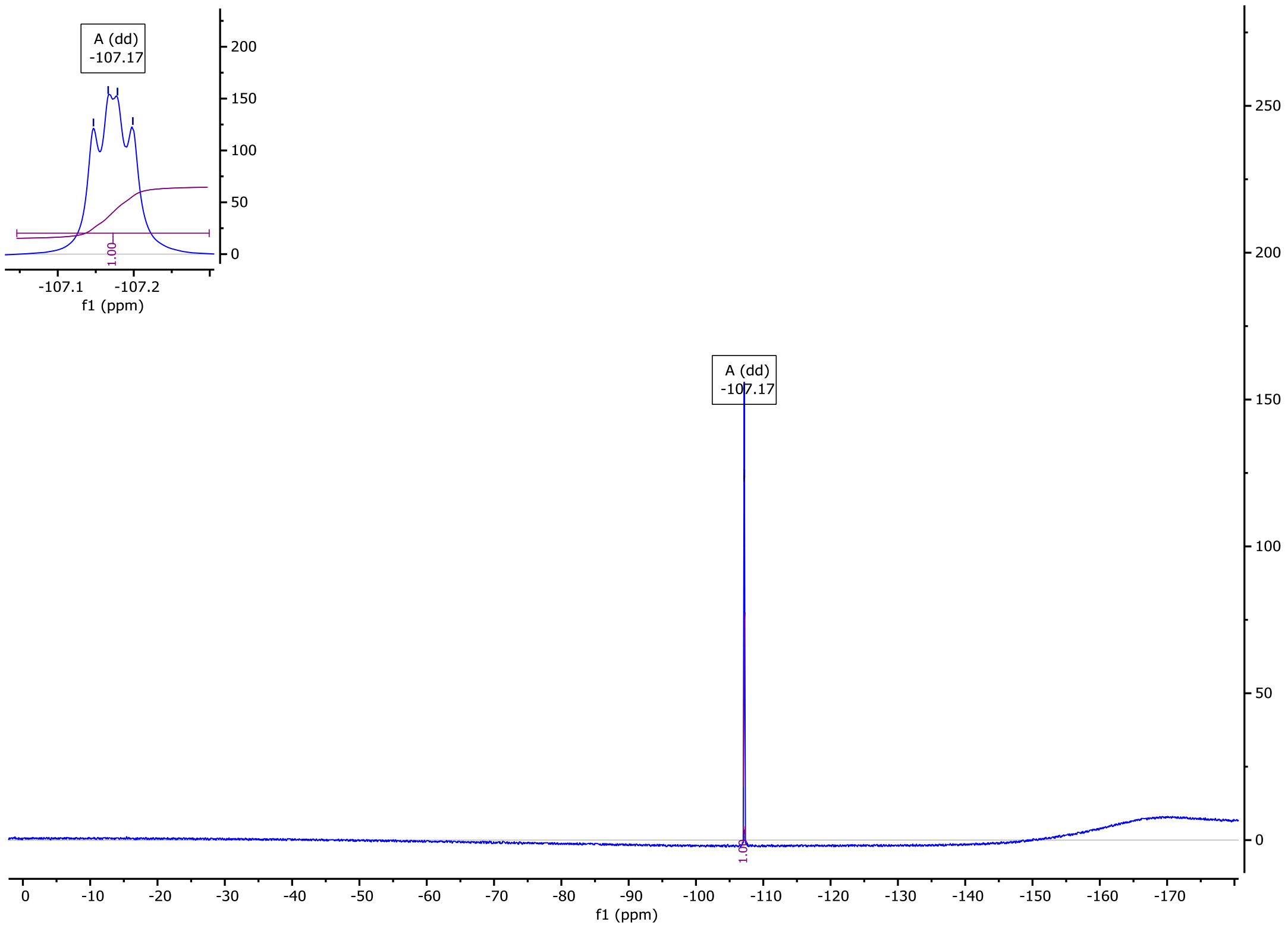

**Fig. S8.** $^{19}F$ NMR of **1a** in $CDCl_3$.

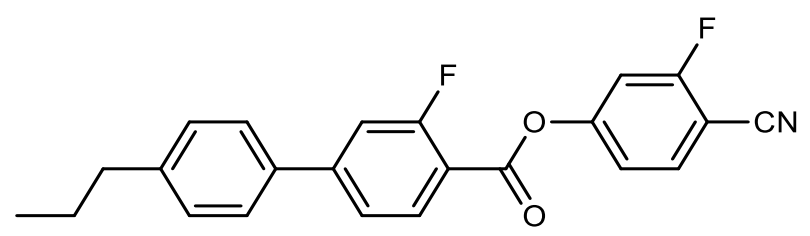

**1b**

*4-cyano-3-fluorophenyl 3-fluoro-4'-propyl-[1,1'-biphenyl]-4-carboxylate*

Yield: (Colourless needles) 140 mg, 74 %

$^{1}$H NMR (400 MHz): 8.12 (t, *J* = 8.0 Hz, 1H, Ar-**H**), 7.71 (t, *J* = 8.4 Hz, 1H, Ar-**H**), 7.56 (ddd, *J* = 8.3, 2.0, 1.9 Hz, 2H, Ar-**H**), 7.52 (dd, *J* = 8.2, 1.7 Hz, 1H, Ar-**H**), 7.45 (dd, *J* = 12.2, 1.7 Hz, 1H, Ar-**H**), 7.31 (ddd, *J* = 8.3, 2.0, 1.9 Hz, 2H, Ar-**H**), 7.28 – 7.22 (m, 2H, Ar-**H**)$^{\dagger}$, 2.66 (t, *J* = 7.4 Hz, 2H, Ar-C**H₂**-CH₂), 1.69 (h, *J* = 7.3 Hz, 2H, CH₂-C**H₂**-CH₃), 0.98 (t, *J* = 7.3 Hz, 3H, CH₂-C**H₃**). $^{\dagger}$ Overlapping $CDCl_3$.

$^{13}$C{$^{1}$H} NMR (101 MHz): 164.95 (d, *J* = 260.4 Hz), 164.16 (d, *J* = 262.5 Hz), 161.37 (d, *J* = 4.3 Hz), 155.29 (d, *J* = 10.7 Hz), 149.55 (d, *J* = 9.1 Hz), 144.25, 135.50, 134.12, 133.00, 129.35, 127.07, 122.63 (d, *J* = 3.3 Hz), 118.77 (d, *J* = 3.7 Hz), 115.36 (d, *J* = 23.1 Hz), 114.64 (d, *J* = 9.3 Hz), 113.47, 111.09 (d, *J* = 22.7 Hz), 99.16 (d, *J* = 15.6 Hz), 37.73, 24.46, 13.83.

$^{19}$F NMR (376 MHz: -103.26 (t, $J_{F-H}$ = 8.1 Hz, 1F, Ar-**F**), -106.86 (dd, $J_{F-H}$ = 12.3, 7.5 Hz, 1F, Ar-**F**).

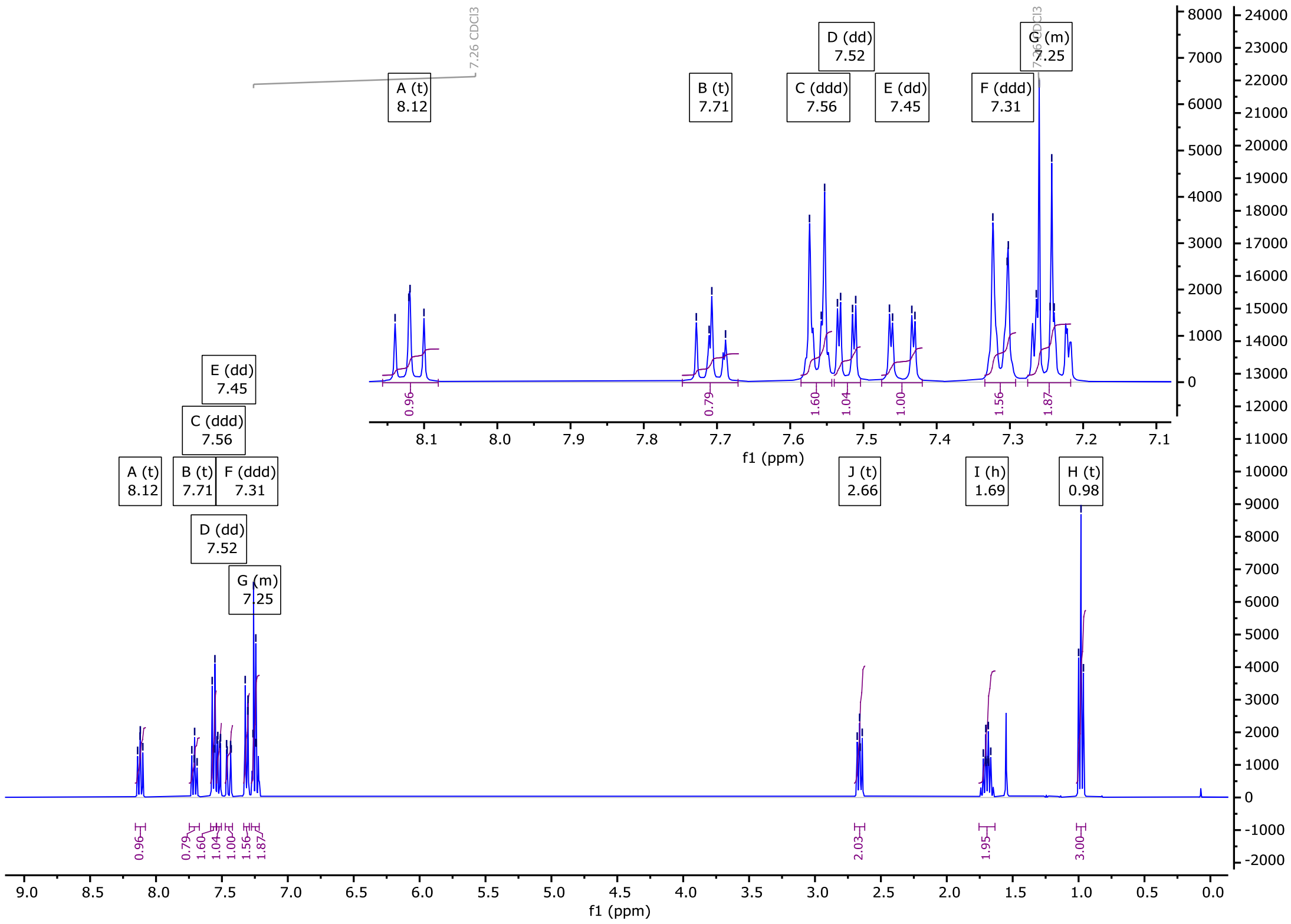

**Fig. S9.** $^{1}H$ NMR of **1b** in $CDCl_3$.

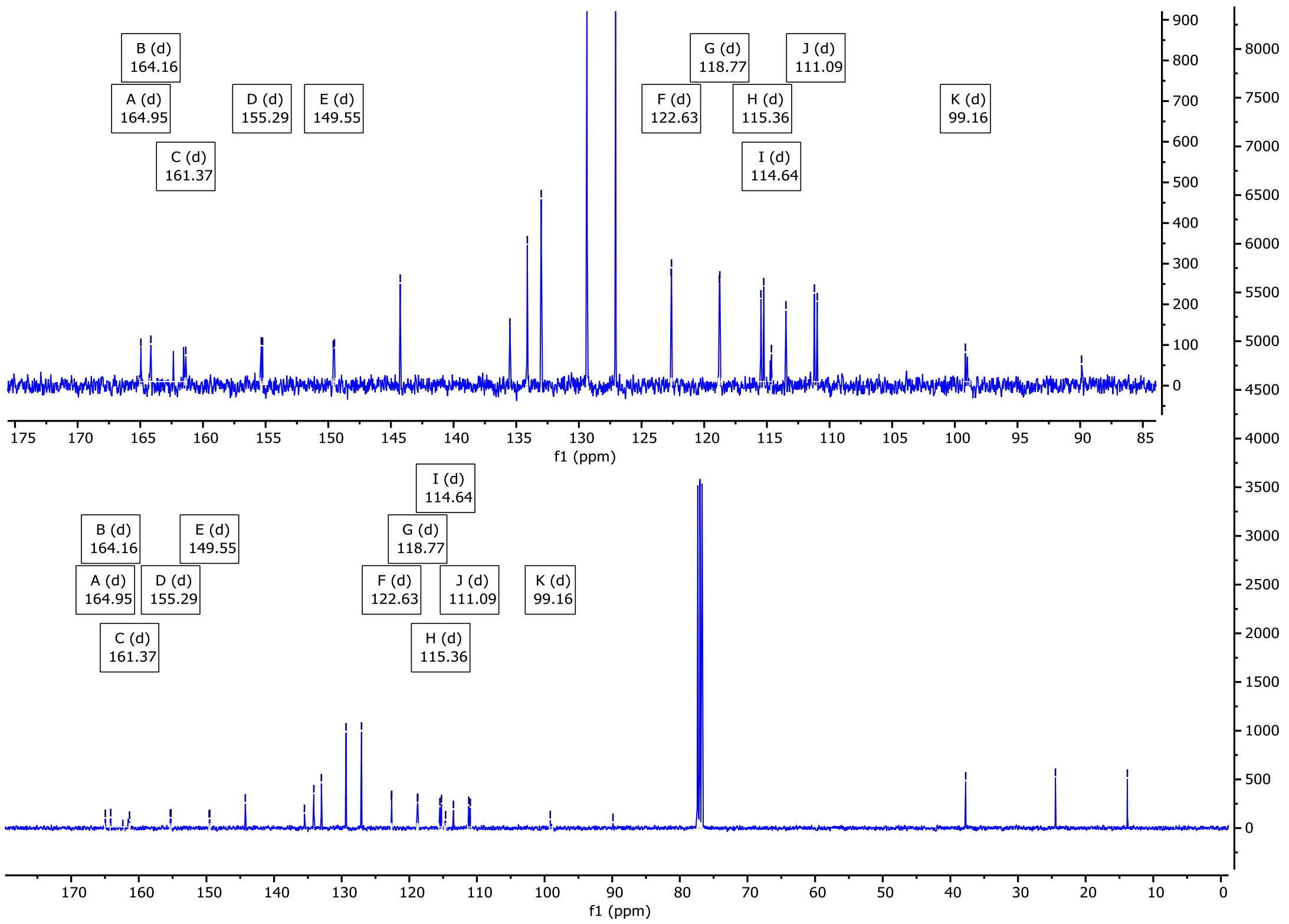

**Fig. S10.** $^{13}C\{^{1}H\}$ NMR of **1b** in $CDCl_3$.

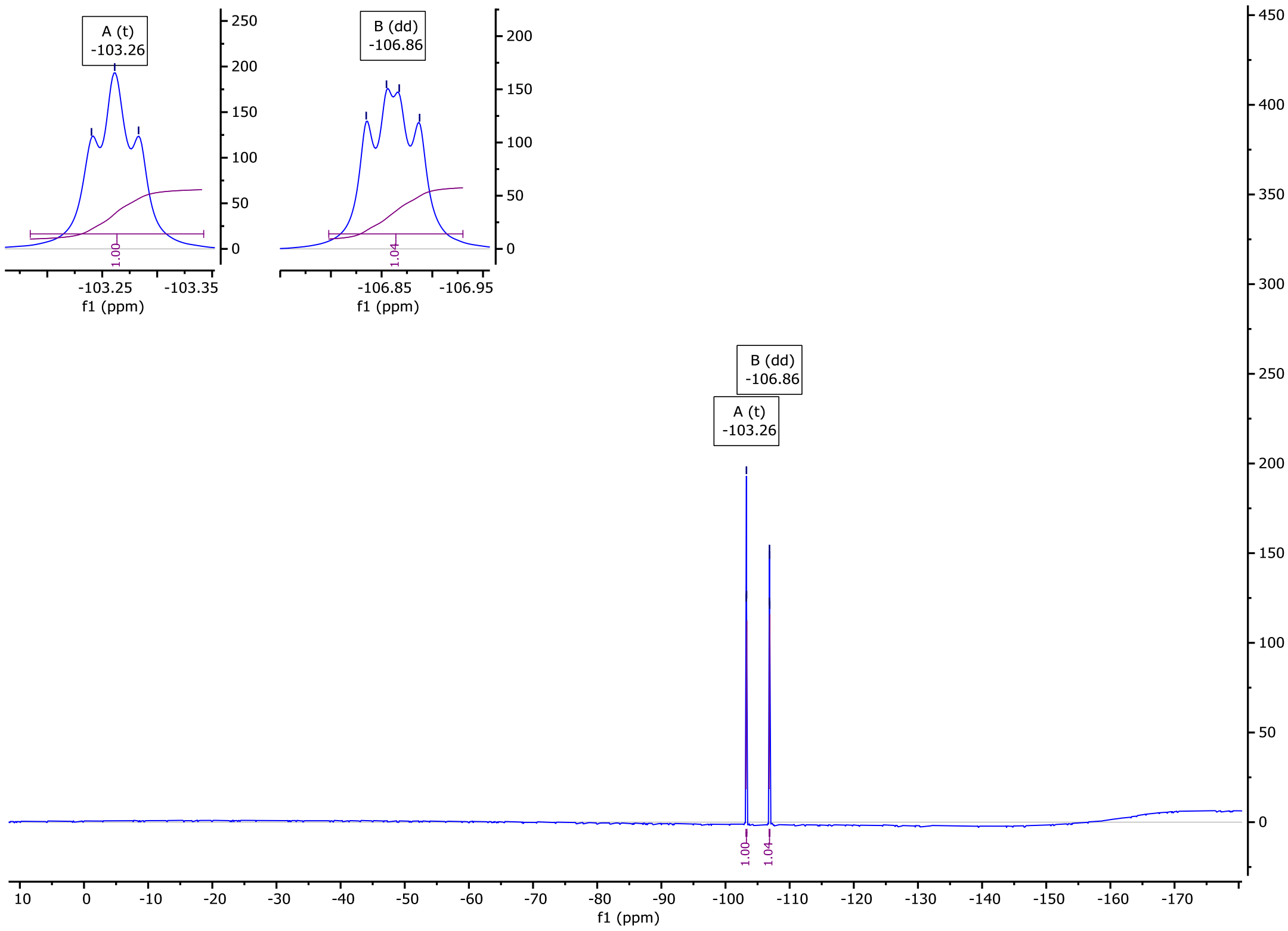

**Fig. S11.** $^{19}F$ NMR of **1b** in $CDCl_3$.

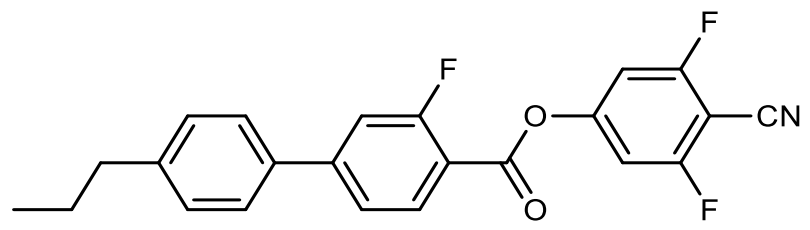

**1c**

*4-cyano-3,5-difluorophenyl 3-fluoro-4'-propyl-[1,1'-biphenyl]-4-carboxylate*

Yield: (White solid) 136 mg, 69 %

$^{1}$H NMR (400 MHz): 8.10 (t, *J* = 7.9 Hz, 1H, Ar-**H**), 7.56 (ddd, *J* = 8.3, 2.0, 1.8 Hz, 2H, Ar-**H**), 7.53 (dd, *J* = 8.3, 1.7 Hz, 1H, Ar-**H**), 7.45 (dd, *J* = 12.2, 1.7 Hz, 1H, Ar-**H**), 7.31 (ddd, *J* = 8.2, 1.9, 1.8 Hz, 2H, Ar-**H**), 7.10 (dd, *J* = 7.7, 2.7 Hz, 2H, Ar-**H**), 2.66 (t, *J* = 7.4 Hz, 2H, Ar-C**H$_2$**-CH$_2$), 1.69 (h, *J* = 7.5 Hz, 2H, CH$_2$-C**H$_2$**-CH$_3$), 0.98 (t, *J* = 7.3 Hz, 3H, CH$_2$-C**H$_3$**).

$^{13}$C{$^{1}$H} NMR (101 MHz): 164.82 (dd, *J* = 261.5, 6.2 Hz), 164.18 (d, *J* = 263.1 Hz), 160.91 (d, *J* = 4.0 Hz), 155.70 (t, *J* = 13.2 Hz), 149.85 (d, *J* = 8.9 Hz), 144.36, 135.40 (d, *J* = 1.4 Hz), 133.03, 129.38, 127.07, 122.69 (d, *J* = 3.3 Hz), 115.40 (d, *J* = 22.8 Hz)*, 114.31 (d, *J* = 9.1 Hz), 108.75, 106.97 (dd, *J* = 23.1, 3.9 Hz), 90.20, 37.73, 24.46, 13.82. * Overlapping signals.

$^{19}$F NMR (376 MHz): -101.92 (d, $J_{F-H}$= 8.8 Hz, 2F, Ar-**F**), -106.54 (dd, $J_{F-H}$ = 12.2, 7.5 Hz, 1F, Ar-**F**).

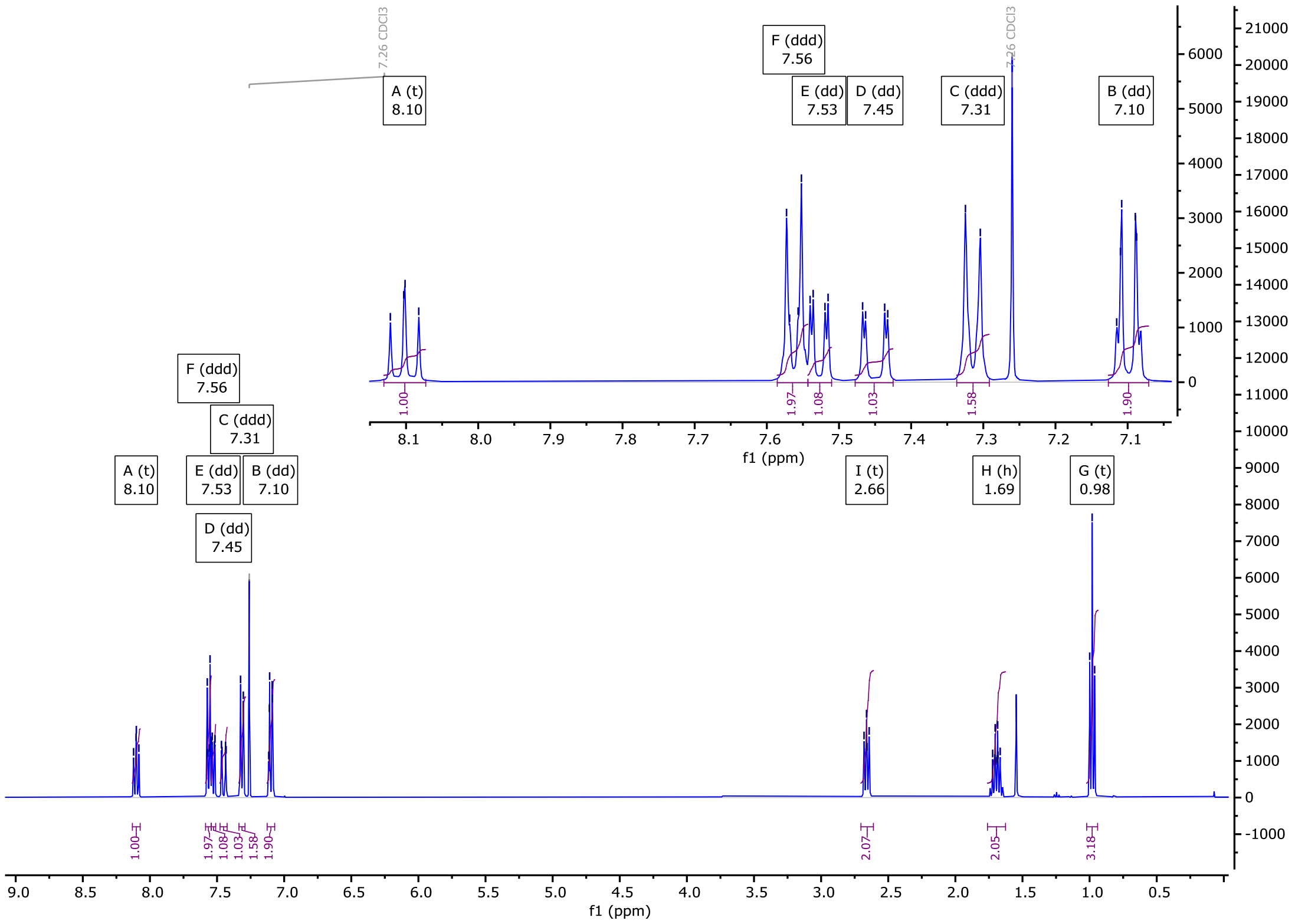

**Fig. S12.** $^{1}H$ NMR of **1c** in $CDCl_3$.

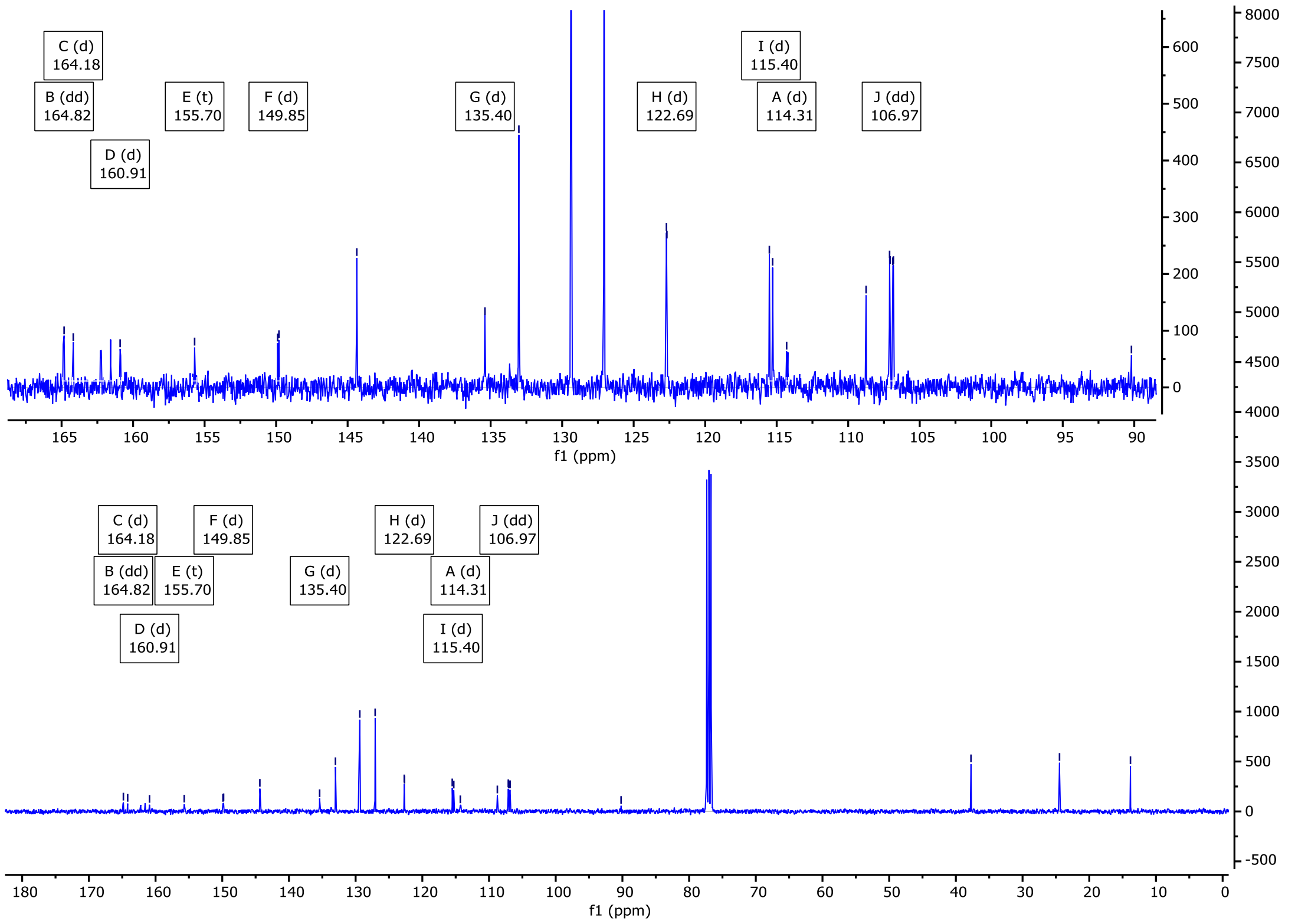

**Fig. S13.** $^{13}C\{^{1}H\}$ NMR of **1c** in $CDCl_3$.

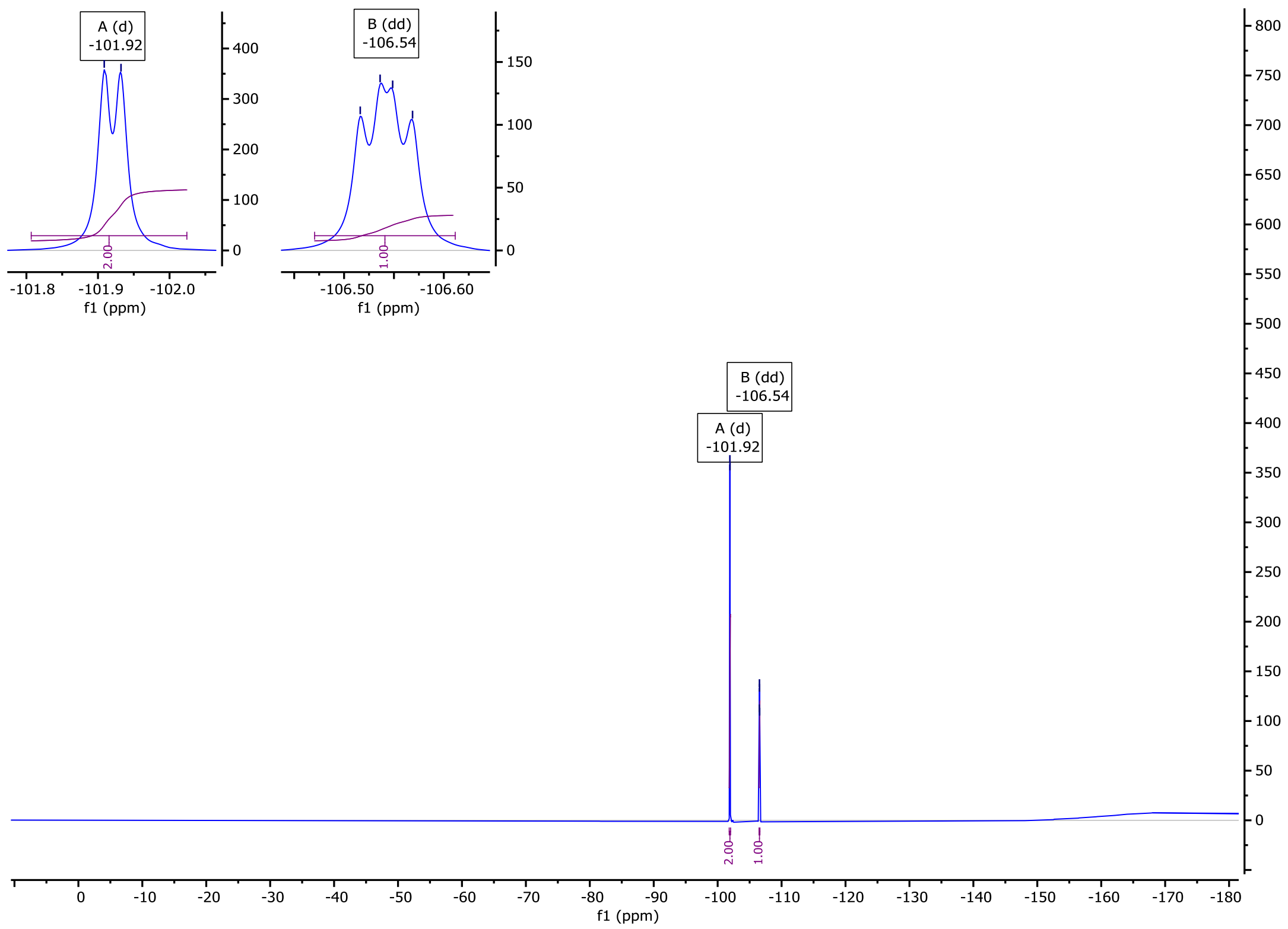

**Fig. S14.** $^{19}F$ NMR of **1c** in $CDCl_3$.

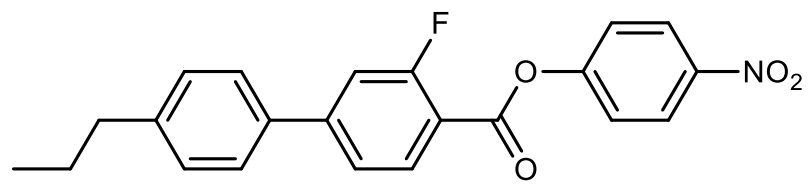

**1d**

*4-nitrophenyl 3-fluoro-4'-propyl-[1,1'-biphenyl]-4-carboxylate*

Yield: (white solid) 133 mg, 70 %

$^{1}$H NMR (400 MHz): 8.34 (ddd, *J* = 9.2, 3.1, 2.2 Hz, 2H, Ar-**H**), 8.15 (dd, *J* = 8.2, 7.5 Hz, 1H, Ar-**H**), 7.57 (ddd, *J* = 8.2, 2.1, 1.8 Hz, 2H, Ar-**H**), 7.53 (dd, *J* = 8.3, 1.7 Hz, 1H, Ar-**H**), 7.48 – 7.42 (m, 3H, Ar-**H**)*, 7.31 (d, *J* = 8.2 Hz, 2H, Ar-**H**), 2.66 (t, *J* = 7.4 Hz, 2H, Ar-C$\mathbf{H_2}$-$CH_2$), 1.69 (h, *J* = 7.4 Hz, 2H, $CH_2$-C$\mathbf{H_2}$-$CH_3$), 0.98 (t, *J* = 7.3 Hz, 3H, $CH_2$-C$\mathbf{H_3}$). * Overlapping signals.

$^{13}$C{$^{1}$H} NMR (126 MHz): 162.96 (d, *J* = 262.0 Hz), 161.85 (d, *J* = 4.3 Hz), 155.54, 149.46 (d, *J* = 9.1 Hz), 145.61, 144.31, 135.68, 133.13, 129.46, 127.19, 125.42, 122.70 (d, *J* = 3.3 Hz), 115.45 (d, *J* = 23.1 Hz), 115.12 (d, *J* = 9.4 Hz), 37.85, 24.60, 13.96.

$^{19}$F NMR (376 MHz): -107.08 (dd, $J_{F\text{-}H}$ = 12.4, 7.2 Hz, 1F, Ar-**F**).

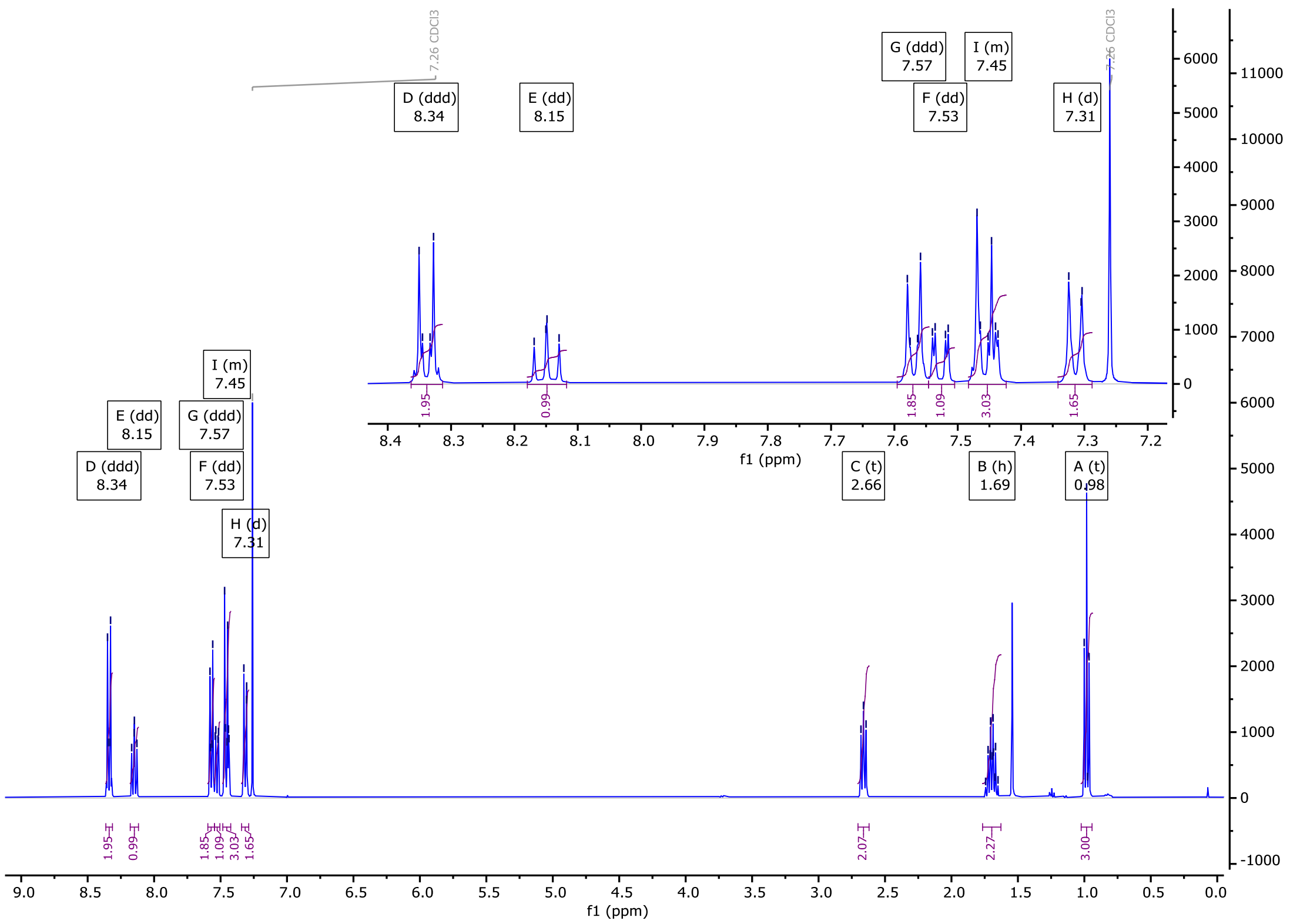

**Fig. S15.** $^{1}H$ NMR of **1d** in $CDCl_3$.

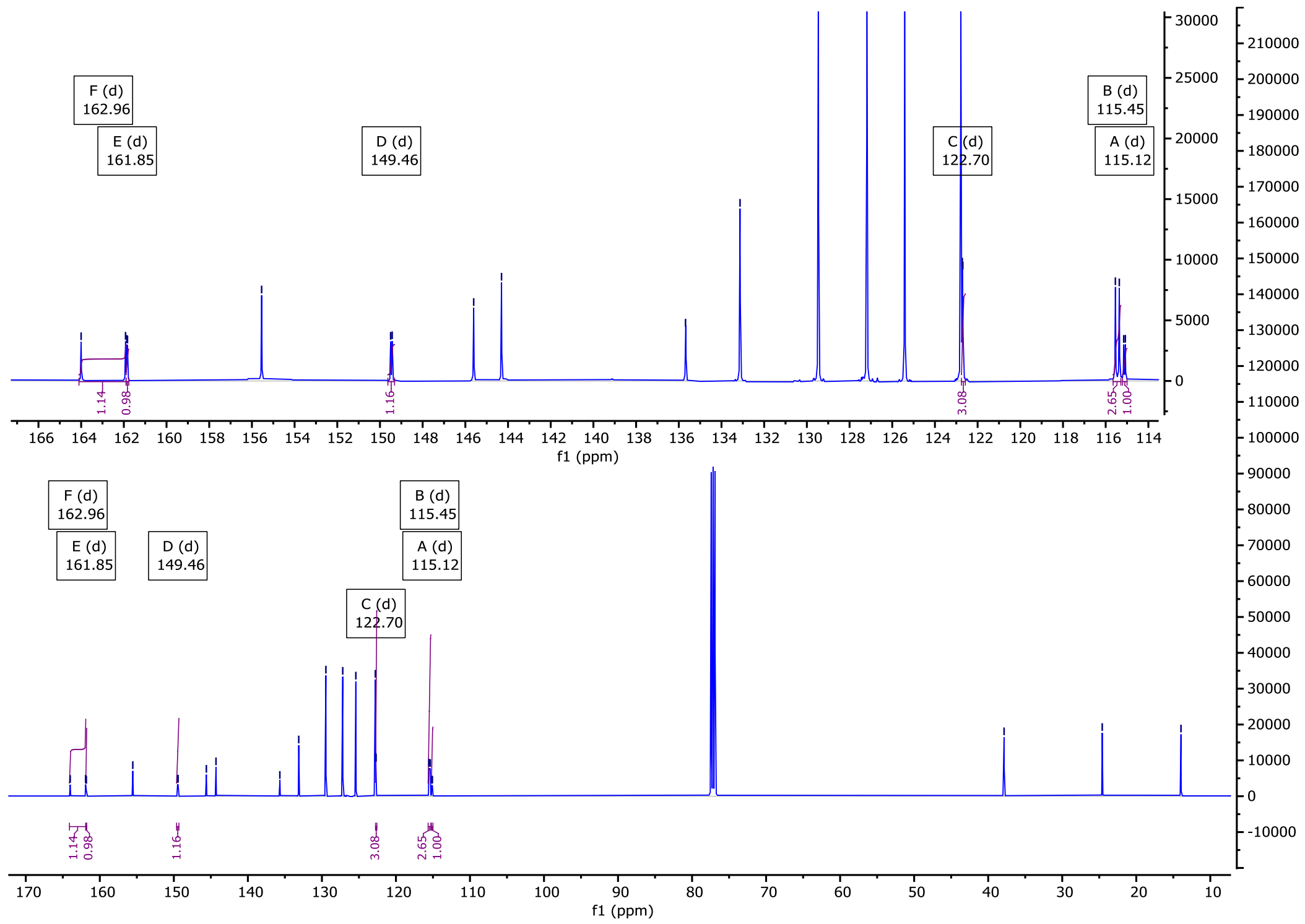

**Fig. S16.** $^{13}C\{^{1}H\}$ NMR of **1d** in $CDCl_3$.

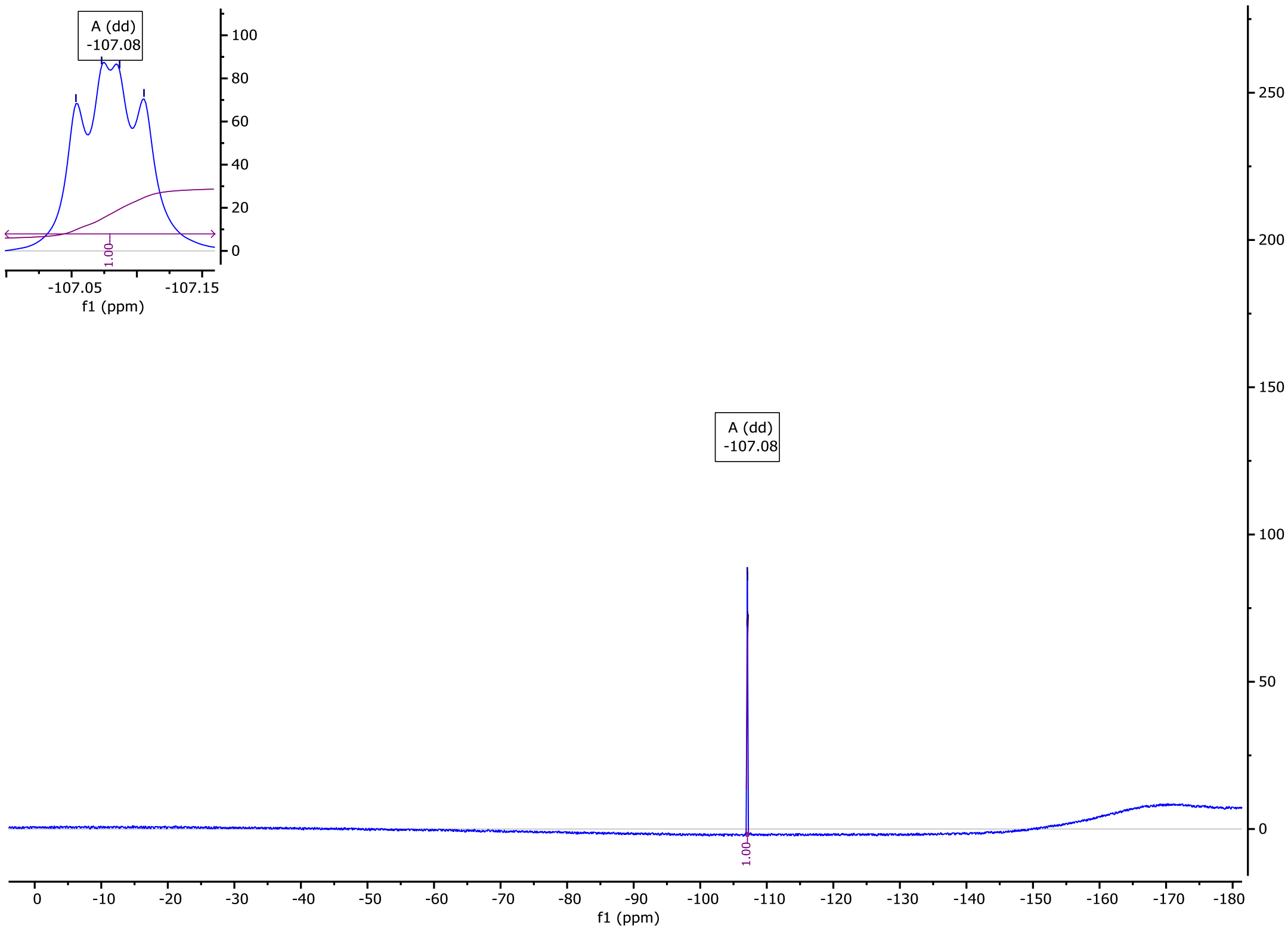

**Fig. S17.** $^{19}F$ NMR of **1d** in $CDCl_3$.

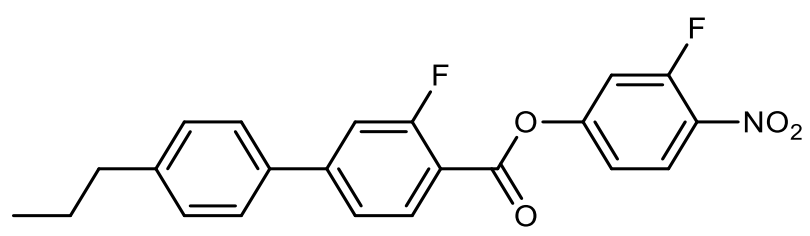

**1e**

*3-fluoro-4-nitrophenyl 3-fluoro-4'-propyl-[1,1'-biphenyl]-4-carboxylate*

Yield: (white solid) 143 mg, 72 %

$^{1}$H NMR (400 MHz): 8.20 (t, *J* = 8.9 Hz, 1H, Ar-**H**), 8.13 (t, *J* = 8.0 Hz, 1H, Ar-**H**), 7.57 (ddd, *J* = 8.2, 2.1, 1.8 Hz, 1H, Ar-**H**), 7.53 (dd, *J* = 8.2, 1.7 Hz, 1H, Ar-**H**), 7.45 (dd, *J* = 12.2, 1.7 Hz, 1H, Ar-**H**), 7.35 – 7.29 (m, 2H, Ar-**H**)*, 7.28 – 7.24 (m, 1H, Ar-**H**)†, 2.66 (t, *J* = 7.7 Hz, 2H, Ar-C**H₂**-CH₂), 1.69 (h, *J* = 7.5 Hz, 2H, $CH_2$-C**H₂**-$CH_3$), 0.98 (t, *J* = 7.3 Hz, 3H, $CH_2$-$CH_3$). * Overlapping peaks, † Overlapping $CDCl_3$.

$^{13}$C{$^{1}$H} NMR (126 MHz): 162.99 (d, *J* = 262.4 Hz), 161.36 (d, *J* = 4.3 Hz), 156.34 (d, *J* = 266.9 Hz), 155.60 (d, *J* = 10.5 Hz), 149.75 (d, *J* = 9.1 Hz), 144.41, 135.02 (d, *J* = 7.1 Hz), 134.99, 133.14, 129.48, 127.37 (d, *J* = 1.8 Hz), 127.19, 122.77 (d, *J* = 3.2 Hz), 118.24 (d, *J* = 4.0 Hz), 115.49 (d, *J* = 22.9 Hz), 114.69 (d, *J* = 9.4 Hz), 112.54 (d, *J* = 23.9 Hz), 37.86, 24.59, 13.95.

$^{19}$F NMR (376 MHz): -106.77 (dd, $J_{F-H}$ = 12.0, 7.5 Hz, 1F, Ar-**F**), -113.06 (t, $J_{F-H}$ = 9.8 Hz, 1F, Ar-**F**).

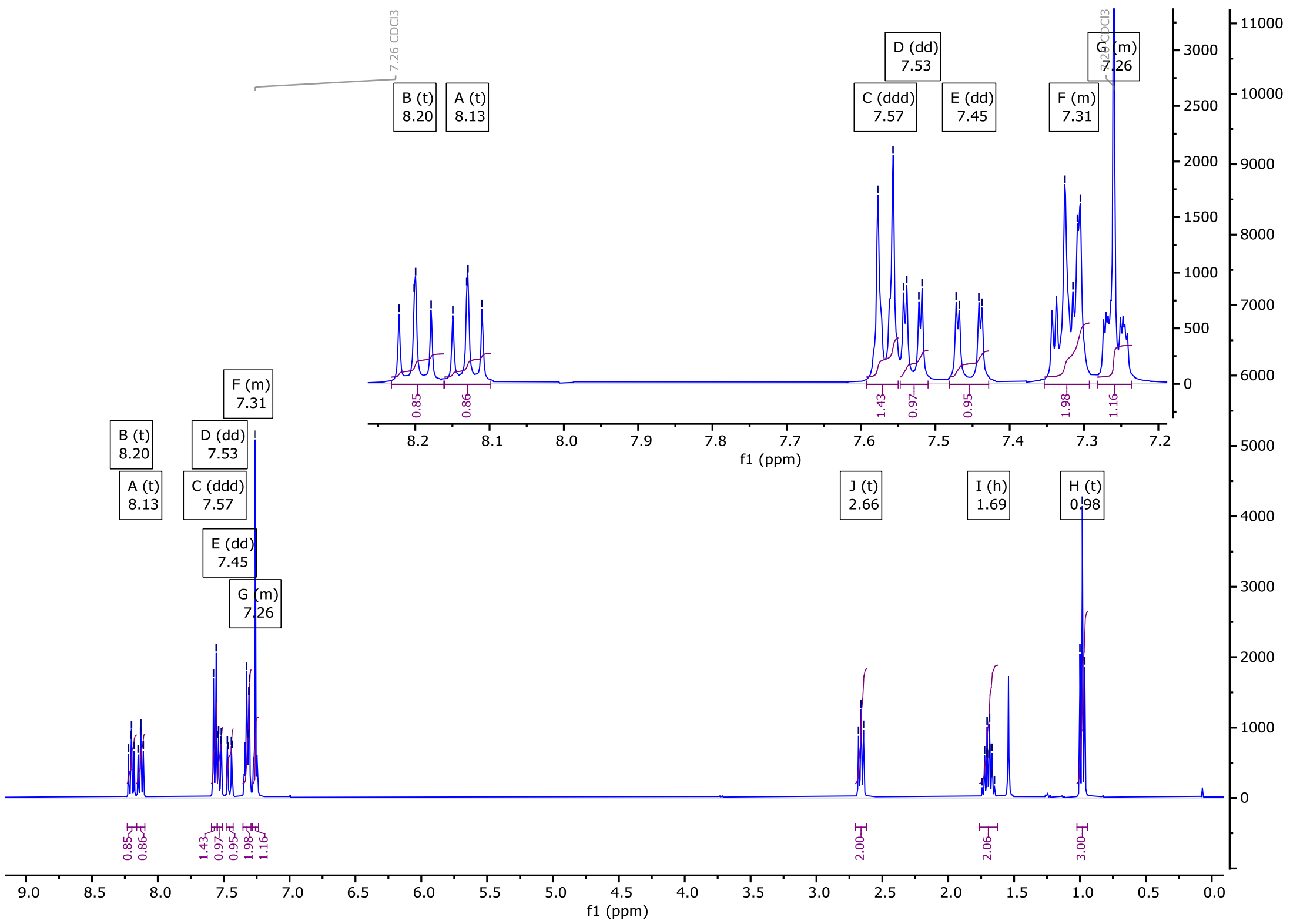

**Fig. S18.** $^{1}H$ NMR of **1e** in $CDCl_3$.

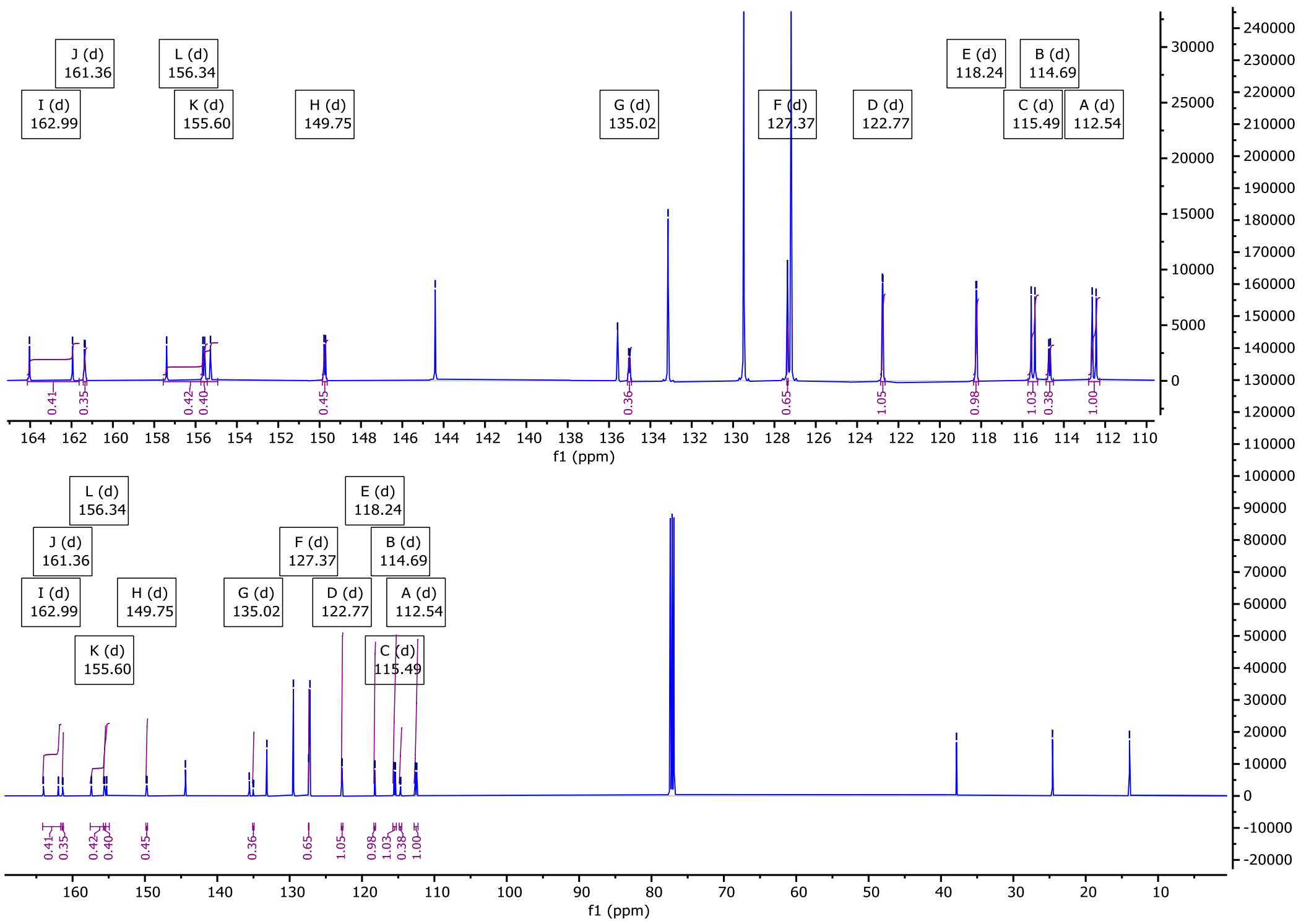

**Fig. S19.** $^{13}C\{^{1}H\}$ NMR of **1e** in $CDCl_3$.

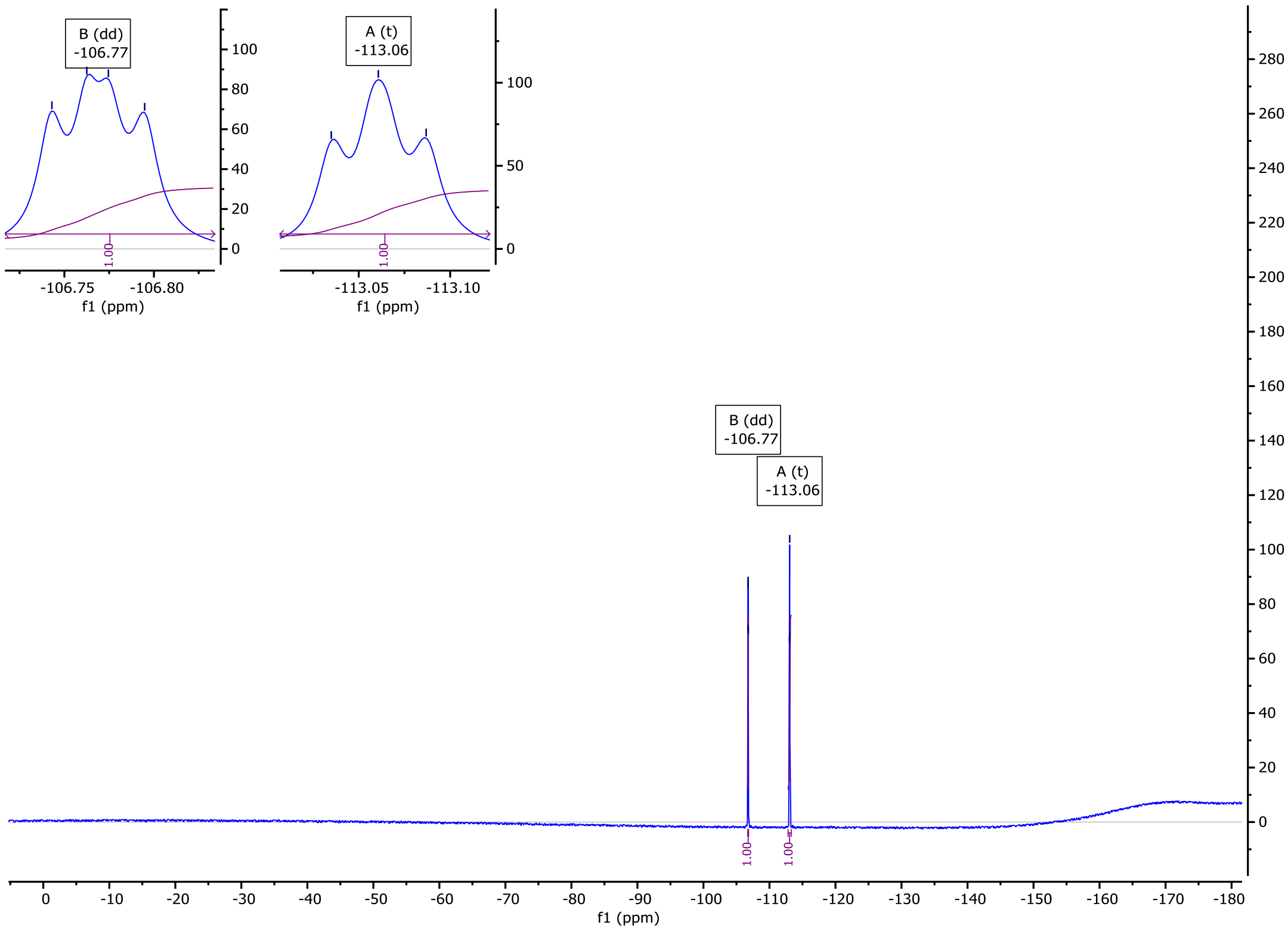

**Fig. S20.** $^{19}F$ NMR of **1e** in $CDCl_3$.

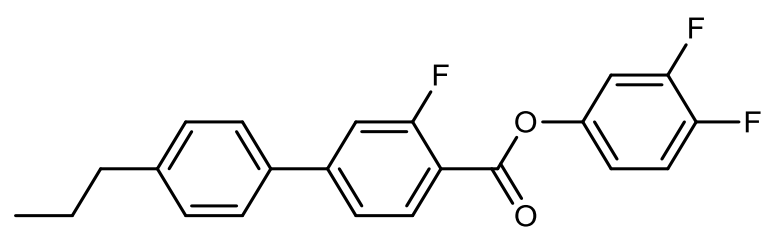

**1f**

*3,4-difluorophenyl 3-fluoro-4'-propyl-[1,1'-biphenyl]-4-carboxylate*

Yield: (colourless needles) 146 mg, 79 %

$^{1}$H NMR (400 MHz): 8.12 (t, *J* = 7.9 Hz, 1H, Ar-**H**), 7.56 (ddd, *J* = 8.3, 2.1, 1.8 Hz, 2H, Ar-**H**), 7.51 (dd, *J* = 8.2, 1.7 Hz, 1H, Ar-**H**), 7.43 (dd, *J* = 12.2, 1.7 Hz, 1H, Ar-**H**), 7.31 (d, *J* = 8.1 Hz, 2H Ar-**H**), 7.22 (d, *J* = 10.0 Hz, 1H, Ar-**H**), 7.16 (dddd, *J* = 13.5, 6.9, 4.0, 0.9 Hz, 1H Ar-**H**), 7.05 – 6.95 (m, 1H, Ar-**H**), 2.66 (t, *J* = 7.3 Hz, 2H, Ar-C**H$_2$**-CH$_2$), 1.70 (h, *J* = 7.3 Hz, 2H, CH$_2$-C**H$_2$**-CH$_3$), 0.99 (t, *J* = 7.3 Hz, 3H, CH$_2$-C**H$_3$**).

$^{13}$C{$^{1}$H} NMR (101 MHz): 162.75 (d, J = 261.5 Hz), 162.32 (d, J = 4.2 Hz), 161.45, 150.57 (dd, J = 250.4, 14.0 Hz), 149.00 (d, J = 8.9 Hz), 148.85 (dd, J = 246.6, 12.2 Hz), 146.23 (dd, J = 8.8, 3.1 Hz), 144.05, 135.67 (d, J = 1.3 Hz), 132.94, 129.31, 127.05, 122.49 (d, J = 3.2 Hz), 117.77 (dd, J = 6.4, 3.8 Hz), 117.35 (d, J = 18.9 Hz), 115.39 (d, J = 9.7 Hz), 115.24 (d, J = 23.2 Hz), 111.86 (d, J = 20.1 Hz), 37.74, 24.48, 13.84.

$^{19}$F NMR (376 MHz): -107.42 (dd, $J_{F-H}$ = 12.2, 7.6 Hz, 1F, Ar-**F**), -134.54 (dt, $J_{F-F}$ = 20.2 Hz, $J_{F-H}$ = 9.6 Hz, 1F, Ar-**F**), -140.23 – -142.19 (m, 1F, Ar-**F**).

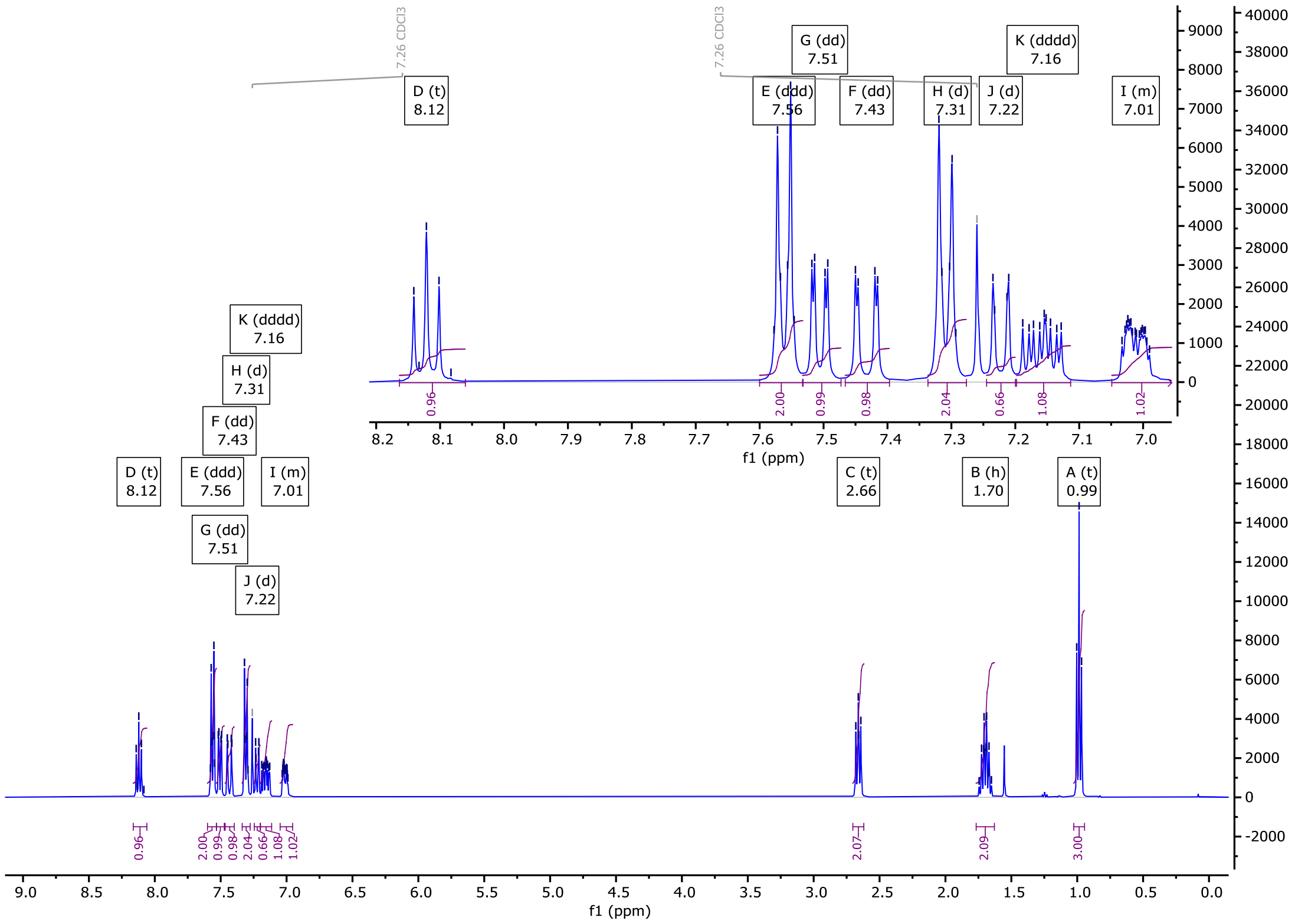

**Fig. S21.** $^{1}H$ NMR of **1f** in $CDCl_3$.

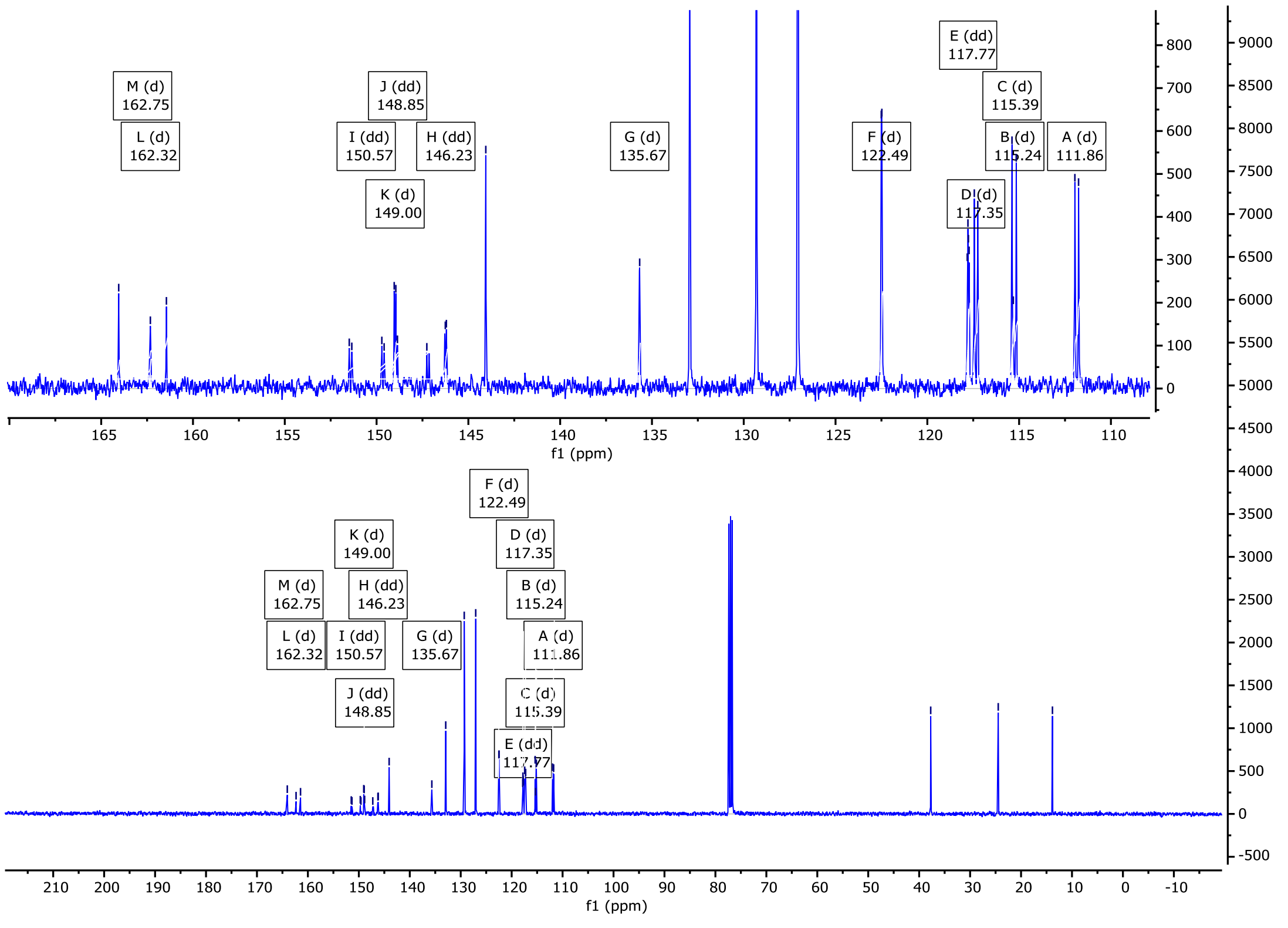

**Fig. S22.** $^{13}C\{^{1}H\}$ NMR of **1f** in $CDCl_3$.

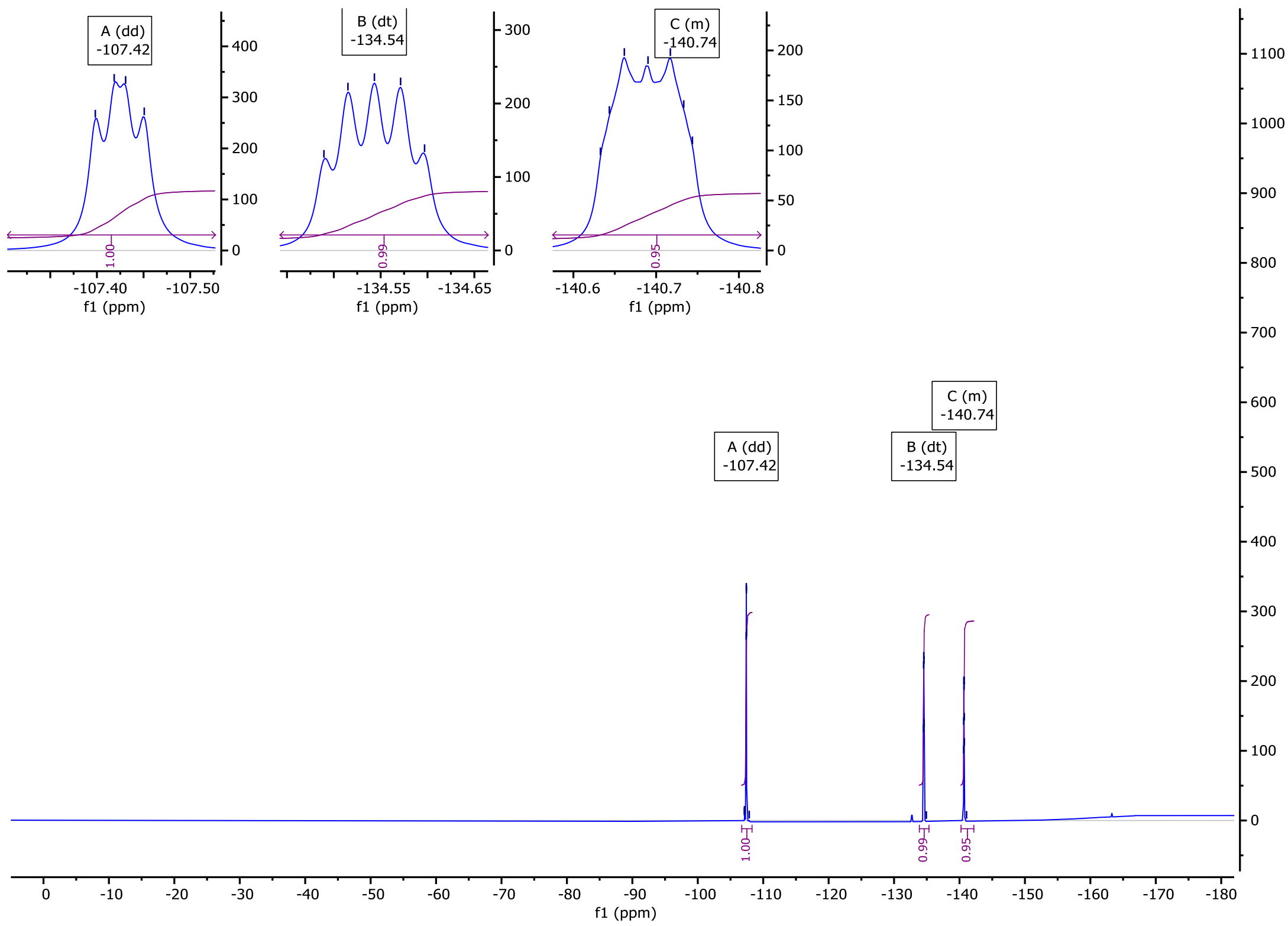

**Fig. S23.** $^{19}F$ NMR of **1f** in $CDCl_3$.

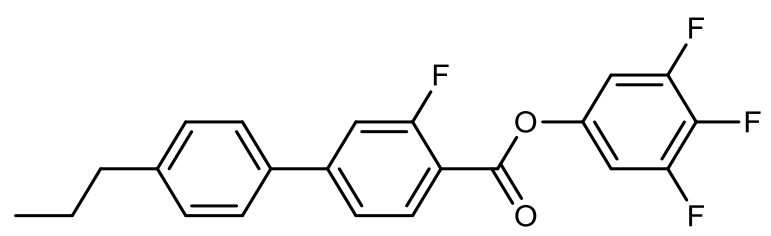

**1g**

*3,4,5-trifluorophenyl 3-fluoro-4'-propyl-[1,1'-biphenyl]-4-carboxylate*

| | |
|---|---|
| Yield: | (colourless needles) 155 mg, 80 % |
| $^{1}$H NMR (400 MHz): | 8.10 (t, *J* = 7.9 Hz, 1H, Ar-**H**), 7.56 (ddd, *J* = 8.2, 2.0, 1.9 Hz, 2H, Ar-**H**), 7.51 (dd, *J* = 8.2, 1.7 Hz, 1H, Ar-**H**), 7.43 (dd, *J* = 12.2, 1.7 Hz, 1H, Ar-**H**), 7.31 (ddd, *J* = 8.3, 1.9, 1.7 Hz, 2H, Ar-**H**), 7.03 – 6.93 (m, 2H, Ar-**H**), 2.66 (t, *J* = 7.4 Hz, 2H, Ar-C$\mathbf{H_2}$-$CH_2$), 1.69 (h, *J* = 7.5 Hz, 2H, $CH_2$-C$\mathbf{H_2}$-$CH_3$), 0.98 (t, *J* = 7.3 Hz, 3H, $CH_2$-C$\mathbf{H_3}$). |
| $^{13}$C{$^{1}$H} NMR (101 MHz): | 162.77 (d, *J* = 262.3 Hz), 161.89 (d, *J* = 4.4 Hz), 151.15 (ddd, *J* = 250.1, 10.6, 5.5 Hz), 149.27 (d, *J* = 9.1 Hz), 144.15, 138.24 (dt, *J* = 250.0, 15.4 Hz), 135.58 (d, *J* = 1.5 Hz), 132.95, 129.33, 127.06, 122.55 (d, *J* = 3.3 Hz), 115.31 (d, *J* = 23.0 Hz), 114.94 (d, *J* = 9.5 Hz), 107.23 (dd, *J* = 24.0, 6.2 Hz), 37.73, 24.47, 13.83. |
| $^{19}$F NMR (376 MHz): | -107.16 (dd, $J_{F\text{-}H}$ = 12.4, 7.6 Hz, 1F, Ar-**F**), -132.74 (dd, $J_{F\text{-}F}$ = 20.7 Hz, $J_{F\text{-}H}$ = 8.7 Hz, 2F, Ar-**F**), -163.29 (tt, $J_{F\text{-}F}$ = 20.9 Hz, $J_{F\text{-}H}$ = 6.0 Hz, 1F, Ar-**F**). |

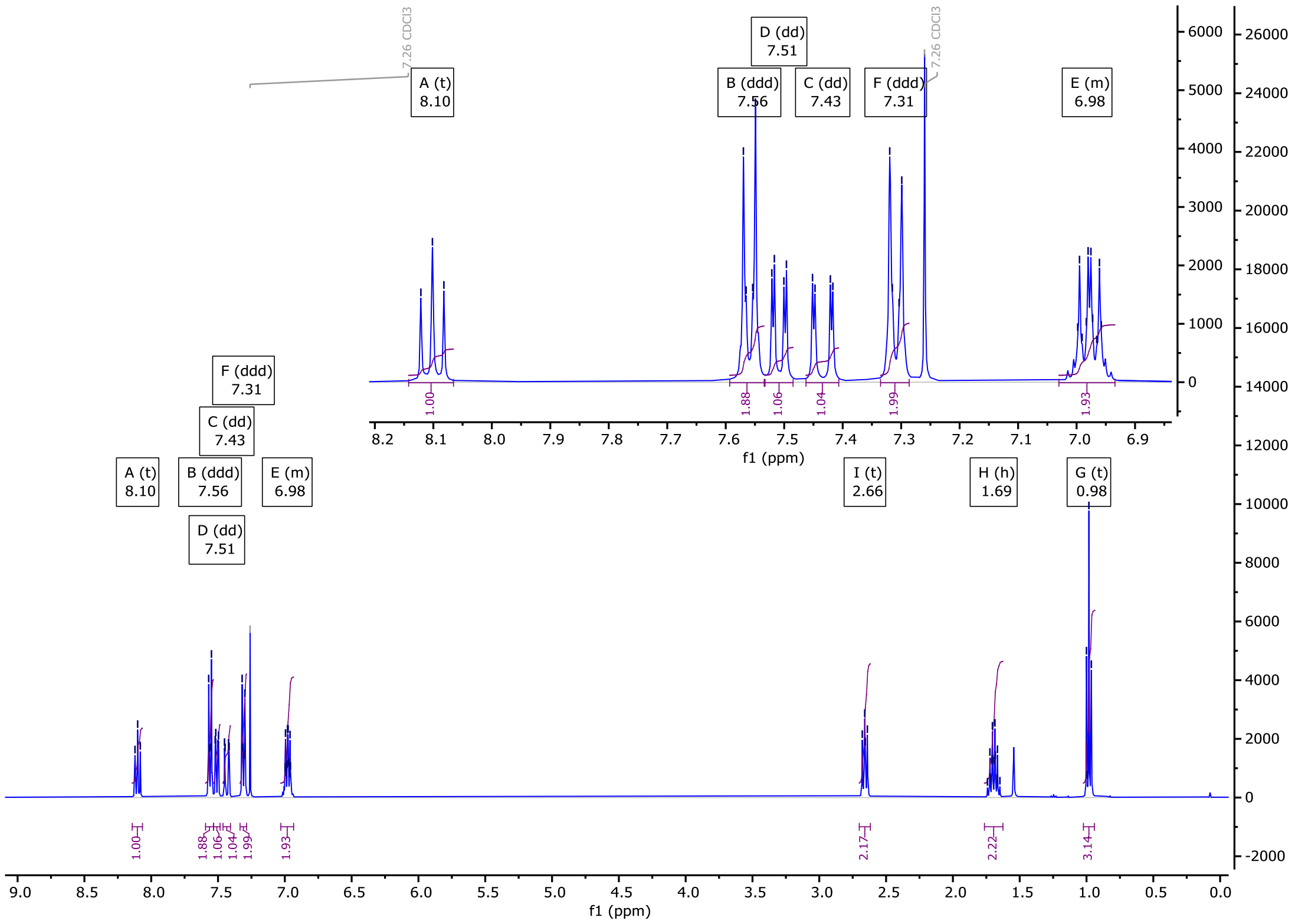

**Fig. S24.** $^{1}H$ NMR of **1g** in $CDCl_3$.

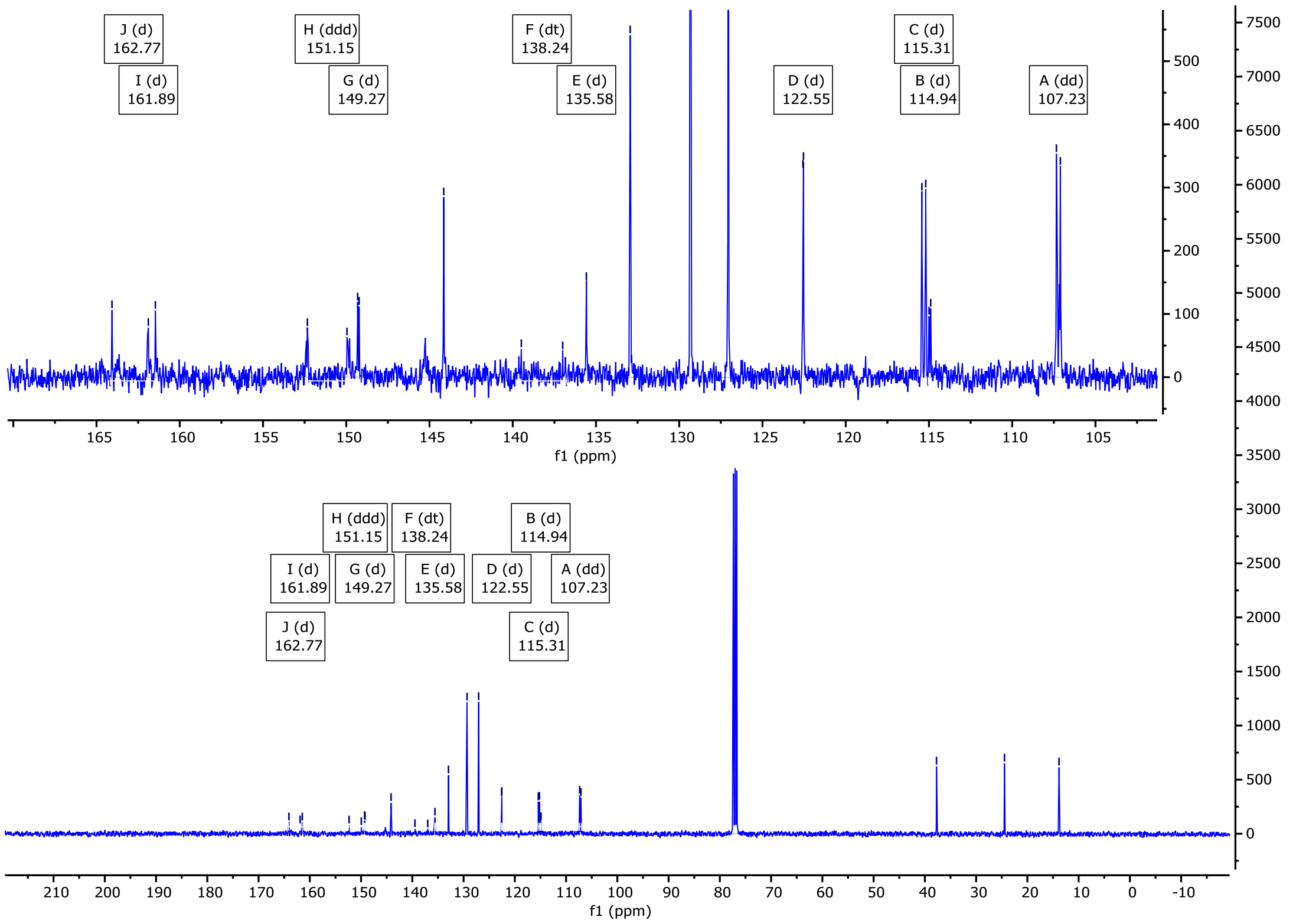

**Fig. S25.** $^{13}C\{^{1}H\}$ NMR of **1g** in $CDCl_3$.

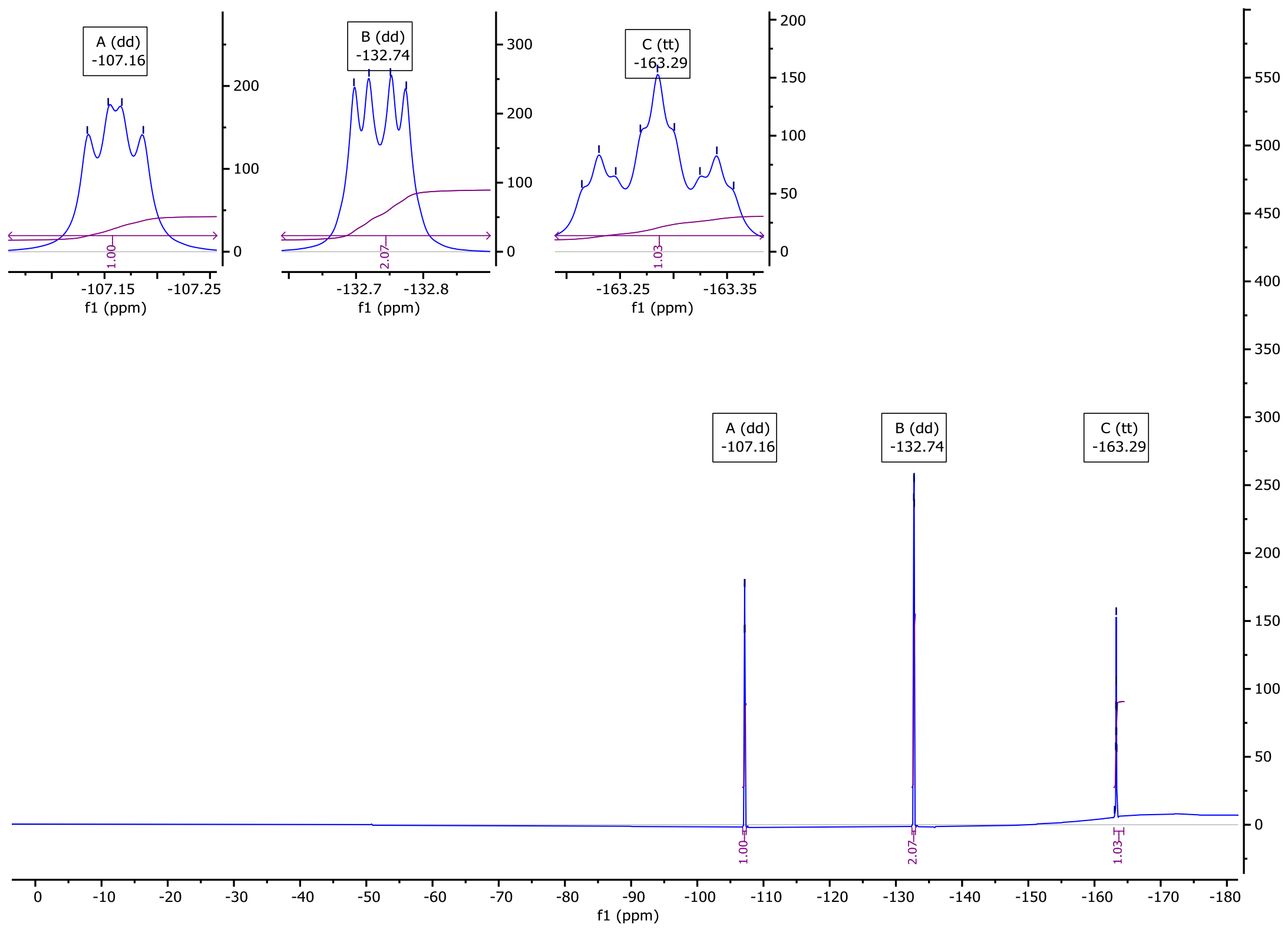

**Fig. S26.** $^{19}$F NMR of **1g** in $CDCl_3$.

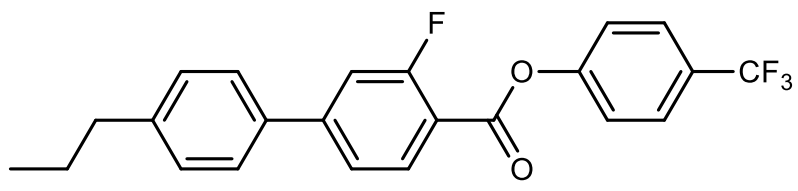

**1h**

*4-(trifluoromethyl)phenyl 3-fluoro-4'-propyl-[1,1'-biphenyl]-4-carboxylate*

Yield: (white solid) 133 mg, 66 %

$^{1}$H NMR (400 MHz): 8.15 (t, *J* = 7.9 Hz, 1H, Ar-**H**), 7.72 (ddd, *J* = 8.5, 2.7, 1.9 Hz, 2H, Ar-**H**), 7.57 (ddd, *J* = 8.2, 2.0, 1.8 Hz, 2H, Ar-**H**), 7.52 (dd, *J* = 8.2, 1.7 Hz, 1H, Ar-**H**), 7.44 (dd, *J* = 12.2, 1.7 Hz, 1H, Ar-**H**), 7.40 (d, *J* = 8.4 Hz, 2H, Ar-**H**), 7.31 (ddd, *J* = 8.2, 2.1, 1.9 Hz, 2H, Ar-**H**), 2.66 (t, *J* = 7.4 Hz, 2H, Ar-C$\mathbf{H_2}$-$CH_2$), 1.70 (h, *J* = 7.4 Hz, 2H, $CH_2$-C$\mathbf{H_2}$-$CH_3$), 0.99 (t, *J* = 7.3 Hz, 3H, $CH_2$-C$\mathbf{H_3}$).

$^{13}$C{$^{1}$H} NMR (101 MHz): 164.09 (d, *J* = 262.1 Hz), 162.20 (d, *J* = 3.9 Hz), 153.20, 149.02 (d, *J* = 8.9 Hz), 144.06, 135.68 (d, *J* = 1.1 Hz), 132.98, 129.31, 128.30 (d, *J* = 32.8 Hz), 127.81 (d, *J* = 15.8 Hz), 126.88 (q, *J* = 3.7 Hz), 125.26, 122.50 (d, *J* = 3.3 Hz), 122.28, 115.40 (d, *J* = 9.4 Hz), 37.74, 24.48, 13.84.

$^{19}$F NMR (376 MHz): -62.20 (s, 3F, C$\mathbf{F_3}$), -107.38 (dd, $J_{F-H}$ = 12.0, 7.6 Hz, 1F, Ar-**F**).

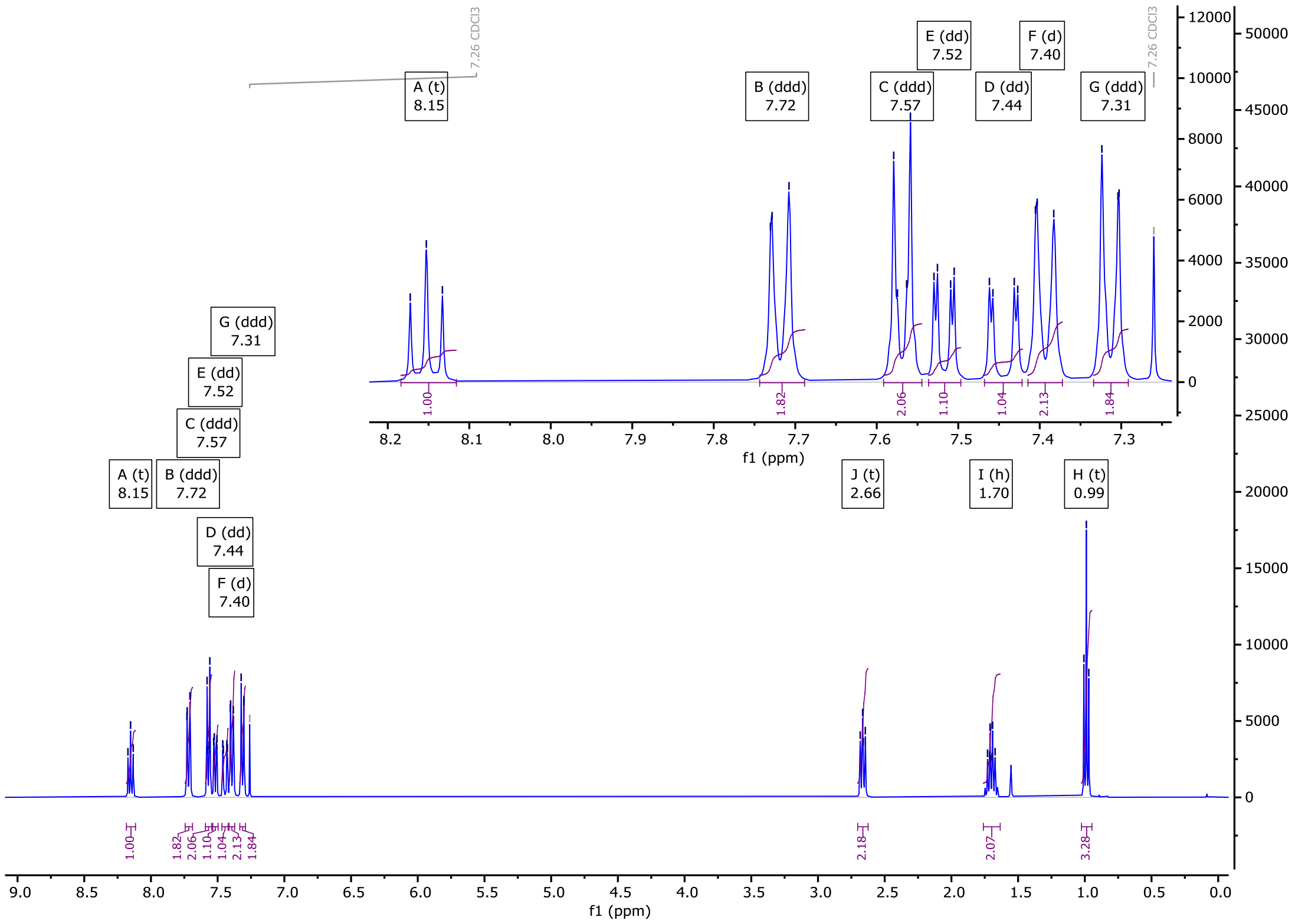

**Fig. S27.** $^1H$ NMR of **1h** in $CDCl_3$.

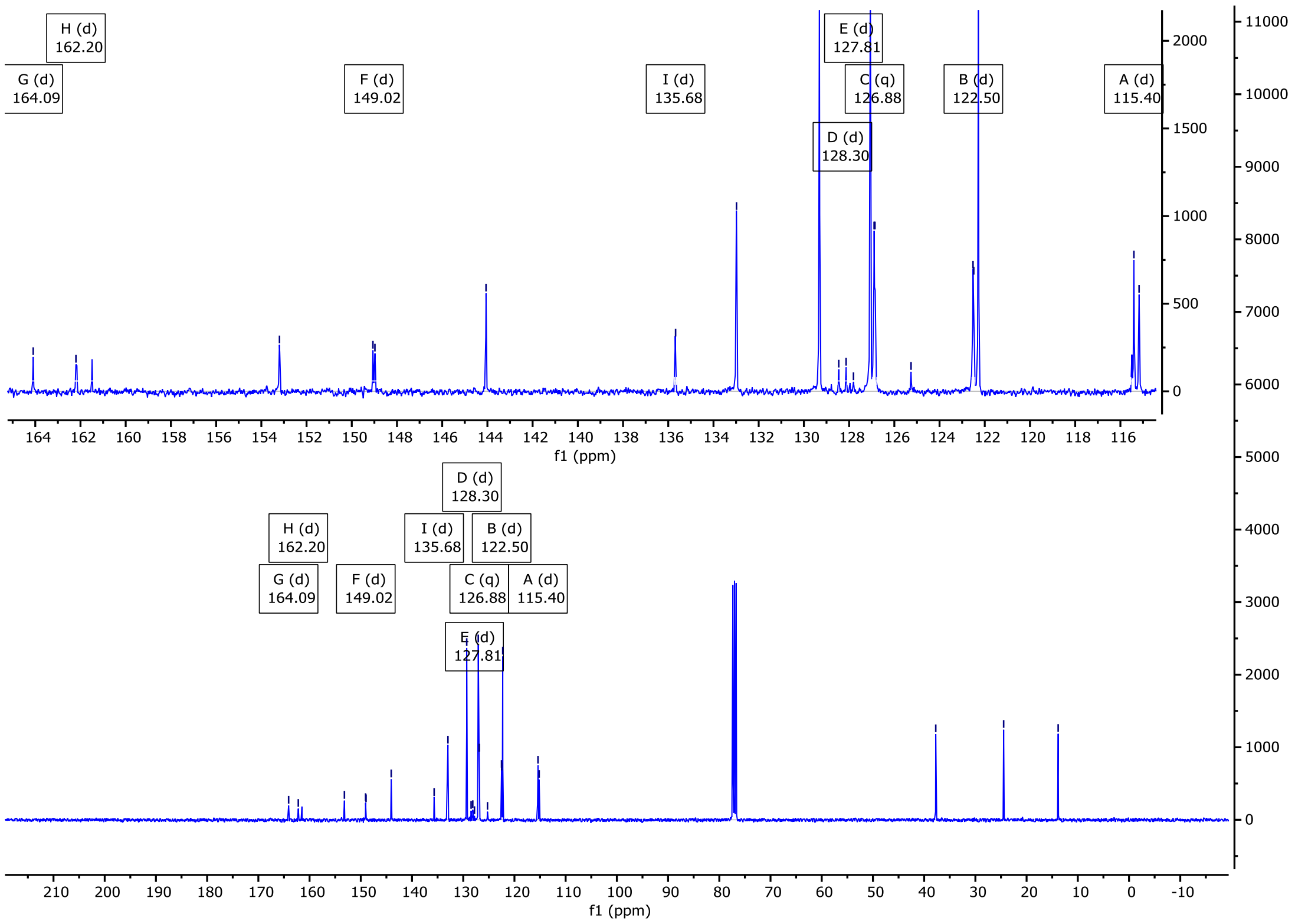

**Fig. S28.** $^{13}C\{^1H\}$ NMR of **1h** in $CDCl_3$.

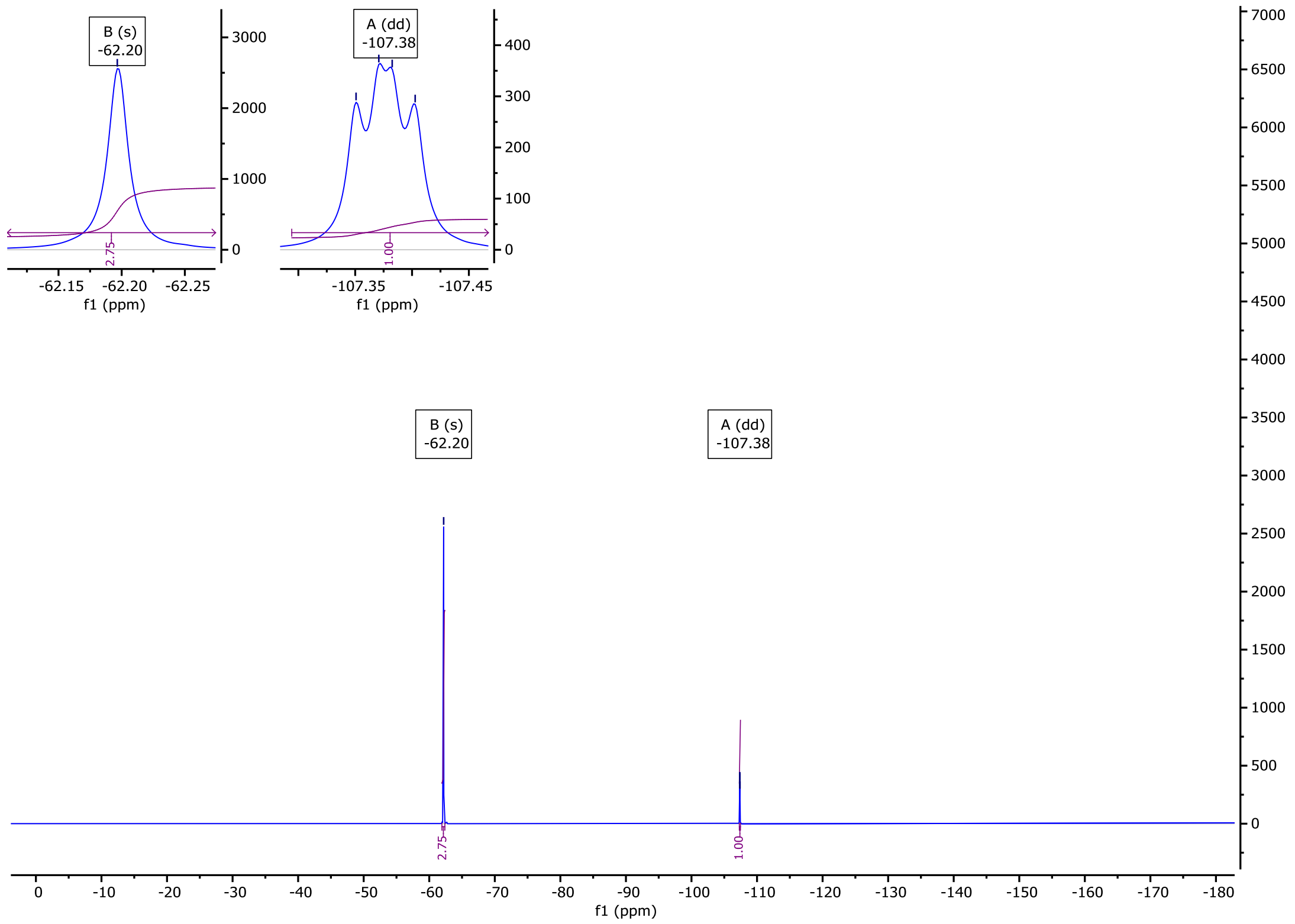

**Fig. S29.** $^{19}$F NMR of **1h** in $CDCl_3$.

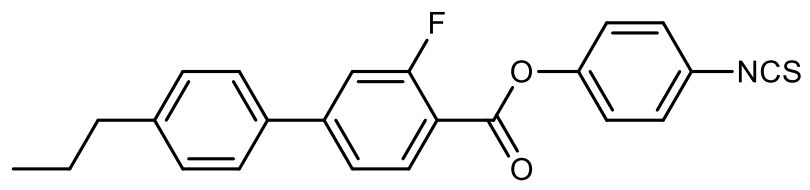

**1i**

*4-isothiocyanatophenyl 3-fluoro-4'-propyl-[1,1'-biphenyl]-4-carboxylate*

| | |
|---|---|
| Yield: | (white solid) 141 mg, 72 % |
| $^{1}$H NMR (400 MHz): | 8.13 (t, *J* = 7.9 Hz, 1H, Ar-**H**), 7.56 (ddd, *J* = 8.2, 1.9, 1.7 Hz, 2H, Ar-**H**), 7.50 (dd, *J* = 8.2, 1.7 Hz, 1H, Ar-**H**), 7.43 (dd, *J* = 12.2, 1.7 Hz, 1H, Ar-**H**), 7.34 – 7.22 (m, 7H, Ar-**H**)*†, 2.66 (t, *J* = 7.4 Hz, 2H, Ar-C**H₂**-CH₂), 1.69 (h, *J* = 7.5 Hz, 2H, CH₂-C**H₂**-CH₃), 0.98 (t, *J* = 7.3 Hz, 3H, CH₂-C**H₃**). * Overlapping peaks, † Overlapping $CDCl_3$. |
| $^{13}$C{$^{1}$H} NMR (126 MHz): | 162.84 (d, *J* = 261.5 Hz), 162.42 (d, *J* = 4.2 Hz), 149.39, 149.01 (d, *J* = 8.9 Hz), 144.13, 136.17, 135.76, 133.05, 129.40, 129.09, 127.15, 126.90, 123.17, 122.57 (d, *J* = 3.3 Hz), 115.59 (d, *J* = 9.5 Hz), 115.35 (d, *J* = 23.1 Hz), 37.83, 24.58, 13.96. |
| $^{19}$F NMR (376 MHz): | -107.49 (dd, $J_{F-H}$ = 11.9, 7.4 Hz, 1F, Ar-**F**). |

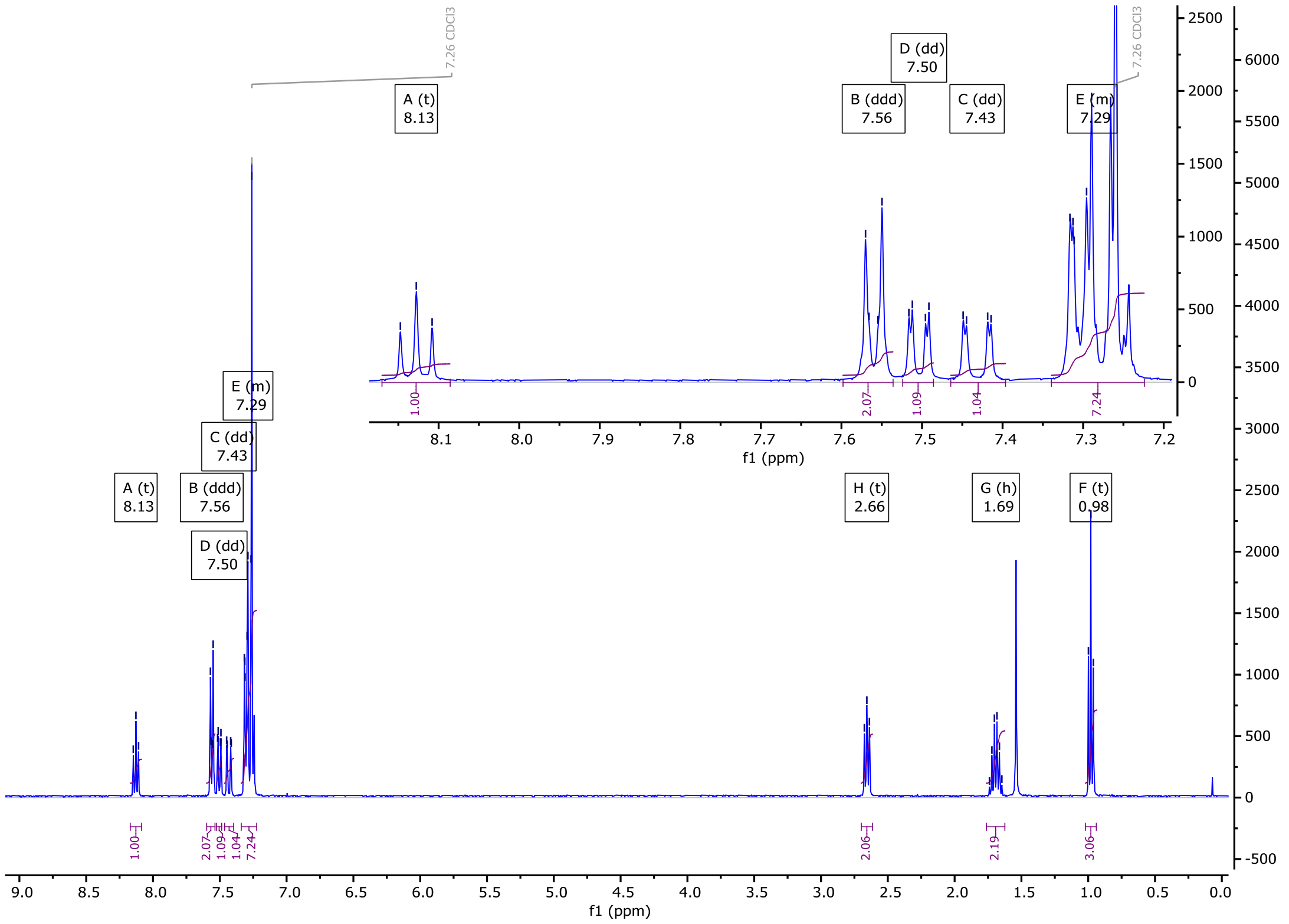

**Fig. S30.** $^{1}H$ NMR of **1i** in $CDCl_3$.

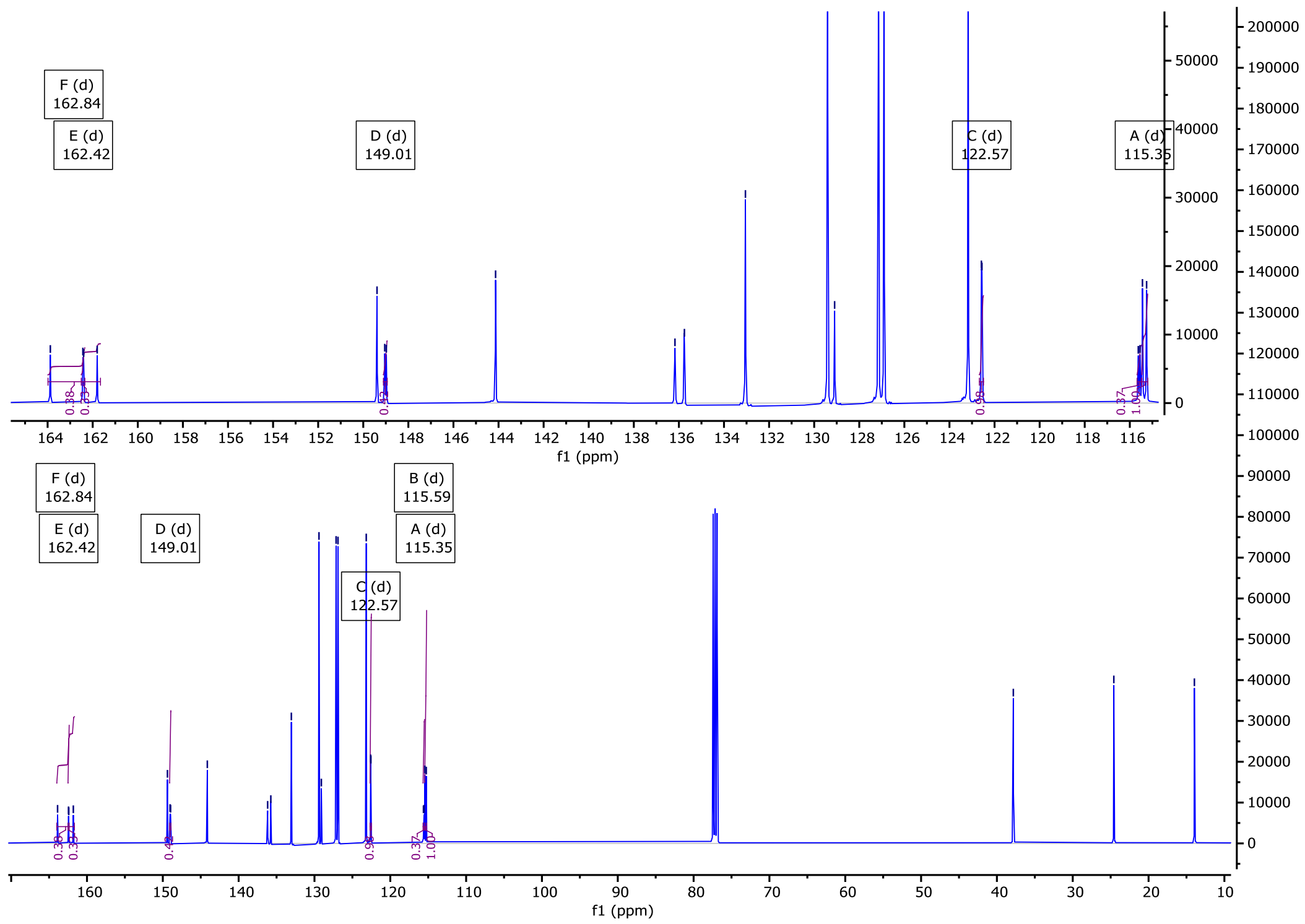

**Fig. S31.** $^{13}C\{^{1}H\}$ NMR of **1i** in $CDCl_3$.

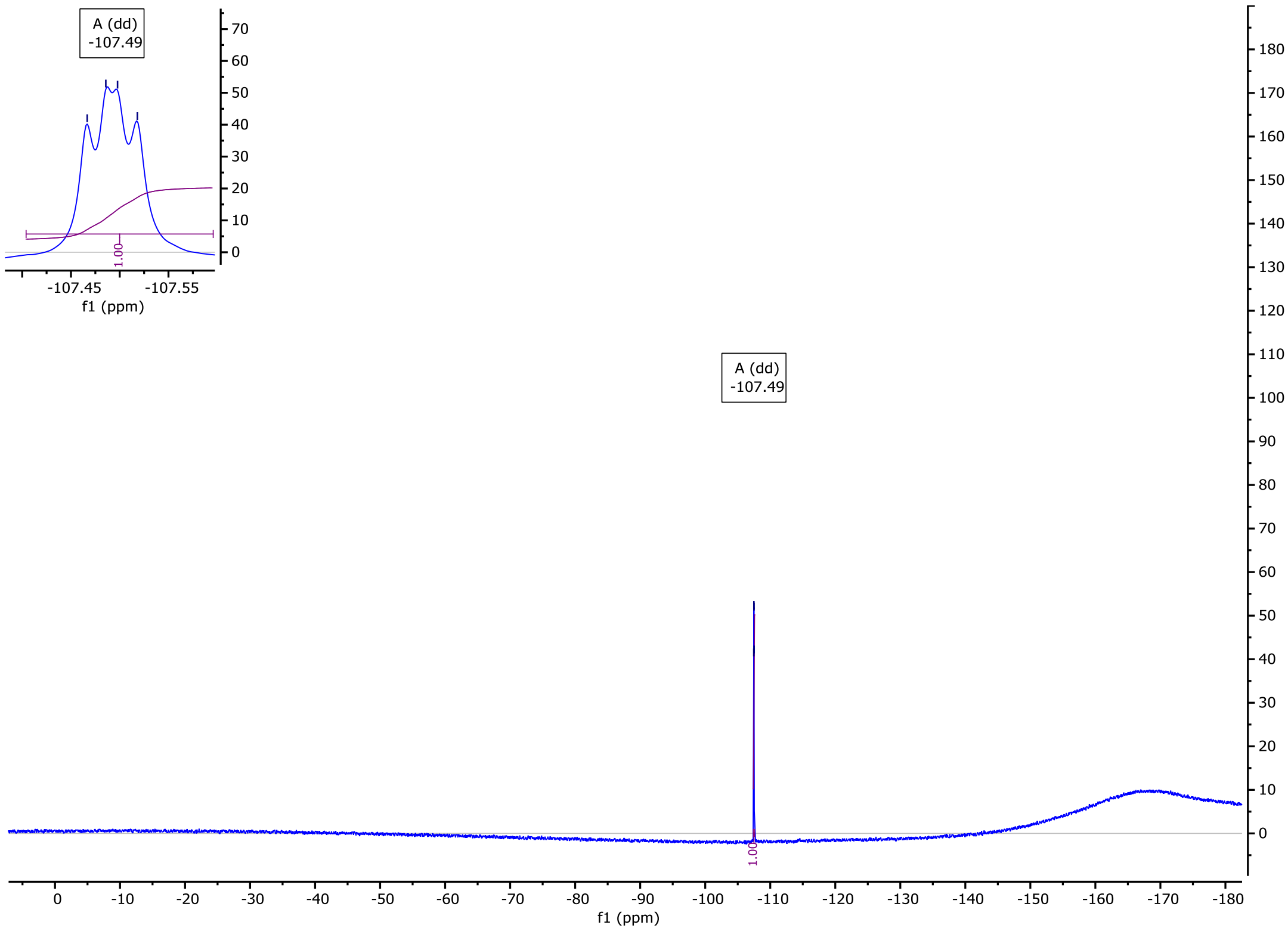

**Fig. S32.** $^{19}$F NMR of **1i** in $CDCl_3$.

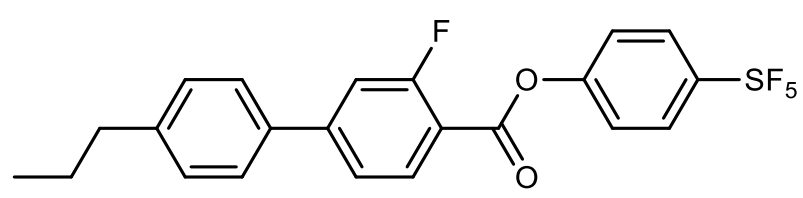

**1j**

*4-(pentafluoro-l6-sulfaneyl)phenyl 3-fluoro-4'-propyl-[1,1'-biphenyl]-4-carboxylate*

Yield: (white solid) 161 mg, 70 %

$^{1}$H NMR (400 MHz): 8.14 (t, *J* = 7.9 Hz, 1H, Ar-**H**), 7.85 (ddd, *J* = 9.0, 3.2, 2.1 Hz, 2H, Ar-**H**), 7.57 (ddd, *J* = 8.2, 2.0, 1.8 Hz, 2H, Ar-**H**), 7.52 (dd, *J* = 8.2, 1.7 Hz, 1H, Ar-**H**), 7.44 (dd, *J* = 12.2, 1.7 Hz, 1H, Ar-**H**), 7.37 (d, *J* = 8.8 Hz, 2H, Ar-**H**), 7.31 (ddd, *J* = 8.3, 2.1, 1.5 Hz, 2H, Ar-**H**), 2.66 (t, *J* = 7.5 Hz, 2H, Ar-C$\mathbf{H_2}$-$CH_2$), 1.69 (h, *J* = 7.4 Hz, 2H, $CH_2$-C$\mathbf{H_2}$-$CH_3$), 0.98 (t, *J* = 7.3 Hz, 3H, $CH_2$-C$\mathbf{H_3}$).

$^{13}$C{$^{1}$H} NMR (101 MHz): 162.80 (d, *J* = 261.9 Hz), 161.98 (d, *J* = 4.2 Hz), 152.53, 151.11 (d, *J* = 18.7 Hz), 149.17 (d, *J* = 8.9 Hz), 144.11, 135.64, 132.99, 129.32, 127.63 (p, *J* = 4.2 Hz), 122.54 (d, *J* = 3.2 Hz), 122.05, 115.31 (d, *J* = 23.3 Hz), 37.74, 24.47, 13.83.

$^{19}$F NMR (376 MHz): 84.16 (p, *J* = 151.1 Hz, 1F, S-$\mathbf{F_{\Lambda}}$), 63.53 (d, *J* = 150.0 Hz, 4F, S-$\mathbf{F_{4\Delta}}$), -107.22 (dd, *J* = 12.2, 7.6 Hz, 1F, Ar-**F**).

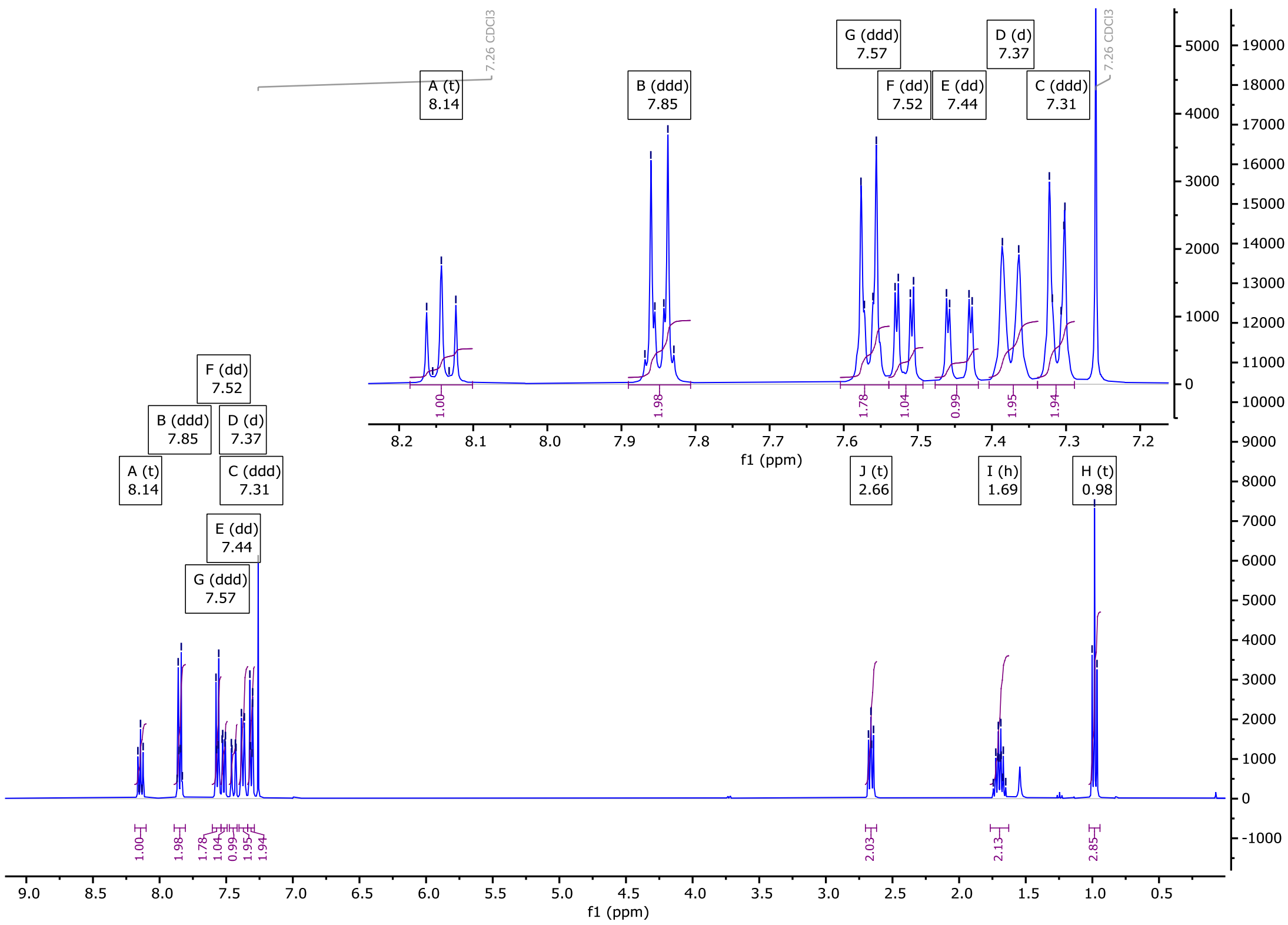

**Fig. S33.** $^{1}H$ NMR of **1j** in $CDCl_3$.

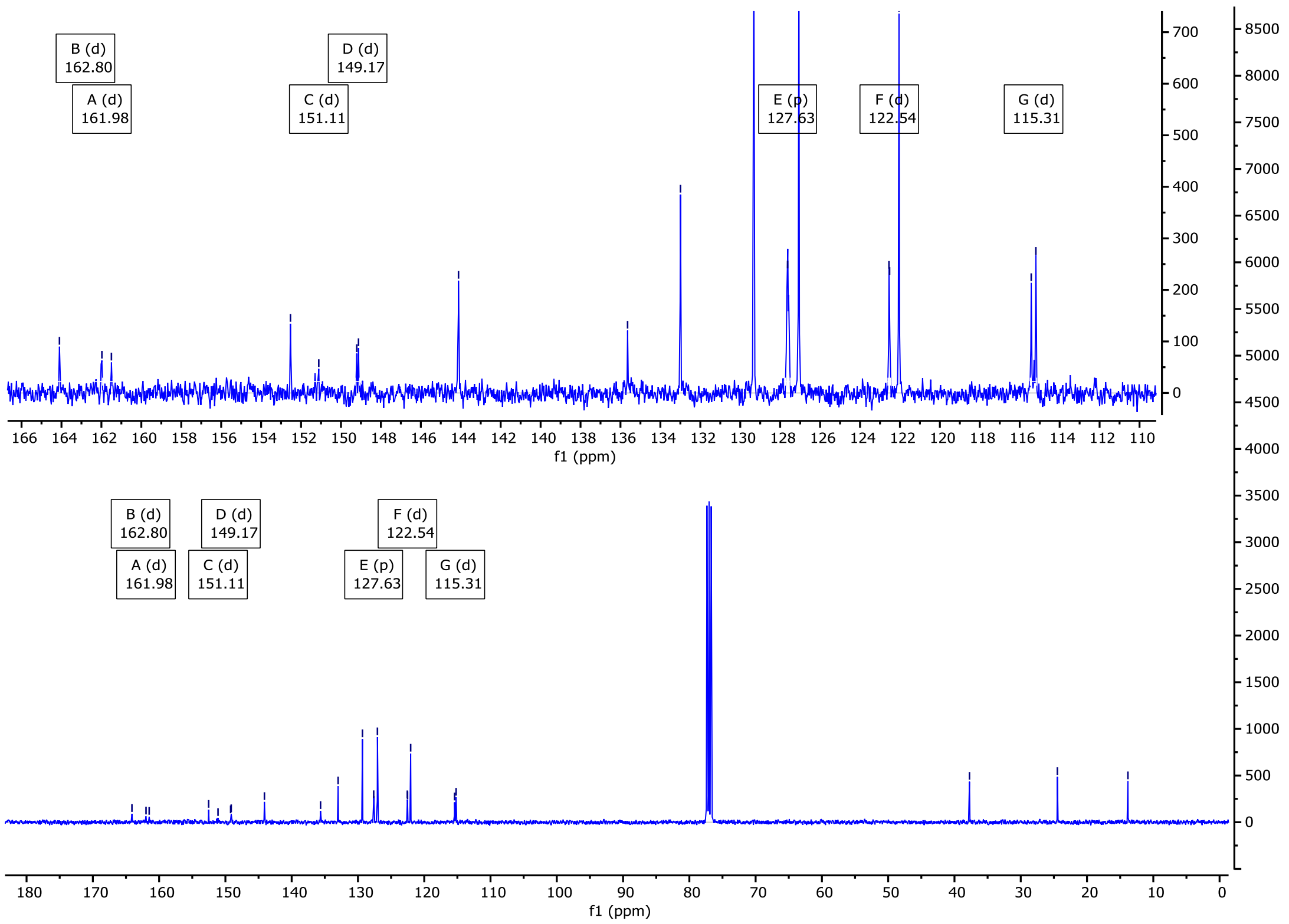

**Fig. S34.** $^{13}C\{^{1}H\}$ NMR of **1j** in $CDCl_3$.

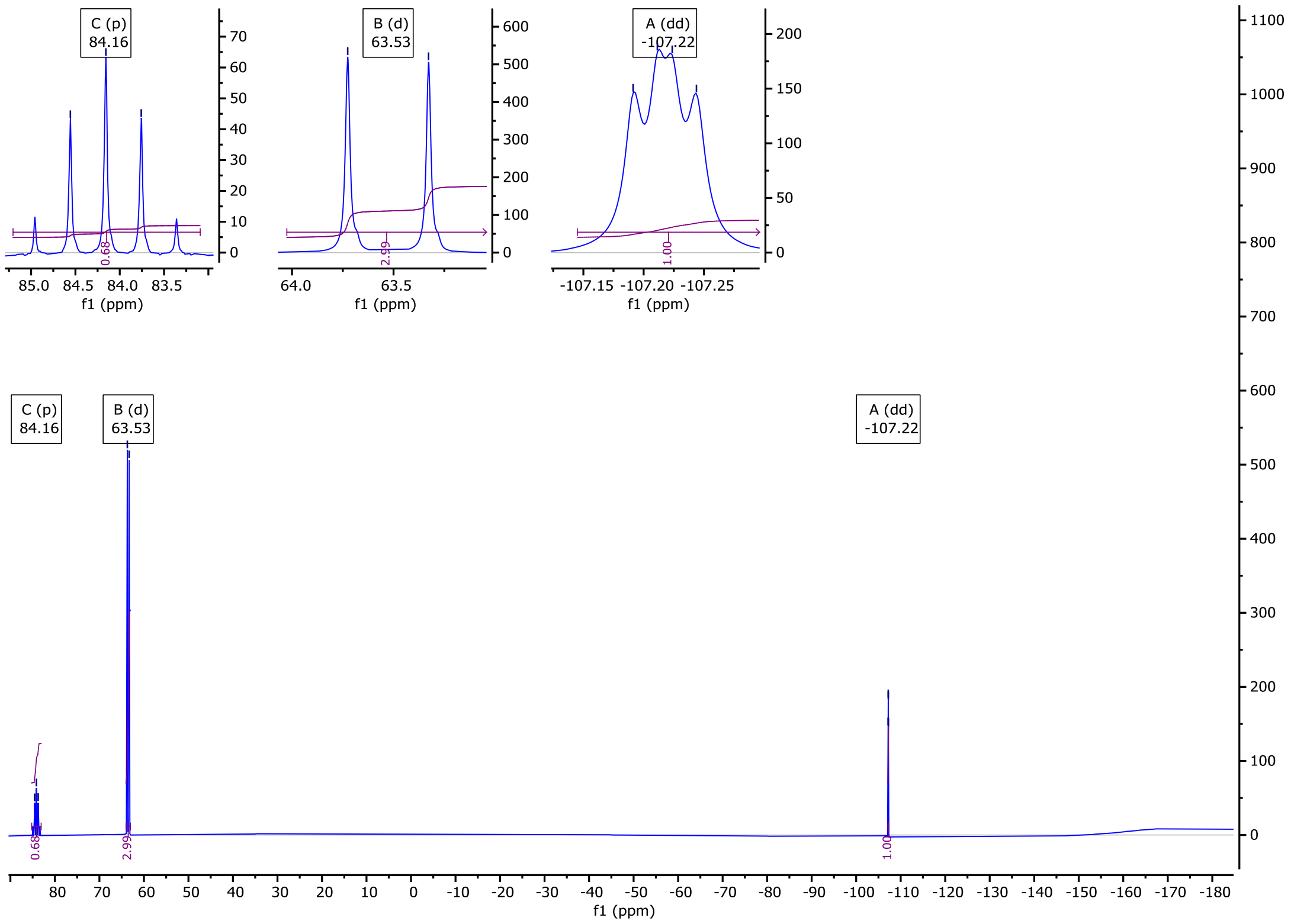

**Fig. S35.** $^{19}F$ NMR of **1j** in $CDCl_3$.

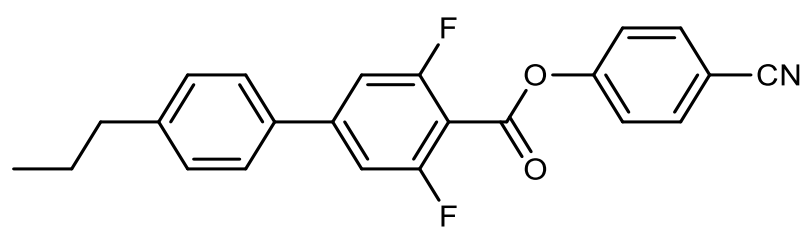

**2a**

*4-cyanophenyl 3,5-difluoro-4'-propyl-[1,1'-biphenyl]-4-carboxylate*

Yield: (white needles) 164 mg, 87 %

$^{1}$H NMR (501 MHz): 7.75 (ddd, *J* = 8.9, 2.4, 2.0 Hz, 2H, Ar-**H**), 7.52 (ddd, *J* = 8.2, 2.3, 1.8 Hz, 2H, Ar-**H**), 7.42 (ddd, *J* = 8.8, 2.4, 2.2 Hz, 2H, Ar-**H**), 7.31 (ddd, *J* = 8.2, 1.9, 1.8 Hz, 2H, Ar-**H**), 7.26 (ddd, *J* = 9.6, 1.7, 1.4 Hz, 2H, Ar-**H**) †, 2.66 (t, *J* = 7.5 Hz, 2H, Ar-C**H₂**-CH₂), 1.69 (h, *J* = 7.3 Hz, 2H, CH₂-C**H₂**-CH₃), 0.98 (t, *J* = 7.3 Hz, 3H, CH₂-C**H₃**). † Overlapping $CDCl_3$.

$^{13}$C{$^{1}$H} NMR (126 MHz): 161.74 (dd, *J* = 258.5, 6.2 Hz), 159.19, 153.71, 148.26 (t, *J* = 10.4 Hz), 144.76, 134.85, 133.92, 129.53, 126.99, 122.97, 118.30, 110.59 (dd, *J* = 22.3, 3.6 Hz), 110.36, 107.13 (t, *J* = 16.4 Hz), 37.83, 24.55, 13.93.

$^{19}$F NMR (376 MHz): -108.05 (d, $J_{F-H}$ = 10.6 Hz, 2F, Ar-**F**).

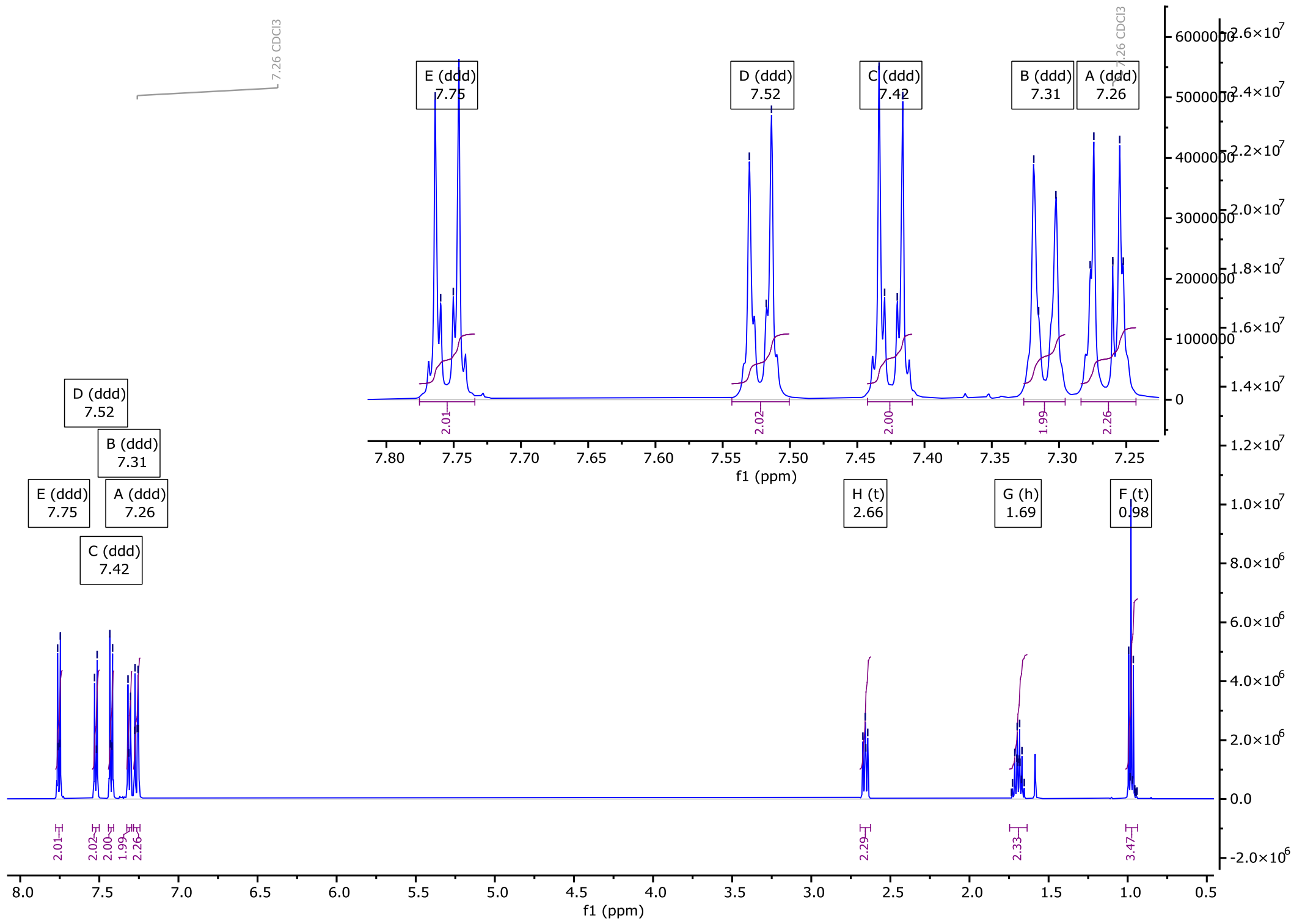

**Fig. S36.** $^{1}$H NMR of **2a** in $CDCl_3$.

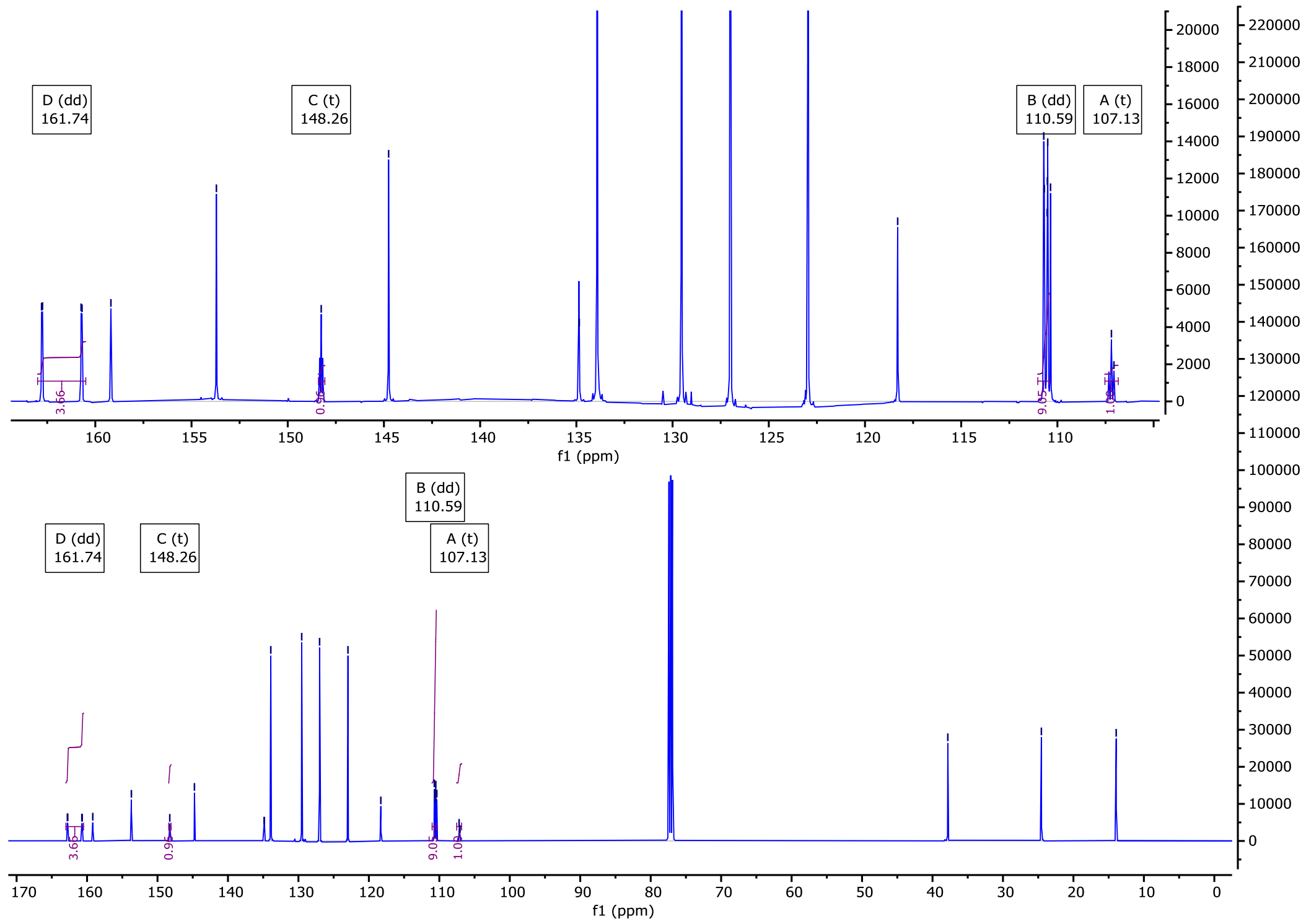

**Fig. S39.** $^{13}C\{^{1}H\}$ NMR of **2a** in $CDCl_3$.

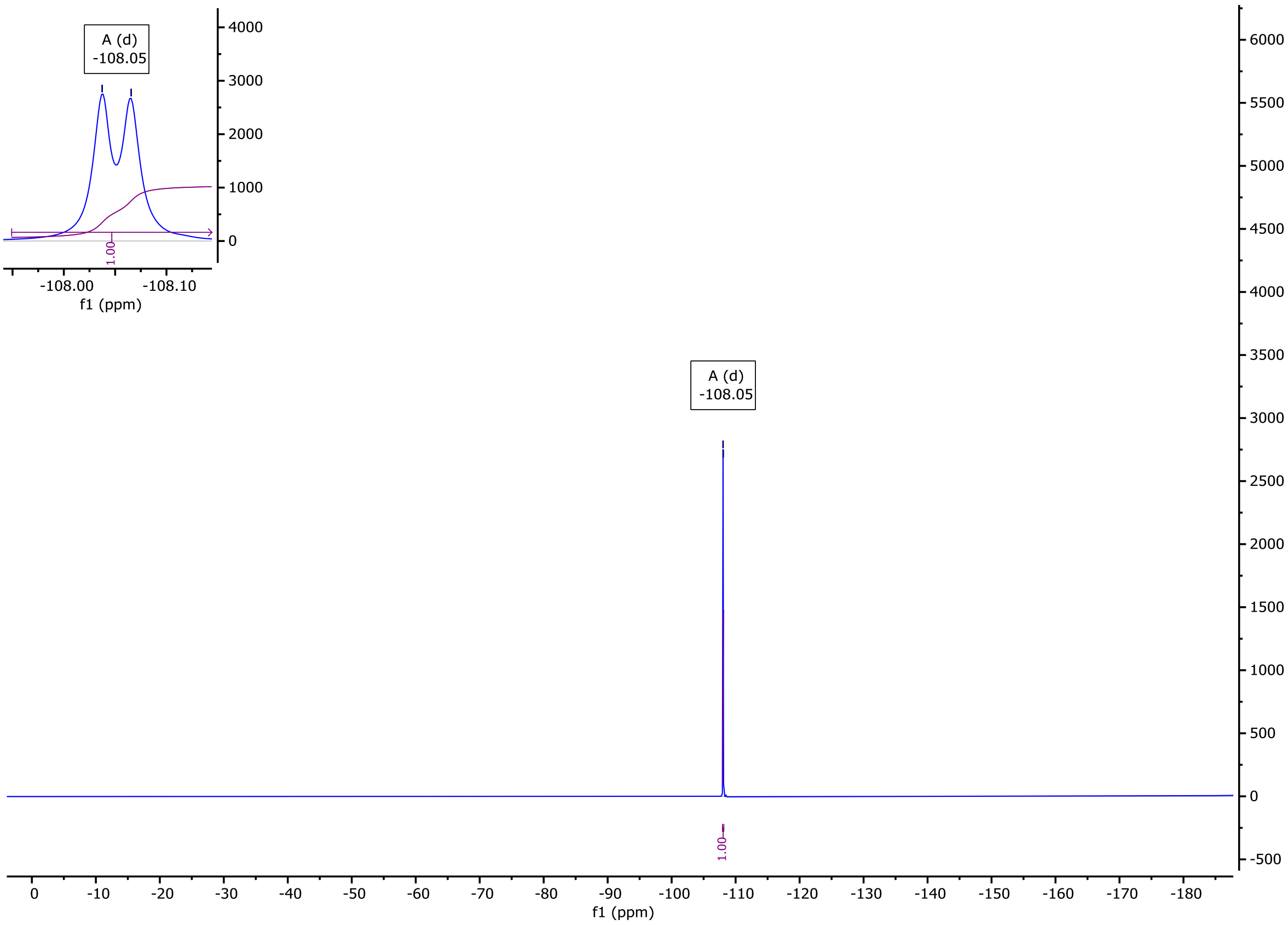

**Fig. S40.** $^{19}F$ NMR of **2a** in $CDCl_3$.

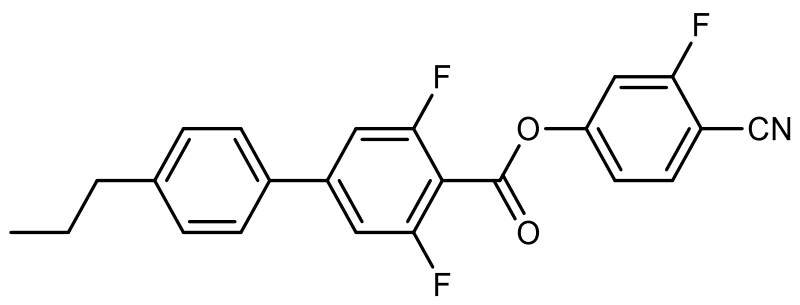

**2b**

*4-cyano-3-fluorophenyl 3,5-difluoro-4'-propyl-[1,1'-biphenyl]-4-carboxylate*

Yield: (white needles) 160 mg, 81 %

$^{1}$H NMR (501 MHz): 7.71 (t, *J* = 8.2 Hz, 1H, Ar-**H**), 7.52 (ddd, *J* = 8.1, 2.4, 1.8 Hz, 2H, Ar-**H**), 7.31 (d, *J* = 8.0 Hz, 2H, Ar-**H**), 7.30 – 7.22 (m, 4H, Ar-**H**) * †, 2.66 (t, *J* = 7.7 Hz, 2H, Ar-C**H₂**-CH₂), 1.69 (h, *J* = 7.4 Hz, 2H, CH₂-C**H₂**-CH₃), 0.98 (t, *J* = 7.3 Hz, 3H, CH₂-C**H₃**). * Overlapping peaks, † Overlapping $CDCl_3$.

$^{13}$C{$^{1}$H} NMR (126 MHz): 164.77 (d, *J* = 261.3 Hz), 161.82 (dd, *J* = 259.1, 6.2 Hz), 158.70, 154.96 (d, *J* = 10.6 Hz), 148.62 (t, *J* = 10.5 Hz), 144.88, 134.76, 134.32, 129.56, 127.00, 118.80 (d, *J* = 3.6 Hz), 113.50, 111.13 (d, *J* = 22.7 Hz), 110.66 (dd, *J* = 22.5, 3.4 Hz), 106.70 (t, *J* = 16.1 Hz), 99.47 (d, *J* = 15.5 Hz), 37.83, 24.54, 13.92.

$^{19}$F NMR (376 MHz): -103.07 (t, $J_{F-H}$ = 8.2 Hz, 1F, Ar-**F**), -107.70 (d, $J_{F-H}$ = 10.7 Hz, 2F, Ar-**F**).

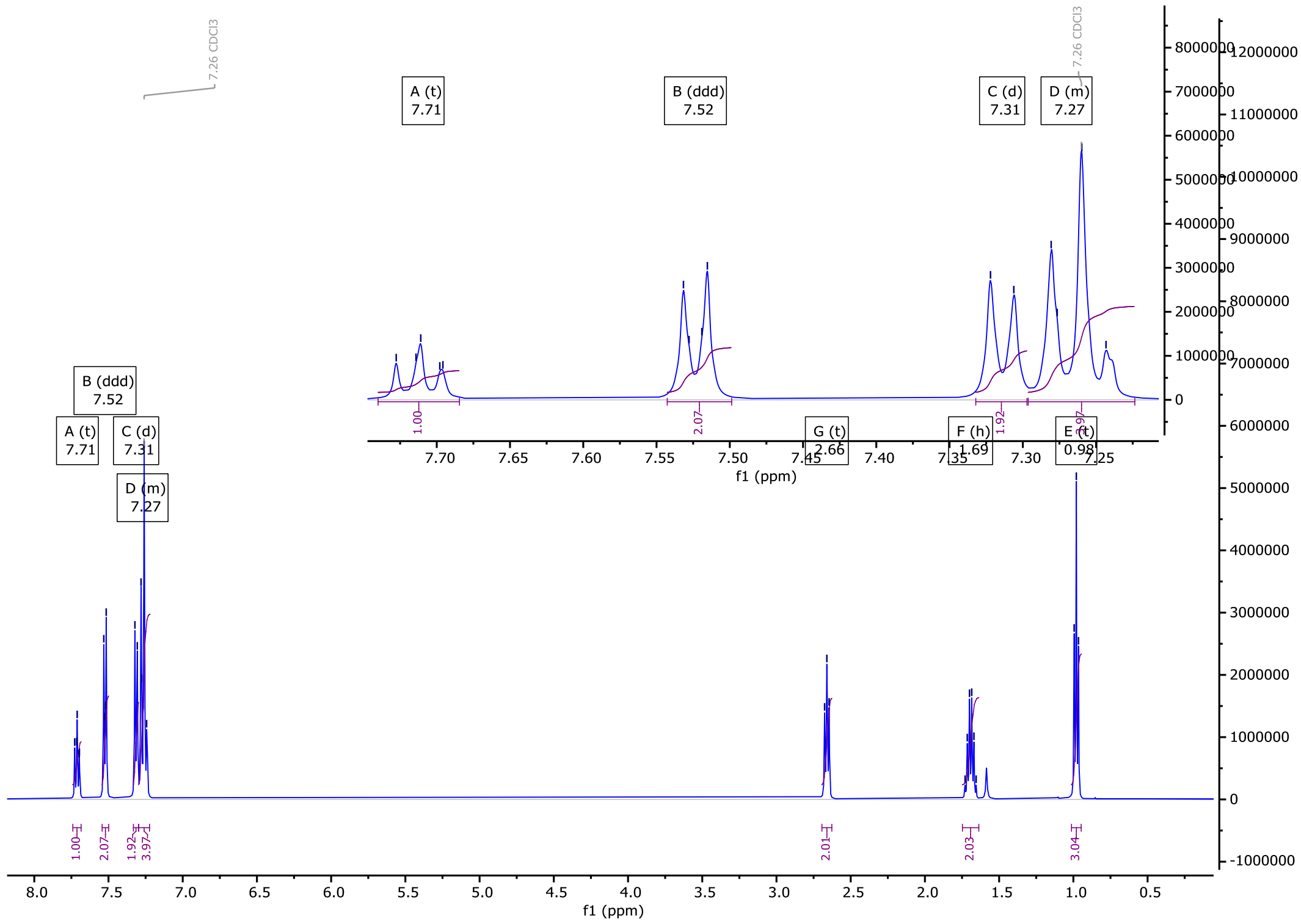

**Fig. S41.** $^{1}H$ NMR of **2b** in $CDCl_3$.

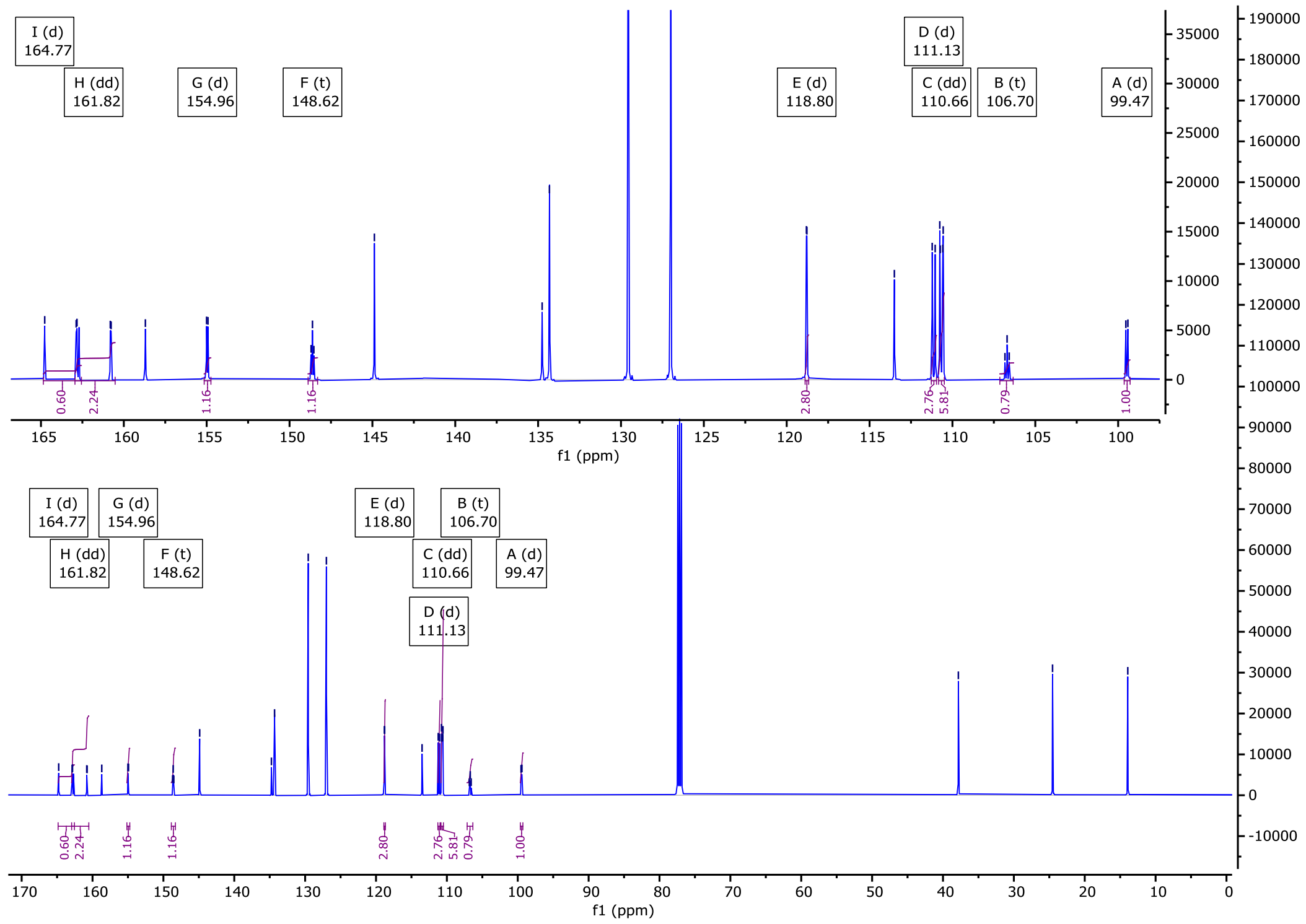

**Fig. S42.** $^{13}C\{^{1}H\}$ NMR of **2b** in $CDCl_3$.

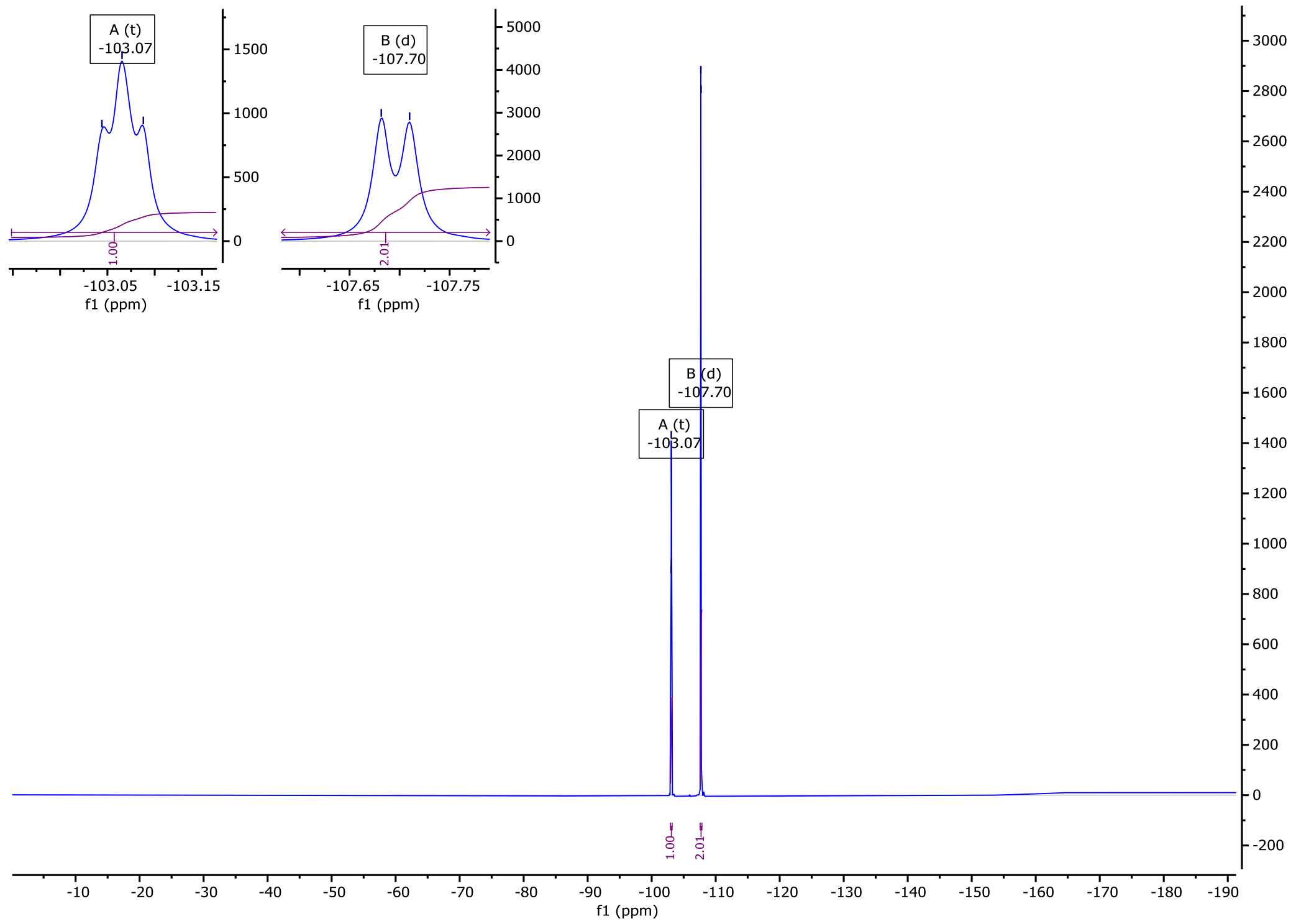

**Fig. S43.** $^{13}C\{^{1}H\}$ NMR of **2b** in $CDCl_3$.

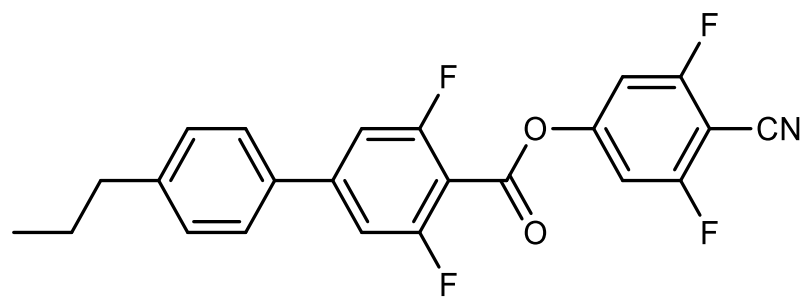

**2c**

*4-cyano-3,5-difluorophenyl 3,5-difluoro-4'-propyl-[1,1'-biphenyl]-4-carboxylate*

| | |
|---|---|
| Yield: | (white solid) 153 mg, 74 % |
| $^1$H NMR (501 MHz): | 7.52 (ddd, *J* = 8.5, 2.0, 1.7 Hz, 2H, Ar-**H**), 7.32 (ddd, *J* = 8.2, 1.8, 1.6 Hz, 2H, Ar-**H**), 7.29 – 7.25 (m, 2H, Ar-**H**) †, 7.11 (ddd, *J* = 7.8, 2.8 Hz, 2H, Ar-**H**), 2.67 (t, *J* = 7.5 Hz, 2H, Ar-C**H₂**-CH₂), 1.69 (h, *J* = 7.5 Hz, 2H, CH₂-C**H₂**-CH₃), 0.98 (t, *J* = 7.3 Hz, 3H, CH₂-C**H₃**).† Overlapping CDCl₃. |
| $^{13}$C{$^1$H} NMR (126 MHz): | 163.65 (dd, *J* = 261.8, 6.6 Hz), 161.89 (dd, *J* = 259.7, 5.9 Hz), 158.21, 155.35 (t, *J* = 13.5 Hz), 148.96 (t, *J* = 10.7 Hz), 145.00, 134.63, 129.58, 127.00, 110.70 (dd, *J* = 22.8, 3.3 Hz), 108.77, 107.02 (dd, *J* = 23.1, 4.0 Hz), 106.22 (t, *J* = 15.8 Hz), 90.56 (t, *J* = 19.4 Hz), 90.72, 90.56, 90.41, 37.83, 24.53, 13.91. |
| $^{19}$F NMR (376 MHz): | -101.70 (d, $J_{F-H}$ = 8.6 Hz, 2F, Ar-**F**), -107.34 (d, $J_{F-H}$ = 10.9 Hz, 2F, Ar-**F**). |

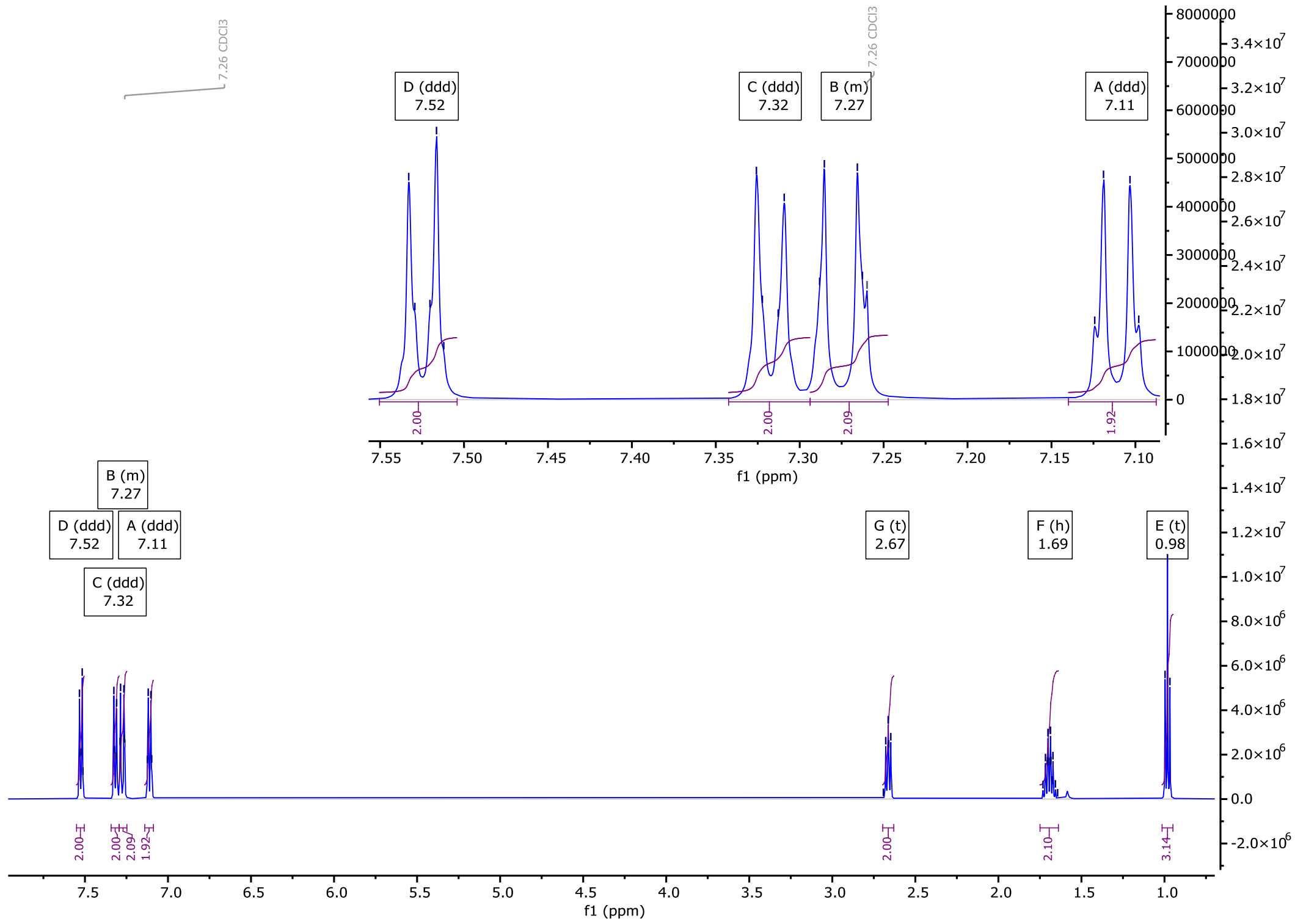

**Fig. S44.** $^{1}H$ NMR of **2c** in $CDCl_3$.

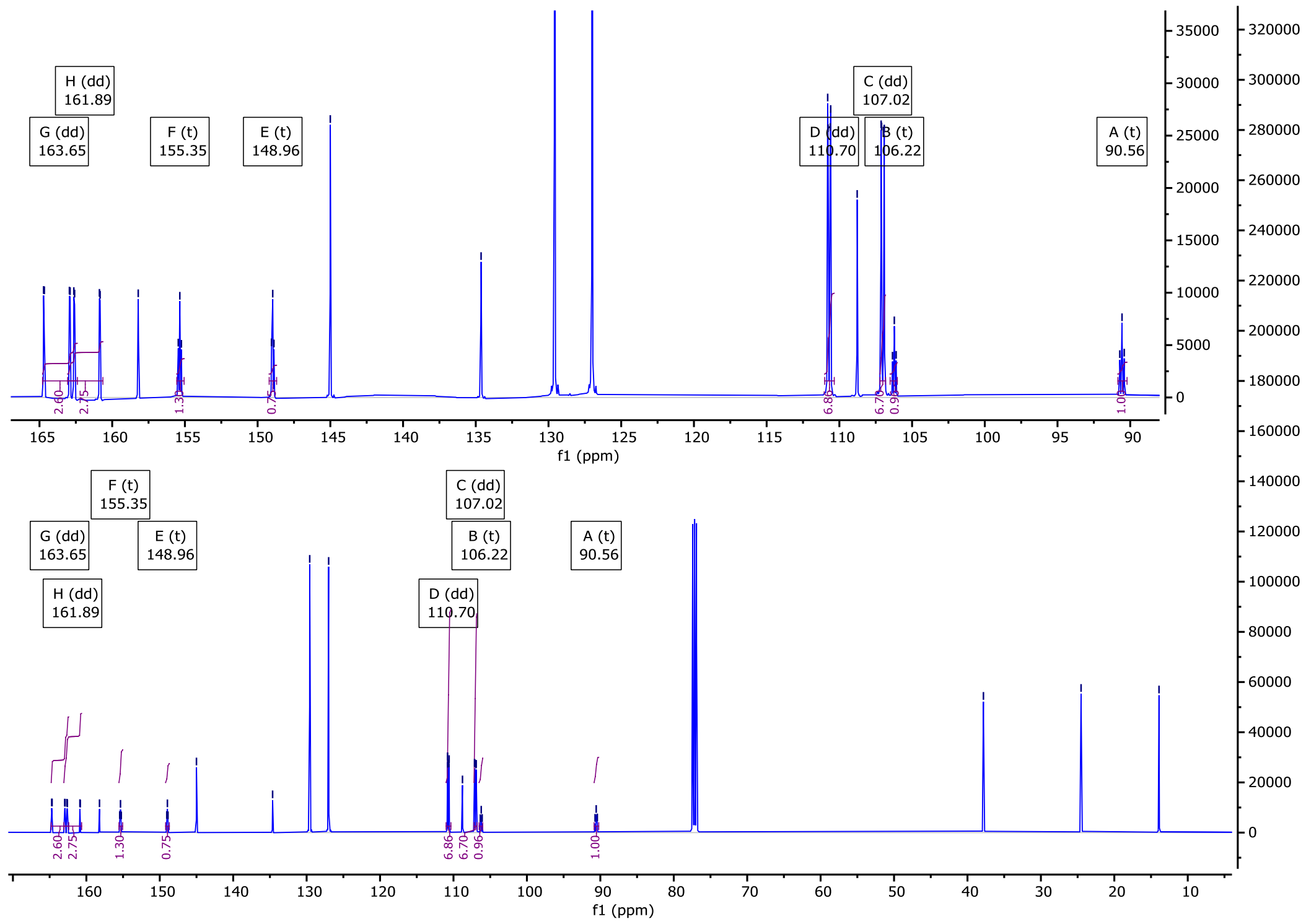

**Fig. S45.** $^{13}C\{^{1}H\}$ NMR of **2c** in $CDCl_3$.

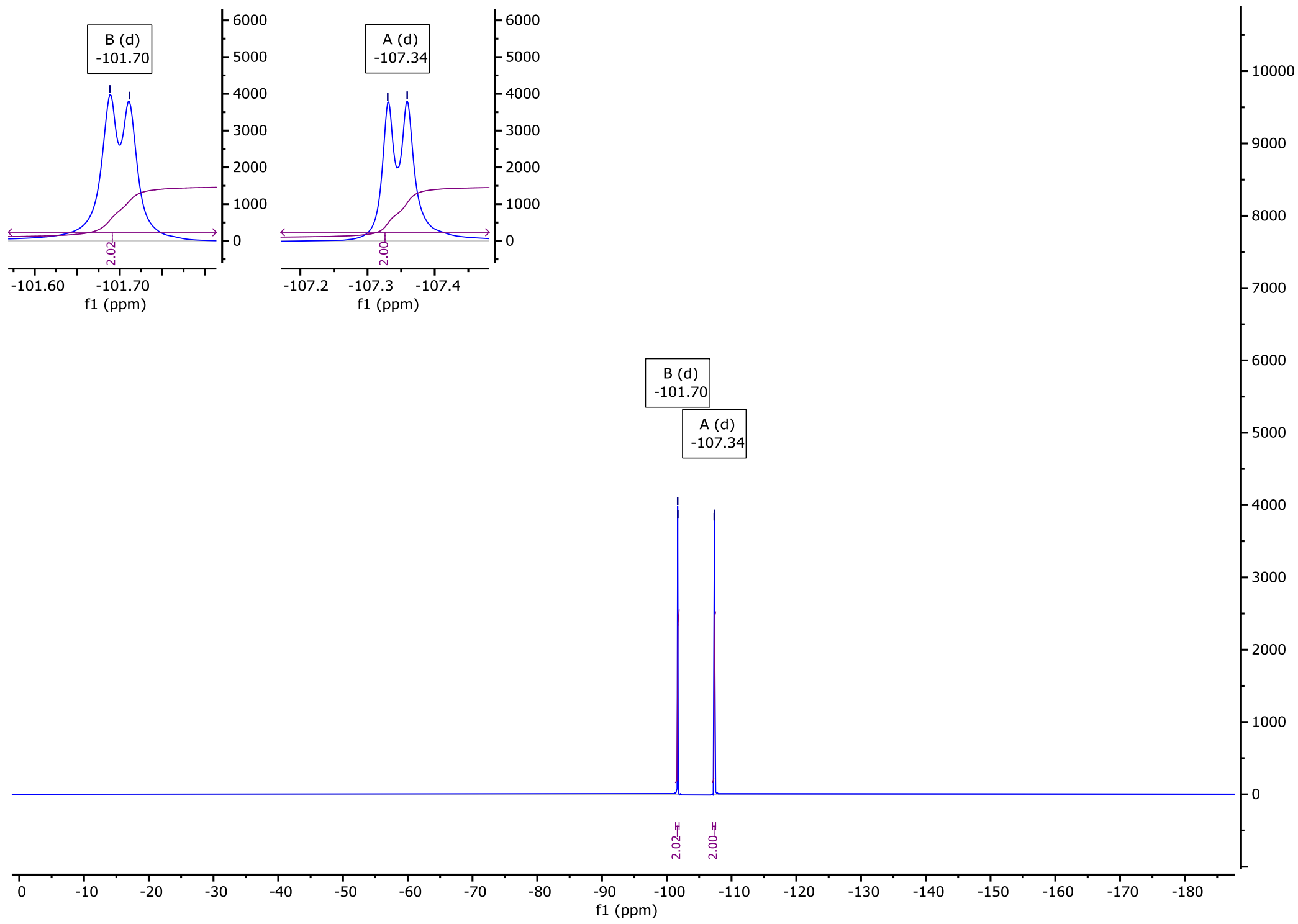

**Fig. S46.** $^{19}F$ NMR of **2c** in $CDCl_3$.

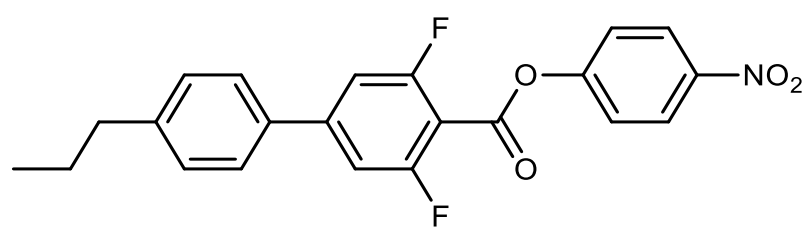

**2d**

*4-nitrophenyl 3,5-difluoro-4'-propyl-[1,1'-biphenyl]-4-carboxylate*

Yield: (white solid) 108 mg, 68 %

$^{1}$H NMR (400 MHz): 8.33 (ddd, *J* = 9.1, 3.1, 2.3 Hz, 2H, Ar-**H**), 7.53 (ddd, *J* = 8.2, 2.2, 1.9 Hz, 2H, Ar-**H**), 7.47 (ddd, *J* = 9.1, 3.1, 2.2 Hz, 2H, Ar-**H**), 7.31 (ddd, *J* = 8.3, 2.0, 1.7 Hz, 2H, Ar-**H**), 7.29 – 7.25 (m, 2H, Ar-**H**) †, 2.66 (t, *J* = 7.5 Hz, 2H, Ar-C**H₂**-CH₂), 1.69 (h, *J* = 7.5 Hz, 2H, CH₂-C**H₂**-CH₃), 0.98 (t, *J* = 7.3 Hz, 3H, CH₂-C**H₃**). † Overlapping $CDCl_3$.

$^{13}$C{$^{1}$H} NMR (101 MHz): 161.68 (dd, *J* = 258.7, 6.3 Hz), 158.98, 155.03, 148.27 (t, *J* = 10.3 Hz), 145.67, 144.69, 134.74, 129.43, 126.89, 125.34, 122.60, 110.52 (dd, *J* = 23.1, 3.0 Hz), 106.97 (t, *J* = 16.3 Hz), 37.72, 24.43, 13.81.

$^{19}$F NMR (376 MHz): -107.92 (d, $J_{F\text{-}H}$ = 10.8 Hz, 2F, Ar-**F**).

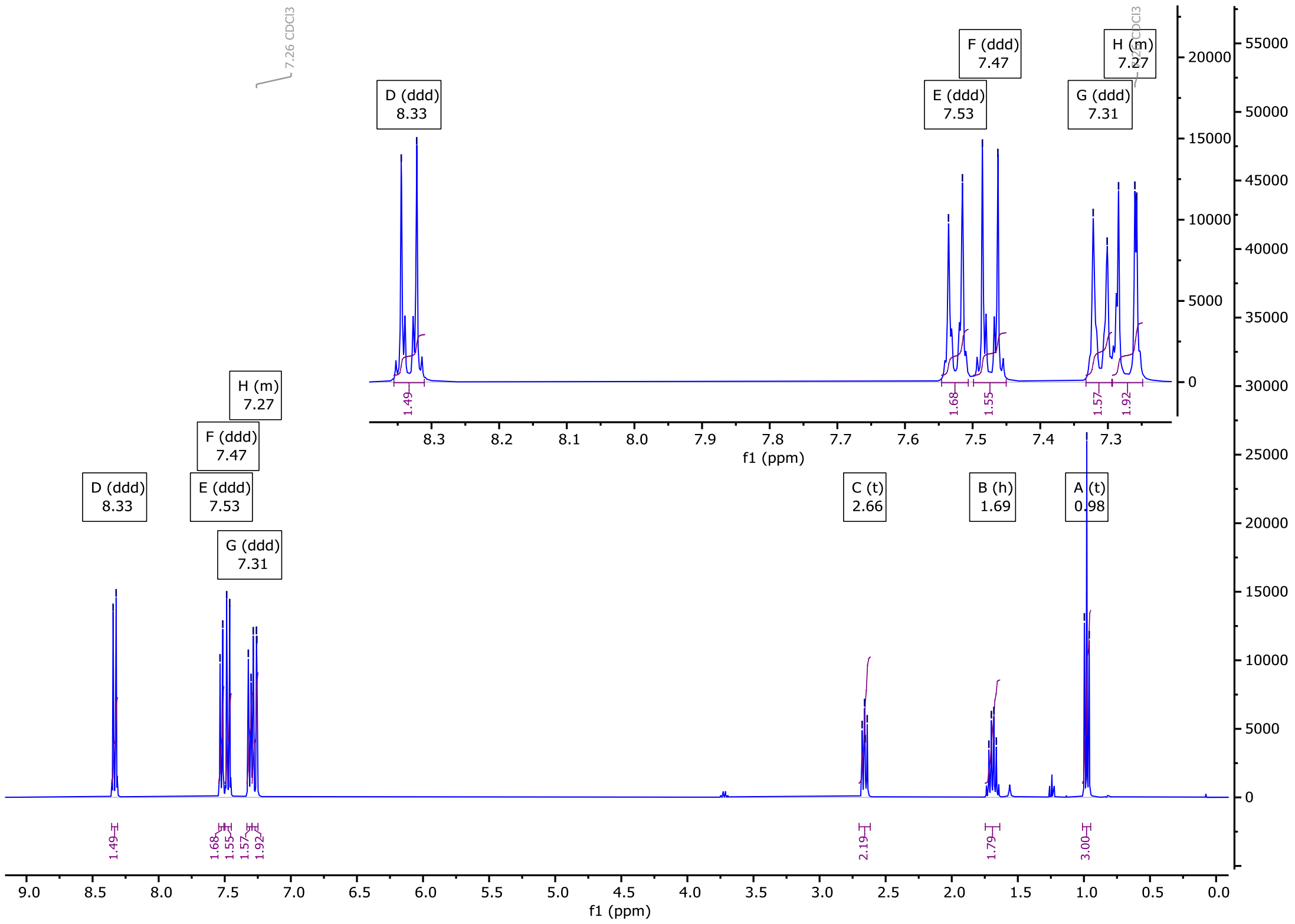

**Fig. S47.** $^{1}H$ NMR of **2d** in $CDCl_3$.

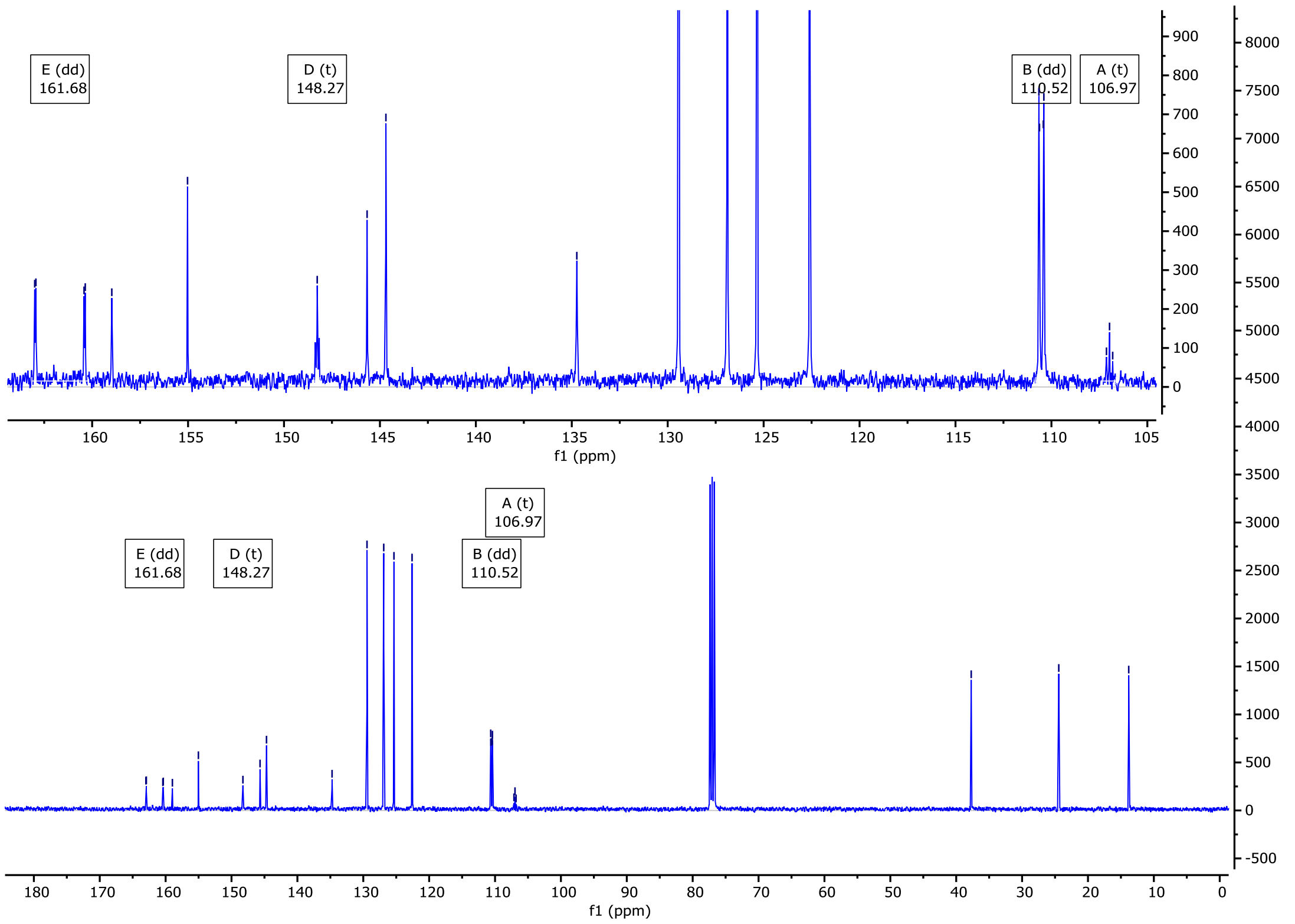

**Fig. S48.** $^{13}C\{^{1}H\}$ NMR of **2d** in $CDCl_3$.

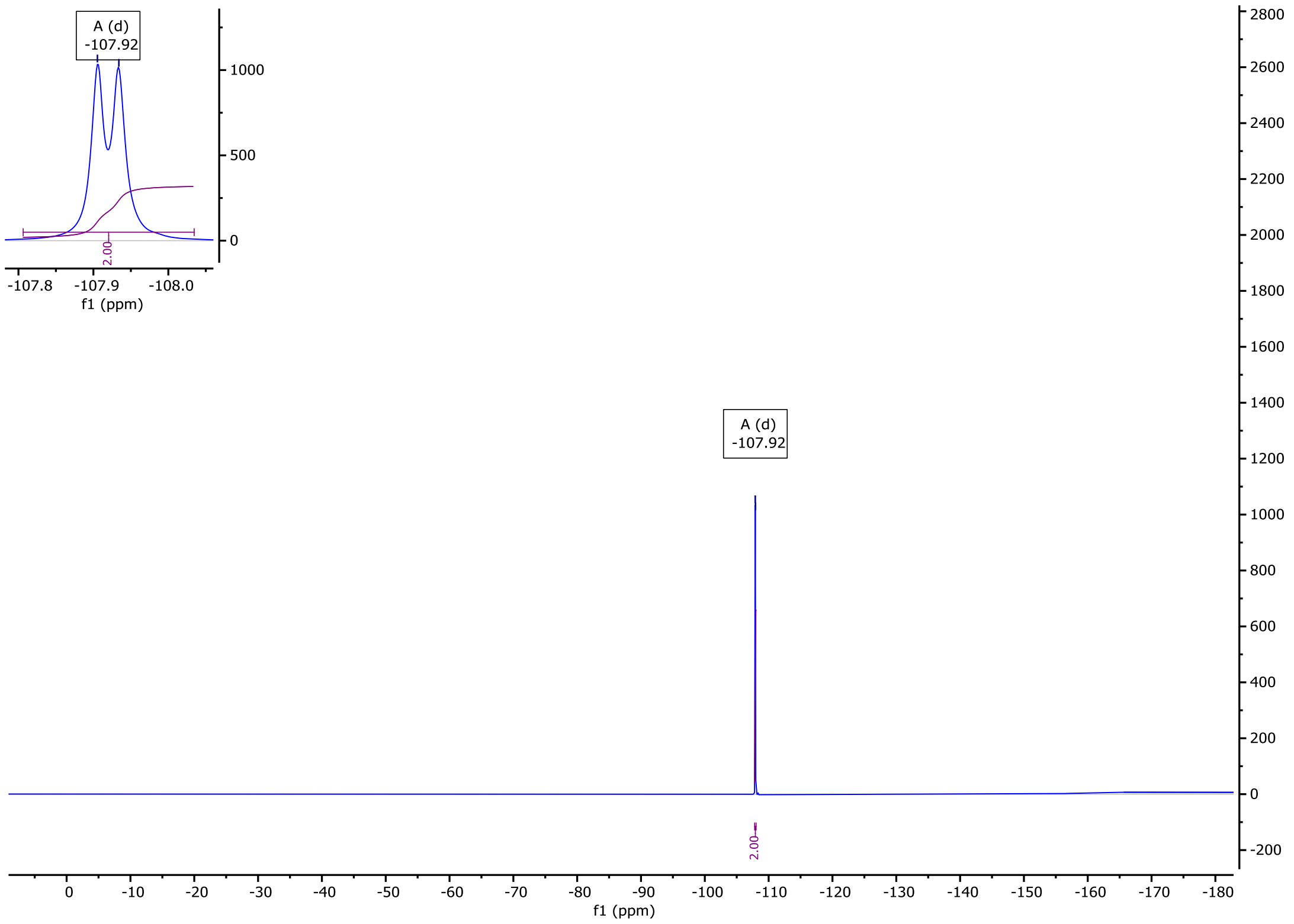

**Fig. S49.** $^{19}F$ NMR of **2d** in $CDCl_3$.

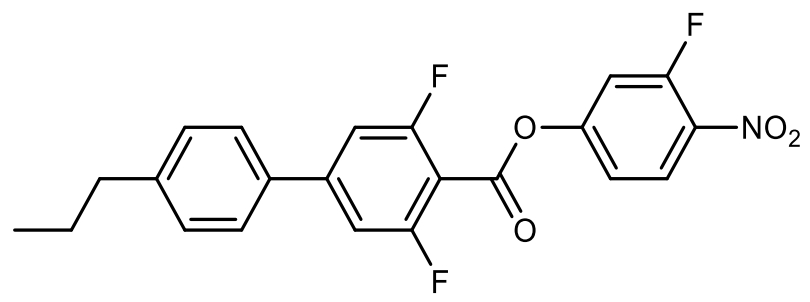

**2e**

*3-fluoro-4-nitrophenyl 3,5-difluoro-4'-propyl-[1,1'-biphenyl]-4-carboxylate*

| | |
|---|---|
| Yield: | (Colourless needles) 143 mg, 69 % |
| $^{1}$H NMR (400 MHz): | 8.20 (t, *J* = 8.8 Hz, 1H, Ar-**H**), 7.53 (ddd, *J* = 8.2, 2.0, 1.8 Hz, 2H, Ar-**H**), 7.37 – 7.24 (m, 6H, Ar-**H**)* †, 2.66 (t, *J* = 7.4 Hz, 2H, Ar-C**H₂**-CH₂), 1.69 (h, *J* = 7.3 Hz, 2H. CH2-C**H₂**-CH₃), 0.98 (t, *J* = 7.3 Hz, 3H, CH₂-C**H₃**). * Overlapping peaks, † Overlapping CDCl₃. |
| $^{13}$C{$^{1}$H} NMR (101 MHz): | 161.74 (dd, *J* = 259.1, 6.2 Hz), 158.48 (d, *J* = 3.0 Hz), 157.53, 154.97 (d, *J* = 19.6 Hz), 148.60 (t, *J* = 10.7 Hz), 144.79, 135.14 (d, *J* = 6.7 Hz), 129.45, 127.30 (d, *J* = 1.6 Hz), 126.90, 118.03 (d, *J* = 4.1 Hz), 112.37 (d, *J* = 24.2 Hz), 110.62 (dd, *J* = 22.9, 3.0 Hz), 37.73, 24.43, 13.81. |
| $^{19}$F NMR (376 MHz): | -107.59 (d, $J_{F-H}$ = 10.8 Hz, 2F, Ar-**F**), -112.88 (t, $J_{F-H}$ = 9.6 Hz, 1F, Ar-**F**). |

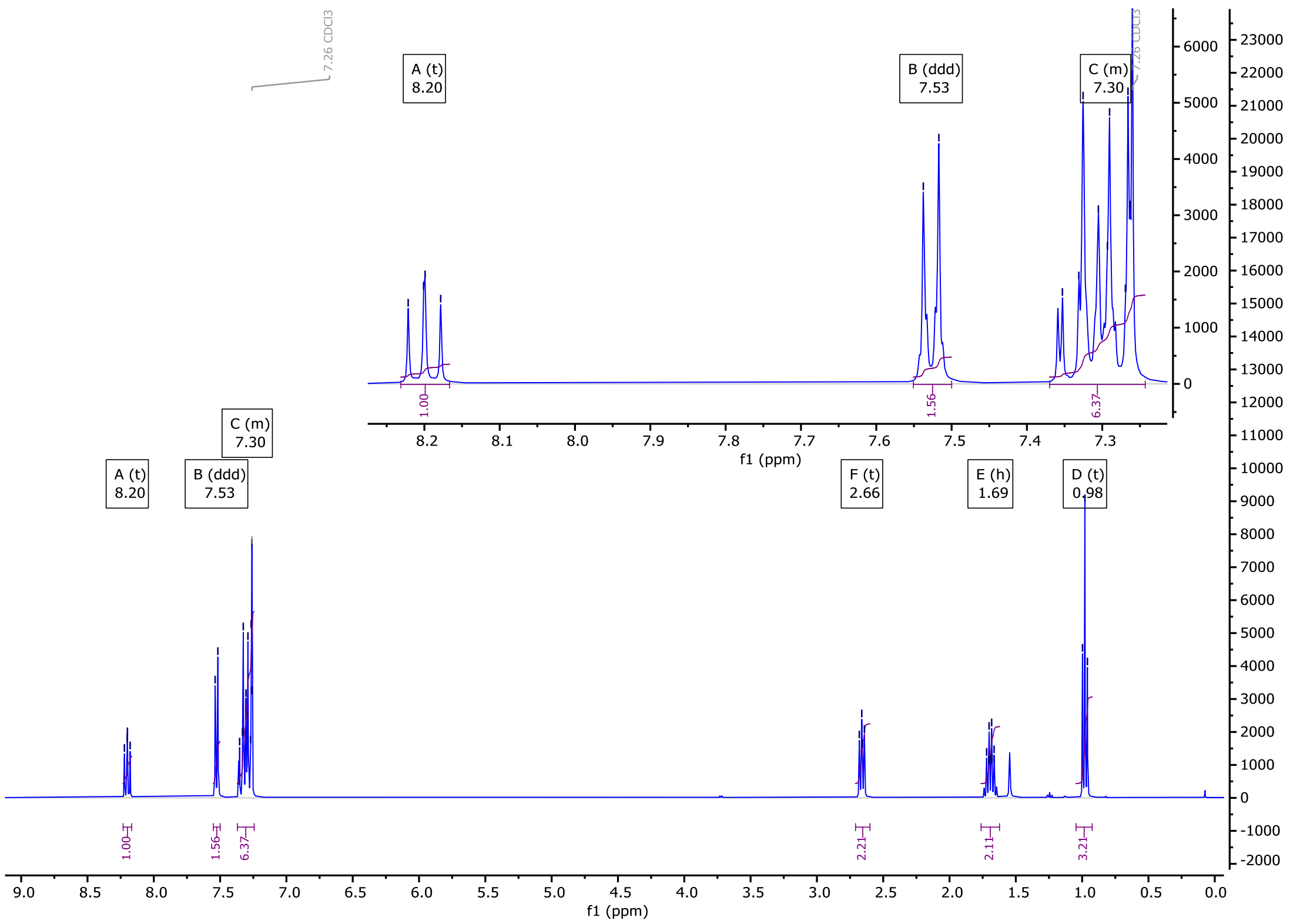

**Fig. S50.** $^{1}H$ NMR of **2e** in $CDCl_3$.

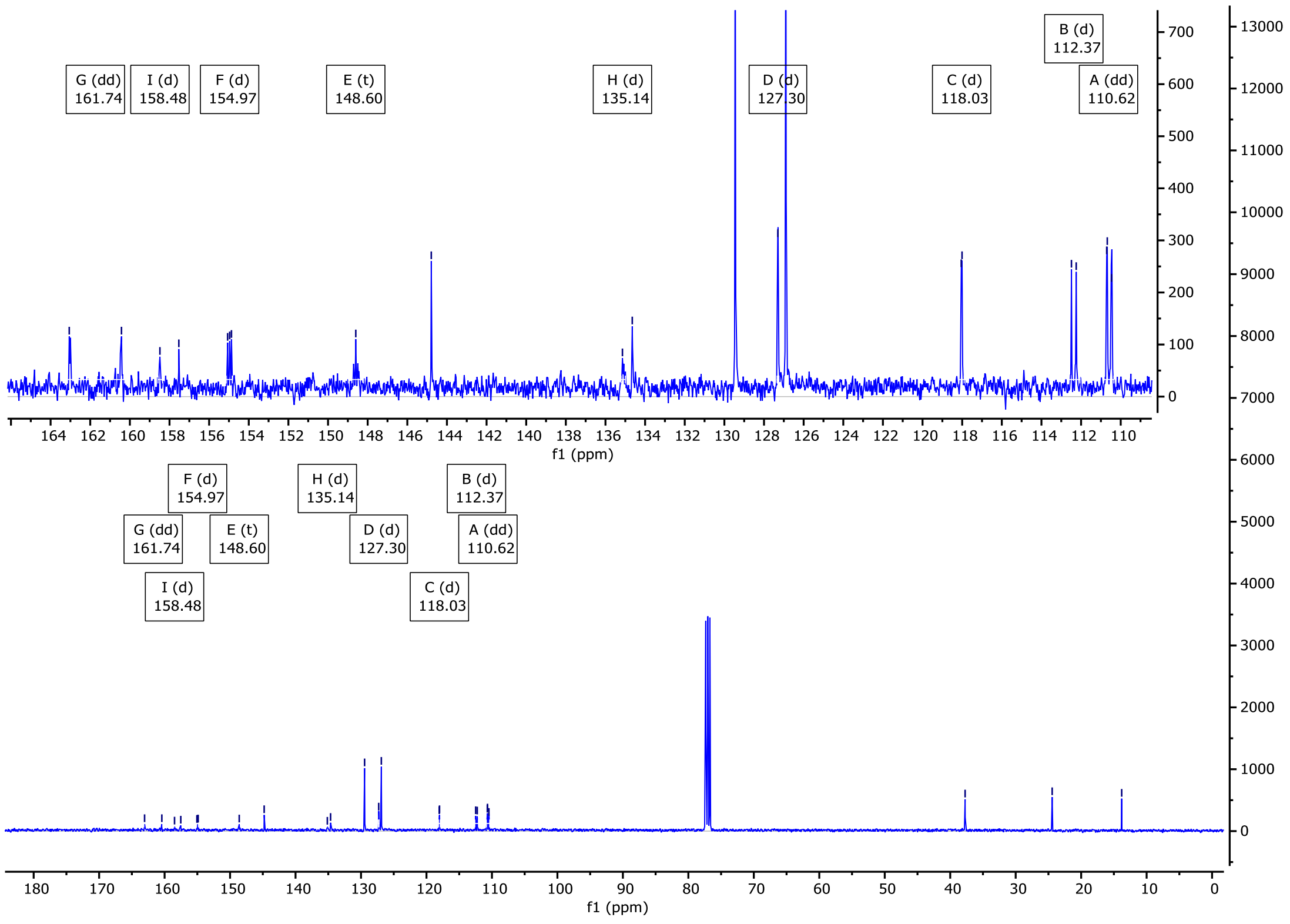

**Fig. S51.** $^{13}C\{^{1}H\}$ NMR of **2e** in $CDCl_3$.

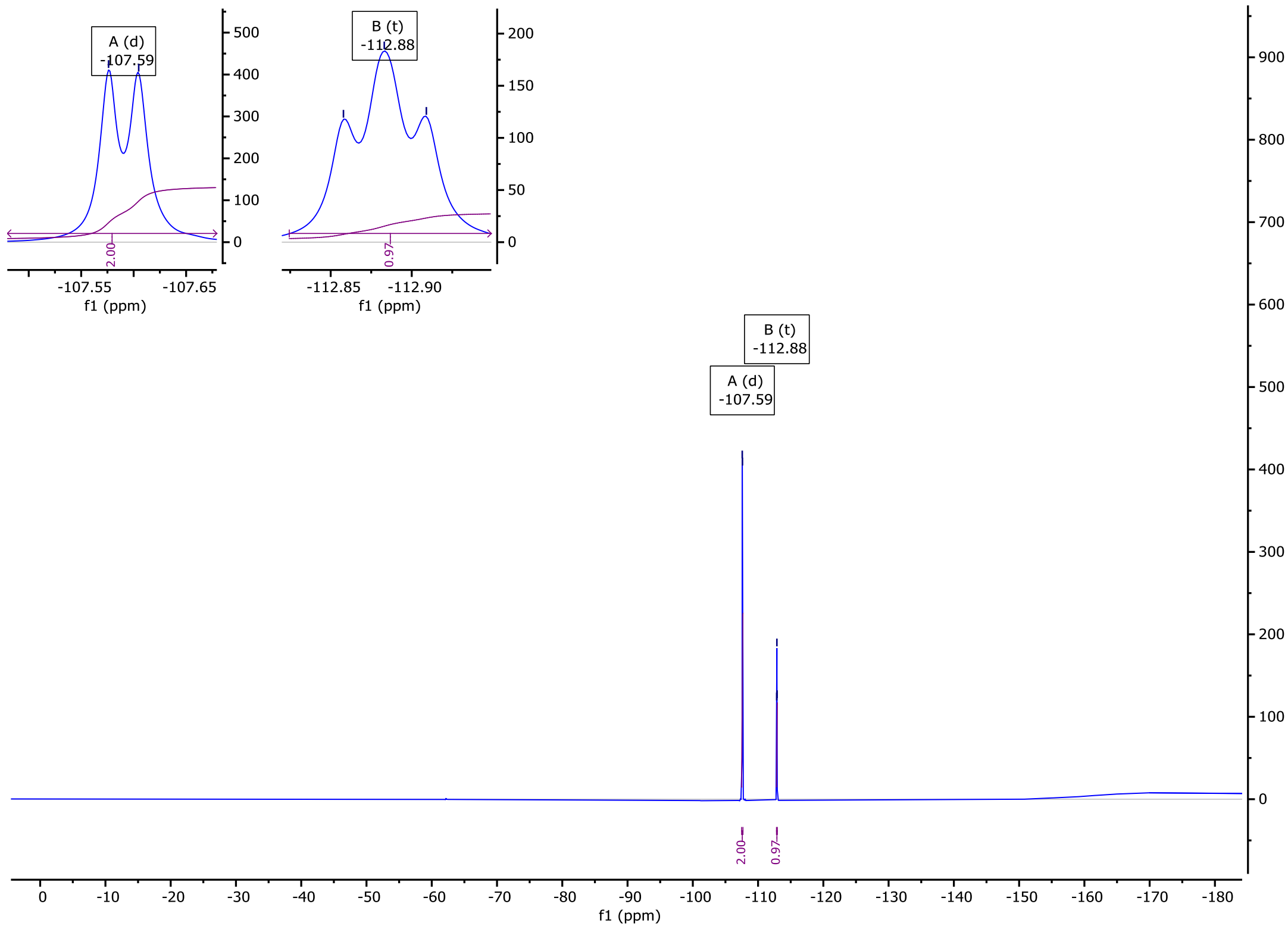

**Fig. S52.** $^{19}F$ NMR of **2e** in $CDCl_3$.

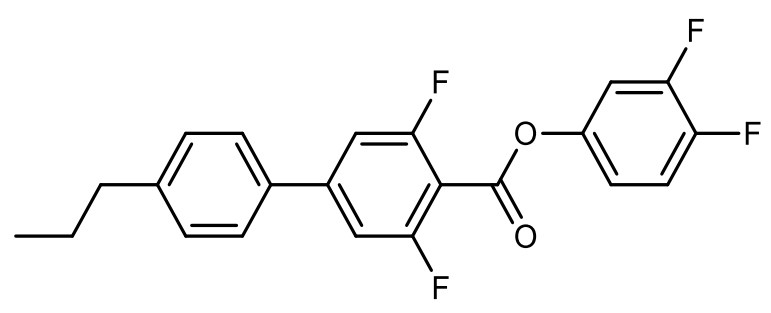

**2f**

*3,4-difluorophenyl 3,5-difluoro-4'-propyl-[1,1'-biphenyl]-4-carboxylate*

Yield: (White Solid) 150 mg, 77 %

$^{1}$H NMR (400 MHz): 7.54 (ddd, *J* = 8.2, 2.1, 1.8 Hz, 2H, Ar-**H**), 7.33 (d, *J* = 8.3 Hz, 2H, Ar-**H**), 7.30 – 7.16 (m, 4H) †, 7.10 – 7.02 (m, 1H, Ar-**H**), 2.68 (t, *J* = 7.4 Hz, 2H, Ar-C**H₂**-CH₂), 1.71 (h, *J* = 7.4 Hz, 2H, CH₂-C**H₂**-CH₃), 1.00 (t, *J* = 7.3 Hz, 3H, CH₂-C**H₃**). † Overlapping $CDCl_3$.

$^{13}$C{$^{1}$H} NMR (101 MHz): 161.54 (dd, *J* = 258.2, 6.6 Hz), 159.60, 150.18 (dd, *J* = 250.5, 13.9 Hz), 148.58 (dd, *J* = 247.0, 12.3 Hz), 147.92 (t, *J* = 10.6 Hz), 145.95 (dd, *J* = 8.6, 2.8 Hz), 144.54, 134.86, 129.40, 126.88, 117.72 (dd, *J* = 6.4, 3.9 Hz), 117.42 (d, *J* = 18.9 Hz), 111.80 (d, *J* = 20.2 Hz), 110.45 (dd, *J* = 23.3, 3.0 Hz), 107.39 (t, *J* = 16.7 Hz), 37.72, 24.44, 13.81.

$^{19}$F NMR (376 MHz): -108.38 (d, $J_{F-H}$ = 10.7 Hz, 2F, Ar-**F**), -134.32 (dt, $J_{F-F}$ = 20.1 Hz, $J_{F-H}$ = 9.6 Hz, 1F, Ar-**F**), -140.13 – -140.35 (m, 1F, Ar-**F**).

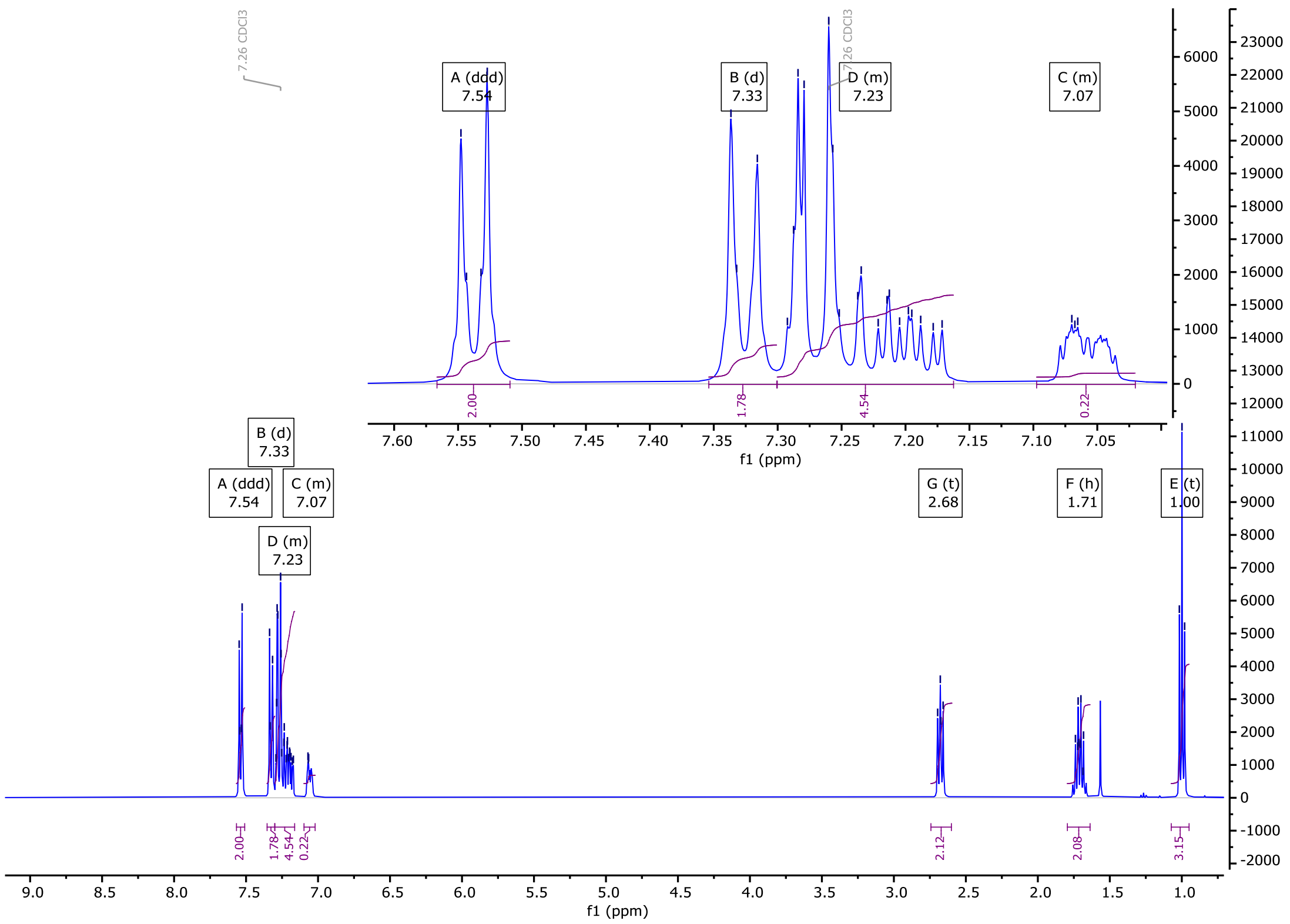

**Fig. S53.** $^{1}H$ NMR of **2f** in $CDCl_3$.

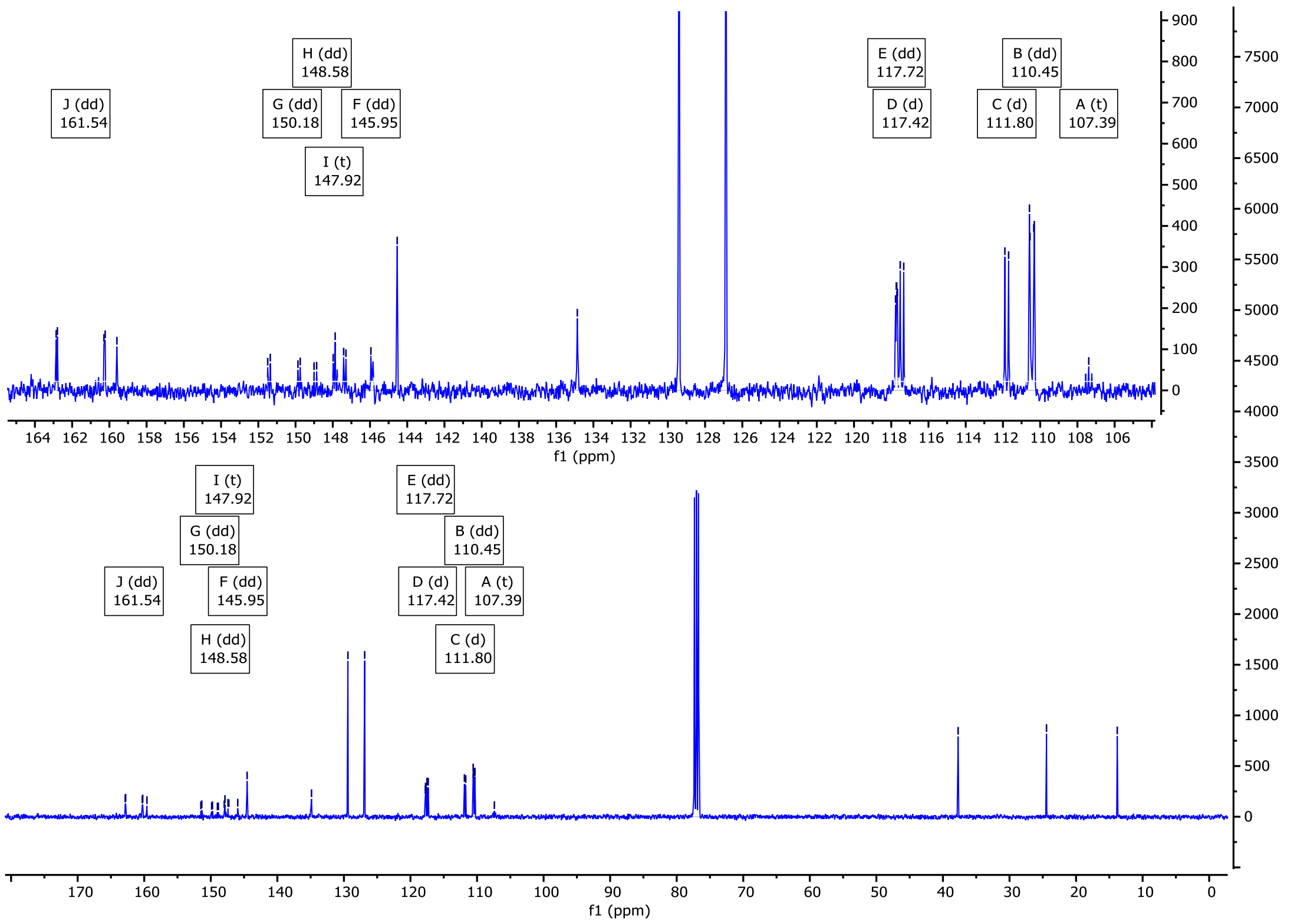

**Fig. S54.** $^{13}C\{^{1}H\}$ NMR of **2f** in $CDCl_3$.

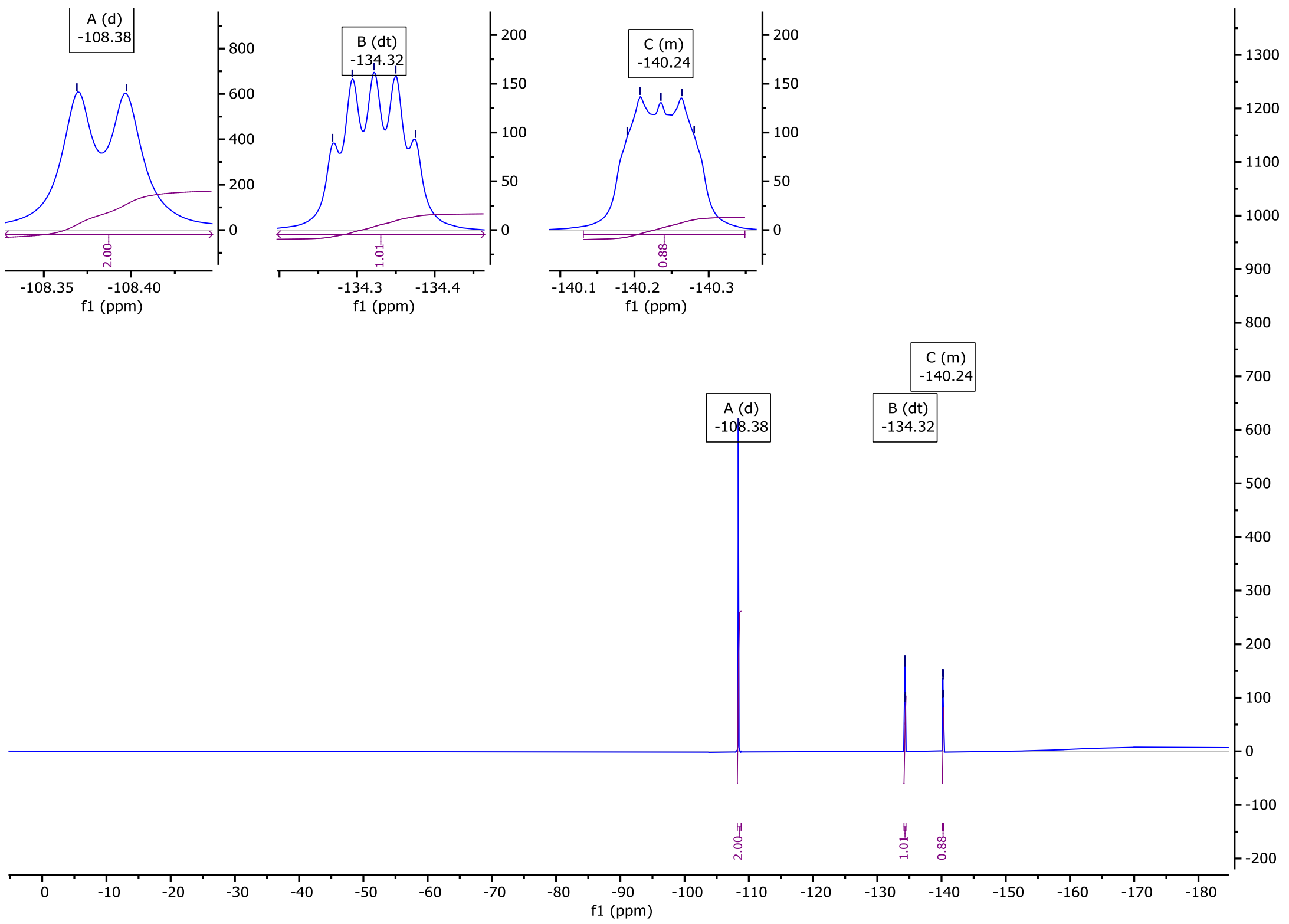

**Fig. S55.** $^{19}F$ NMR of **2f** in $CDCl_3$.

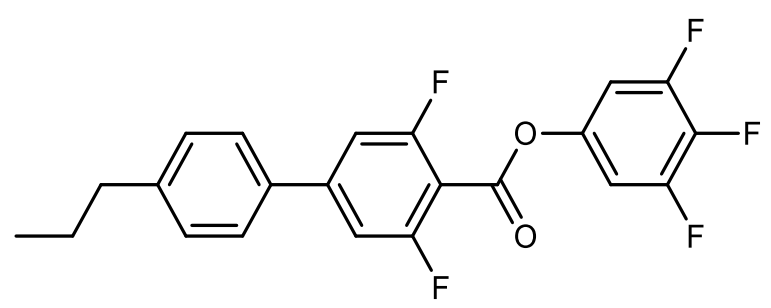

**2g**

*3,4,5-trifluorophenyl 3,5-difluoro-4'-propyl-[1,1'-biphenyl]-4-carboxylate*

| | |
|---|---|
| Yield: | (colourless needles) 150 mg, 74 % |
| $^{1}$H NMR (400 MHz): | 7.52 (ddd, *J* = 8.2, 2.2, 1.9 Hz, 2H, Ar-**H**), 7.31 (d, *J* = 8.1 Hz, 2H, Ar-**H**), 7.26 (d, *J* = 9.7 Hz, 2H, Ar-**H**) †, 7.05 – 6.96 (m, 2H, Ar-**H**), 2.66 (t, *J* = 7.4 Hz, 2H, Ar-C**H₂**-CH₂), 1.69 (h, *J* = 7.2 Hz, 2H, CH₂-C**H₂**-CH₃), 0.98 (t, *J* = 7.3 Hz, 3H, CH₂-C**H₃**). † Overlapping $CDCl_3$. |
| $^{13}$C{$^{1}$H} NMR (101 MHz): | 161.68 (dd, *J* = 258.6, 6.3 Hz), 151.13 (ddd, *J* = 251.0, 10.5, 5.1 Hz), 148.27 (t, *J* = 10.5 Hz), 144.98 (td, *J* = 11.3, 4.1 Hz), 144.73, 137.24 (dt, *J* = 250.9, 15.6 Hz), 134.84, 129.50, 126.96, 110.57 (dd, *J* = 23.2, 3.2 Hz), 107.48 – 106.96 (m), 37.80, 24.51, 13.88. |
| $^{19}$F NMR (376 MHz): | -108.00 (d, $J_{F-H}$ = 10.7 Hz, 2F, Ar-**F**), -132.40 (dd, $J_{F-F}$ = 20.7 Hz, $J_{F-H}$ = 8.5 Hz), -162.76 (tt, $J_{F-F}$ = 20.6 Hz, $J_{F-H}$ = 5.8 Hz). |

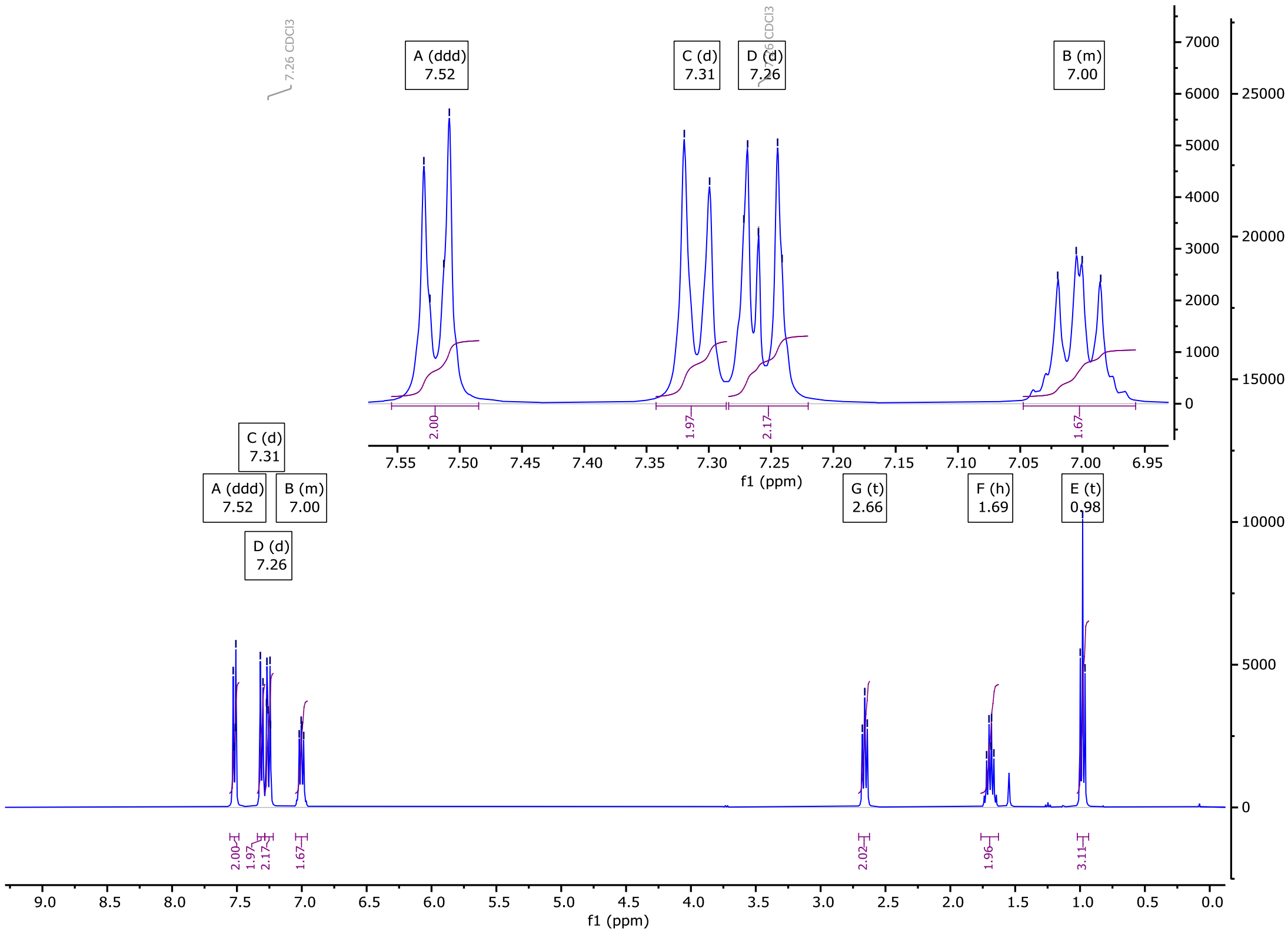

**Fig. S56.** $^{1}H$ NMR of **2g** in $CDCl_3$.

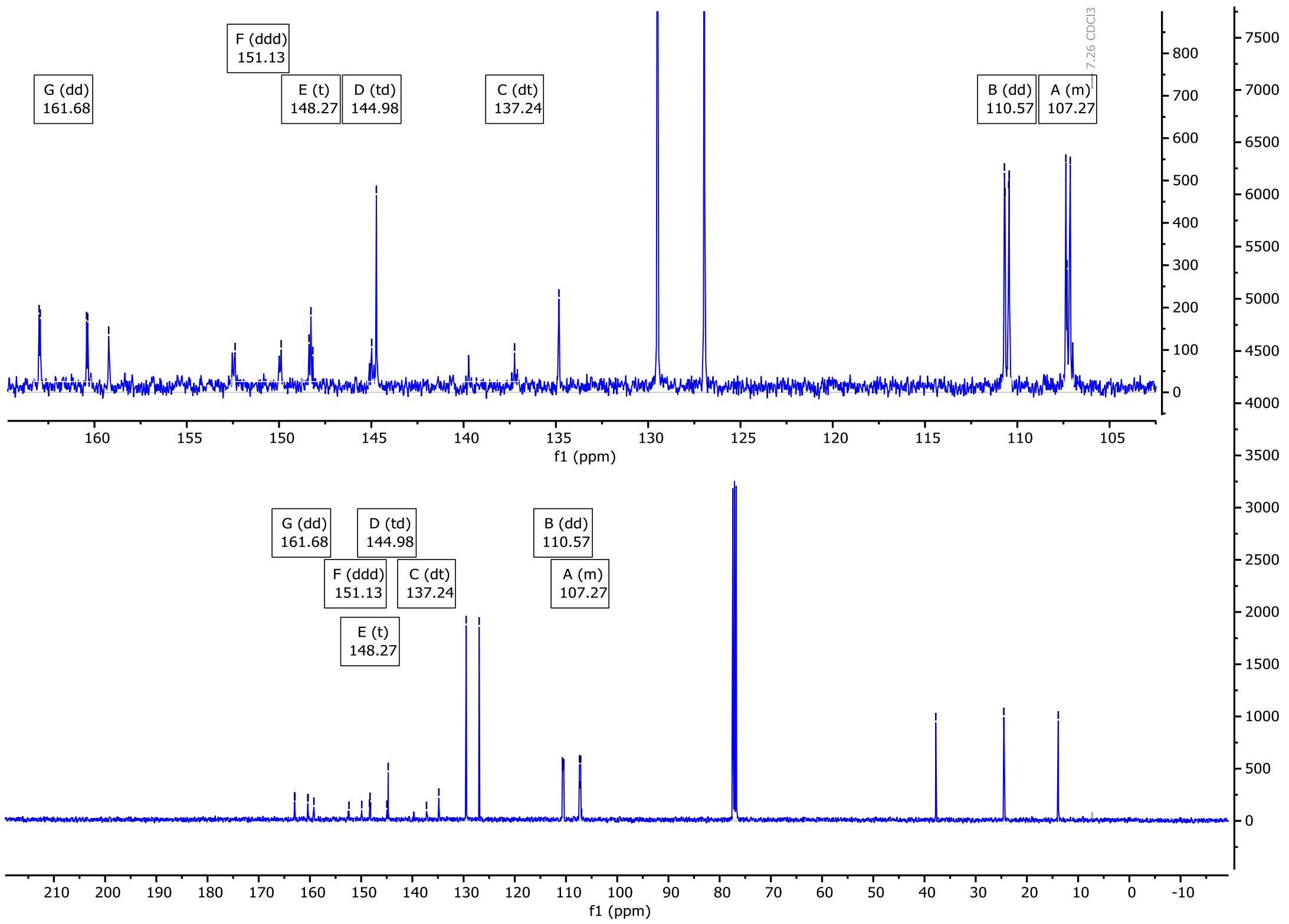

**Fig. S57.** $^{13}C\{^{1}H\}$ NMR of **2g** in $CDCl_3$.

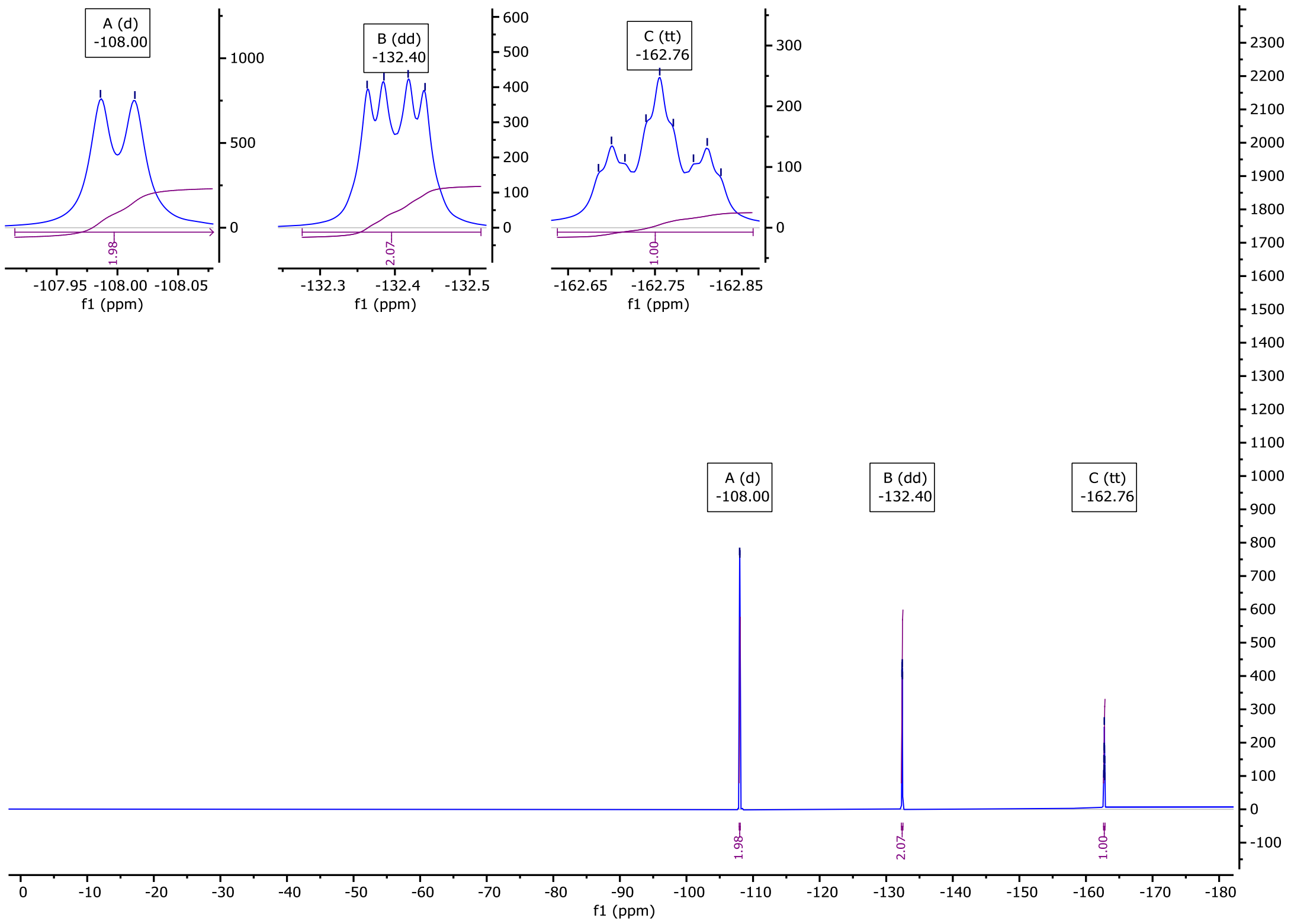

**Fig. S58.** $^{19}F$ NMR of **2g** in $CDCl_3$.

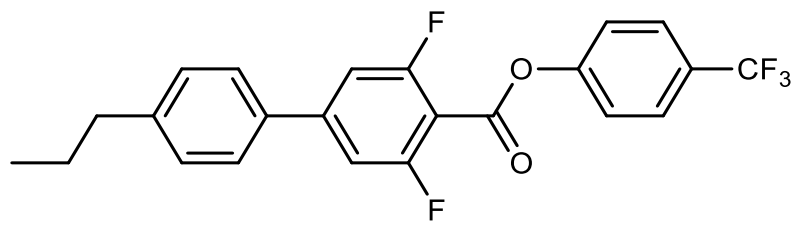

**2h**

*4-(trifluoromethyl)phenyl 3,5-difluoro-4'-propyl-[1,1'-biphenyl]-4-carboxylate*

Yield: (White Solid) 139 mg, 66 %

$^1$H NMR (501 MHz): 7.72 (ddd, *J* = 8.4, 2.5, 1.8 Hz, 2H, Ar-**H**), 7.52 (ddd, *J* = 8.2, 1.9, 1.9 Hz, 2H, Ar-**H**), 7.41 (ddd, *J* = 8.4, 2.4, 2.0 Hz, 2H, Ar-**H**), 7.31 (ddd, *J* = 8.2, 1.8, 1.8 Hz, 2H, Ar-**H**), 7.26 (ddd, *J* = 9.6, 1.8, 1.2 Hz, 2H, Ar-**H**) †, 2.66 (t, *J* = 7.5 Hz, 2H, Ar-C**H₂**-CH₂), 1.69 (h, *J* = 7.5 Hz, 2H, CH₂-C**H₂**-CH₃), 0.98 (t, *J* = 7.3 Hz, 3H, CH₂-C**H₃**). † Overlapping $CDCl_3$.

$^{13}$C{$^1$H} NMR (126 MHz): 161.71 (dd, *J* = 258.2, 6.3 Hz), 159.58, 148.02 (t, *J* = 10.4 Hz), 147.94, 144.68, 135.00, 128.70 (q, *J* = 32.9 Hz), 127.15 – 127.05 (m), 125.06, 122.89, 122.33, 110.58 (dd, *J* = 22.4, 3.6 Hz), 110.49, 107.59 (t, *J* = 17.0 Hz), 37.85, 24.58, 13.95.

$^{19}$F NMR (376 MHz): -62.25 (s, 3F, Ar-C**F₃**), -108.33 (d, $J_{F-H}$ = 10.4 Hz, 2F, Ar-**F**).

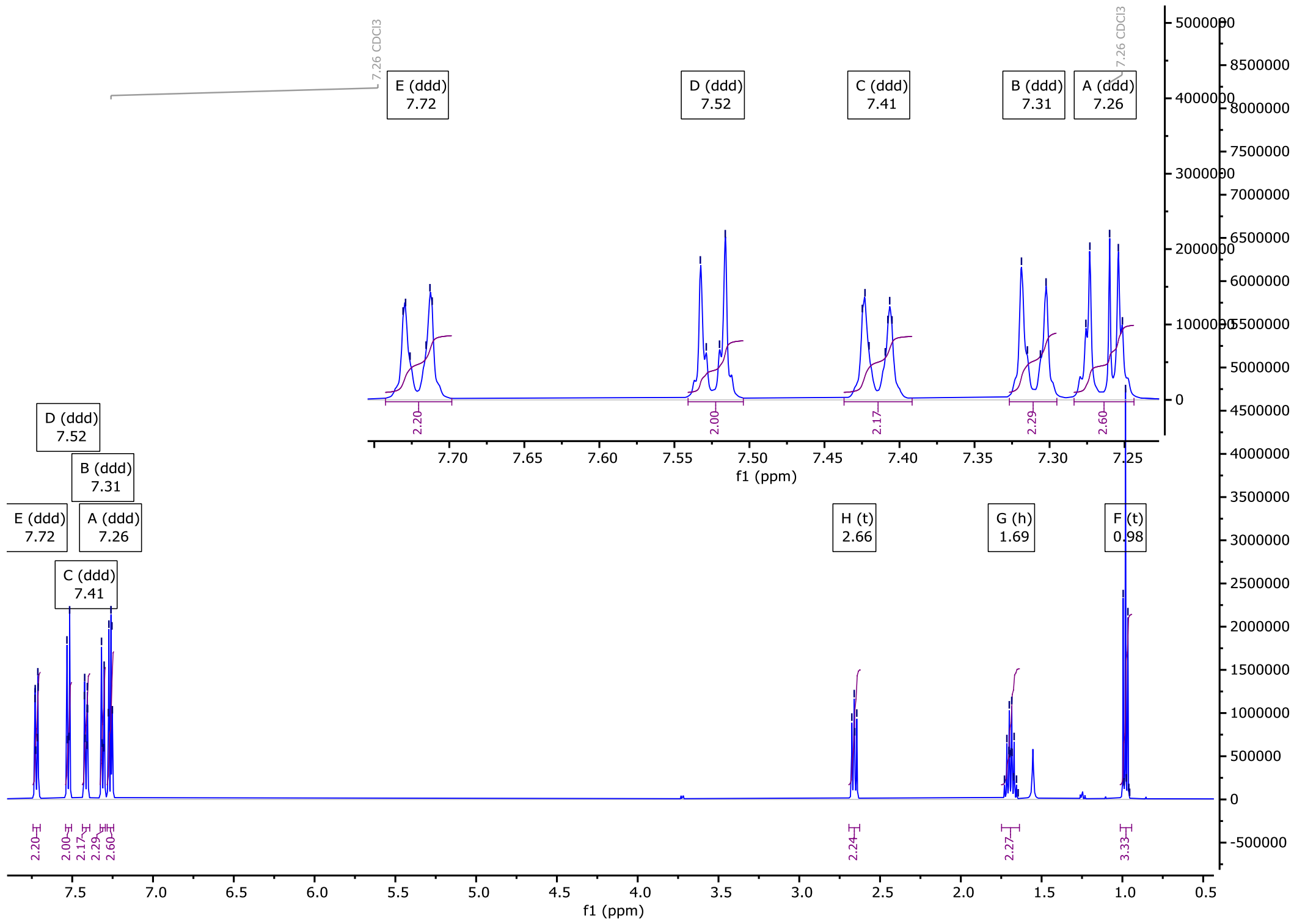

**Fig. S59.** $^1$H NMR of **2h** in $CDCl_3$.

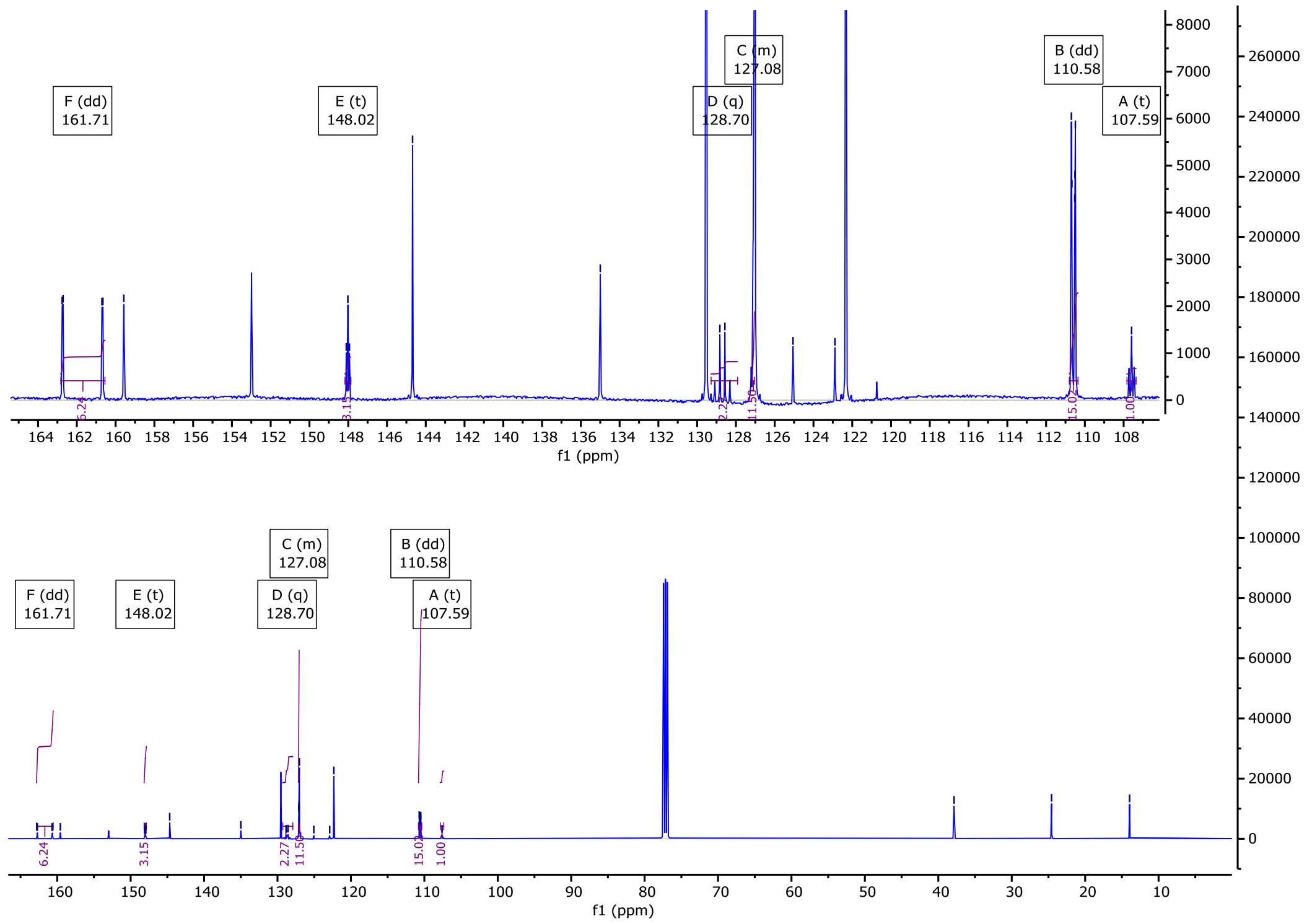

**Fig. S60.** $^{13}C\{^1H\}$ NMR of **2h** in $CDCl_3$.

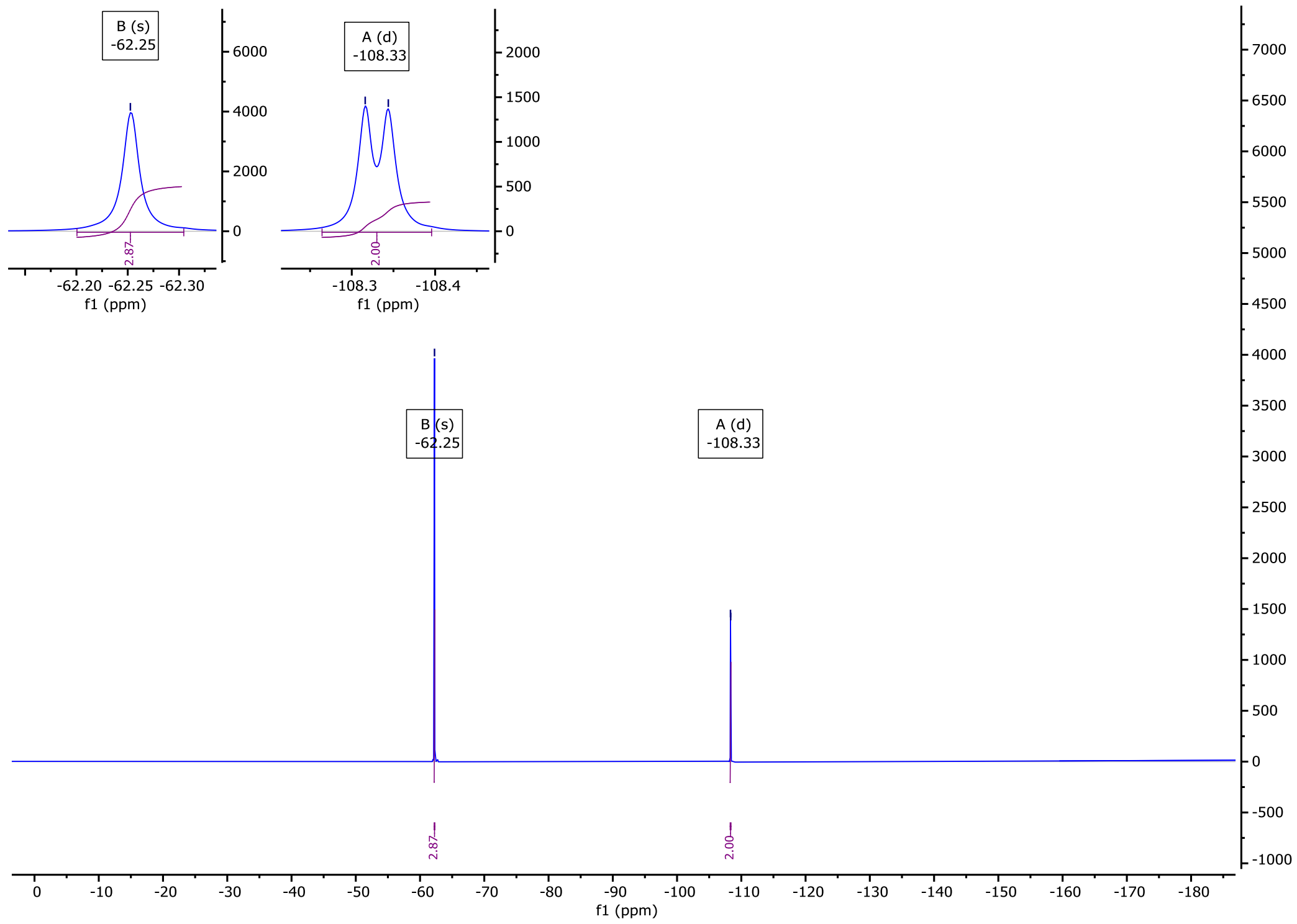

**Fig. S61.** $^{19}F$ NMR of **2h** in $CDCl_3$.

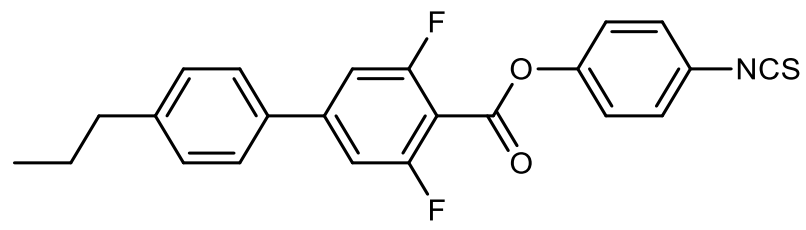

**2i**

*4-isothiocyanatophenyl 3,5-difluoro-4'-propyl-[1,1'-biphenyl]-4-carboxylate*

Yield: (White Solid) 117 mg, 57 %

$^{1}$H NMR (400 MHz): 7.49 (ddd, *J* = 8.2, 1.9, 1.8 Hz, 2H, Ar-**H**), 7.30 – 7.20 (m, 6H, Ar-**H**) †*, 2.63 (t, *J* = 7.4 Hz, 2H, Ar-C**H₂**-CH₂), 1.66 (h, *J* = 7.4 Hz, 2H, CH₂-C**H₂**-CH₃), 0.95 (t, *J* = 7.3 Hz, 3H, CH₂-C**H₃**). * Overlapping peaks, † Overlapping $CDCl_3$.

$^{13}$C{$^{1}$H} NMR (101 MHz): 161.54 (dd, *J* = 258.0, 6.4 Hz), 159.60, 148.97, 147.80 (t, *J* = 10.3 Hz), 136.24, 134.88, 129.39, 129.30, 122.99, 110.44 (dd, *J* = 23.3, 2.8 Hz), 107.54 (t, *J* = 17.0 Hz), 37.72, 24.44, 13.82.

$^{19}$F NMR (376 MHz): -108.39 (d, $J_{F-H}$ = 10.5 Hz, 2F, Ar-**F**).

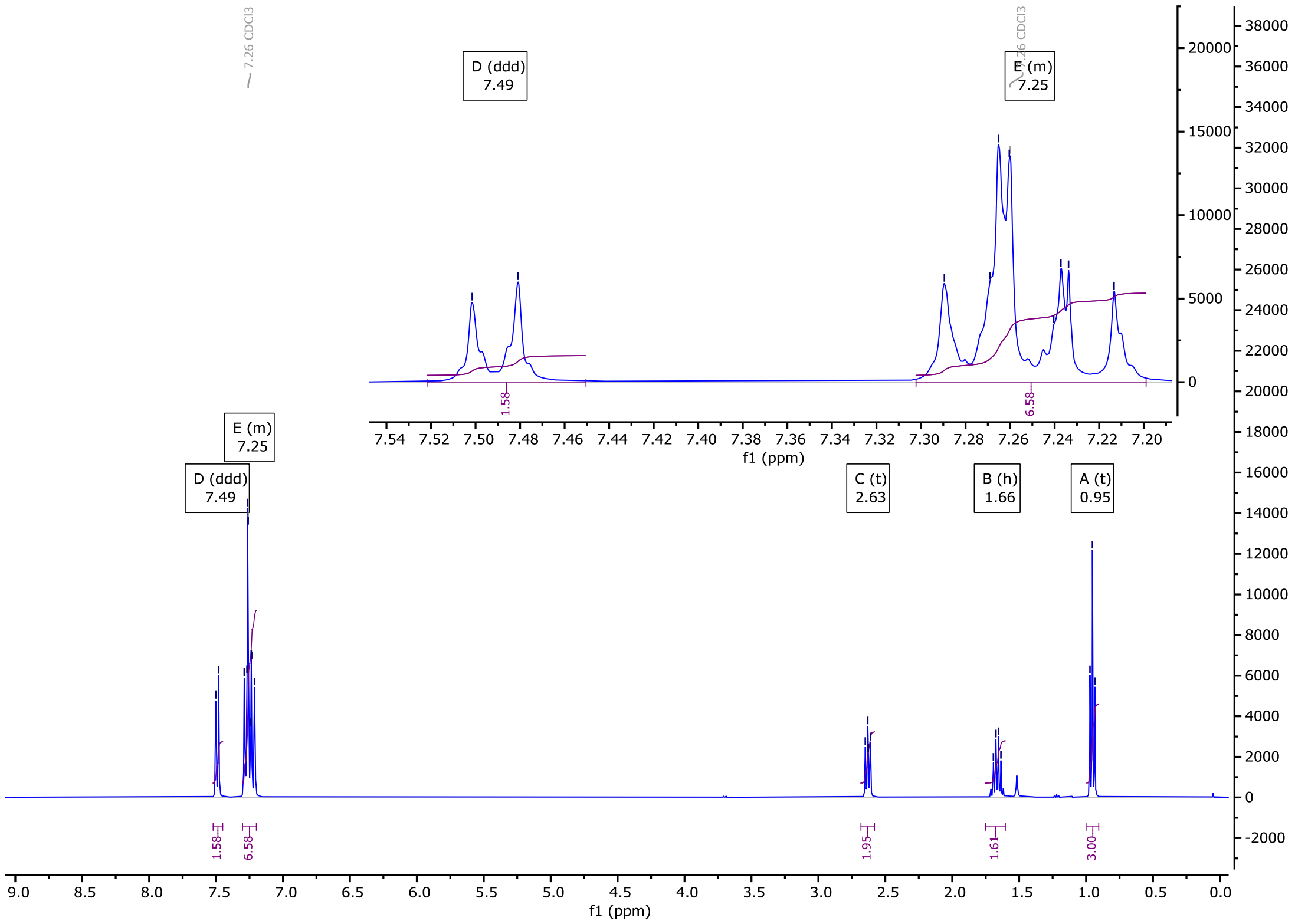

**Fig. S62.** $^{1}H$ NMR of **2i** in $CDCl_3$.

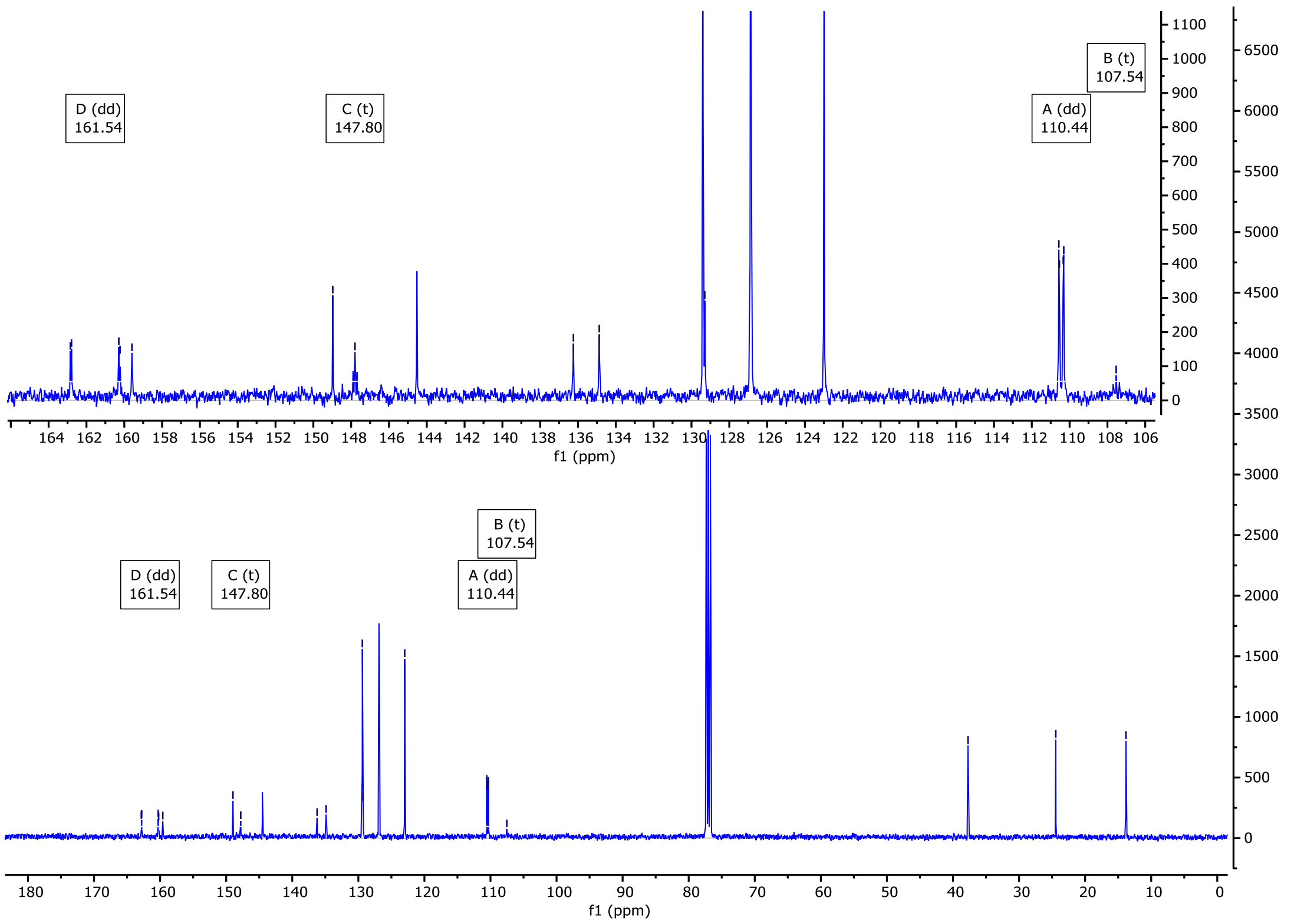

**Fig. S63.** $^{13}C\{^{1}H\}$ NMR of **2i** in $CDCl_3$.

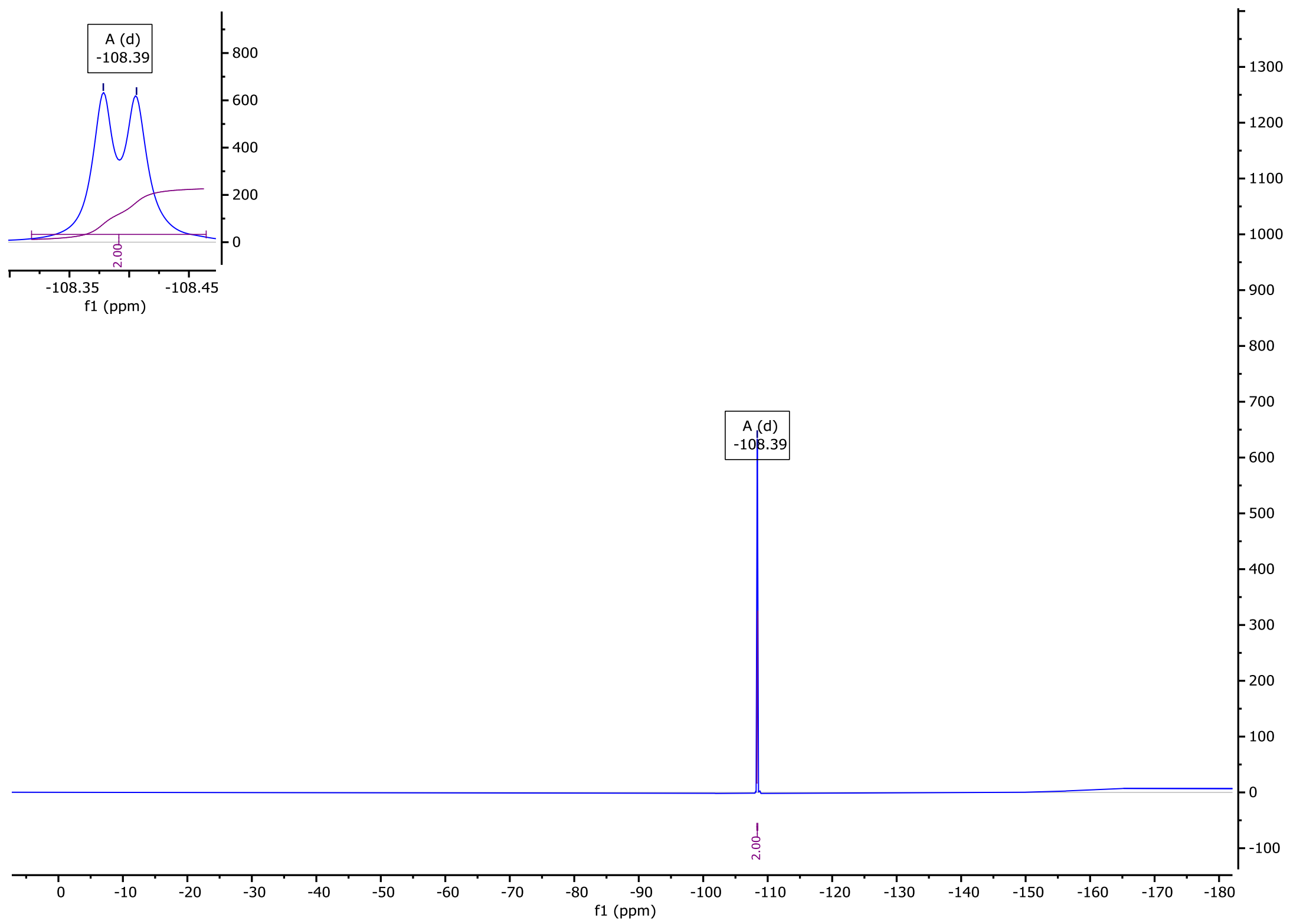

**Fig. S64.** $^{19}$F NMR of **2i** in $CDCl_3$.

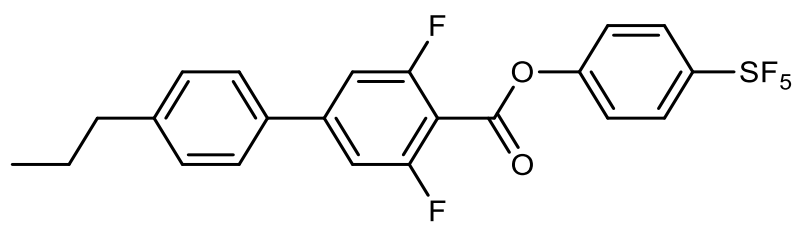

**2j**

*4-(pentafluoro-l6-sulfaneyl)phenyl 3,5-difluoro-4'-propyl-[1,1'-biphenyl]-4-carboxylate*

Yield: (White Solid) 143 mg, 60 %

$^{1}$H NMR (400 MHz): 7.85 (ddd, *J* = 9.1, 3.2, 2.1 Hz, 2H, Ar-**H**), 7.53 (ddd, *J* = 8.3, 2.2, 1.8 Hz, 2H, Ar-**H**), 7.40 (d, *J* = 8.9 Hz, 2H, Ar-**H**), 7.31 (ddd, *J* = 8.2, 1.8, 1.7 Hz, 2H, Ar-**H**), 7.29 – 7.24 (m, 2H) †, 2.66 (t, *J* = 7.4 Hz, 2H, Ar-C$\mathbf{H_2}$-$CH_2$), 1.70 (h, *J* = 7.4 Hz, 2H, $CH_2$-C$\mathbf{H_2}$-$CH_3$), 0.98 (t, *J* = 7.3 Hz, 3H, $CH_2$-C$\mathbf{H_3}$). † Overlapping $CDCl_3$.

$^{13}$C{$^{1}$H} NMR (101 MHz): 161.61 (dd, *J* = 258.2, 6.2 Hz), 159.24, 152.19, 151.31 (t, *J* = 18.0 Hz), 148.05 (t, *J* = 10.5 Hz), 148.16, 148.05, 147.95, 144.61, 134.82, 129.42, 127.69 (p, *J* = 4.6 Hz), 126.89, 121.98, 110.48 (dd, *J* = 23.3, 3.0 Hz), 107.24 (t, *J* = 16.4 Hz), 37.72, 24.44, 13.81.

$^{19}$F NMR (376 MHz): 84.01 (p, *J* = 149.8 Hz, 1F, S-$\mathbf{F_{\Lambda}}$), 63.48 (d, *J* = 150.0 Hz, 4F, S-$\mathbf{F_{4\Delta}}$), -108.15 (d, *J* = 10.6 Hz, 2F, Ar-**F**).

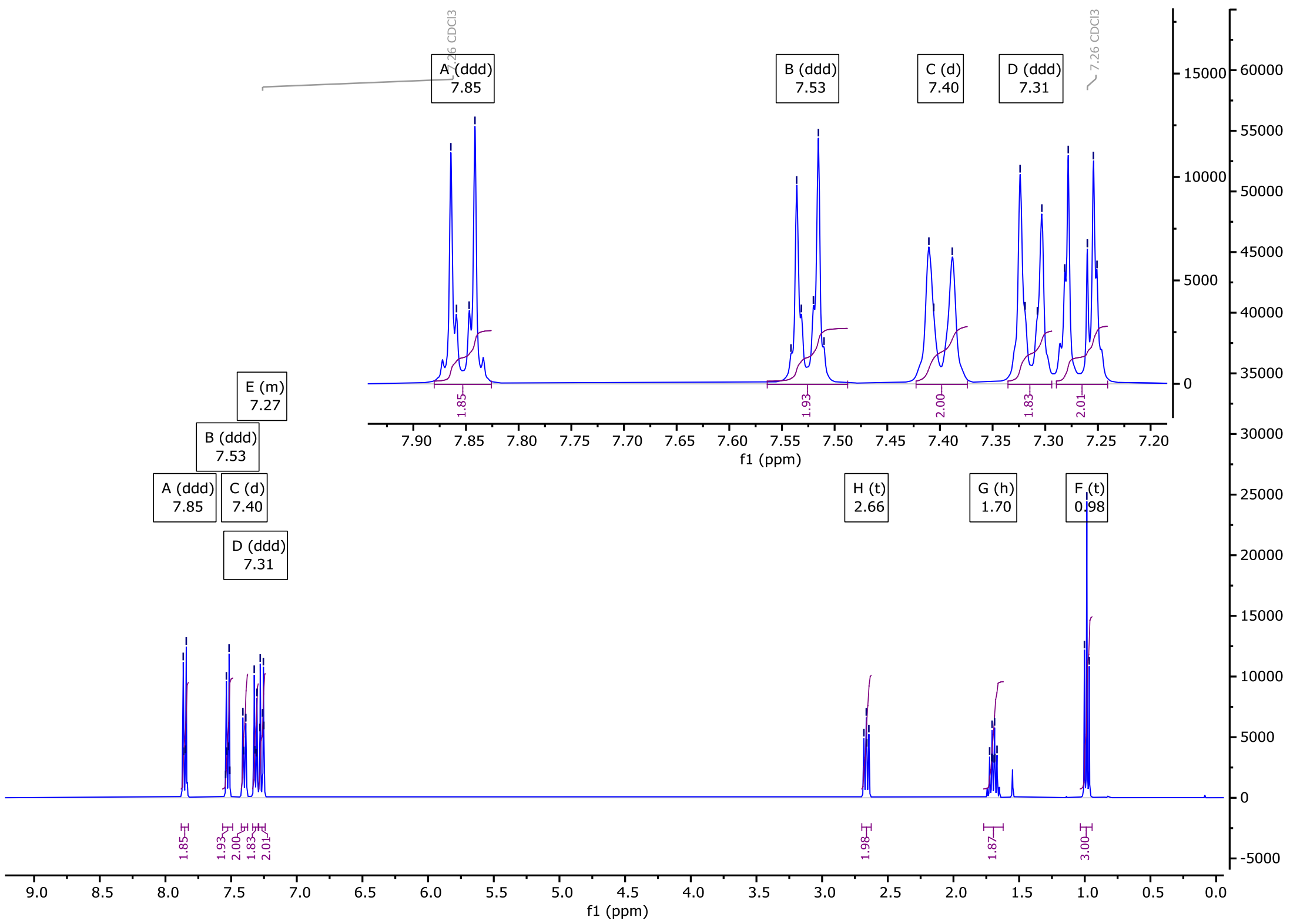

**Fig. S65.** $^{1}H$ NMR of **2j** in $CDCl_3$.

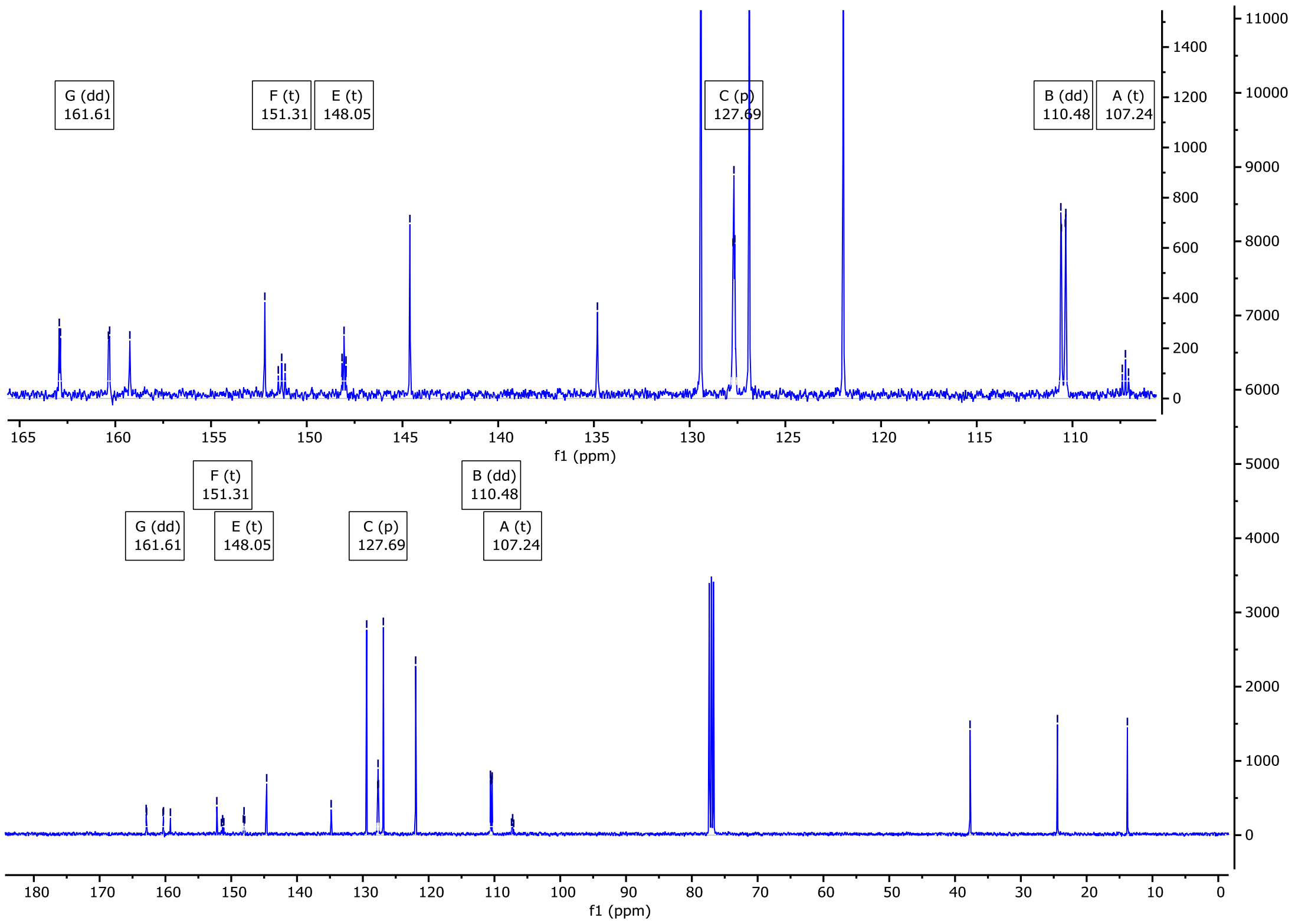

**Fig. S66.** $^{13}C\{^{1}H\}$ NMR of **2j** in $CDCl_3$.

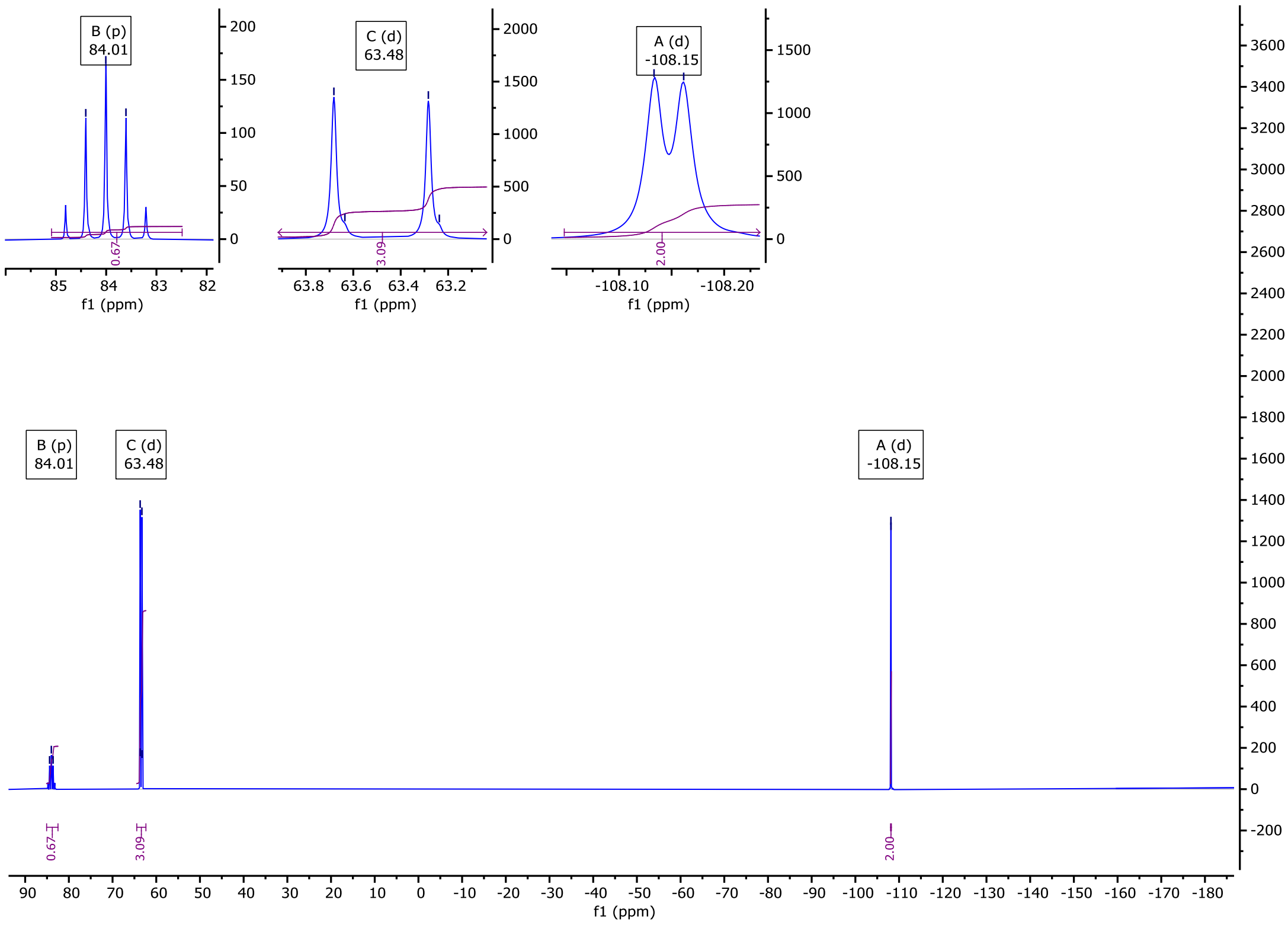

**Fig. S67.** $^{19}$F NMR of **2j** in $CDCl_3$.

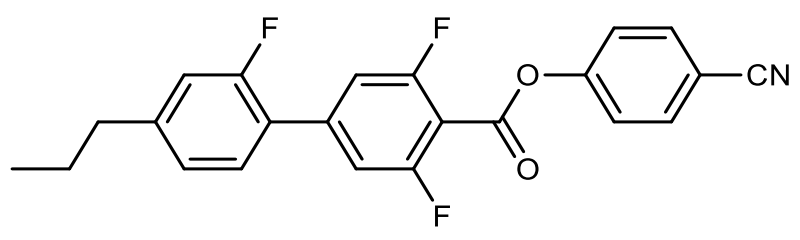

**3a**

*4-cyanophenyl 2',3,5-trifluoro-4'-propyl-[1,1'-biphenyl]-4-carboxylate*

Yield: (White Solid) 146 mg, 74 %

$^1$H NMR (400 MHz): 7.76 (ddd, *J* = 8.7, 2.5, 2.0 Hz, 2H, Ar-**H**), 7.43 (ddd, *J* = 8.7, 2.5, 2.0 Hz, 2H, Ar-**H**), 7.36 (t, *J* = 8.0 Hz, 1H, Ar-**H**), 7.30 – 7.23 (m, 2H, Ar-**H**)†, 7.08 (dd, *J* = 7.9, 1.7 Hz, 1H, Ar-**H**), 7.03 (dd, *J* = 11.9, 1.6 Hz, 1H, Ar-**H**), 2.65 (t, *J* = 7.5 Hz, 2H, Ar-C**$H_2$**-$CH_2$), 1.69 (h, *J* = 7.4 Hz, 2H, $CH_2$-$CH_2$-$CH_3$), 0.98 (t, *J* = 7.3 Hz, 3H, $CH_2$-C**$H_3$**). † Overlapping $CDCl_3$.

$^{13}$C{$^1$H} NMR (101 MHz): 161.13 (dd, *J* = 258.2, 6.3 Hz), 159.58 (d, *J* = 250.3 Hz), 158.96 (d, *J* = 1.9 Hz)4, 153.57, 147.09 (d, *J* = 7.7 Hz), 142.75 (t, *J* = 11.2 Hz), 133.82, 129.74 (d, *J* = 2.9 Hz), 125.05 (d, *J* = 3.1 Hz), 122.84, 118.17, 116.47 (d, *J* = 22.0 Hz), 112.72 (dt, *J* = 23.7, 3.9 Hz), 110.31, 37.54, 24.13, 13.72.

$^{19}$F NMR (376 MHz): -108.62 (d, $J_{F-H}$ = 10.4 Hz, 2F, Ar-**F**), -117.54 (dd, $J_{F-H}$ = 12.0 Hz, $J_{F-H}$= 8.1 Hz, 1F, Ar-**F**).

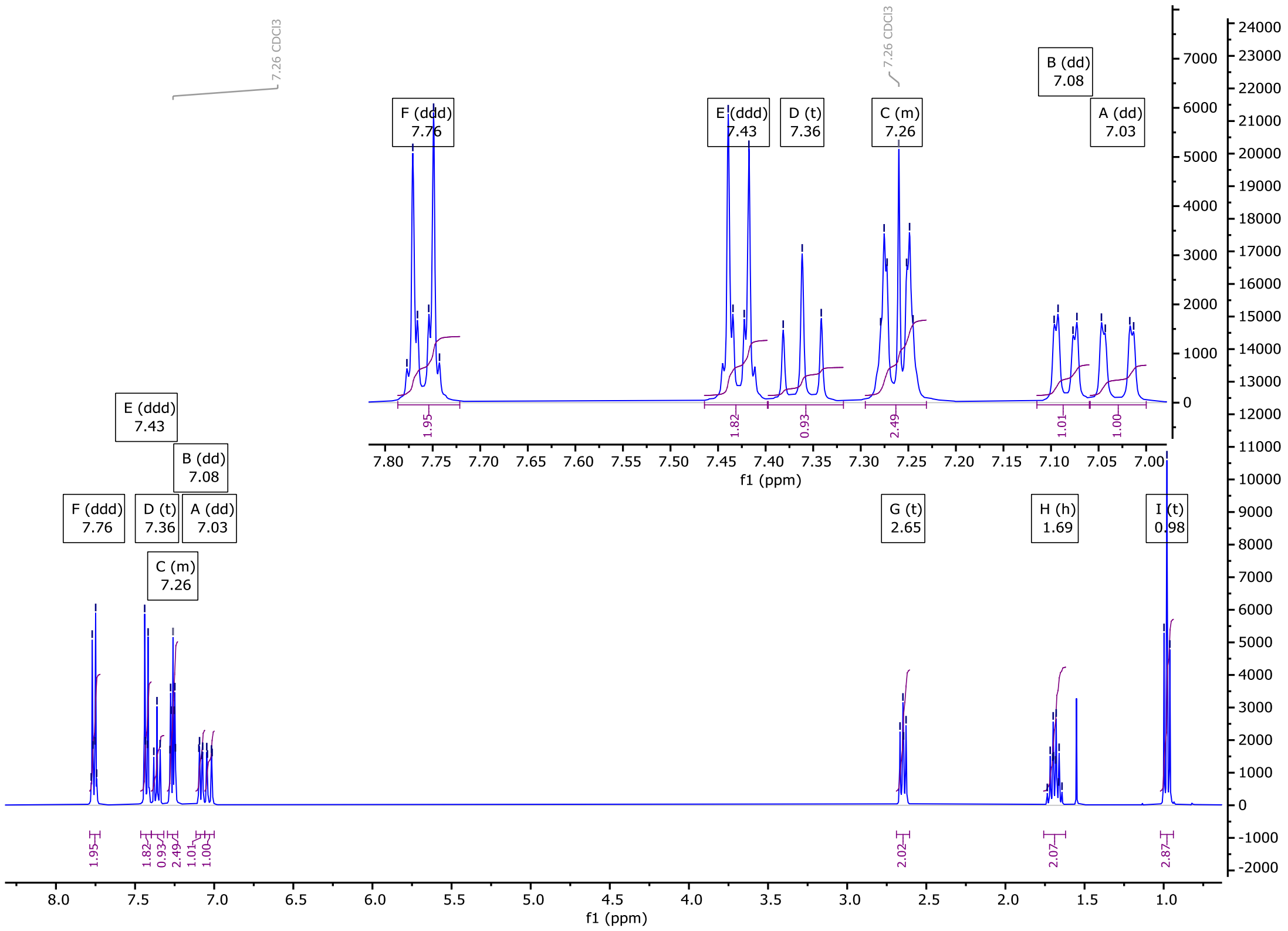

**Fig. S68.** $^{1}H$ NMR of **3a** in $CDCl_3$.

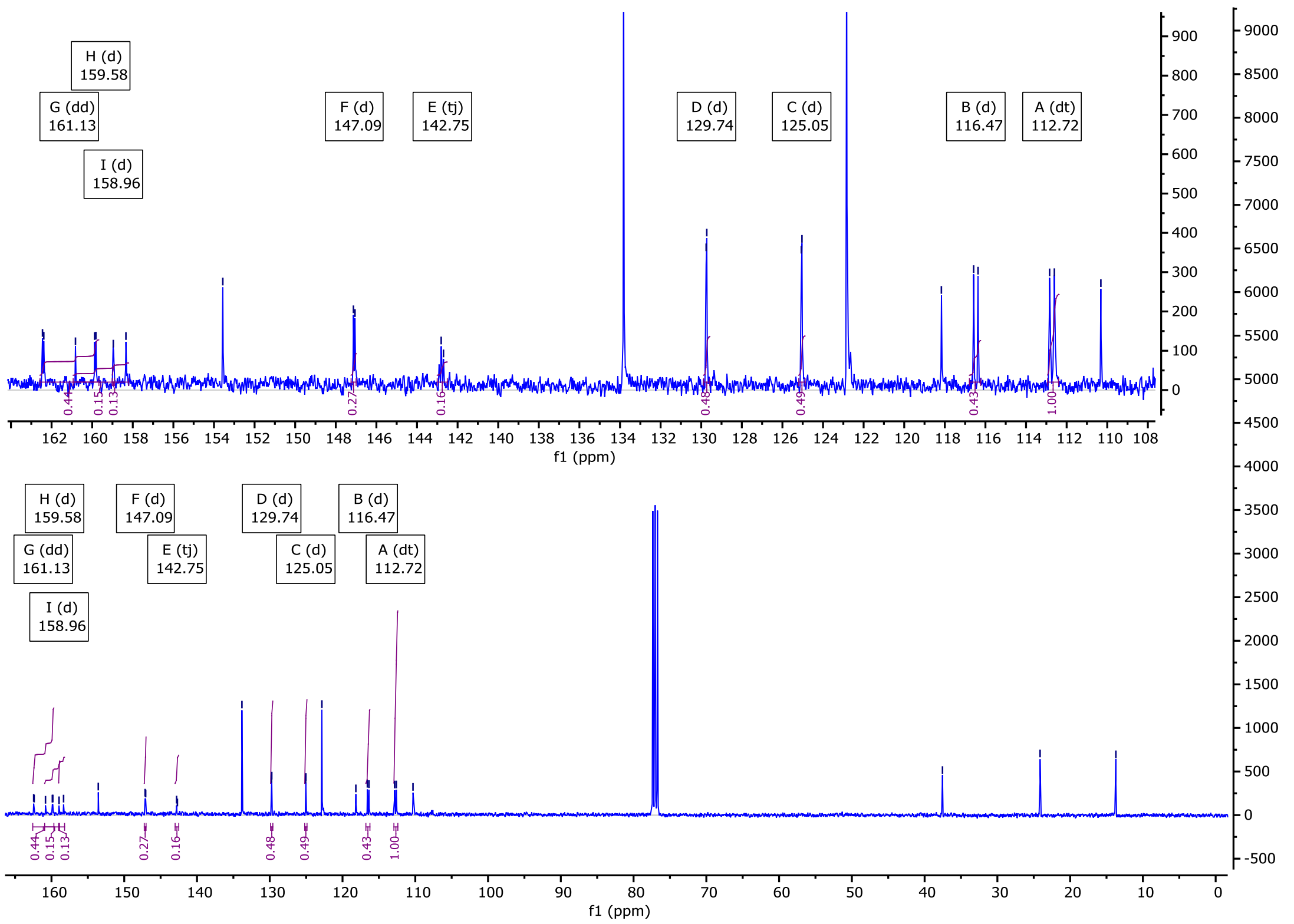

**Fig. S69.** $^{13}C\{^{1}H\}$ NMR of **3a** in $CDCl_3$.

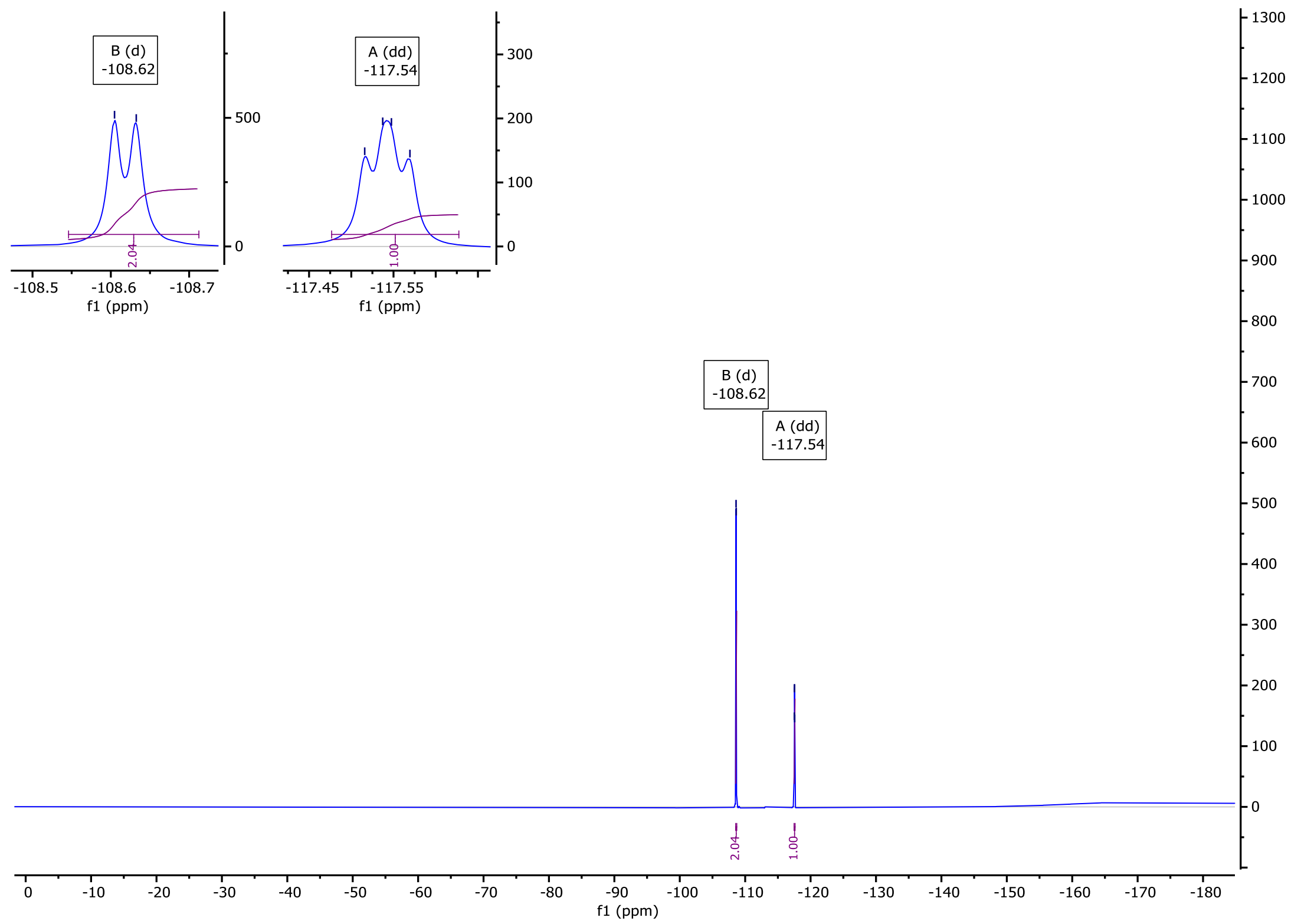

**Fig. S70.** $^{19}F$ NMR of **3a** in $CDCl_3$.

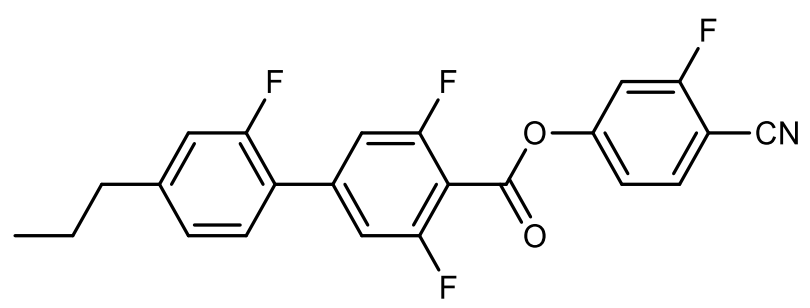

**3b**

*4-cyano-3-fluorophenyl 2',3,5-trifluoro-4'-propyl-[1,1'-biphenyl]-4-carboxylate*

Yield: (White Solid) 140 mg, 68 %

$^{1}$H NMR (400 MHz): 7.72 (t, *J* = 8.7 Hz, 1H, Ar-**H**), 7.36 (t, *J* = 8.0 Hz, 1H, Ar-**H**), 7.31 – 7.23 (m, 4H, Ar-**H**)$^{*†}$, 7.09 (dd, *J* = 7.9, 1.7 Hz, 1H, Ar-**H**), 7.03 (dd, *J* = 12.0, 1.6 Hz, 1H, Ar-**H**), 2.65 (t, *J* = 7.4 Hz, 2H, Ar-C**H₂**-CH₂), 1.69 (h, *J* = 7.3 Hz, 2H, $CH_2$-C$\mathbf{H_2}$-$CH_3$), 0.98 (t, *J* = 7.3 Hz, 3H, $CH_2$-C$\mathbf{H_3}$). * Overlapping peaks, † Overlapping $CDCl_3$.

$^{13}$C{$^{1}$H} NMR (101 MHz): 163.63 (d, *J* = 261.1 Hz), 162.50 (d, *J* = 6.2 Hz), 160.83, 158.41 (d, *J* = 13.8 Hz), 157.36 (dd, *J* = 517.5, 5.6 Hz), 147.21 (d, *J* = 7.9 Hz), 143.17 (t, *J* = 11.3 Hz), 134.22, 129.73 (d, *J* = 2.9 Hz), 125.08 (d, *J* = 3.1 Hz), 122.54 (d, *J* = 12.6 Hz), 118.65 (d, *J* = 3.6 Hz), 116.50 (d, *J* = 21.8 Hz), 112.73 (dt, *J* = 23.4, 3.1 Hz), 111.01 (d, *J* = 22.8 Hz), 107.25 (d, *J* = 17.3 Hz), 99.44 (d, *J* = 15.5 Hz), 37.55, 24.12, 13.71.

$^{19}$F NMR (376 MHz): -102.99 (t, $J_{F-H}$ = 8.2 Hz, 1F, Ar-**F**), -108.27 (d, $J_{F-H}$ = 10.7 Hz, 2F, Ar-**F**), -117.50 (dd, $J_{F-H}$ = 12.1 Hz, $J_{F-H}$ = 8.3 Hz, 1F, Ar-**H**).

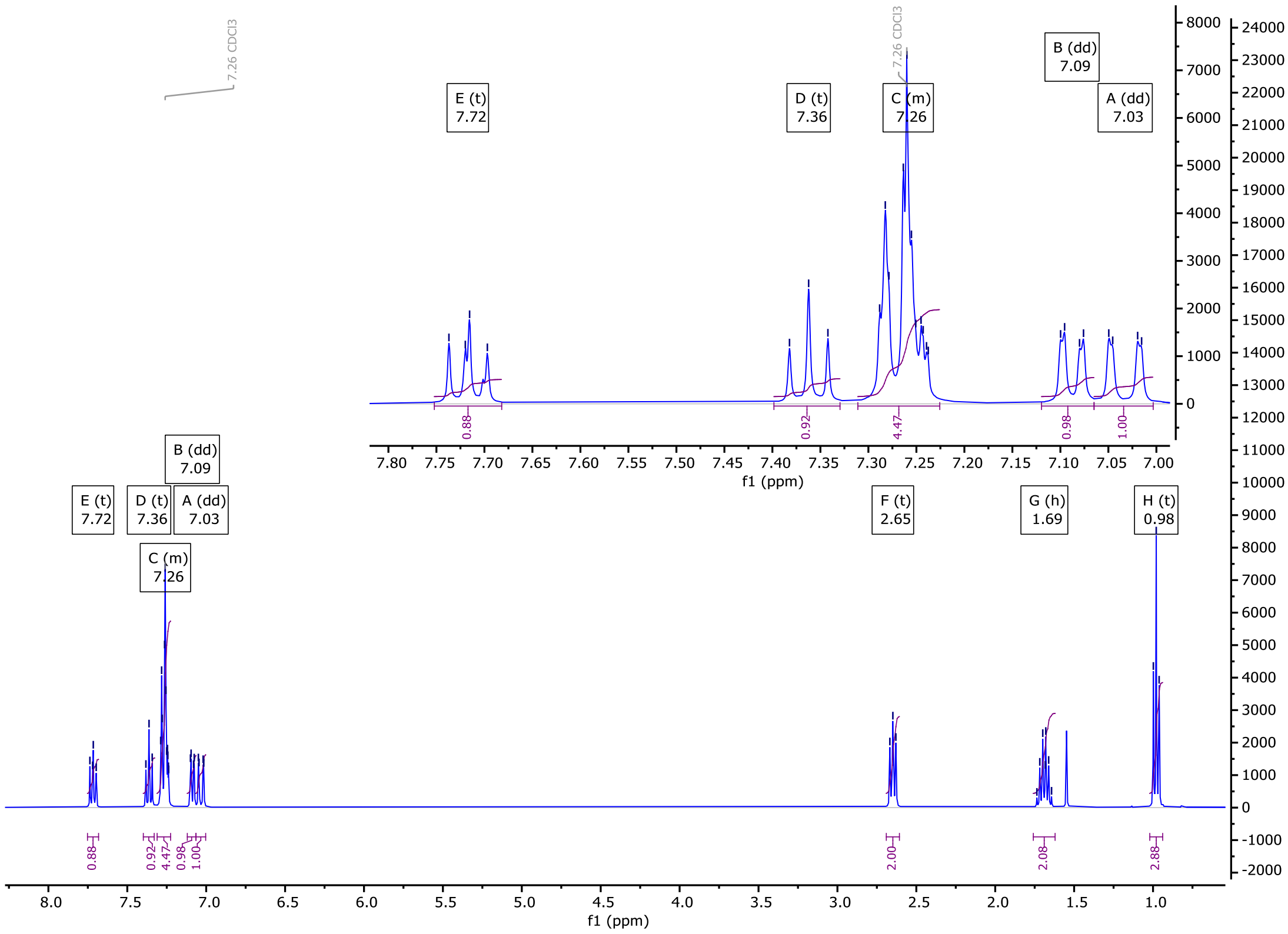

**Fig. S71.** $^{1}H$ NMR of **3b** in $CDCl_3$.

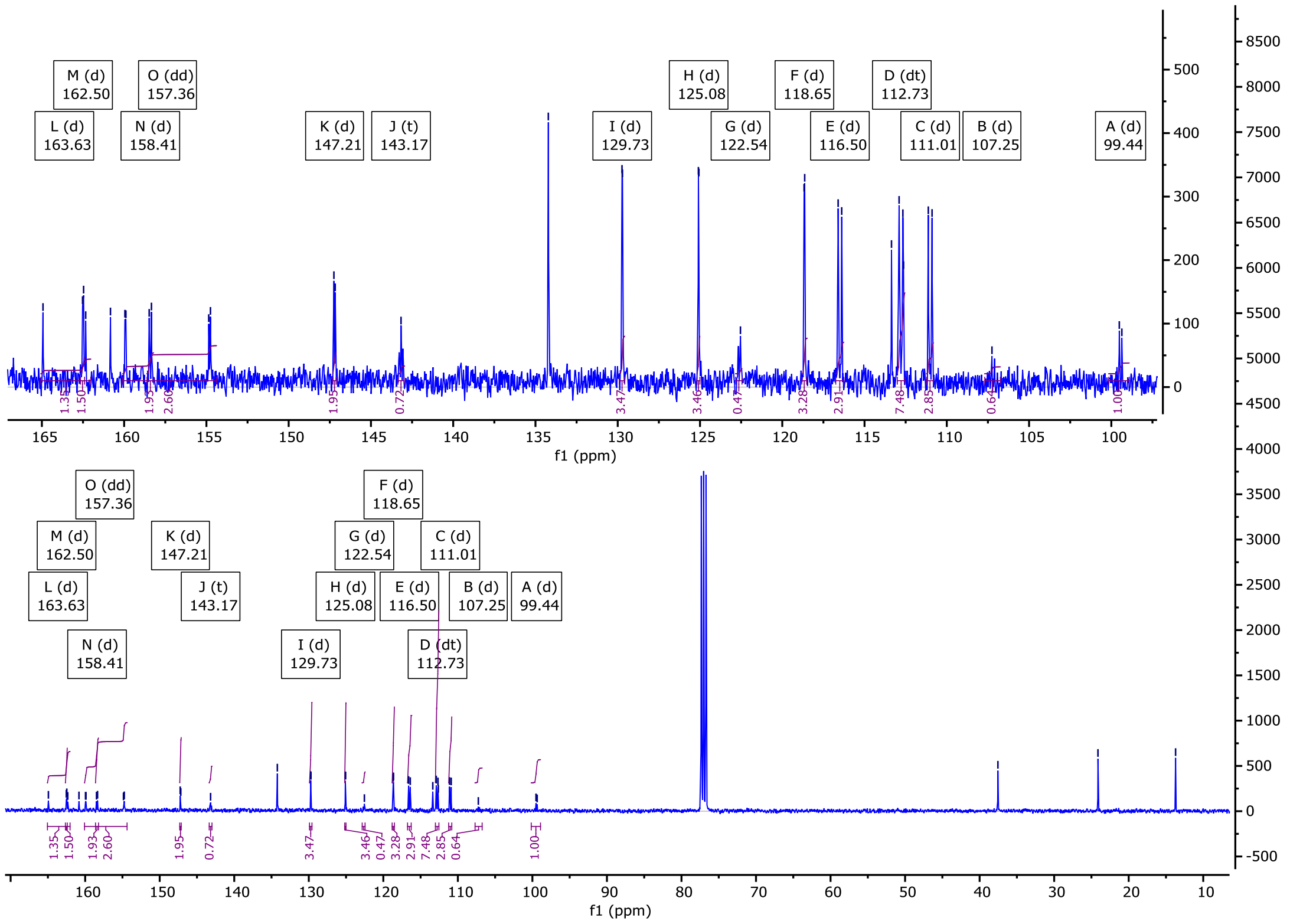

**Fig. S72.** $^{13}C\{^{1}H\}$ NMR of **3b** in $CDCl_3$.

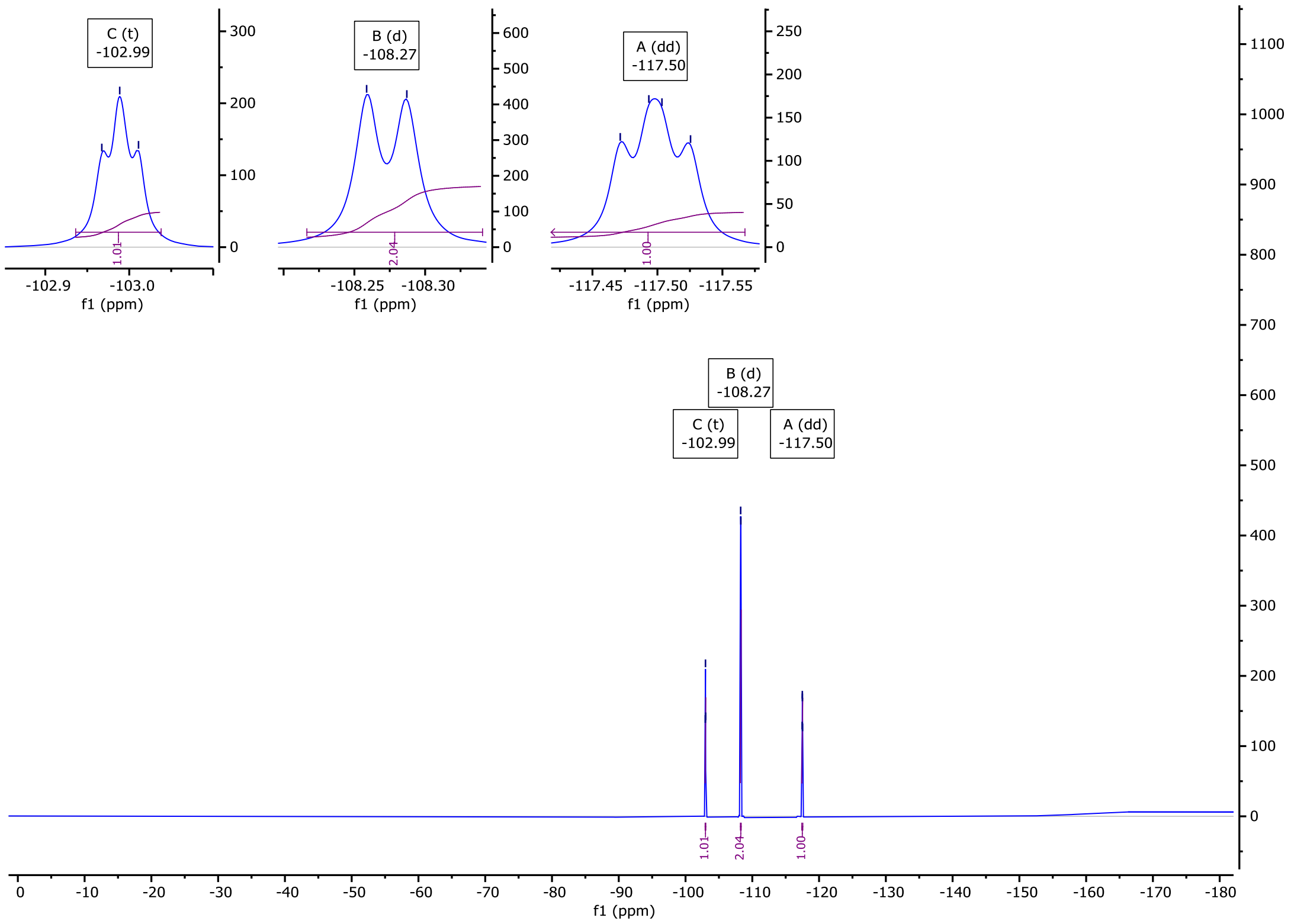

**Fig. S73.** $^{19}$F NMR of **3b** in $CDCl_3$.

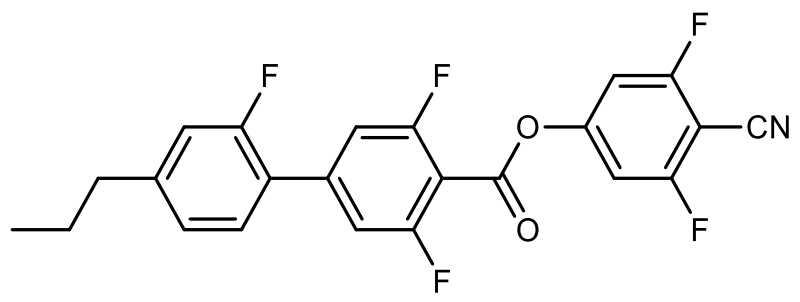

**3c**

*4-cyano-3,5-difluorophenyl 2',3,5-trifluoro-4'-propyl-[1,1'-biphenyl]-4-carboxylate*

Yield: (White Solid) 125 mg, 58 %

$^{1}$H NMR (400 MHz): 7.36 (t, *J* = 8.0 Hz, 1H, Ar-**H**), 7.30 – 7.24 (m, 2H, Ar-**H**) †, 7.15 – 7.06 (m, 3H, Ar-**H**) *, 7.03 (dd, *J* = 12.1, 1.7 Hz, 1H, Ar-**H**), 2.65 (t, *J* = 7.4 Hz, 2H, Ar-C**H₂**-CH₂), 1.69 (h, *J* = 7.5 Hz, 2H, CH₂-C**H₂**-CH₃), 0.98 (t, *J* = 7.3 Hz, 3H, CH₂-C**H₃**).* Overlapping peaks, † Overlapping $CDCl_3$.

$^{13}$C{$^{1}$H} NMR (101 MHz): 163.56 (dd, *J* = 261.8, 6.3 Hz), 161.28 (dd, *J* = 259.3, 5.9 Hz), 160.83, 159.59 (d, *J* = 250.5 Hz), 155.25 (t, *J* = 13.6 Hz), 147.33 (d, *J* = 7.9 Hz), 143.54 (t, *J* = 11.1 Hz), 129.71 (d, *J* = 3.2 Hz), 125.11 (d, *J* = 3.2 Hz), 122.45 (d, *J* = 11.3 Hz), 116.52 (d, *J* = 21.9 Hz), 112.84 (dt, *J* = 23.5, 3.8 Hz), 108.64, 106.90 (dd, *J* = 23.2, 3.9 Hz), 90.54 (t, *J* = 19.5 Hz), 37.55, 24.12, 13.70.

$^{19}$F NMR (376 MHz): -101.59 (d, $J_{F-H}$= 8.7 Hz, 2F, Ar-**F**), -107.93 (d, $J_{F-H}$ = 10.7 Hz, 2F, Ar-**F**), -117.48 (t, $J_{F-H}$ = 10.4 Hz, 1F, Ar-**F**)

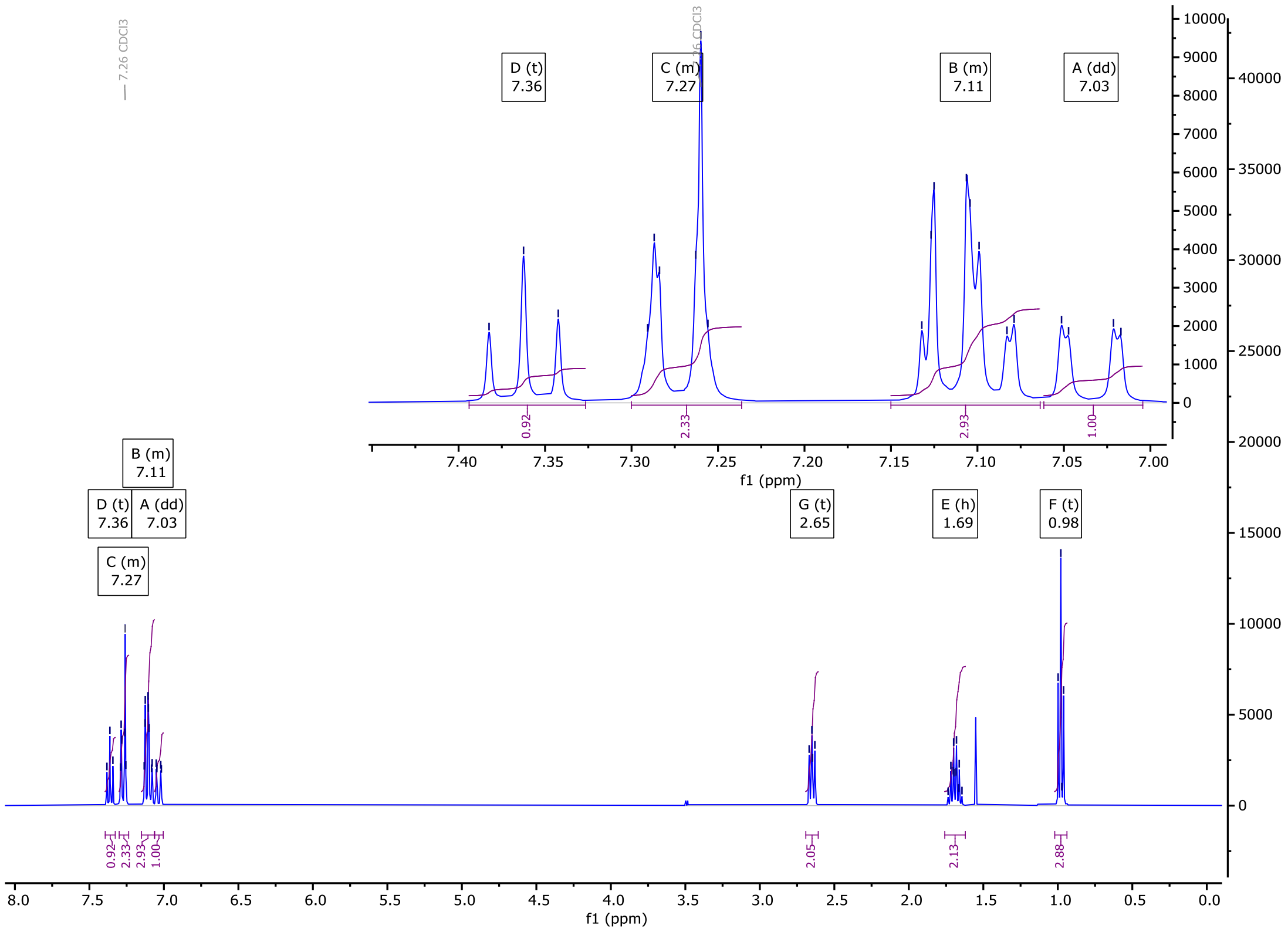

**Fig. S74.** $^1H$ NMR of **3c** in $CDCl_3$.

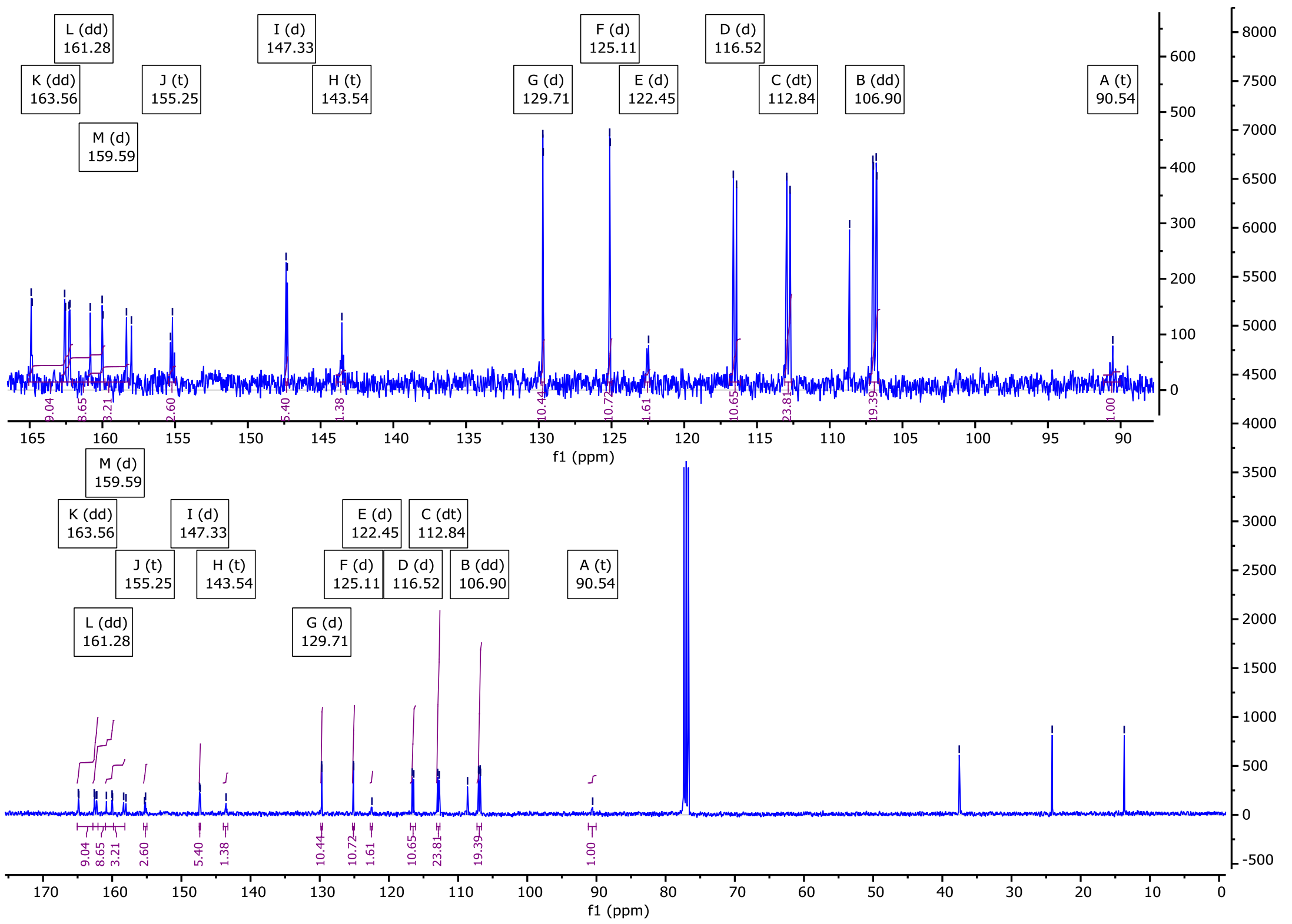

**Fig. S75.** $^{13}C\{^1H\}$ NMR of **3c** in $CDCl_3$.

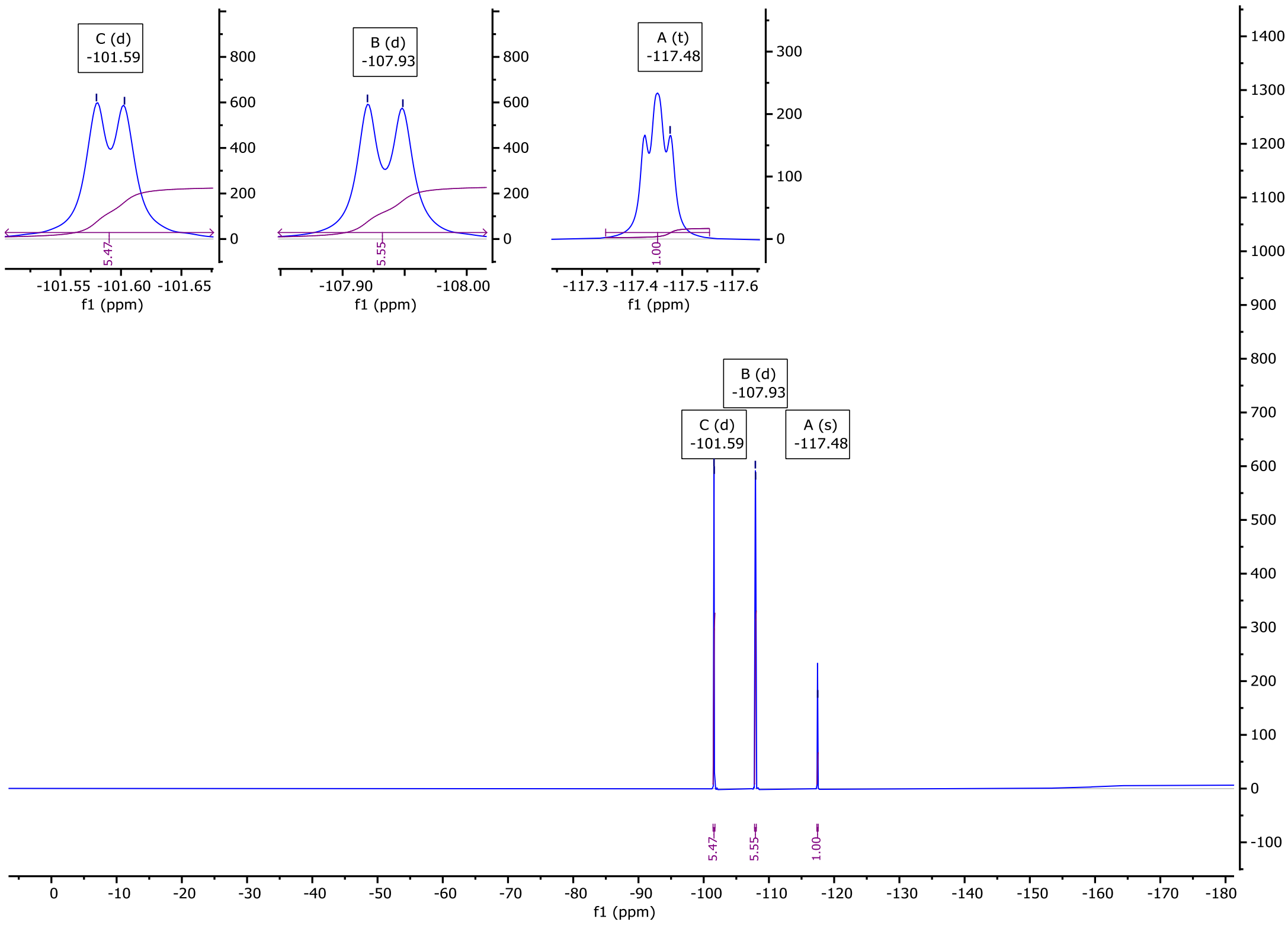

**Fig. S76.** $^{19}$F NMR of **3c** in $CDCl_3$.

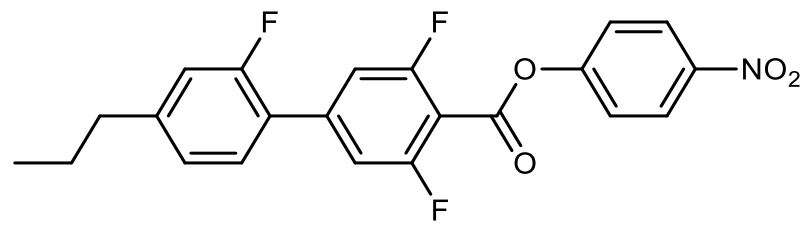

**3d**

*4-nitrophenyl 2',3,5-trifluoro-4'-propyl-[1,1'-biphenyl]-4-carboxylate*

Yield: (White Solid) 116 mg, 56 %

$^1$H NMR (400 MHz): 8.34 (ddd, *J* = 9.4, 3.1, 2.2 Hz, 2H, Ar-**H**), 7.48 (ddd, *J* = 9.1, 3.0, 2.2 Hz, 2H, Ar-**H**), 7.37 (t, *J* = 8.0 Hz, 1H, Ar-**H**), 7.30 – 7.23 (m, 2H, Ar-**H**)$^†$, 7.09 (dd, *J* = 7.9, 1.7 Hz, 1H, Ar-**H**), 7.03 (dd, *J* = 12.0, 1.6 Hz, 1H, Ar-**H**), 2.65 (t, *J* = 7.4 Hz, 2H, Ar-C**H$_2$**-CH$_2$), 1.69 (h, *J* = 7.4 Hz, 2H, CH$_2$-C**H$_2$**-CH$_3$), 0.98 (t, *J* = 7.3 Hz, 3H, CH$_2$-C**H$_3$**). $^†$ Overlapping $CDCl_3$.

$^{13}$C{$^1$H} NMR (101 MHz): 161.18 (dd, *J* = 258.5, 6.4 Hz), 159.85 (d, *J* = 250.8 Hz), 158.87, 154.98, 147.14 (d, *J* = 7.8 Hz), 145.70, 142.98 (t, *J* = 10.9 Hz), 129.74 (d, *J* = 3.2 Hz), 125.36, 125.07 (d, *J* = 3.2 Hz), 122.58, 116.48 (d, *J* = 22.0 Hz), 112.75 (dt, *J* = 23.3, 3.4 Hz), 107.61 (t, *J* = 16.3 Hz), 37.54, 24.13, 13.72.

$^{19}$F NMR (376 MHz): -108.48 (d, $J_{F-H}$ = 10.6 Hz, 2F, Ar-**F**), -117.52 (dd, $J_{F-H}$ = 11.9 Hz, $J_{F-H}$ = 8.2 Hz, 1F, Ar-**F**).

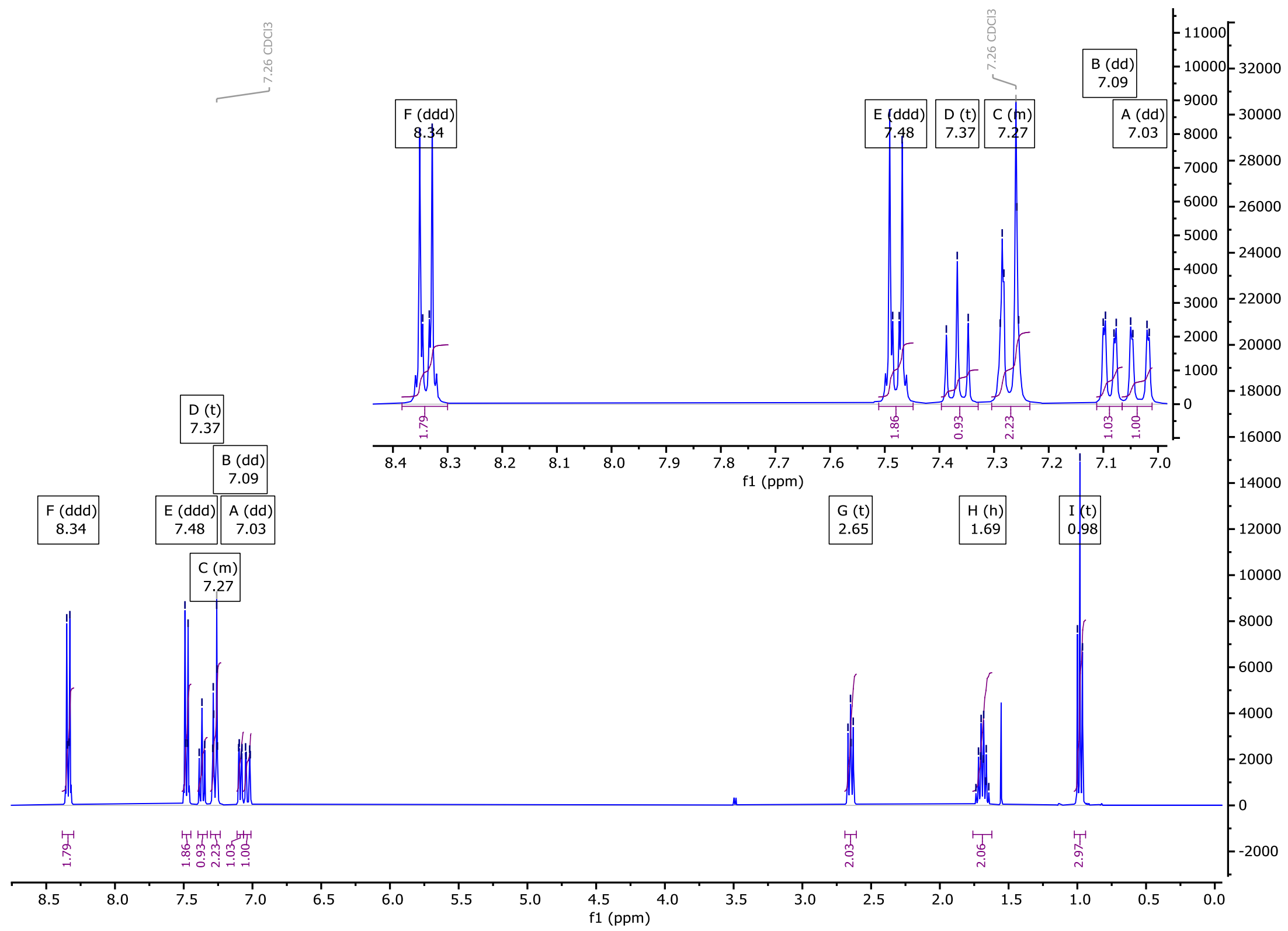

**Fig. S77.** $^{1}H$ NMR of **3d** in $CDCl_3$.

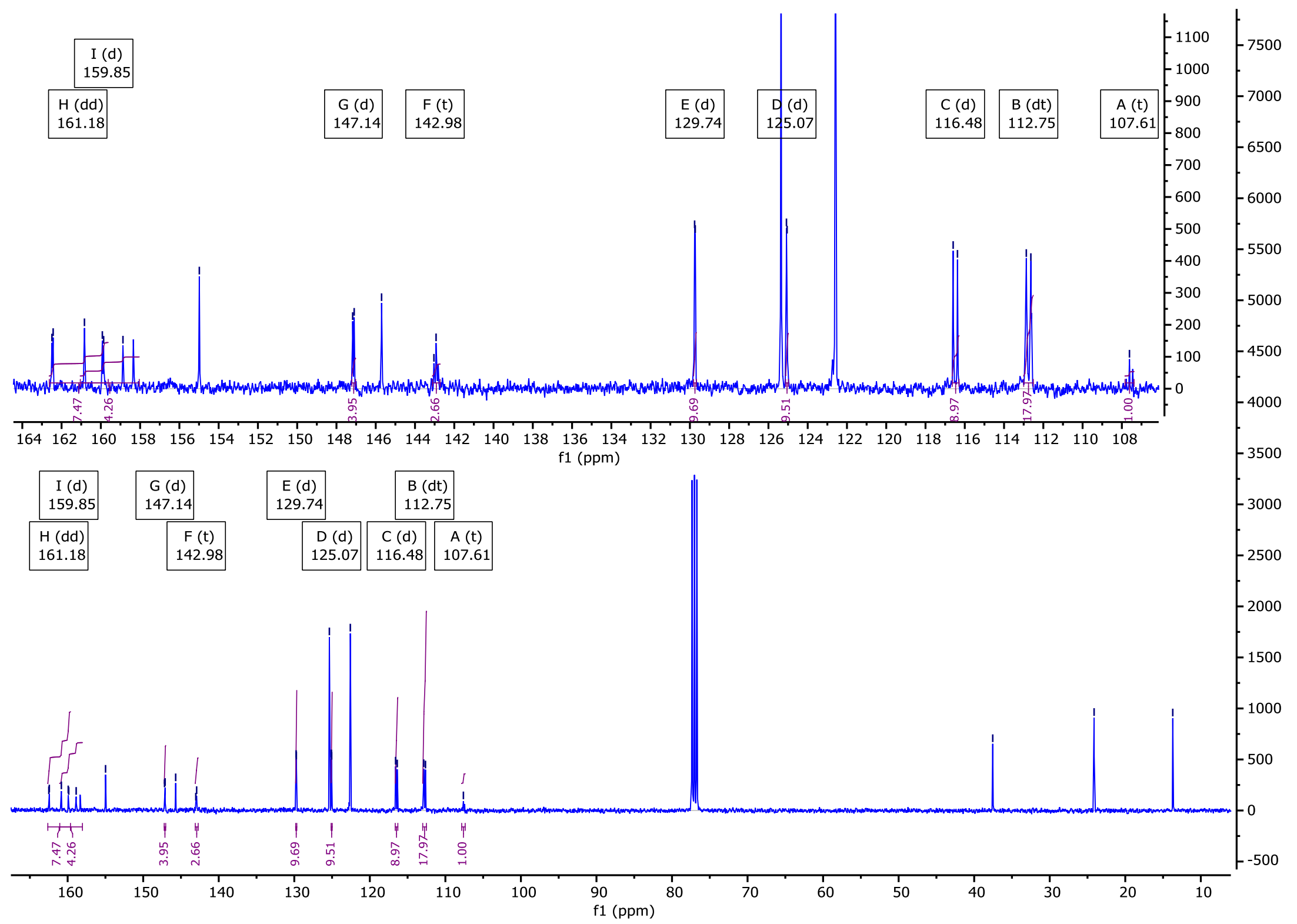

**Fig. S78.** $^{13}C\{^{1}H\}$ NMR of **3d** in $CDCl_3$.

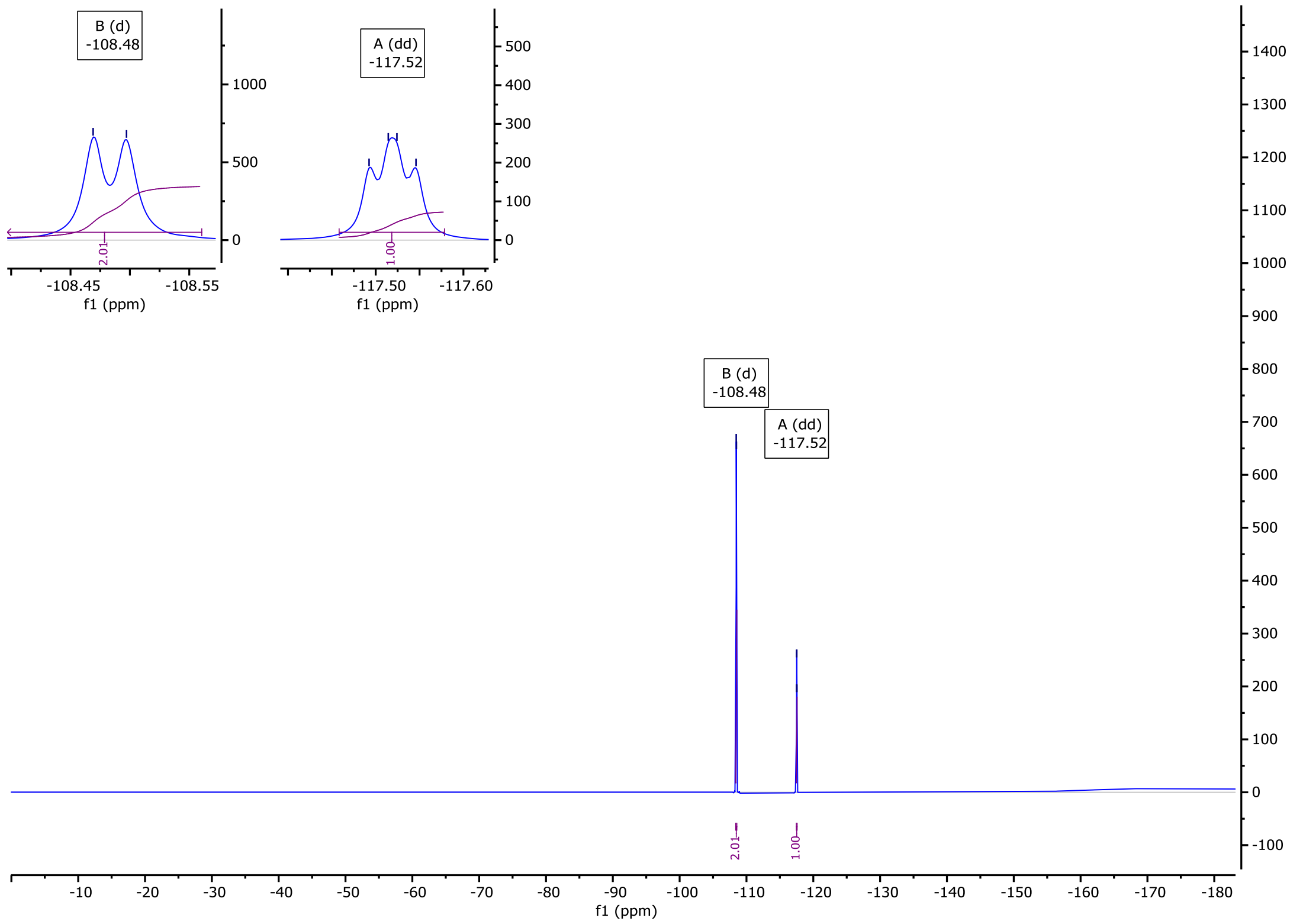

**Fig. S79.** $^{19}F$ NMR of **3d** in $CDCl_3$.

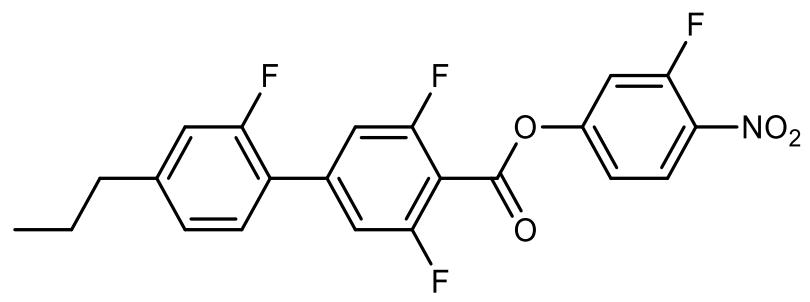

**3e**

*3-fluoro-4-nitrophenyl 2',3,5-trifluoro-4'-propyl-[1,1'-biphenyl]-4-carboxylate*

Yield: (White Solid) 117 mg, 54 %

$^{1}$H NMR (400 MHz): 8.20 (t, *J* = 8.4 Hz, 1H, Ar-**H**), 7.39 – 7.32 (m, 2H, Ar-**H**)*, 7.31 – 7.25 (m, 3H, Ar-**H**)†, 7.09 (dd, *J* = 7.9, 1.6 Hz, 1H, Ar-**H**), 7.03 (dd, *J* = 12.0, 1.6 Hz, 1H, Ar-**H**), 2.65 (t, *J* = 7.4 Hz, 2H, Ar-C**H₂**-CH₂), 1.69 (h, *J* = 7.3 Hz, 2H, CH₂-C**H₂**-CH₃), 0.98 (t, *J* = 7.3 Hz, 3H, CH₂-C**H₃**). * Overlapping signals, † Overlapping $CDCl_3$.

$^{13}$C{$^{1}$H} NMR (101 MHz): 161.08 (dd, *J* = 258.6, 6.5 Hz), 158.37 (d, *J* = 2.7 Hz), 155.01 (d, *J* = 253.8 Hz), 154.90 (d, *J* = 3.7 Hz), 147.24 (d, *J* = 8.0 Hz), 143.27 (t, *J* = 10.5 Hz), 135.18 (t, *J* = 7.6 Hz), 129.73 (d, *J* = 2.9 Hz), 127.32, 125.09 (d, *J* = 3.1 Hz), 122.59 (d, *J* = 12.1 Hz), 118.01 (d, *J* = 4.1 Hz), 116.50 (d, *J* = 22.0 Hz), 112.80 (dt, *J* = 23.3, 3.4 Hz), 112.36 (d, *J* = 24.0 Hz), 37.55, 24.12, 13.71.

$^{19}$F NMR (376 MHz): -108.15 (d, $J_{F-H}$ = 10.7 Hz, 2F, Ar-**F**), -112.84 (t, $J_{F-H}$ = 9.6 Hz, 1F, Ar-**F**), -117.48 (t, $J_{F-H}$ = 9.8 Hz, 1F, Ar-**F**).

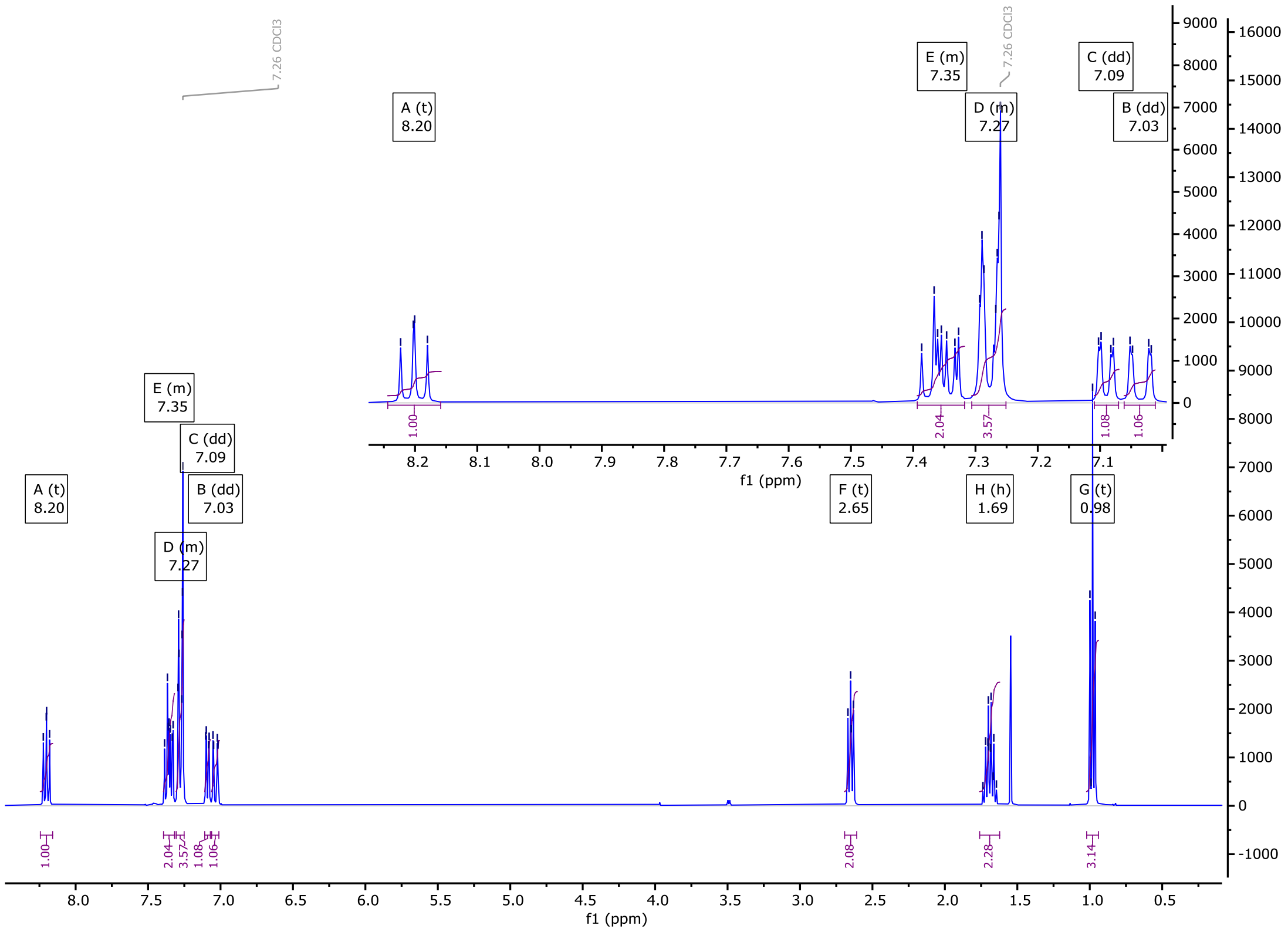

**Fig. S80.** $^{1}H$ NMR of **3e** in $CDCl_3$.

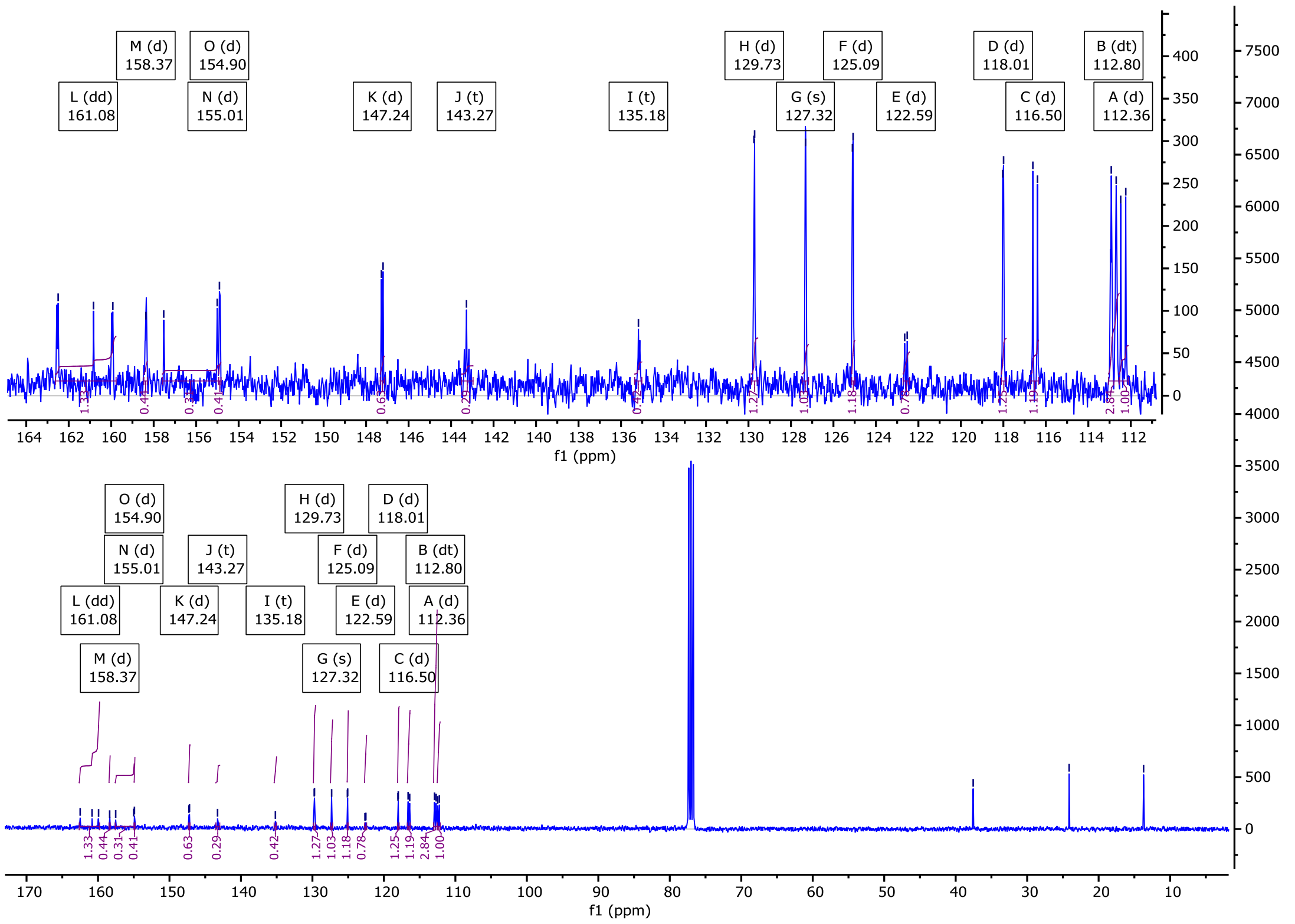

**Fig. S81.** $^{13}C\{^{1}H\}$ NMR of **3e** in $CDCl_3$.

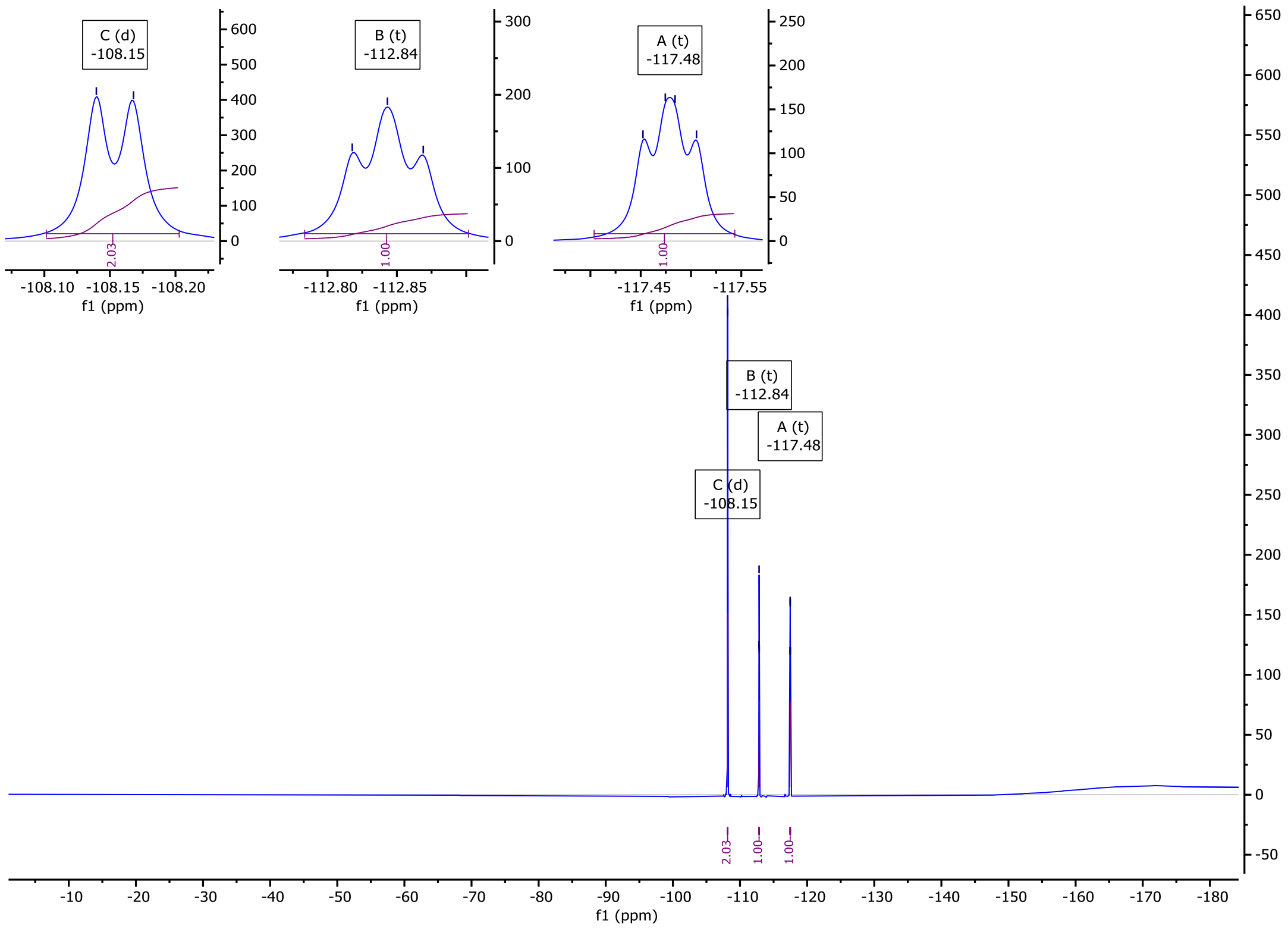

**Fig. S82.** $^{19}$F NMR of **3e** in $CDCl_3$.

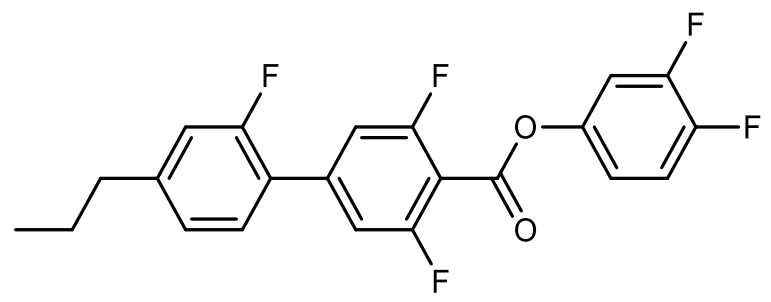

**3f**

*3,4-difluorophenyl 2',3,5-trifluoro-4'-propyl-[1,1'-biphenyl]-4-carboxylate*

| | |
|---|---|
| Yield: | (White Solid) 130 mg, 64 % |
| $^{1}$H NMR (400 MHz): | 7.36 (t, *J* = 8.0 Hz, 1H, Ar-**H**), 7.28 – 7.14 (m, 4H, Ar-**H**)*†, 7.08 (dd, *J* = 7.9, 1.6 Hz, 1H, Ar-**H**), 7.06 – 7.00 (m, 2H, Ar-**H**)*, 2.64 (t, *J* = 7.5 Hz, 2H, Ar-C**H₂**-CH₂), 1.69 (h, *J* = 7.5 Hz, 2H, CH₂-C**H₂**-CH₃), 0.98 (t, *J* = 7.4 Hz, 3H, CH₂-C**H₃**). * Overlapping signals, † Overlapping $CDCl_3$. |
| $^{13}$C{$^{1}$H} NMR (101 MHz): | 162.33 (dd, *J* = 257.9, 6.3 Hz), 159.59 (d, *J* = 251.1 Hz), 158.33, 146.99 (d, *J* = 7.7 Hz), 145.89 (dd, *J* = 7.8, 3.0 Hz), 142.51 (t, *J* = 10.6 Hz), 129.75 (d, *J* = 3.0 Hz), 125.02 (d, *J* = 3.2 Hz), 122.72 (d, *J* = 10.2 Hz), 117.68 (dd, *J* = 6.1, 4.1 Hz), 117.44 (d, *J* = 18.9 Hz), 116.45 (d, *J* = 22.0 Hz), 112.66 (dt, *J* = 23.1, 3.2 Hz), 111.78 (d, *J* = 20.3 Hz), 108.03 (t, *J* = 17.2 Hz), 37.54, 24.13, 13.72. |
| $^{19}$F NMR (376 MHz): | -108.93 (d, $J_{F-H}$ = 10.4 Hz, 2F, Ar-**F**), -117.58 (dd, $J_{F-H}$ = 12.0 Hz, $J_{F-H}$ = 8.3 Hz, 1F, Ar-**F**), -134.19 – -134.37 (m, 1F, Ar-**F**), -140.06 – -140.25 (m, 1F, Ar-**F**). |

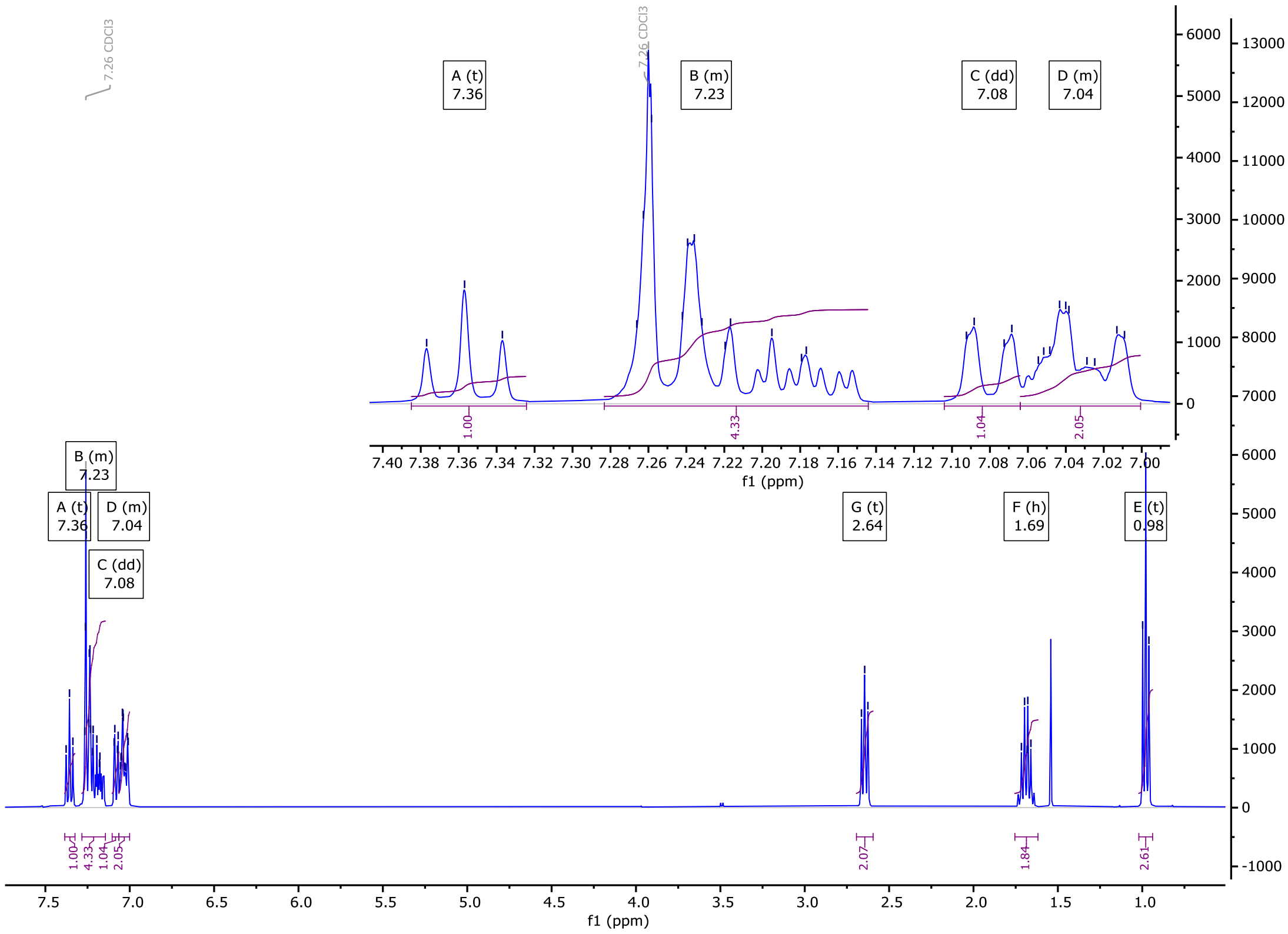

**Fig. S83.** $^{1}H$ NMR of **3f** in $CDCl_3$.

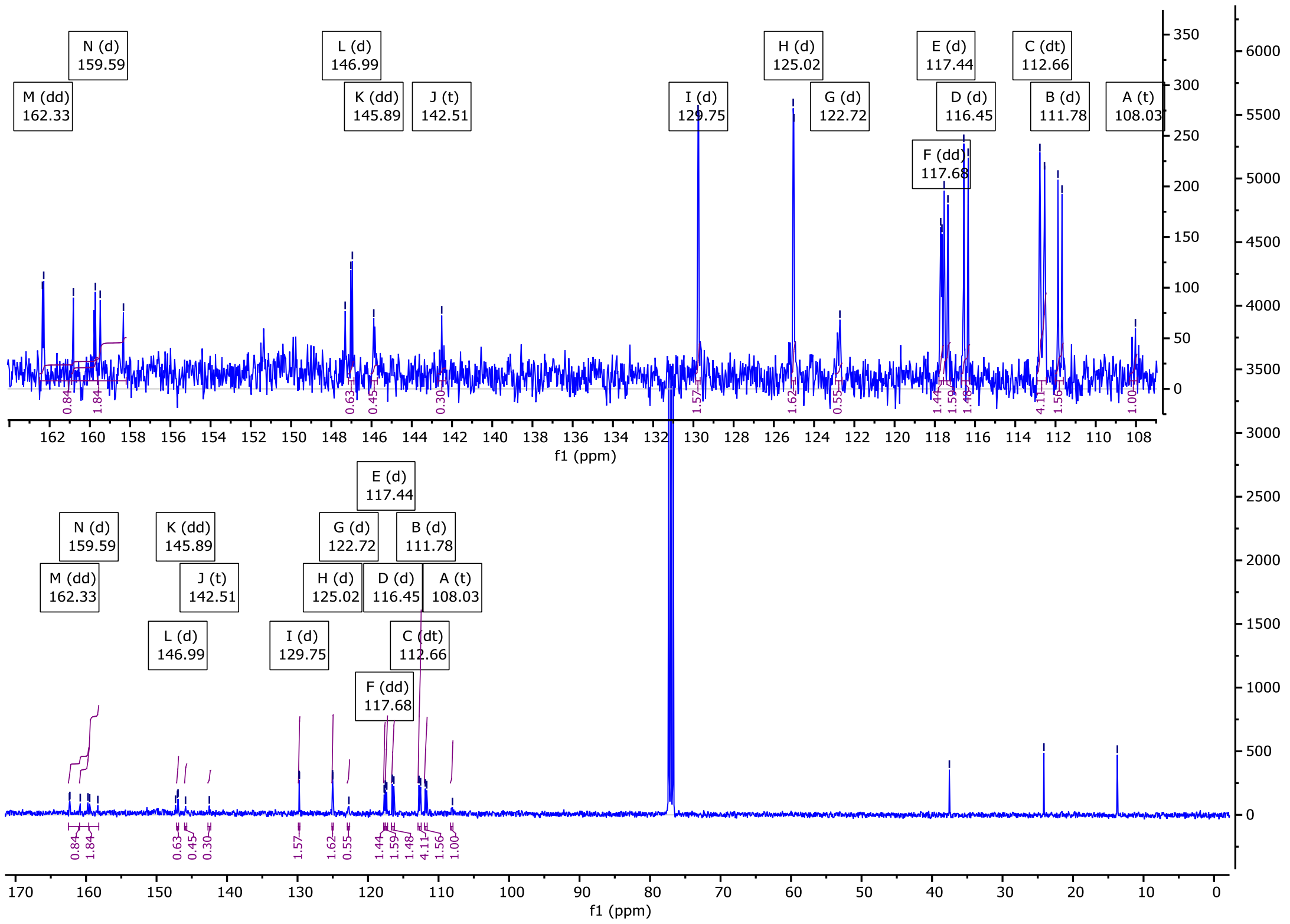

**Fig. S84.** $^{13}C\{^{1}H\}$ NMR of **3f** in $CDCl_3$.

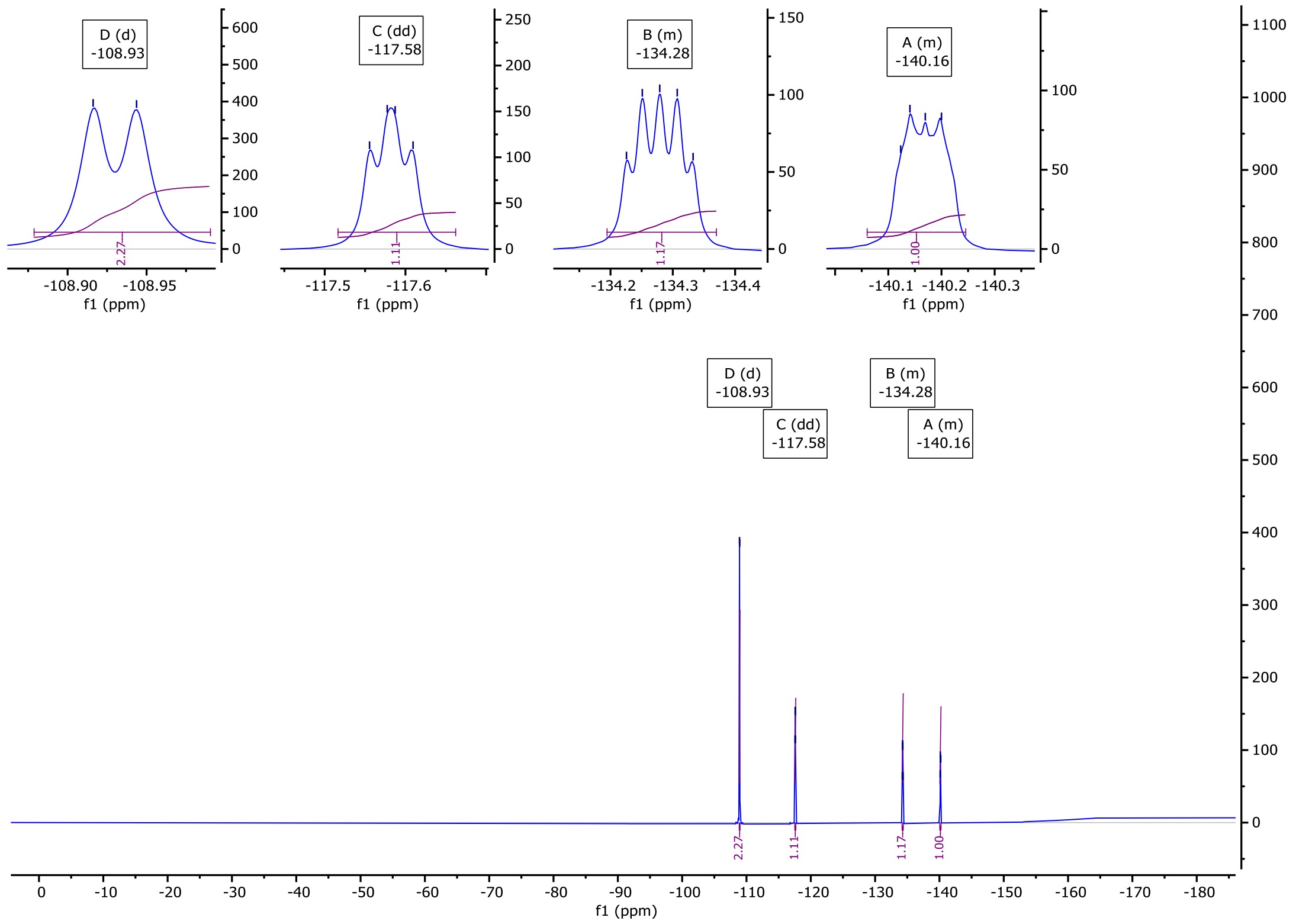

**Fig. S85.** $^{19}$F NMR of **3f** in $CDCl_3$.

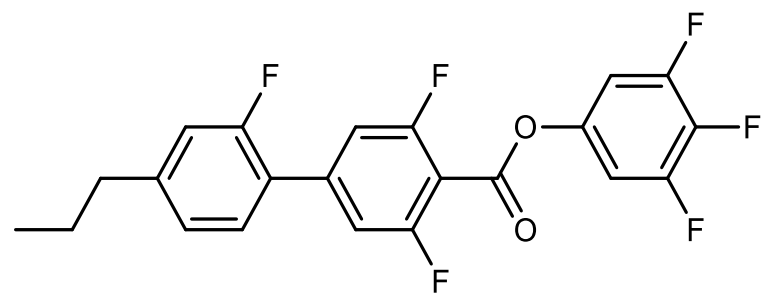

**3g**

*3,4,5-trifluorophenyl 2',3,5-trifluoro-4'-propyl-[1,1'-biphenyl]-4-carboxylate*

Yield: (Colourless Needles) 111 mg, 52 %

$^{1}$H NMR (400 MHz): 7.36 (t, *J* = 8.0 Hz, 1H, Ar-**H**), 7.29 – 7.22 (m, 2H, Ar-**H**)$^{†}$, 7.08 (dd, *J* = 8.0, 1.7 Hz, 1H, Ar-**H**), 7.05 – 6.96 (m, 3H)$^{*}$, 2.65 (t, *J* = 7.4 Hz, 2H, Ar-C**H₂**-CH₂), 1.69 (h, *J* = 7.2 Hz, 2H, CH₂-C**H₂**, CH₃), 0.98 (t, *J* = 7.3 Hz, 3H, CH₂-C**H₃**).$^{*}$ Overlapping signals,$^{†}$ Overlapping $CDCl_3$.

$^{13}$C{$^{1}$H} NMR (126 MHz): 161.24 (dd, *J* = 258.4, 6.2 Hz), 159.53 (d, *J* = 250.5 Hz), 159.18, 151.27 (ddd, *J* = 251.1, 10.8, 5.1 Hz), 147.24 (d, *J* = 7.9 Hz), 145.01 (td, *J* = 11.5, 3.9 Hz), 142.98 (t, *J* = 11.0 Hz), 138.56 (dt, *J* = 250.8, 15.2 Hz), 129.87 (d, *J* = 3.1 Hz), 125.18 (d, *J* = 3.2 Hz), 122.81 (dt, *J* = 12.4, 2.4 Hz), 116.60 (d, *J* = 22.1 Hz), 112.84 (dt, *J* = 22.8, 3.9 Hz), 107.68 (t, *J* = 16.6 Hz), 107.29 (dd, *J* = 18.7, 6.1 Hz), 37.68, 24.26, 13.84.

$^{19}$F NMR (376 MHz): -108.63 (d, $J_{F-H}$ = 10.5 Hz, 2F, Ar-**F**), -117.54 (t, $J_{F-H}$ = 9.5 Hz, 1F, Ar-**F**), -132.42 (dd, $J_{F-F}$ = 20.8 Hz, $J_{F-H}$ = 8.4 Hz, 2F, Ar-**F**), -162.75 (tt, $J_{F-F}$ = 20.7 Hz, $J_{F-H}$ = 6.7 Hz, 1F, Ar-**F**).

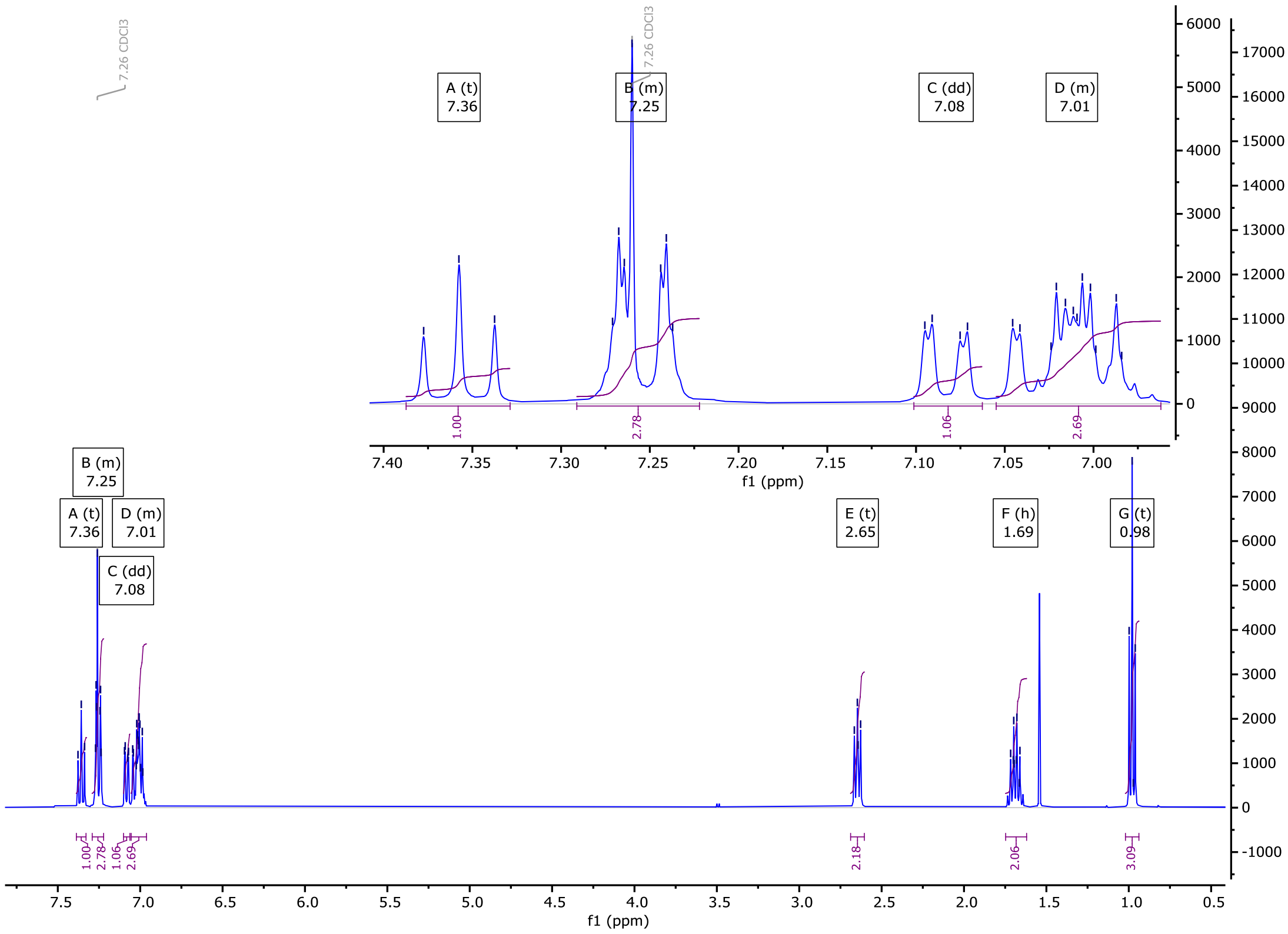

**Fig. S86.** $^{1}H$ NMR of **3g** in $CDCl_3$.

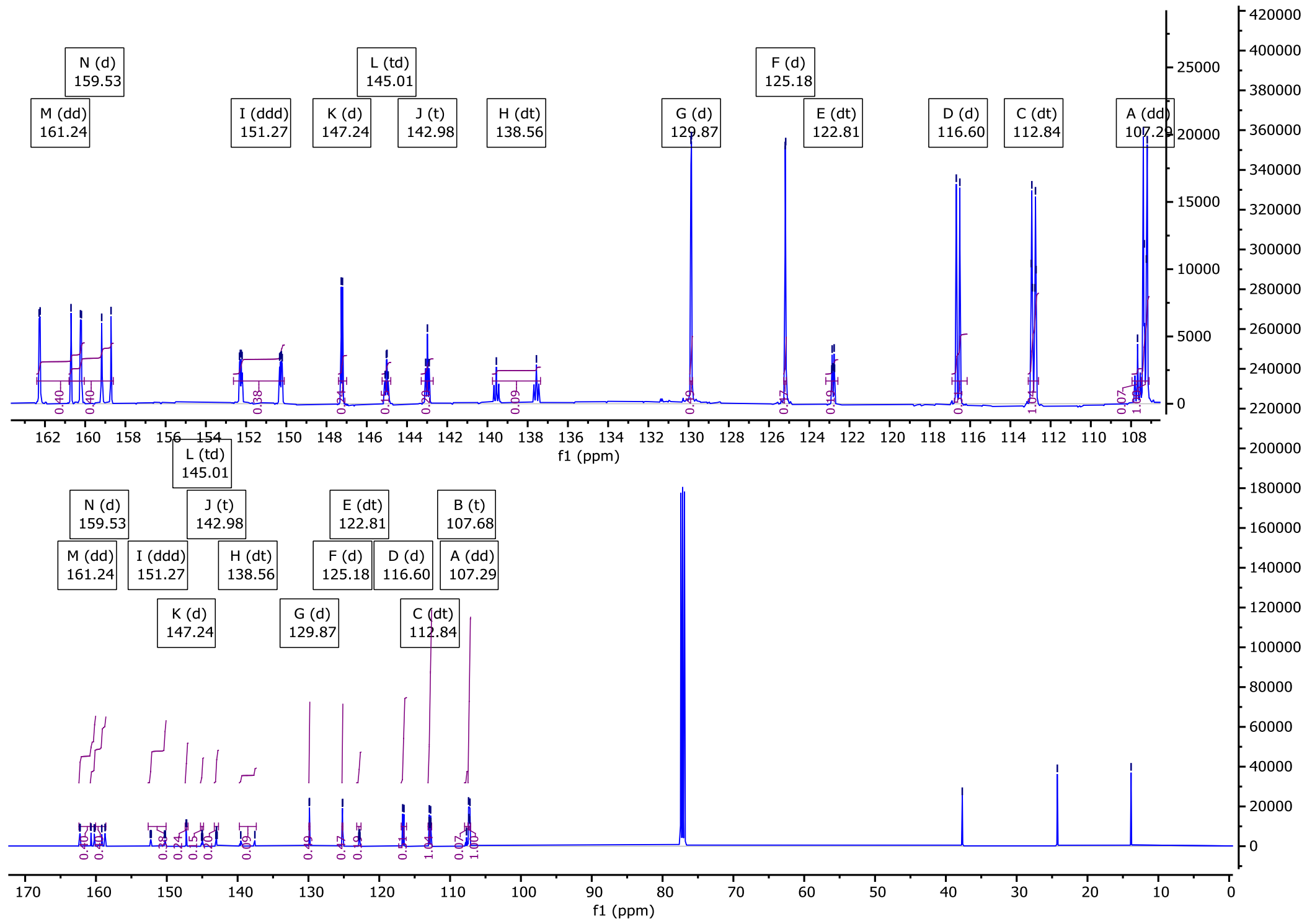

**Fig. S87.** $^{13}C\{^{1}H\}$ NMR of **3g** in $CDCl_3$.

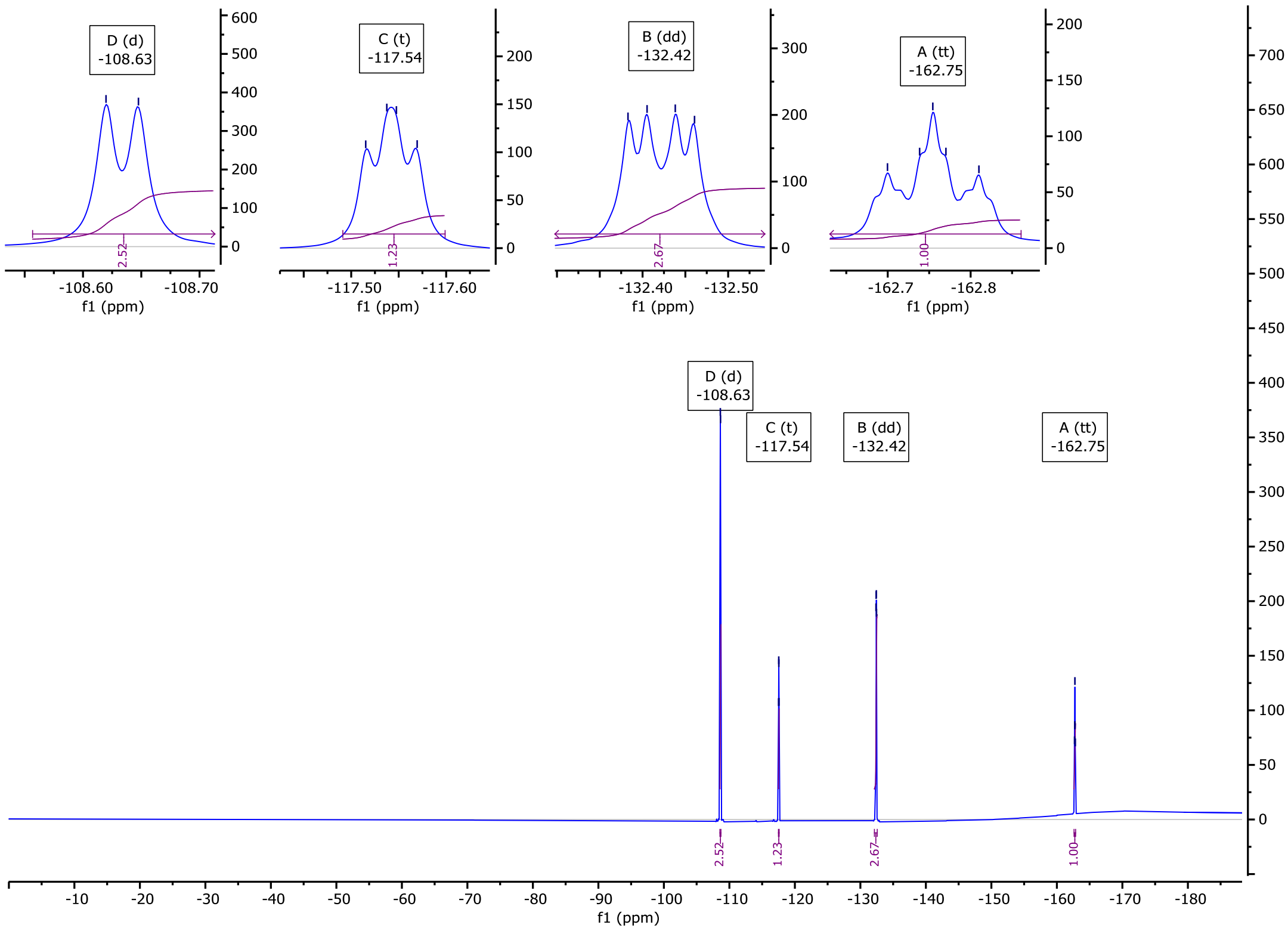

**Fig. S88.** $^{19}F$ NMR of **3g** in $CDCl_3$.

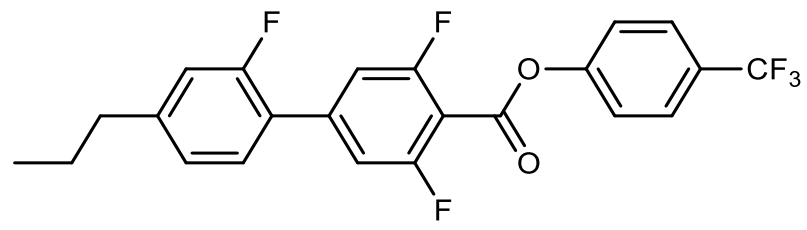

**3h**

*4-(trifluoromethyl)phenyl 2',3,5-trifluoro-4'-propyl-[1,1'-biphenyl]-4-carboxylate*

Yield: (White Solid) 112 mg, 51 %

$^1$H NMR (400 MHz): 7.72 (d, *J* = 8.5 Hz, 2H, Ar-**H**), 7.42 (d, *J* = 8.3 Hz, 2H, Ar-**H**), 7.36 (t, *J* = 8.0 Hz, 1H), 7.28 – 7.23 (m, 2H, Ar-**H**) †, 7.08 (dd, *J* = 7.9, 1.7 Hz, 1H, Ar-**H**), 7.03 (dd, *J* = 12.0, 1.6 Hz, 1H, Ar-**H**), 2.65 (t, *J* = 7.4 Hz, 2H, Ar-C$\mathbf{H_2}$-$CH_2$), 1.69 (h, *J* = 7.5 Hz, 2H, $CH_2$-C$\mathbf{H_2}$-$CH_3$), 0.98 (t, *J* = 7.3 Hz, 3H, $CH_2$-C$\mathbf{H_3}$). † Overlapping $CDCl_3$.

$^{13}$C{$^1$H} NMR (126 MHz): 161.21 (dd, *J* = 257.9, 6.4 Hz), 159.64 (d, *J* = 250.1 Hz), 159.47, 152.96, 147.13 (d, *J* = 7.9 Hz), 142.67 (t, *J* = 10.9 Hz), 129.90 (d, *J* = 3.0 Hz), 128.75 (q, *J* = 32.9 Hz), 127.10 (m), 125.16 (d, *J* = 3.1 Hz), 125.05, 122.93 (d, *J* = 10.6 Hz), 122.31, 116.59 (d, *J* = 21.9 Hz), 112.81 (dd, *J* = 22.6, 3.6 Hz), 108.23 (t, *J* = 16.8 Hz), 37.69, 24.28, 13.85.

f$^{19}$F NMR (376 MHz): -62.26 (s, 3F, Ar-**F**), -108.88 (d, $J_{F-H}$ = 10.4 Hz, 2F, Ar-**F**), -117.58 (t, $J_{F-H}$ = 9.6 Hz, 1F, Ar-**F**).

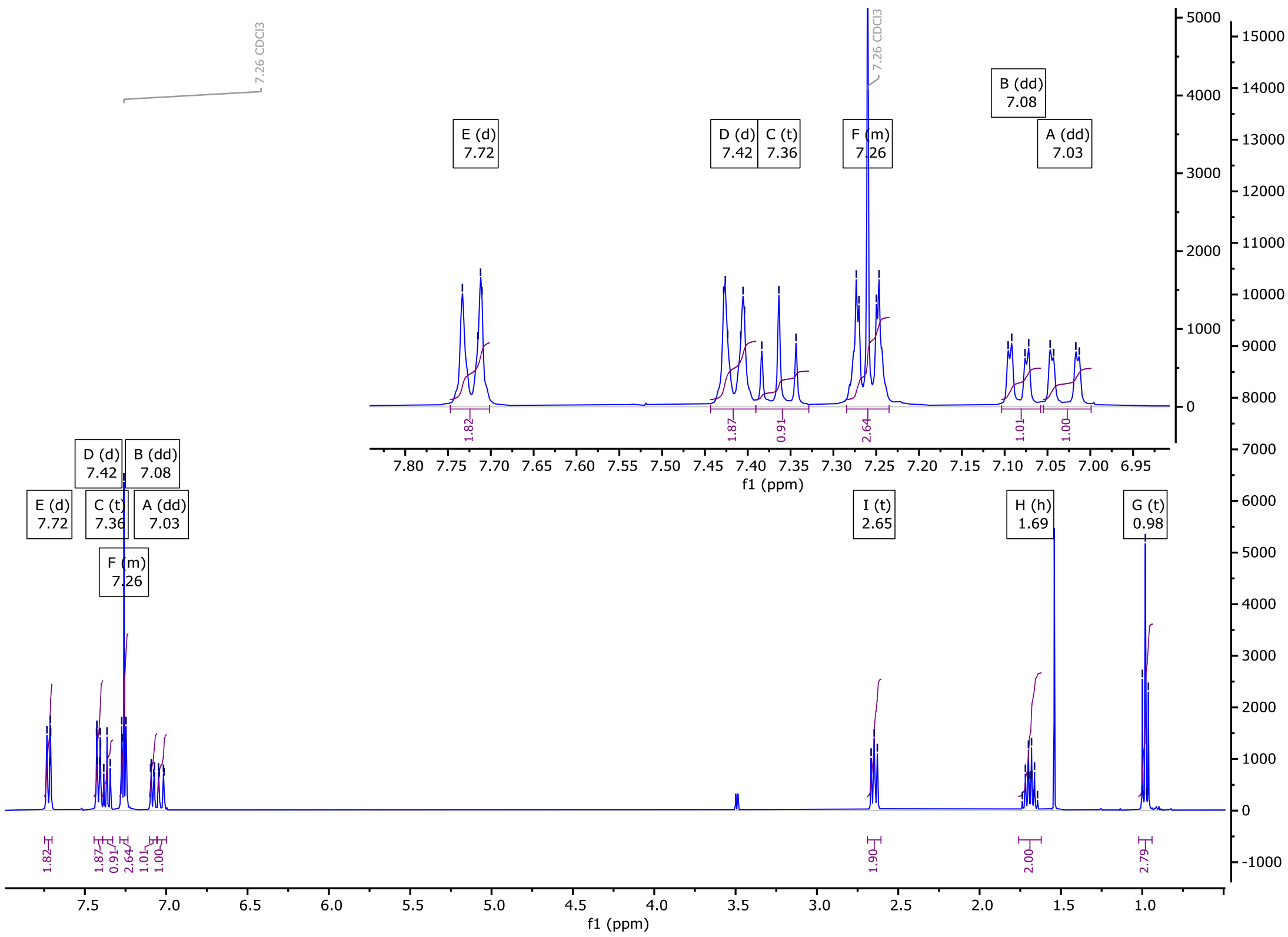

**Fig. S89.** $^{1}H$ NMR of **3h** in $CDCl_3$.

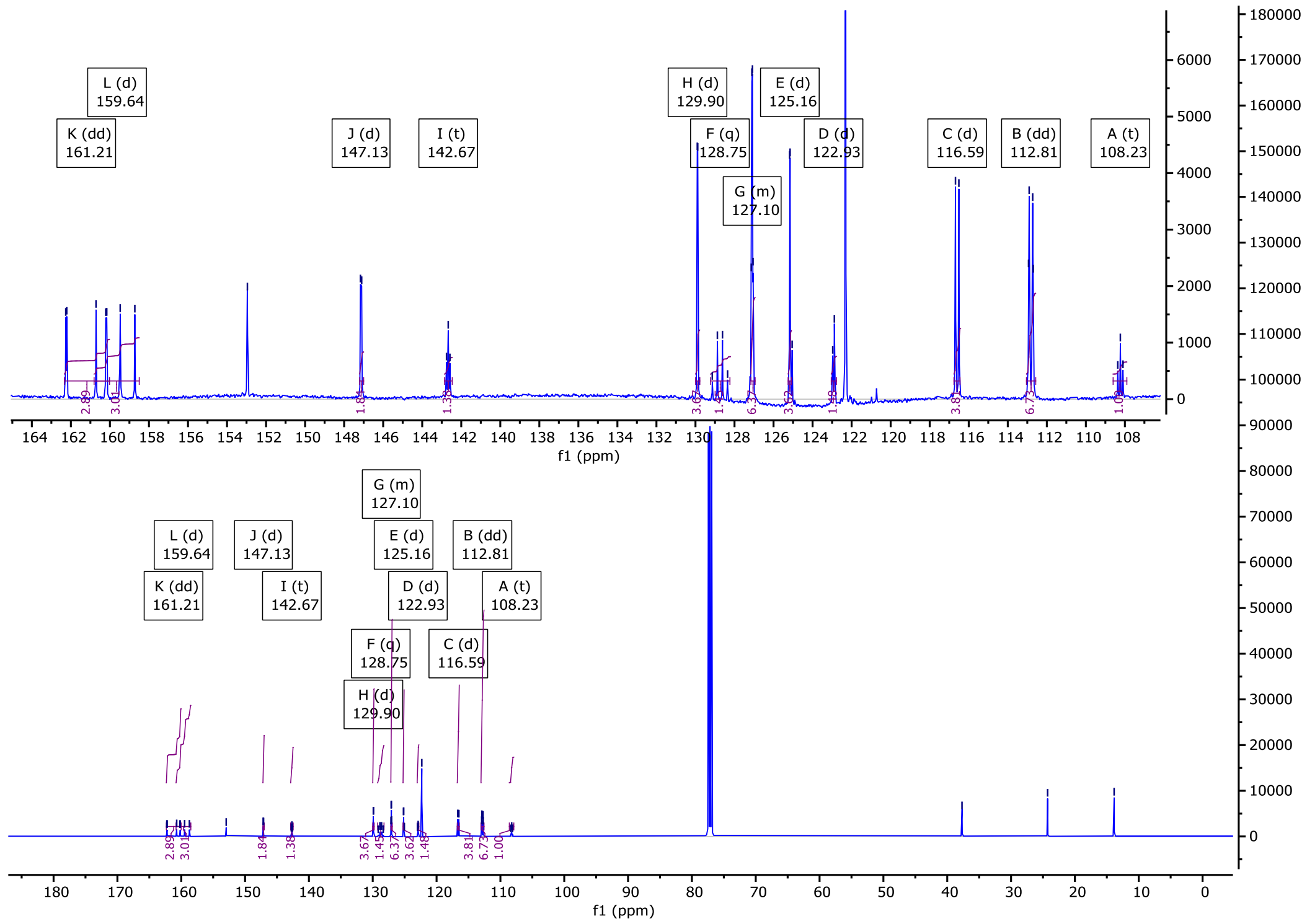

**Fig. S90.** $^{13}C\{^{1}H\}$ NMR of **3h** in $CDCl_3$.

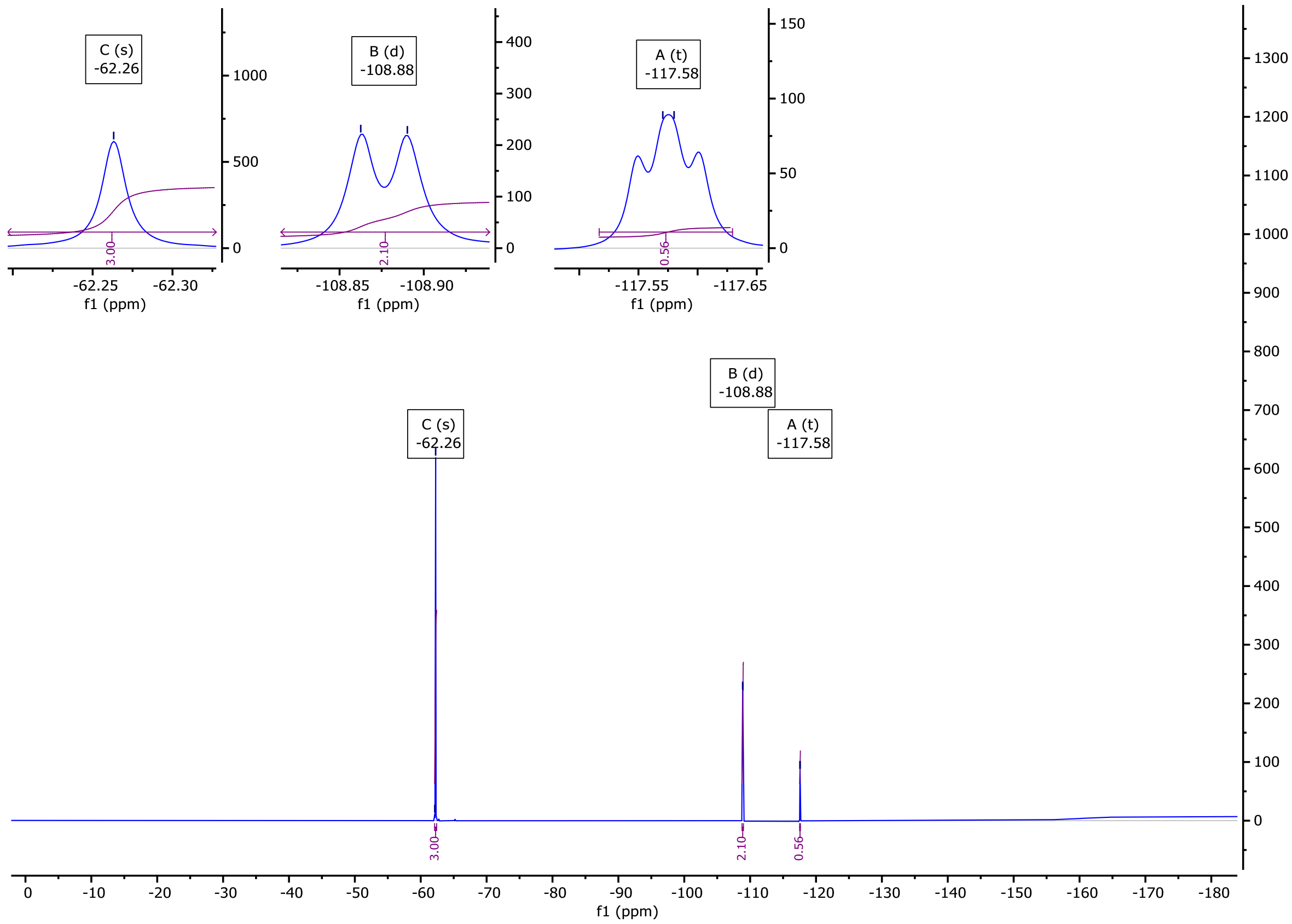

**Fig. S91.** $^{19}F$ NMR of **3h** in $CDCl_3$.

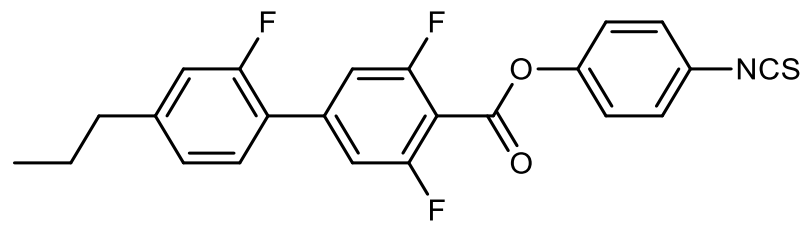

**3i**

*4-isothiocyanatophenyl 2',3,5-trifluoro-4'-propyl-[1,1'-biphenyl]-4-carboxylate*

Yield: (White Solid) 107 mg, 50 %

$^{1}$H NMR (501 MHz): 7.35 (d, *J* = 8.0 Hz, 1H, Ar-**H**), 7.32 – 7.23 (m, 6H, Ar-**H**)*†, 7.08 (dd, *J* = 7.8, 1.7 Hz, 1H, Ar-**H**), 7.03 (dd, *J* = 11.9, 1.6 Hz, 1H, Ar-**H**), 2.64 (t, *J* = 7.5 Hz, 2H, Ar-C**H₂**-CH₂), 1.69 (h, *J* = 7.5 Hz, 2H, CH₂-C**H₂**-CH₃), 0.98 (t, *J* = 7.3 Hz, 3H, CH₂-C**H₃**). * Overlapping signals, † Overlapping $CDCl_3$.

$^{13}$C{$^{1}$H} NMR (126 MHz): 161.17 (dd, *J* = 257.8, 6.4 Hz), 159.68 (d, *J* = 250.4 Hz), 158.72, 149.07, 147.11 (d, *J* = 7.9 Hz), 142.58 (t, *J* = 10.9 Hz), 136.37, 129.89 (d, *J* = 3.0 Hz), 129.47, 127.00, 125.16 (d, *J* = 3.1 Hz), 123.10, 122.93 (d, *J* = 12.7 Hz), 116.59 (d, *J* = 22.0 Hz), 112.79 (dt, *J* = 23.0, 4.0 Hz), 108.30 (t, *J* = 17.0 Hz), 37.68, 24.28, 13.86.

$^{19}$F NMR (376 MHz): -108.95 (d, $J_{F\text{-}H}$ = 10.2 Hz, 2F, Ar-**F**), -117.59 (t, $J_{F\text{-}H}$ = 10.1 Hz, 1F, Ar-**F**).

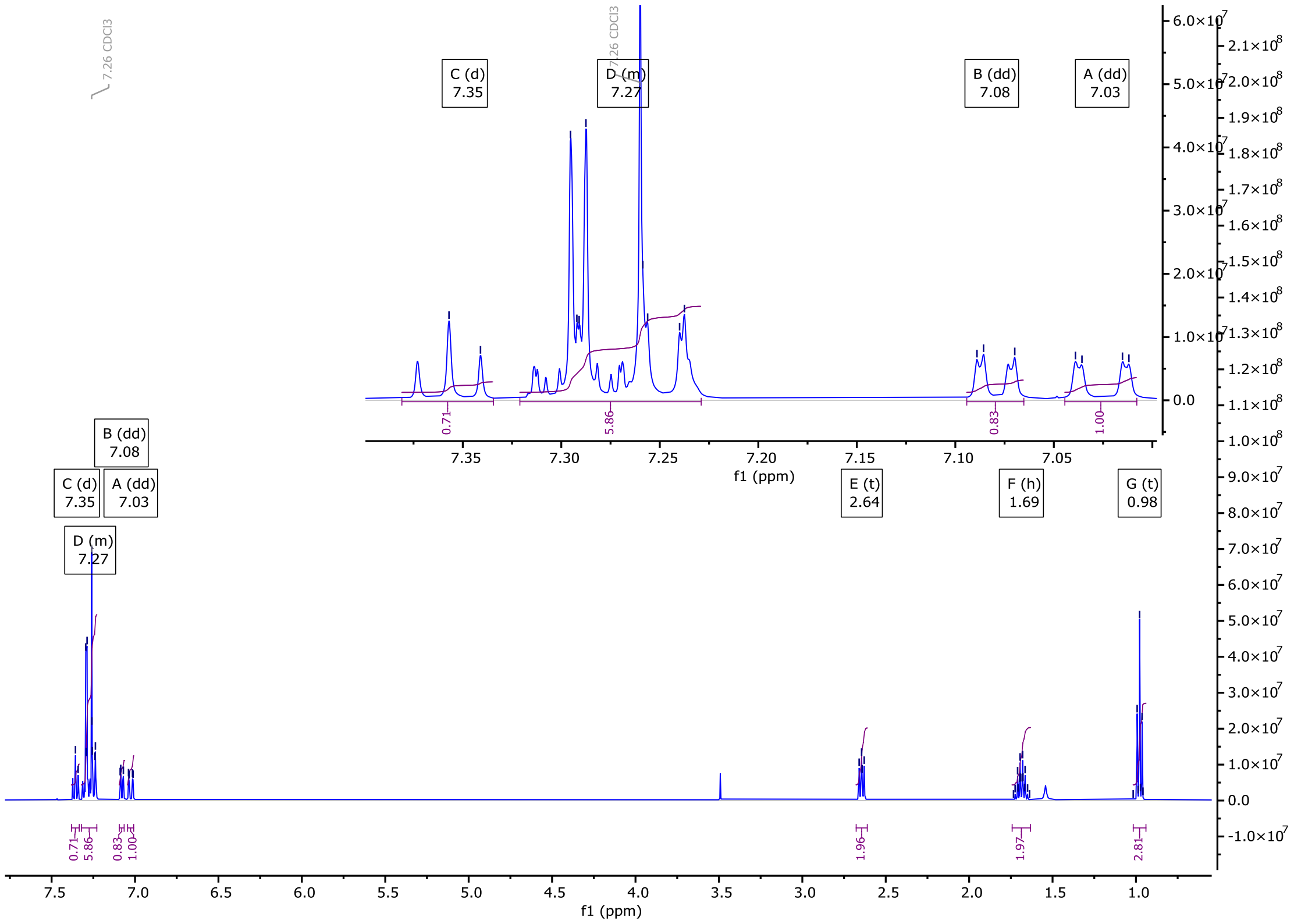

**Fig. S92.** $^{1}H$ NMR of **3i** in $CDCl_3$

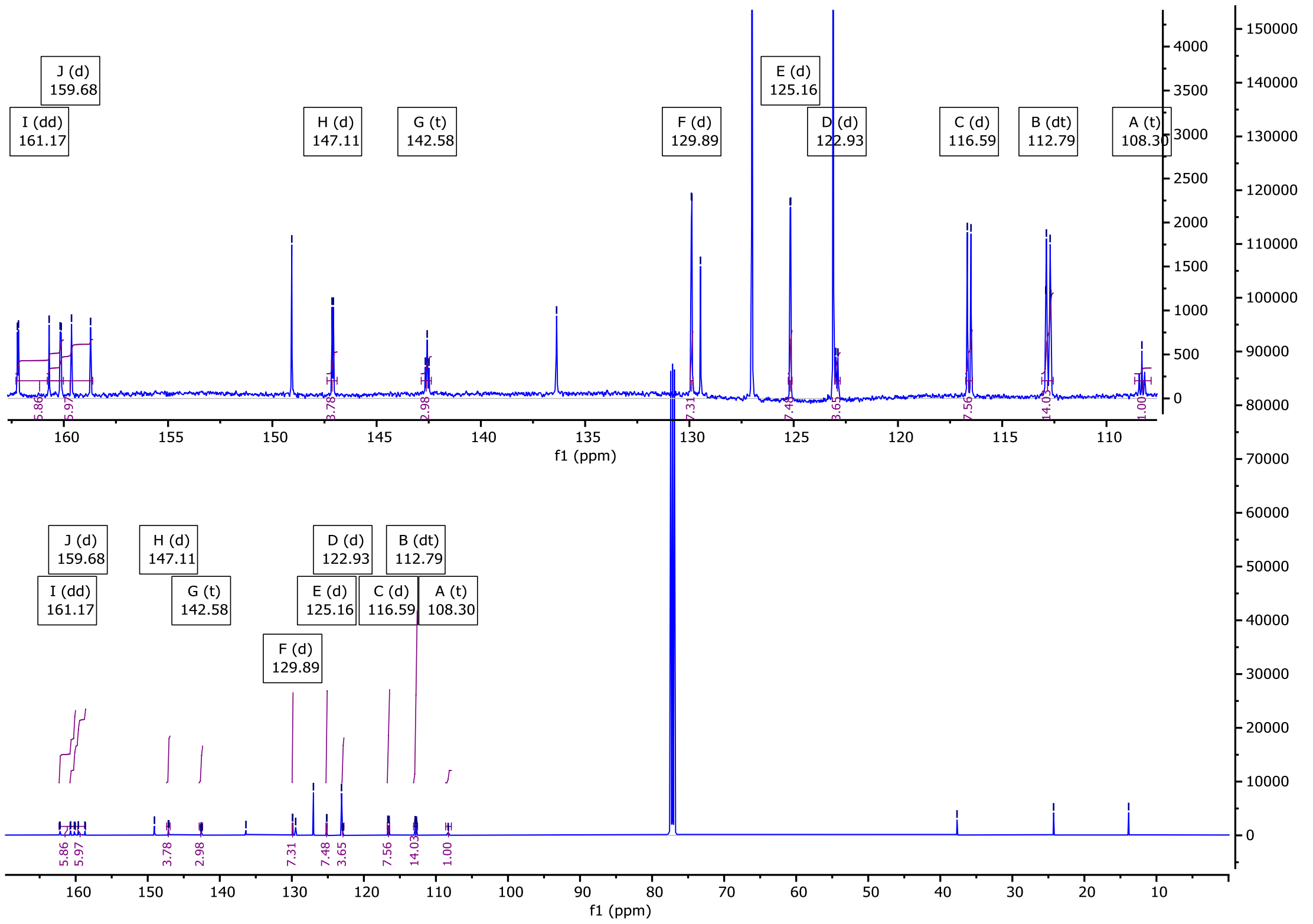

**Fig. S93.** $^{13}C\{^{1}H\}$ NMR of **3i** in $CDCl_3$.

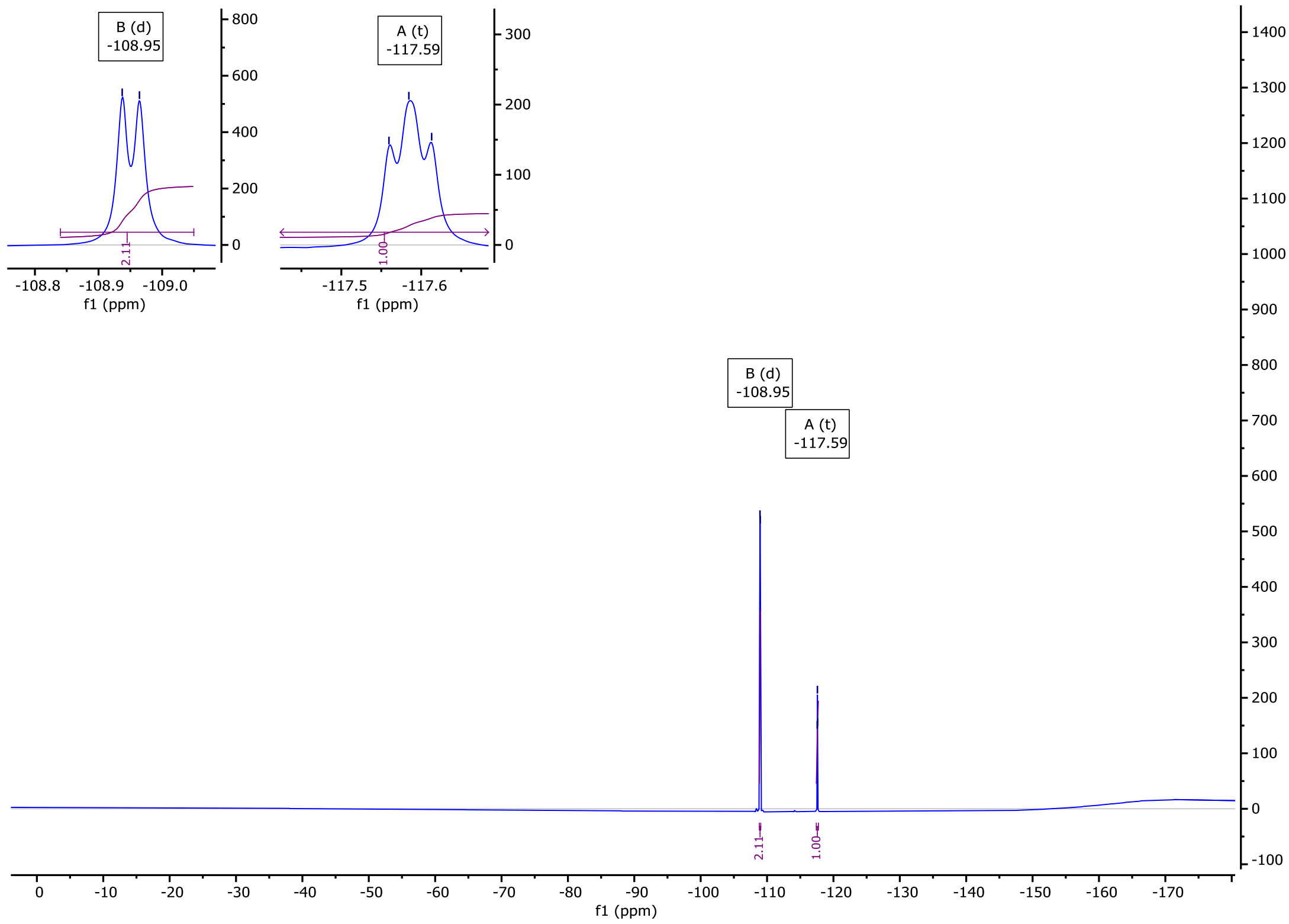

**Fig. S94.** $^{19}$F NMR of **3i** in $CDCl_3$

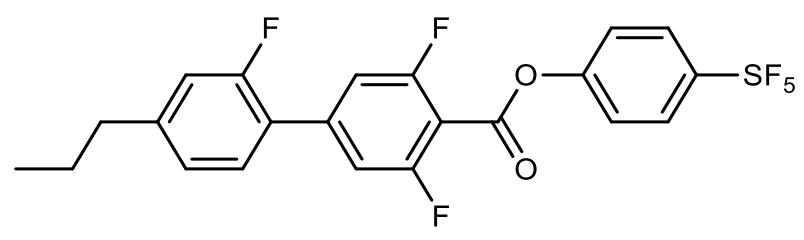

**3j**

*4-(pentafluoro-l6-sulfaneyl)phenyl 2',3,5-trifluoro-4'-propyl-[1,1'-biphenyl]-4-carboxylate*

| | |
|---|---|
| Yield: | (White Solid) 120 mg, 48 % |
| $^1$H NMR (400 MHz): | 7.85 (ddd, *J* = 9.1, 3.3, 2.1 Hz, 2H, Ar-**H**), 7.40 (d, *J* = 8.9 Hz, 2H, Ar-**H**), 7.36 (t, *J* = 7.9 Hz, 1H, Ar-**H**), 7.29 – 7.23 (m, 2H, Ar-**H**) †, 7.09 (dd, *J* = 7.9, 1.7 Hz, 1H, Ar-**H**), 7.03 (dd, *J* = 12.0, 1.6 Hz, 1H, Ar-**H**), 2.65 (t, *J* = 7.5 Hz, 2H, Ar-C**H₂**-CH₂), 1.69 (h, *J* = 7.5 Hz, 2H, CH₂-C**H₂**-CH₃), 0.98 (t, *J* = 7.3 Hz, 3H, CH₂-C**H₃**). † Overlapping $CDCl_3$. |
| $^{13}$C{$^1$H} NMR (101 MHz): | 161.08 (dd, *J* = 258.2, 6.1 Hz), 159.43 (d, *J* = 250.2 Hz), 158.34, 152.15, 147.05 (d, *J* = 7.9 Hz), 142.70 (t, *J* = 10.8 Hz), 129.75 (d, *J* = 3.0 Hz), 127.71 (t, *J* = 4.7 Hz), 125.04 (d, *J* = 3.2 Hz), 121.96, 116.47 (d, *J* = 21.9 Hz), 112.71 (dt, *J* = 23.8, 3.6 Hz), 37.54, 24.13, 13.71. |
| $^{19}$F NMR (376 MHz): | 83.96 (p, $J_{F\text{-}S}$ = 151.7 Hz, 1F, S-$\mathbf{F_{\Lambda}}$), 63.47 (d, $J_{F\text{-}S}$ = 150.0 Hz, 4F, S-$\mathbf{F_{4\Delta}}$), -108.70 (d, $J_{F\text{-}H}$ = 10.4 Hz, 2F, Ar-**F**), -117.55 (t, $J_{F\text{-}H}$ = 9.9 Hz, 1F, Ar-**F**). |

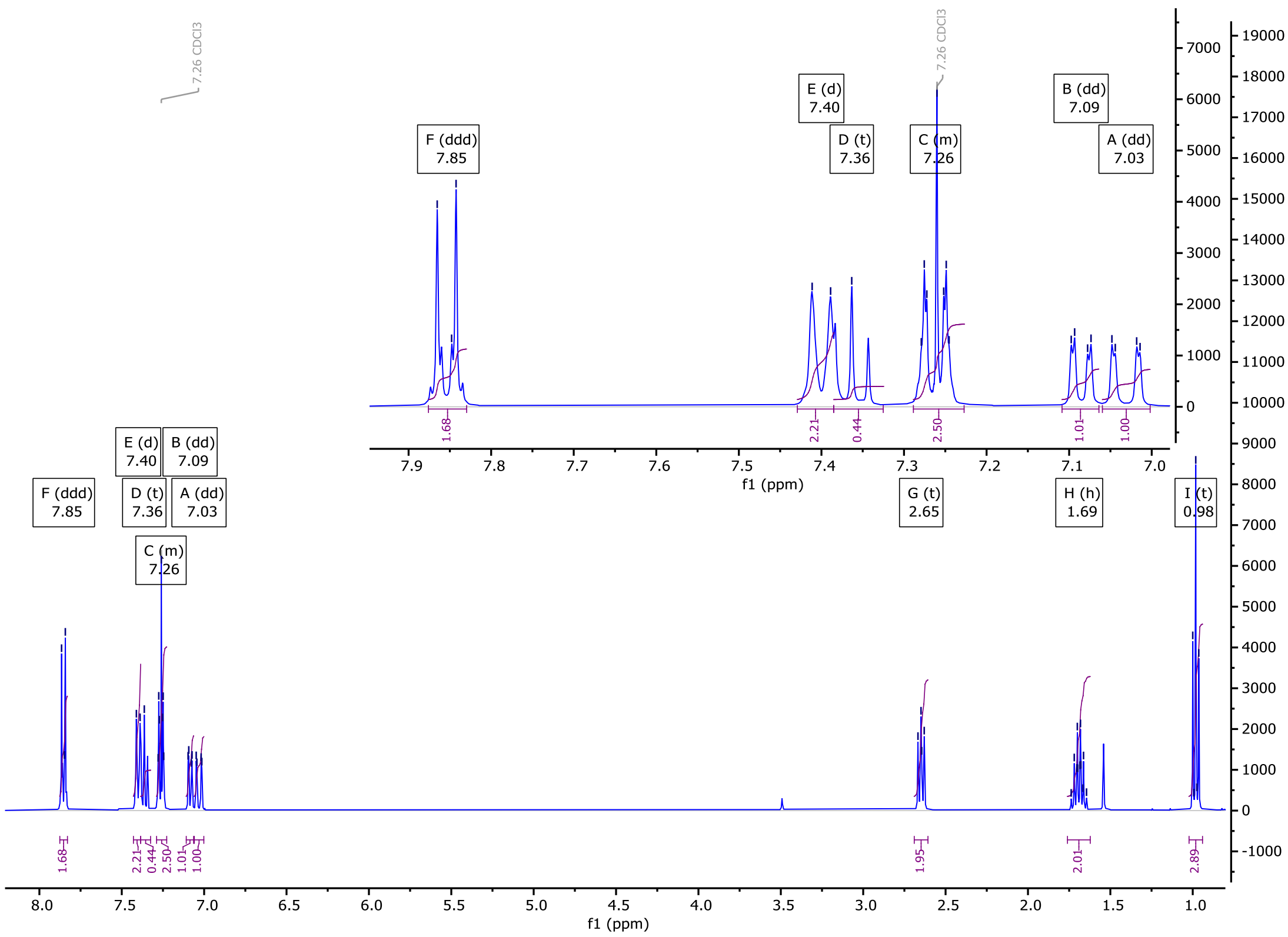

**Fig. S95.** $^{1}H$ NMR of **3j** in $CDCl_3$.

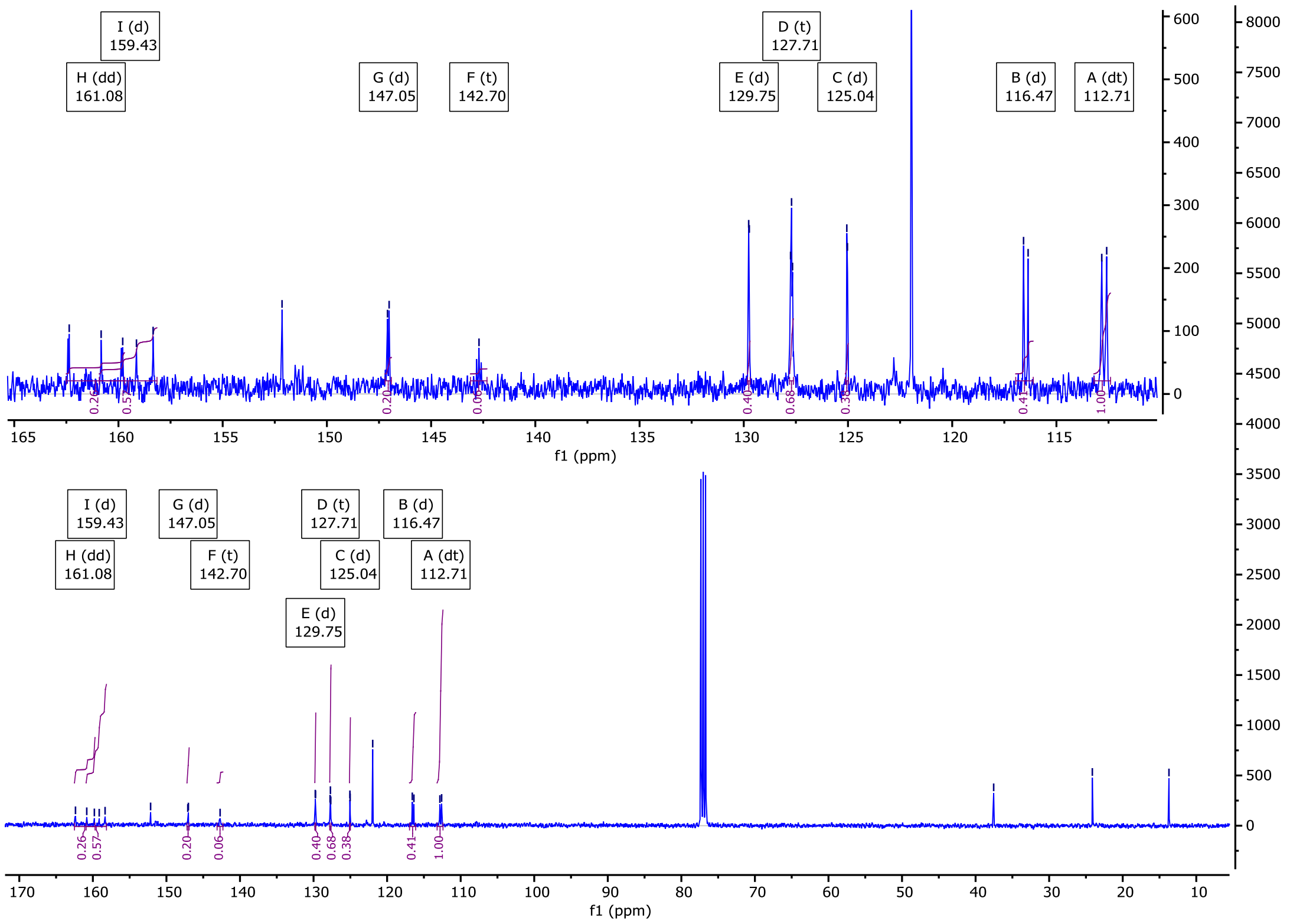

**Fig. S96.** $^{13}C\{^{1}H\}$ NMR of **3j** in $CDCl_3$.

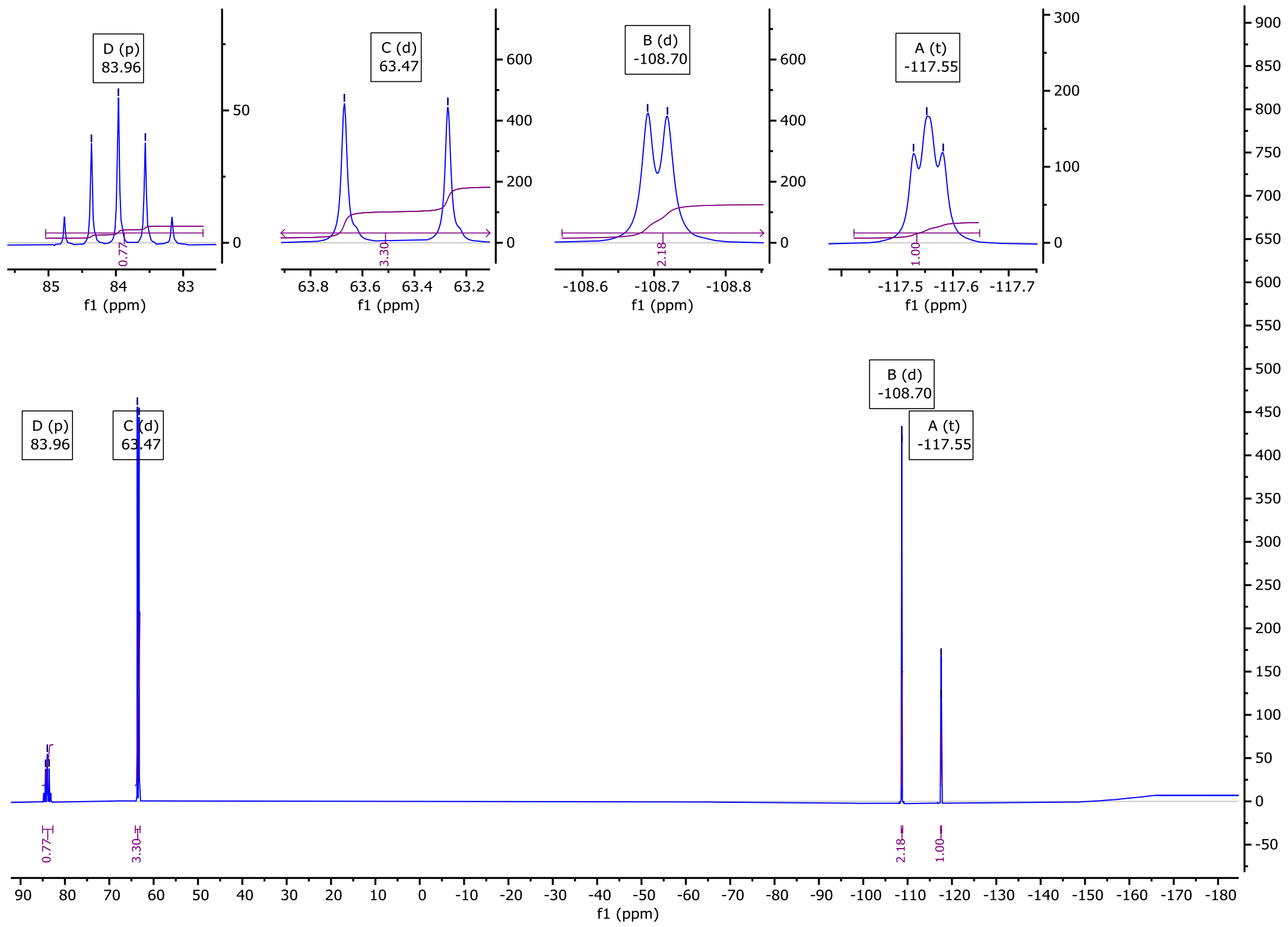

**Fig. S97.** $^{19}F$ NMR of **3j** in $CDCl_3$.

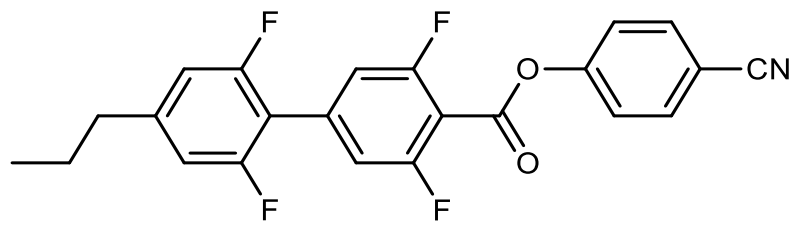

**4a**

*4-cyanophenyl 2',3,5,6'-tetrafluoro-4'-propyl-[1,1'-biphenyl]-4-carboxylate*

Yield: (Whie Solid) 114 mg, 55 %

$^{1}$H NMR (501 MHz): 7.76 (ddd, *J* = 8.8, 2.5, 1.9 Hz, 2H, Ar-**H**), 7.43 (ddd, *J* = 8.7, 2.4, 2.0 Hz, 2H, Ar-**H**), 7.21 – 7.17 (m, 2H, Ar-**H**), 6.86 (ddd, *J* = 9.0, 2.9, 2.0 Hz, 2H, Ar-**H**), 2.63 (t, *J* = 7.4 Hz, 2H, Ar-C**H₂**-CH₂), 1.68 (h, *J* = 7.3 Hz, 2H, CH₂-C**H₂**-CH₃), 0.98 (t, *J* = 7.4 Hz, 3H, CH₂-C**H₃**).

$^{13}$C{$^{1}$H} NMR (126 MHz): 160.90 (dd, *J* = 258.3, 6.3 Hz), 159.45 (dd, *J* = 250.9, 6.5 Hz), 158.98, 153.65, 147.45 (t, *J* = 9.5 Hz), 136.53 (t, *J* = 11.2 Hz), 133.97, 122.94, 118.28, 114.62 (dd, *J* = 24.2, 2.3 Hz), 112.48 (t, *J* = 17.9 Hz), 112.13 (dd, *J* = 20.1, 4.9 Hz), 110.48, 108.66 (t, *J* = 16.7 Hz), 37.82, 23.97, 13.75.

$^{19}$F NMR (376 MHz): -109.09 (d, $J_{F-H}$ = 10.1 Hz, 2F, Ar-**F**), -114.94 (d, $J_{F-H}$ = 9.7 Hz, 2F, Ar-**F**).

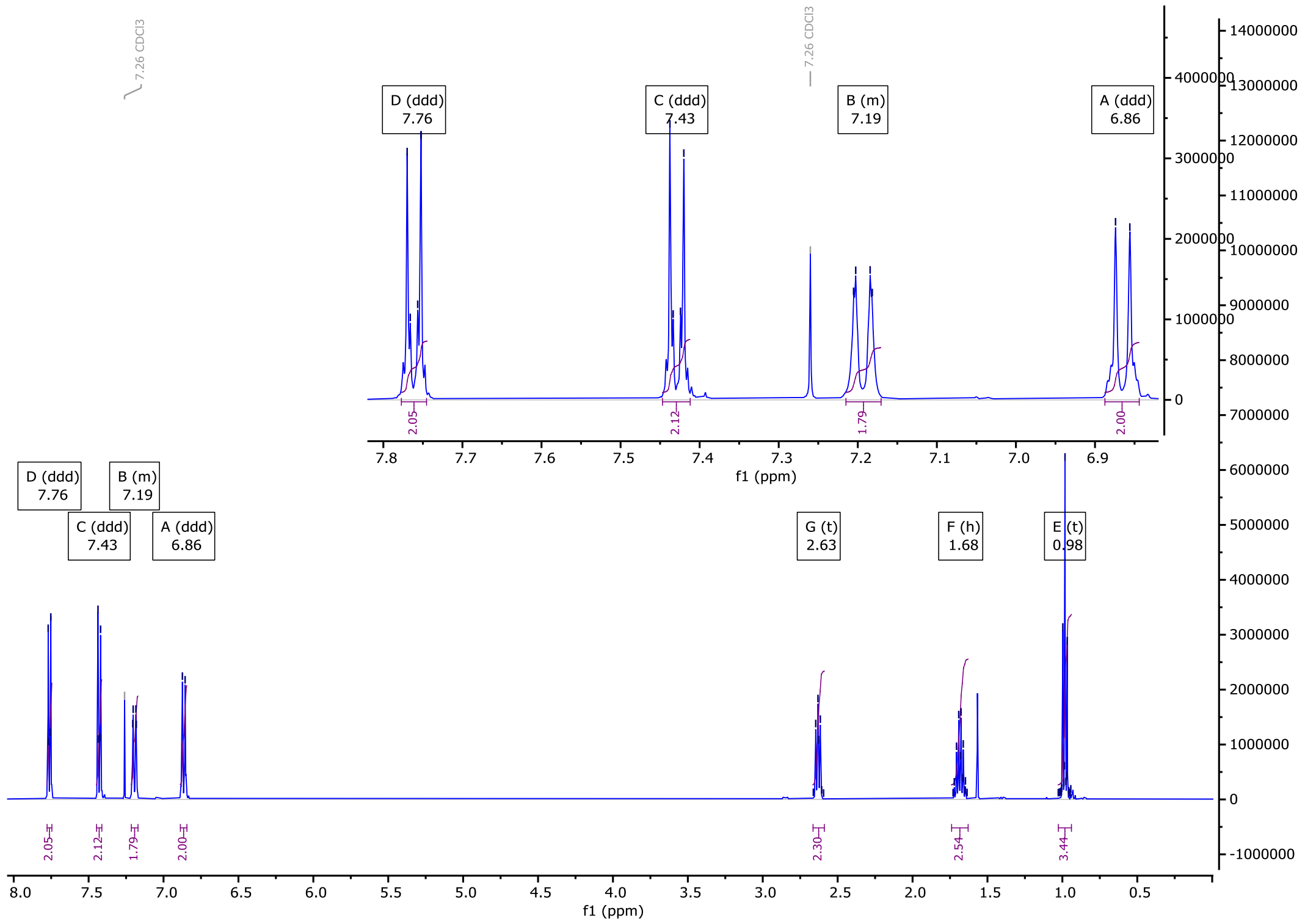

**Fig. S98.** $^{1}H$ NMR of **4a** in $CDCl_3$.

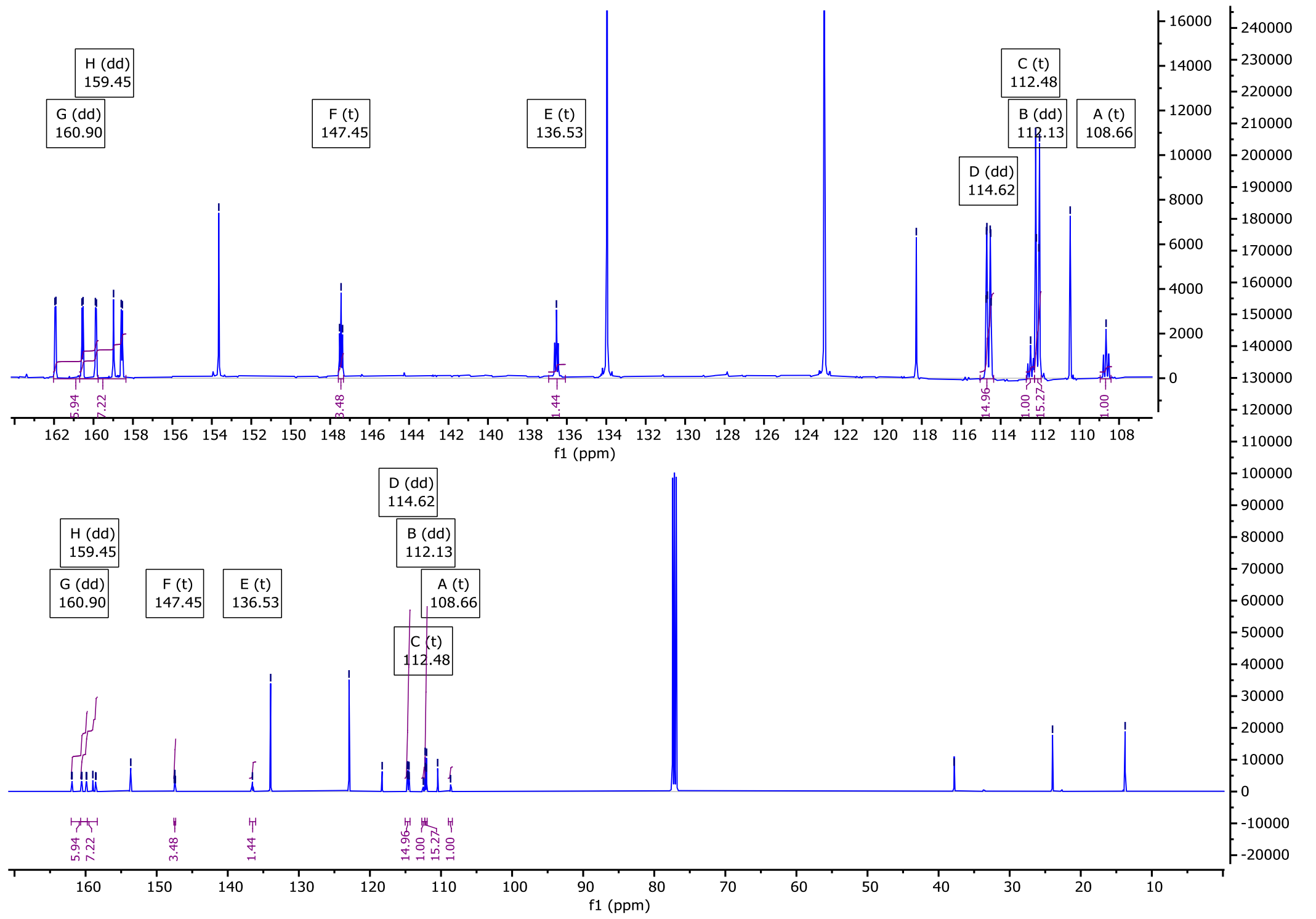

**Fig. S99.** $^{13}C\{^{1}H\}$ NMR of **4a** in $CDCl_3$.

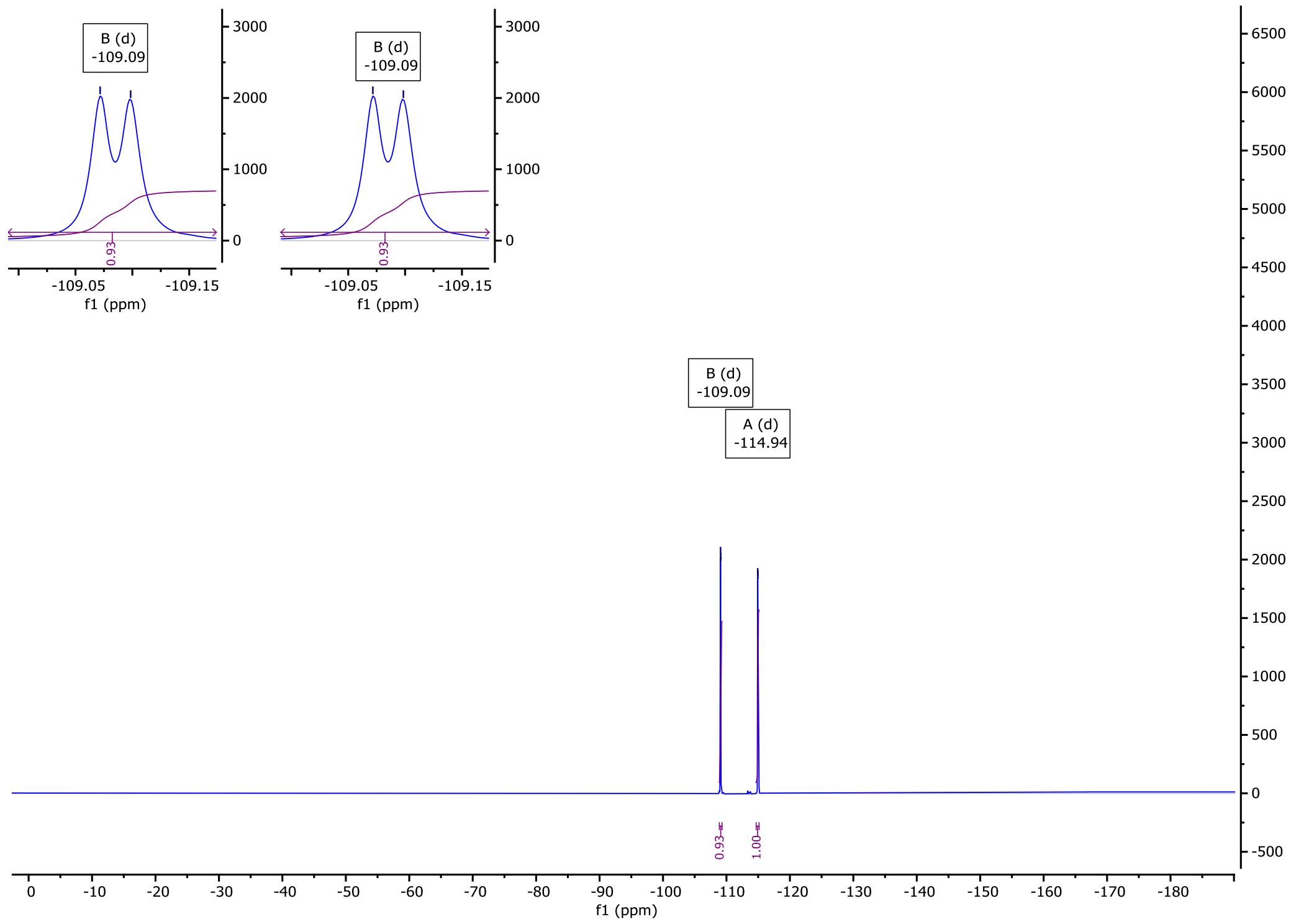

**Fig. S100.** $^{19}$F NMR of **4a** in $CDCl_3$.

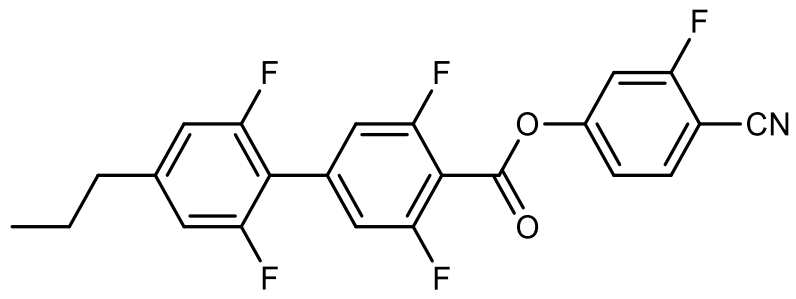

**4b**

*4-cyano-3-fluorophenyl 2',3,5,6'-tetrafluoro-4'-propyl-[1,1'-biphenyl]-4-carboxylate*

Yield: (White Solid) 95 mg, 44 %

$^1$H NMR (501 MHz): 7.72 (t, *J* = 7.7 Hz, 1H, Ar-**H**), 7.30 – 7.23 (m, 2H, Ar-**H**)$^†$, 7.20 (d, *J* = 10.1 Hz, 2H, Ar-**H**), 6.87 (d, *J* = 9.1 Hz, 2H, Ar-**H**), 2.63 (t, *J* = 7.6 Hz, 2H, Ar-C**H₂**-CH₂), 1.68 (h, *J* = 7.4 Hz, 2H, CH₂-C**H₂**-CH₃), 0.98 (t, *J* = 7.3 Hz, 3H, CH₂-C**H₃**). $^†$ Overlapping $CDCl_3$.

$^{13}$C{$^1$H} NMR (126 MHz): 163.77 (d, *J* = 261.1 Hz), 160.98 (dd, *J* = 258.8, 6.1 Hz), 159.56 (dd, *J* = 250.6, 6.9 Hz), 158.50, 154.88 (d, *J* = 10.7 Hz), 147.56 (t, *J* = 9.5 Hz), 136.91 (t, *J* = 11.4 Hz), 134.38, 118.77 (d, *J* = 3.7 Hz), 114.69 (dd, *J* = 23.6, 2.4 Hz), 112.48 (t, *J* = 17.5 Hz), 112.16 (dd, *J* = 20.5, 4.6 Hz), 111.13 (d, *J* = 22.9 Hz), 108.19 (t, *J* = 16.3 Hz), 99.62 (d, *J* = 15.6 Hz), 37.82, 23.98, 13.76.

$^{19}$F NMR (376 MHz): -102.94 (t, $J_{F-H}$ = 8.2 Hz, 1F, Ar-**F**), -108.75 (d, $J_{F-H}$ = 10.4 Hz, 2F, Ar-**F**), -114.93 (d, $J_{F-H}$ = 9.6 Hz, 2F, Ar-**F**).

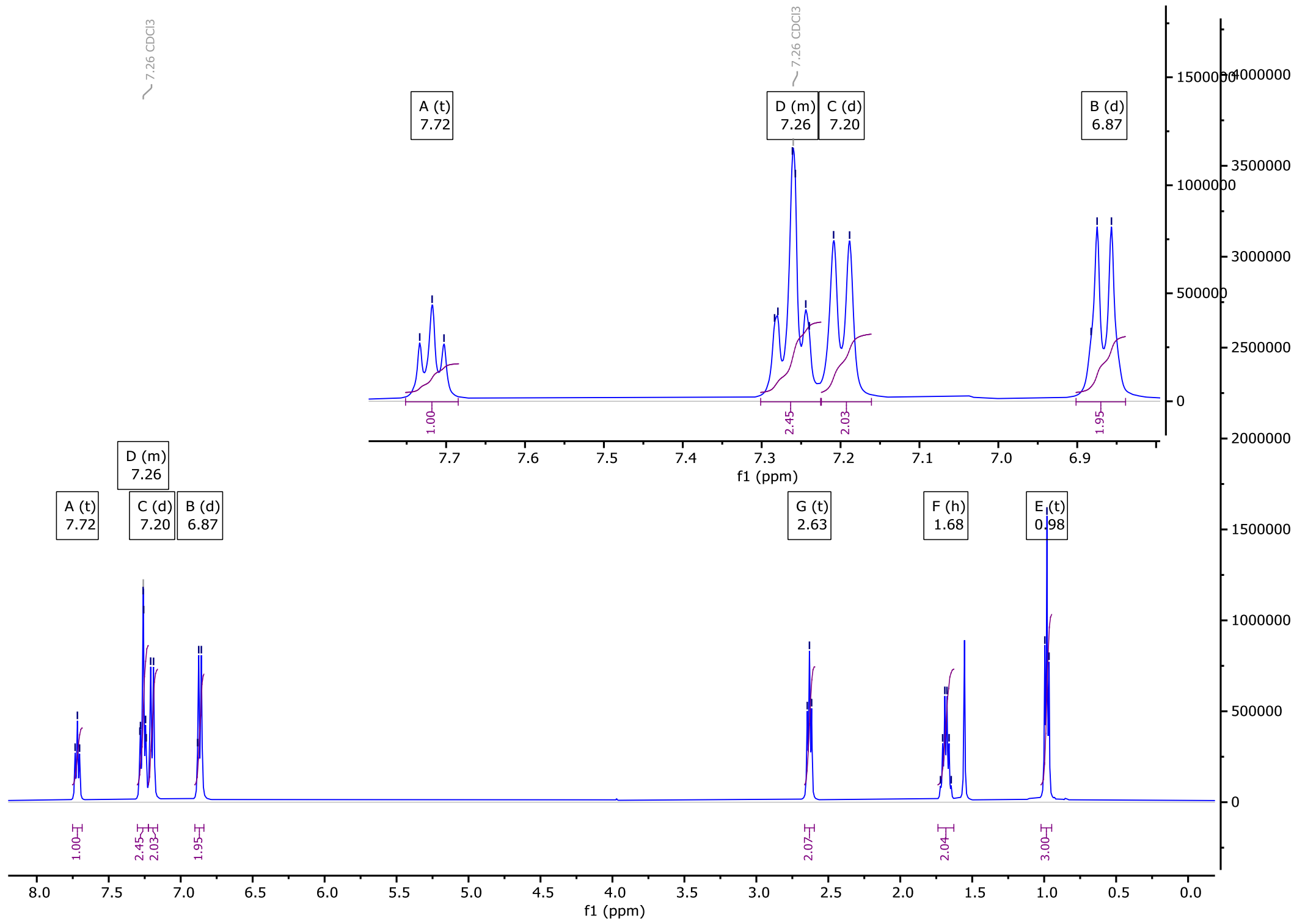

**Fig. S101.** $^{1}H$ NMR of **4b** in $CDCl_3$.

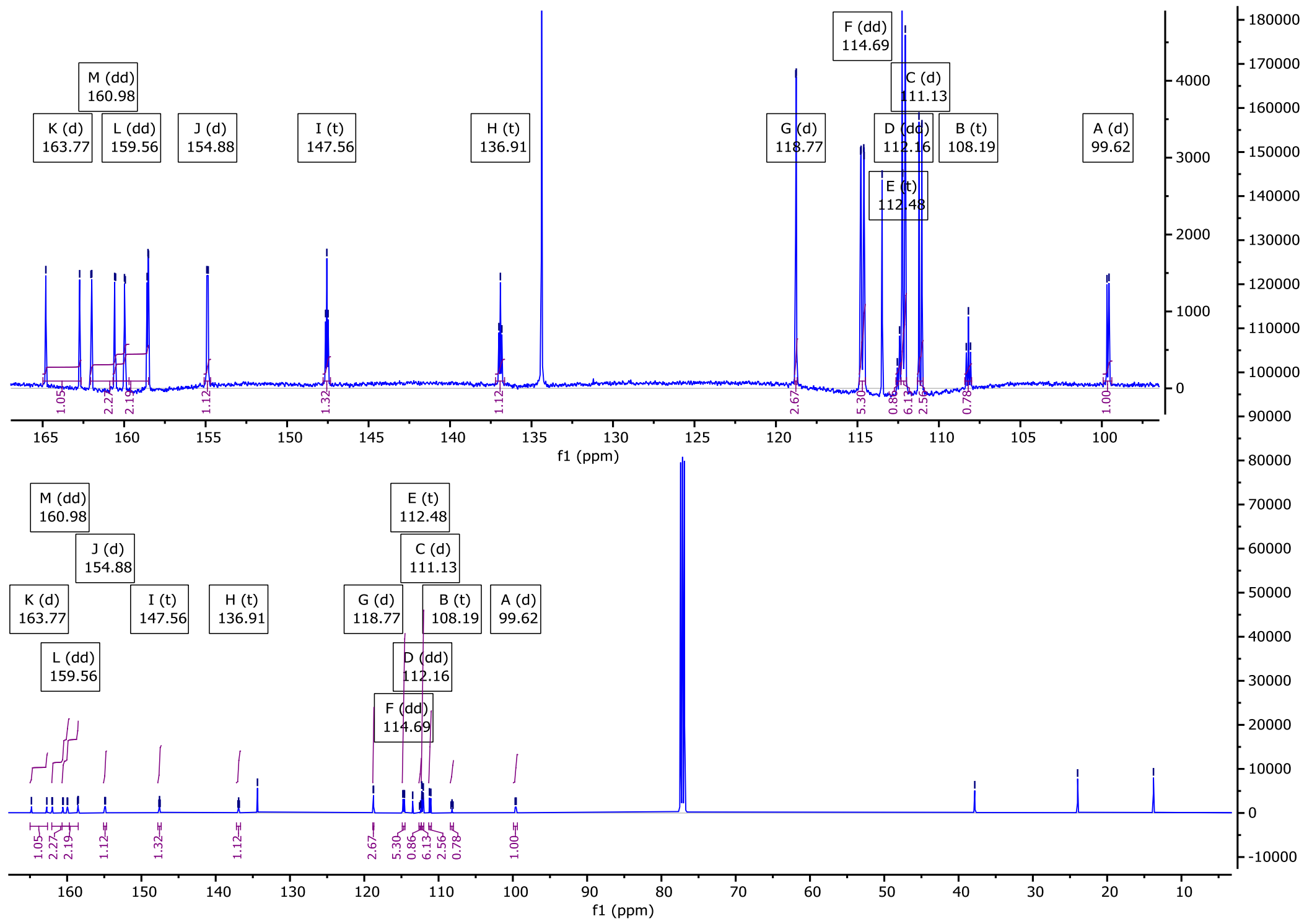

**Fig. S102.** $^{13}C\{^{1}H\}$ NMR of **4b** in $CDCl_3$.

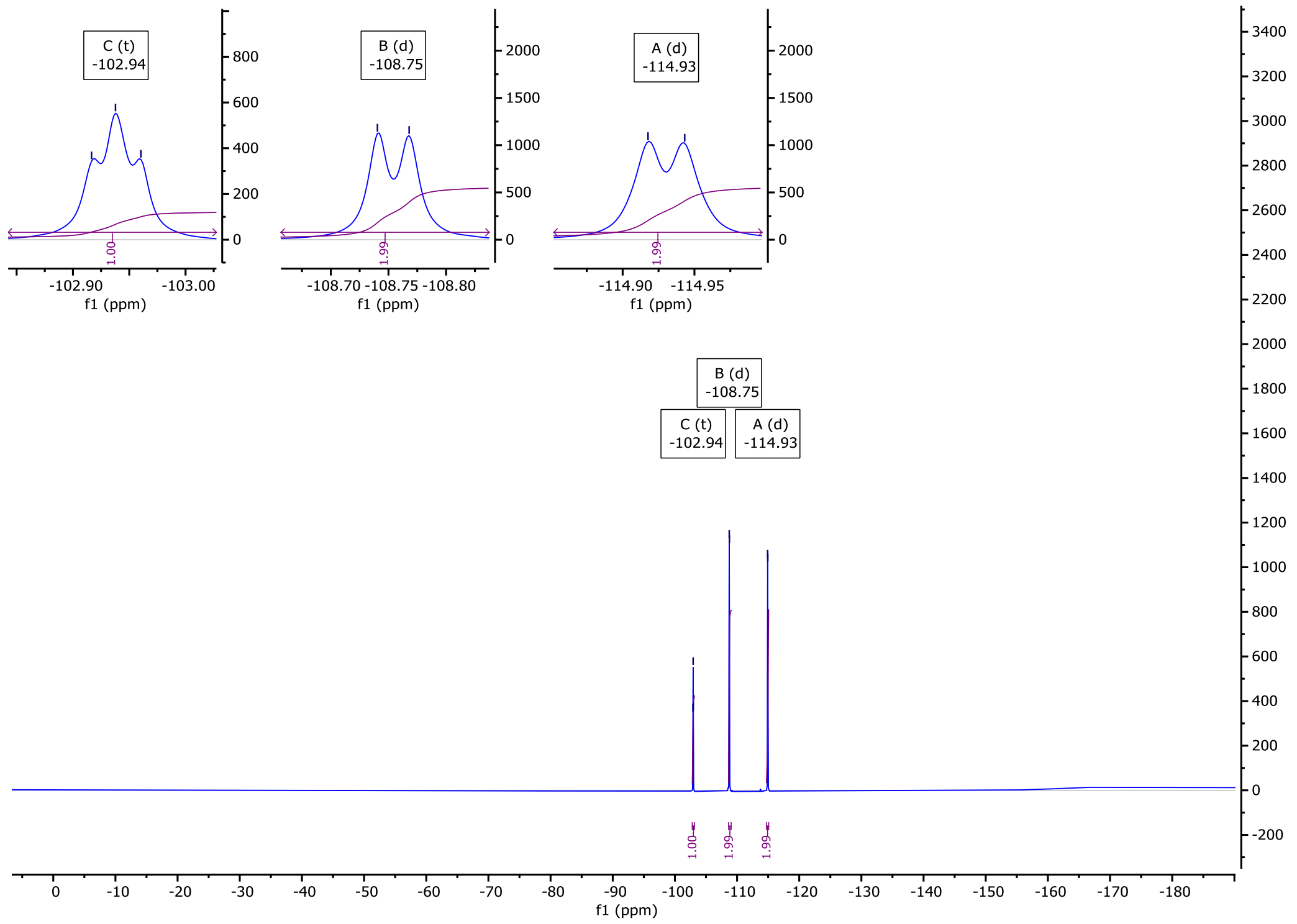

**Fig. S103.** $^{19}F$ NMR of **4b** in $CDCl_3$.

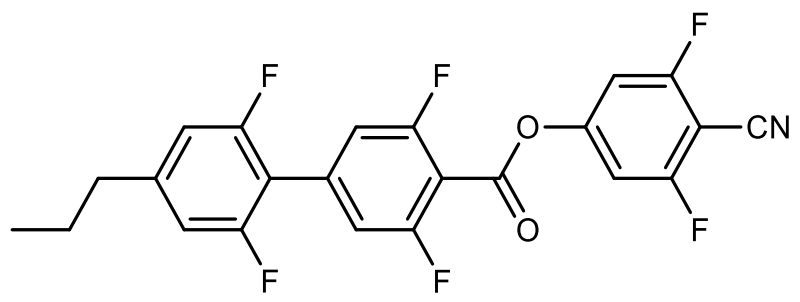

**4c**

*4-cyano-3,5-difluorophenyl 2',3,5,6'-tetrafluoro-4'-propyl-[1,1'-biphenyl]-4-carboxylate*

Yield: (White Solid) 95 mg, 42 %

$^1$H NMR (501 MHz): 7.21 (ddd, *J* = 9.2, 1.4, 1.2 Hz, 2H, Ar-**H**), 7.14 – 7.10 (m, 2H, Ar-**H**), 6.87 (ddd, *J* = 9.2, 2.7, 2.3 Hz, 2H, Ar-**H**), 2.63 (t, *J* = 7.3 Hz, 2H, Ar-C**H₂**-CH₂), 1.68 (h, *J* = 7.5 Hz, 2H, CH₂-C**H₂**-CH₃), 0.98 (t, *J* = 7.3 Hz, 3H, CH₂-C**H₃**).

$^{13}$C{$^1$H} NMR (126 MHz): 163.71 (dd, *J* = 261.9, 6.5 Hz), 161.40 (dd, *J* = 259.7, 5.7 Hz), 159.55 (dd, *J* = 250.7, 7.0 Hz), 158.04, 155.25 (t, *J* = 13.5 Hz), 147.67 (t, *J* = 9.5 Hz), 137.29 (t, *J* = 11.3 Hz), 114.73 (dd, *J* = 24.1, 2.6 Hz), 112.40 – 111.95 (m), 108.77, 107.74 (t, *J* = 16.7 Hz), 107.03 (dd, *J* = 23.2, 4.0 Hz), 90.75 (t, *J* = 19.4 Hz), 37.83, 23.98, 13.77.

$^{19}$F NMR (376 MHz): -101.50 (d, $J_{F\text{-}H}$ = 8.7 Hz, 2F, Ar-**F**), -108.43 (d, $J_{F\text{-}H}$ = 10.3 Hz, 2F, Ar-**F**), -114.91 (d, $J_{F\text{-}H}$ = 9.7 Hz, 2F, Ar-**F**).

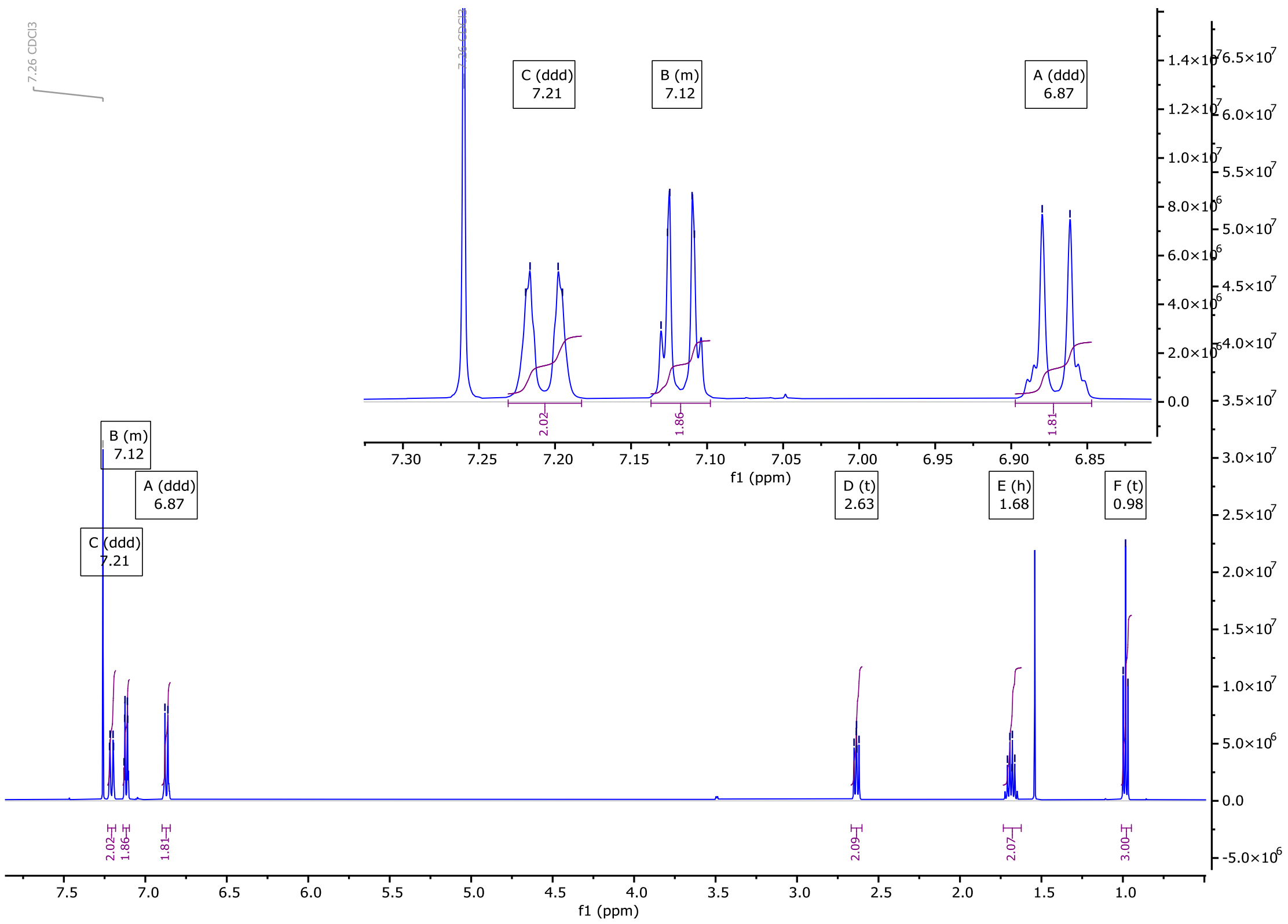

**Fig. S104.** $^1H$ NMR of **4c** in $CDCl_3$.

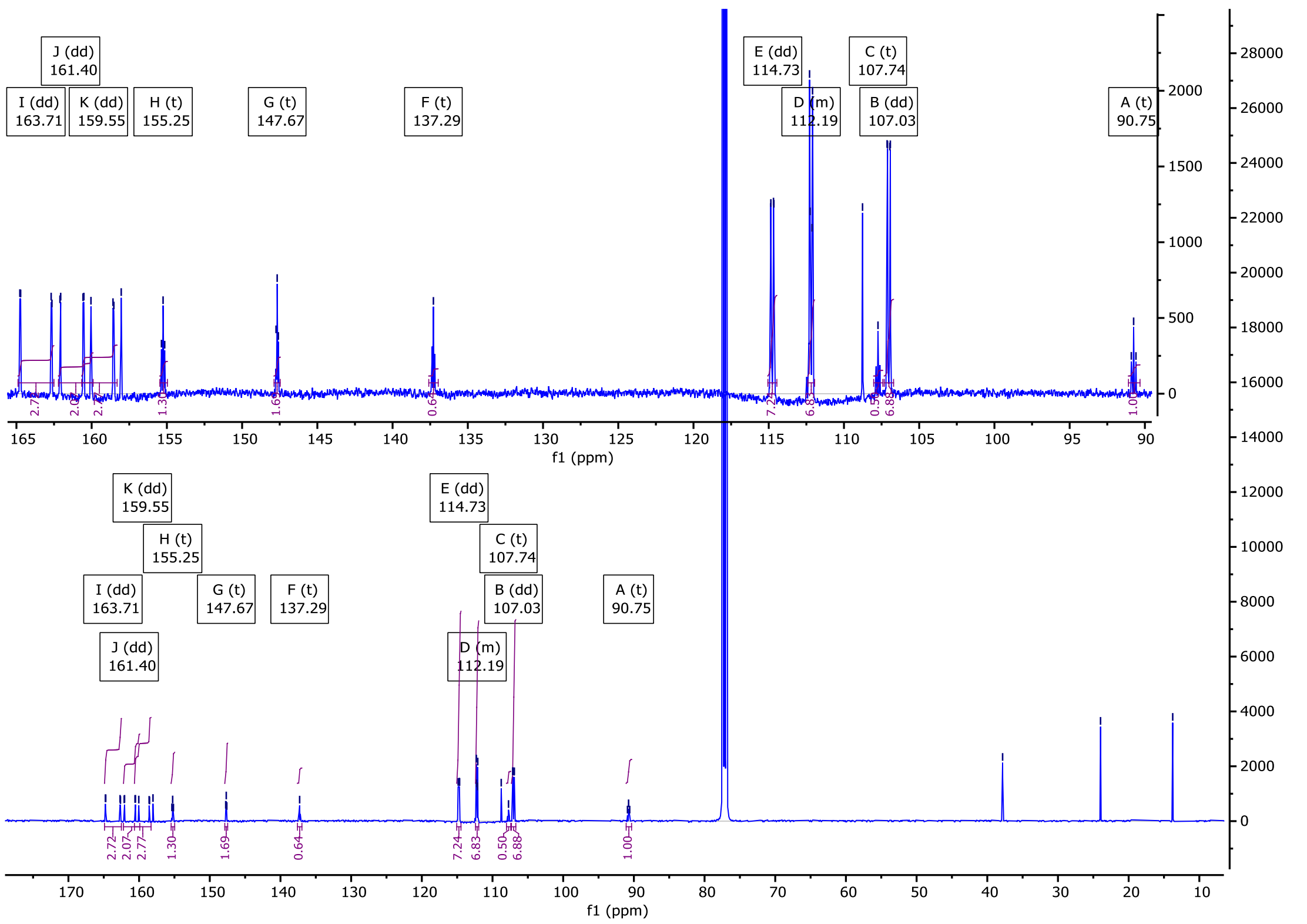

**Fig. S105.** $^{13}C\{^1H\}$ NMR of **4c** in $CDCl_3$.

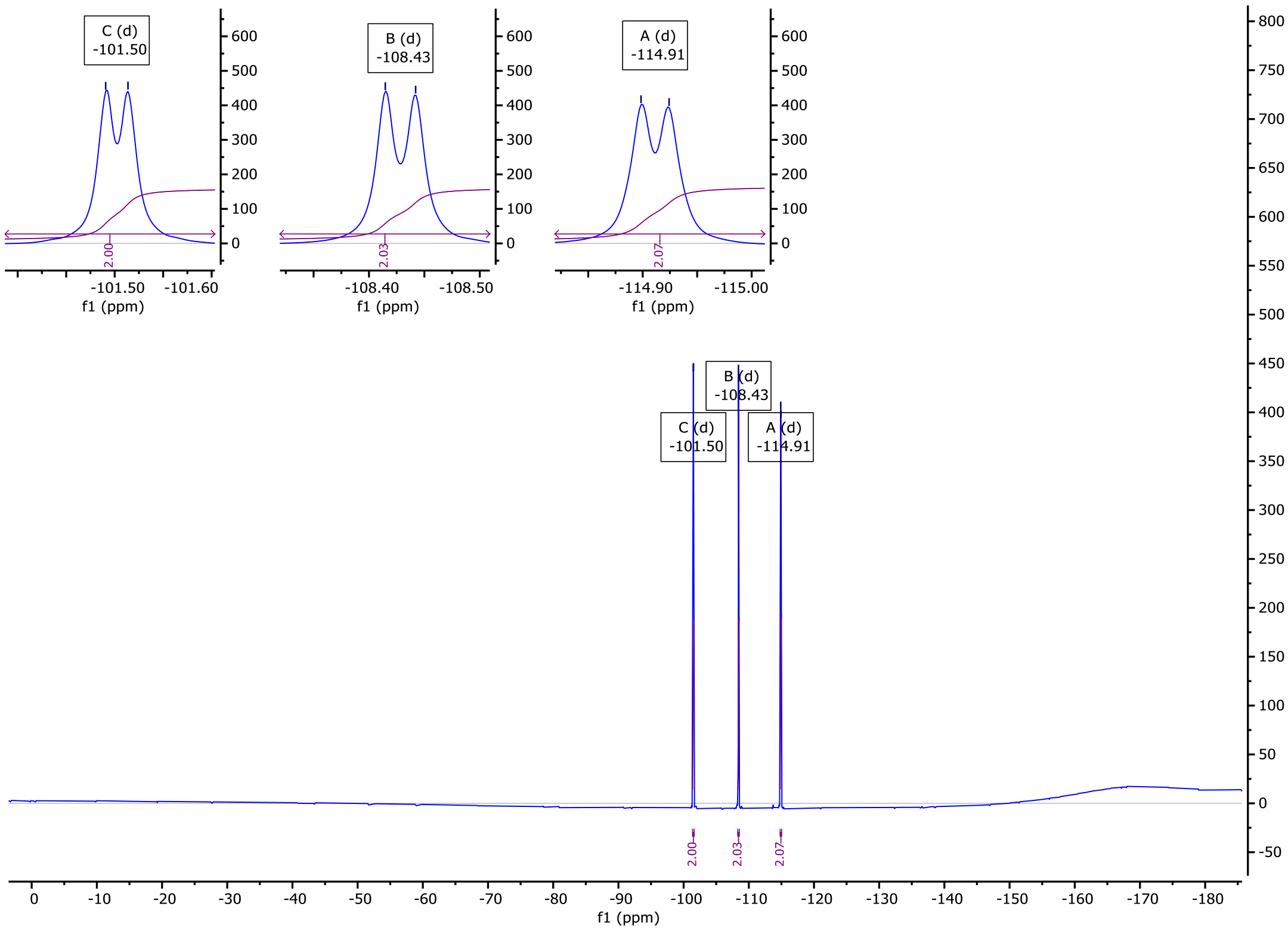

**Fig. S106.** $^{19}F$ NMR of **4c** in $CDCl_3$.

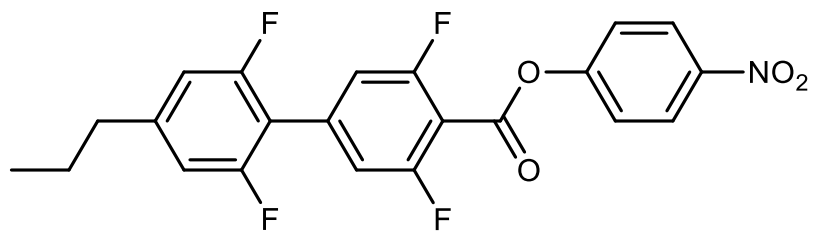

**4d**

*4-nitrophenyl 2',3,5,6'-tetrafluoro-4'-propyl-[1,1'-biphenyl]-4-carboxylate*

| | |
|---|---|
| Yield: | (White Solid) 82 mg, 38 % |
| $^{1}$H NMR (501 MHz): | 8.34 (ddd, *J* = 9.1, 2.9, 2.2 Hz, 2H, Ar-**H**), 7.48 (ddd, *J* = 9.2, 3.3, 2.1 Hz, 2H, Ar-**H**), 7.23 – 7.18 (m, 2H, Ar-**H**), 6.87 (ddd, *J* = 9.0, 2.8, 2.4 Hz, 2H, Ar-**H**), 2.63 (t, *J* = 7.5 Hz, 2H, Ar-C**H₂**-CH₂), 1.69 (h, *J* = 7.4 Hz, 2H, CH₂-C**H₂**-CH₃), 0.99 (t, *J* = 7.4 Hz, 3H, CH₂-C**H₃**). |
| $^{13}$C{$^{1}$H} NMR (126 MHz): | 160.95 (dd, *J* = 258.5, 6.2 Hz), 159.44 (dd, *J* = 250.5, 7.0 Hz), 158.90, 158.61, 158.55, 155.07, 147.49 (t, *J* = 9.5 Hz), 145.86, 136.66 (t, *J* = 11.4 Hz), 125.52, 122.70, 114.66 (d, *J* = 2.5 Hz), 112.41 (t, *J* = 17.8 Hz), 112.15 (dd, *J* = 20.4, 5.0 Hz), 108.55 (t, *J* = 16.5 Hz), 37.82, 23.98, 13.77. |
| $^{19}$F NMR (376 MHz): | -108.96 (d, $J_{F-H}$ = 10.2 Hz, 2F, Ar-**F**), -114.93 (d, $J_{F-H}$ = 9.8 Hz, 2F, Ar-**F**). |

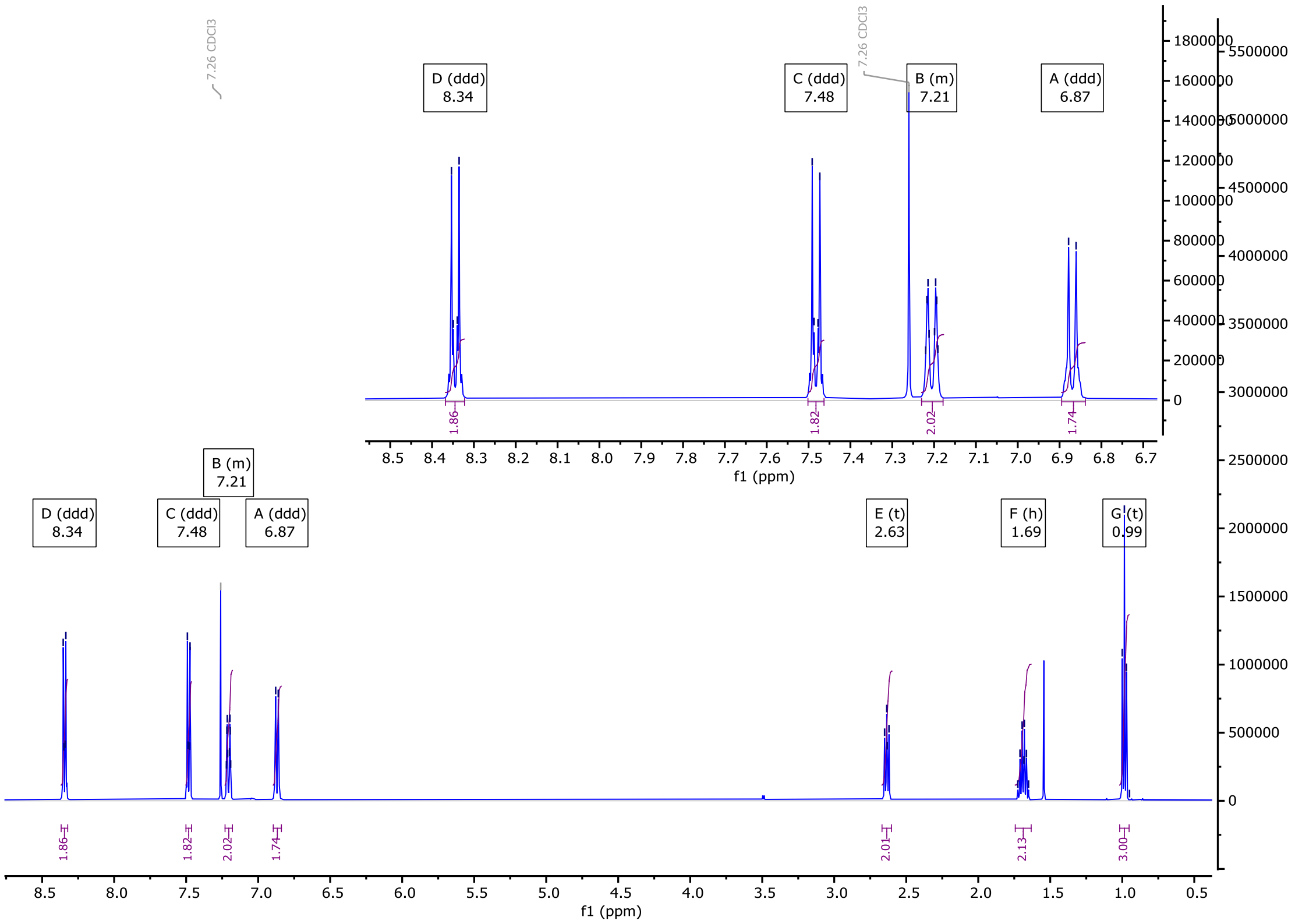

**Fig. S107.** $^{1}H$ NMR of **4d** in $CDCl_3$.

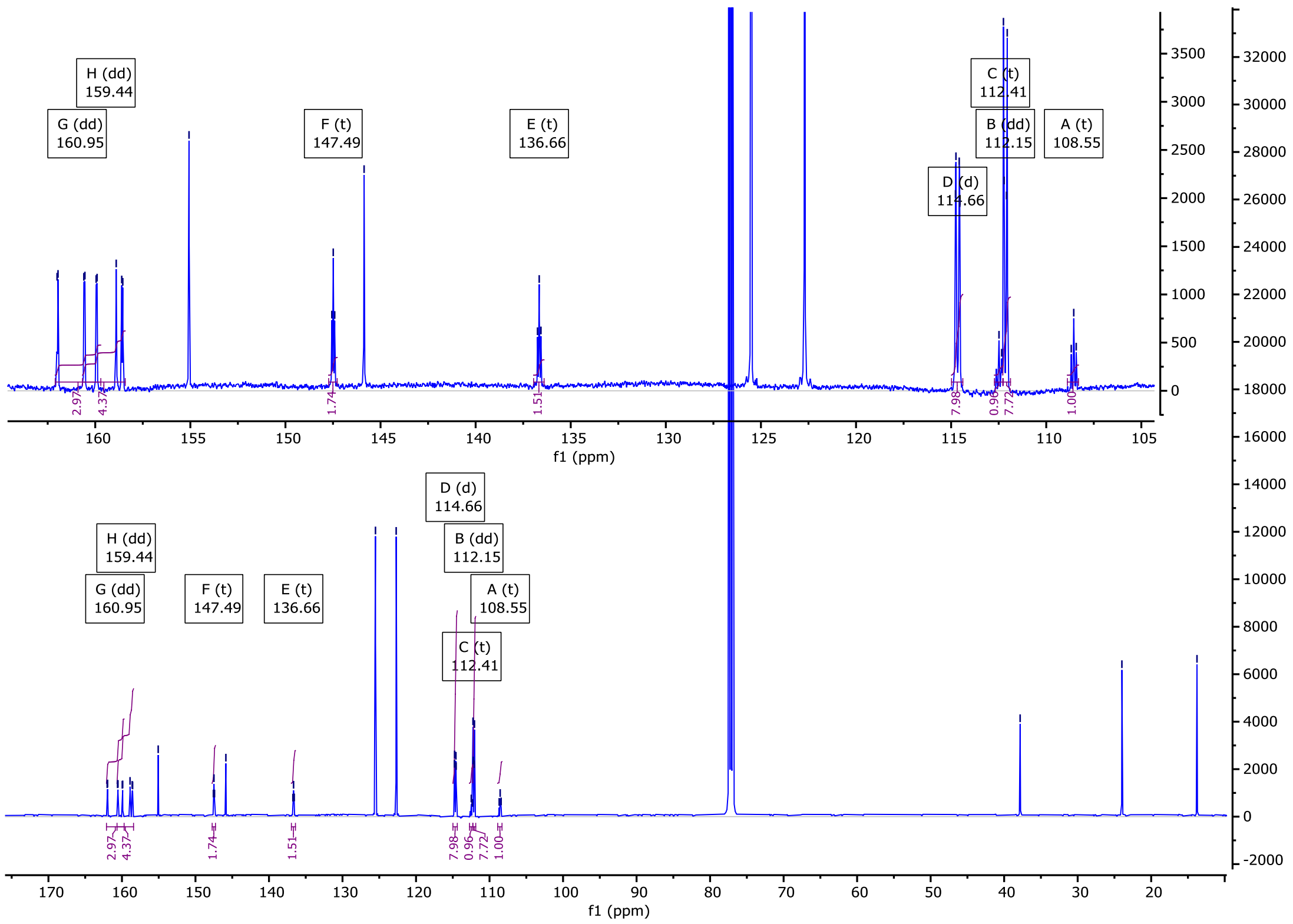

**Fig. S108.** $^{13}C\{^{1}H\}$ NMR of **4d** in $CDCl_3$.

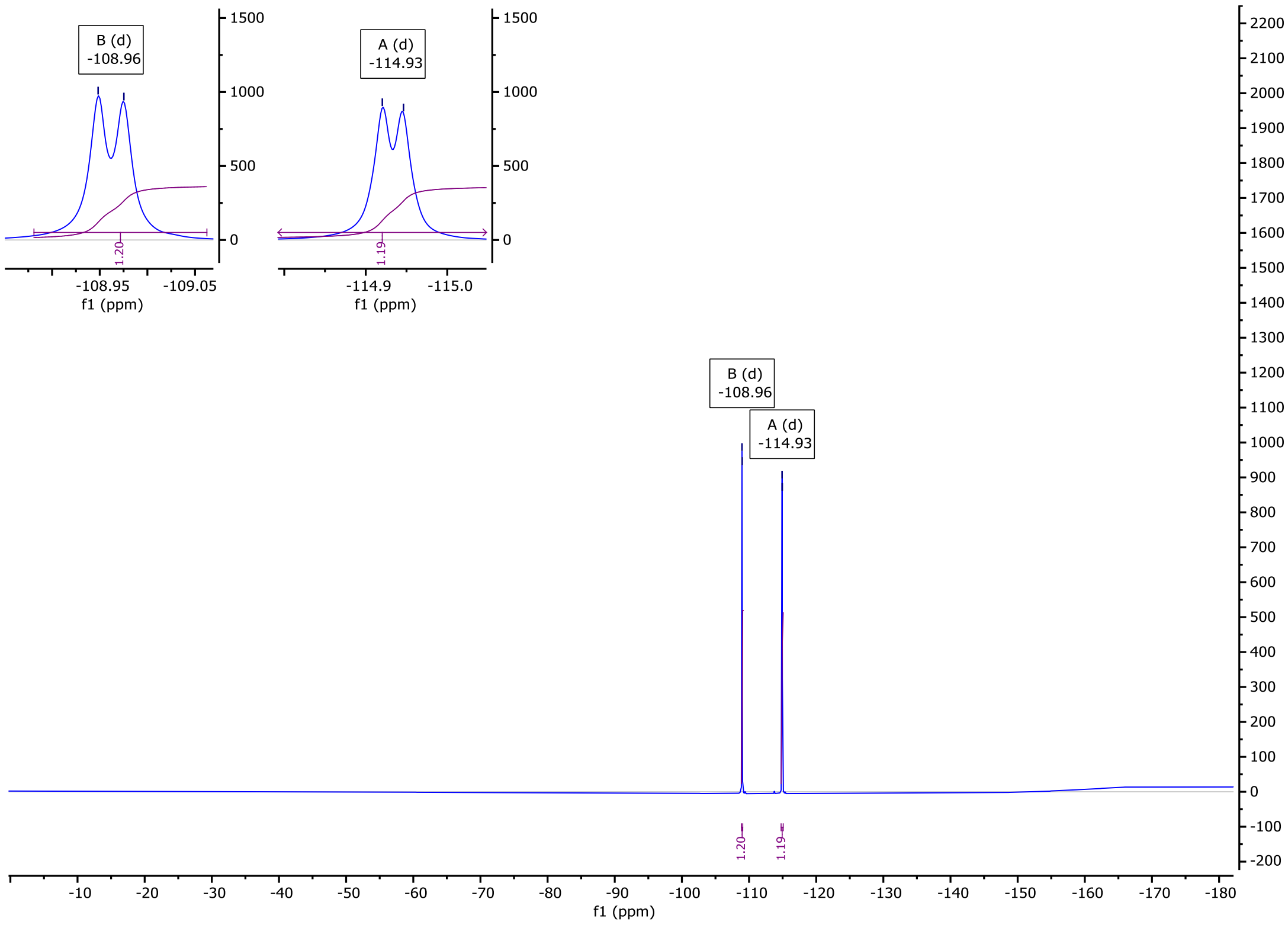

**Fig. S109.** $^{19}F$ NMR of **4d** in $CDCl_3$.

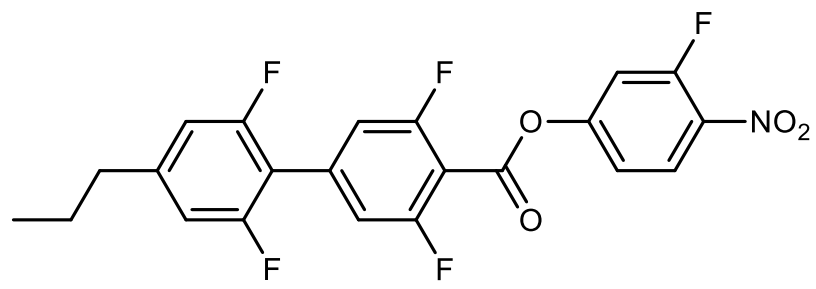

**4e**

*3-fluoro-4-nitrophenyl 2',3,5,6'-tetrafluoro-4'-propyl-[1,1'-biphenyl]-4-carboxylate*

Yield: (White Solid) 74 mg, 33 %

$^{1}$H NMR (501 MHz): 8.21 (t, *J* = 8.9 Hz, 1H, Ar-**H**), 7.35 (dd, *J* = 11.1, 2.4 Hz, 1H, Ar-**H**), 7.28 (ddd, *J* = 9.0, 2.5, 1.5 Hz, 1H, Ar-**H**), 7.21 (ddd, *J* = 9.2, 1.3 Hz, 2H, Ar-**H**), 6.87 (ddd, *J* = 9.2, 2.9, 2.2 Hz, 2H, Ar-**H**), 2.64 (t, *J* = 7.4 Hz, 2H, Ar-C**H₂**-CH₂), 1.68 (h, *J* = 7.4 Hz, 2H, CH₂-C**H₂**-CH₃), 0.99 (t, *J* = 7.3 Hz, 3H, CH₂-C**H₃**).

$^{13}$C{$^{1}$H} NMR (126 MHz): 161.02 (dd, *J* = 259.0, 6.1 Hz), 159.57 (dd, *J* = 250.6, 6.9 Hz), 156.34 (d, *J* = 267.3 Hz), 155.03 (d, *J* = 10.4 Hz), 147.59 (t, *J* = 9.5 Hz), 137.02 (t, *J* = 11.4 Hz), 135.32 (d, *J* = 7.1 Hz), 127.48 (d, *J* = 2.1 Hz), 118.13 (d, *J* = 4.2 Hz), 114.72 (dd, *J* = 23.9, 2.4 Hz), 112.49 (d, *J* = 24.0 Hz), 112.17 (dd, *J* = 20.3, 4.6 Hz), 108.09 (t, *J* = 16.3 Hz), 37.83, 23.98, 13.77.

$^{19}$F NMR (376 MHz): -108.64 (d, $J_{F-H}$ = 10.1 Hz, 2F, Ar-**F**), -112.79 (t, $J_{F-H}$ = 9.6 Hz, 1F, Ar-**F**), -114.92 (d, $J_{F-H}$ = 9.7 Hz, 2F, Ar-**F**).

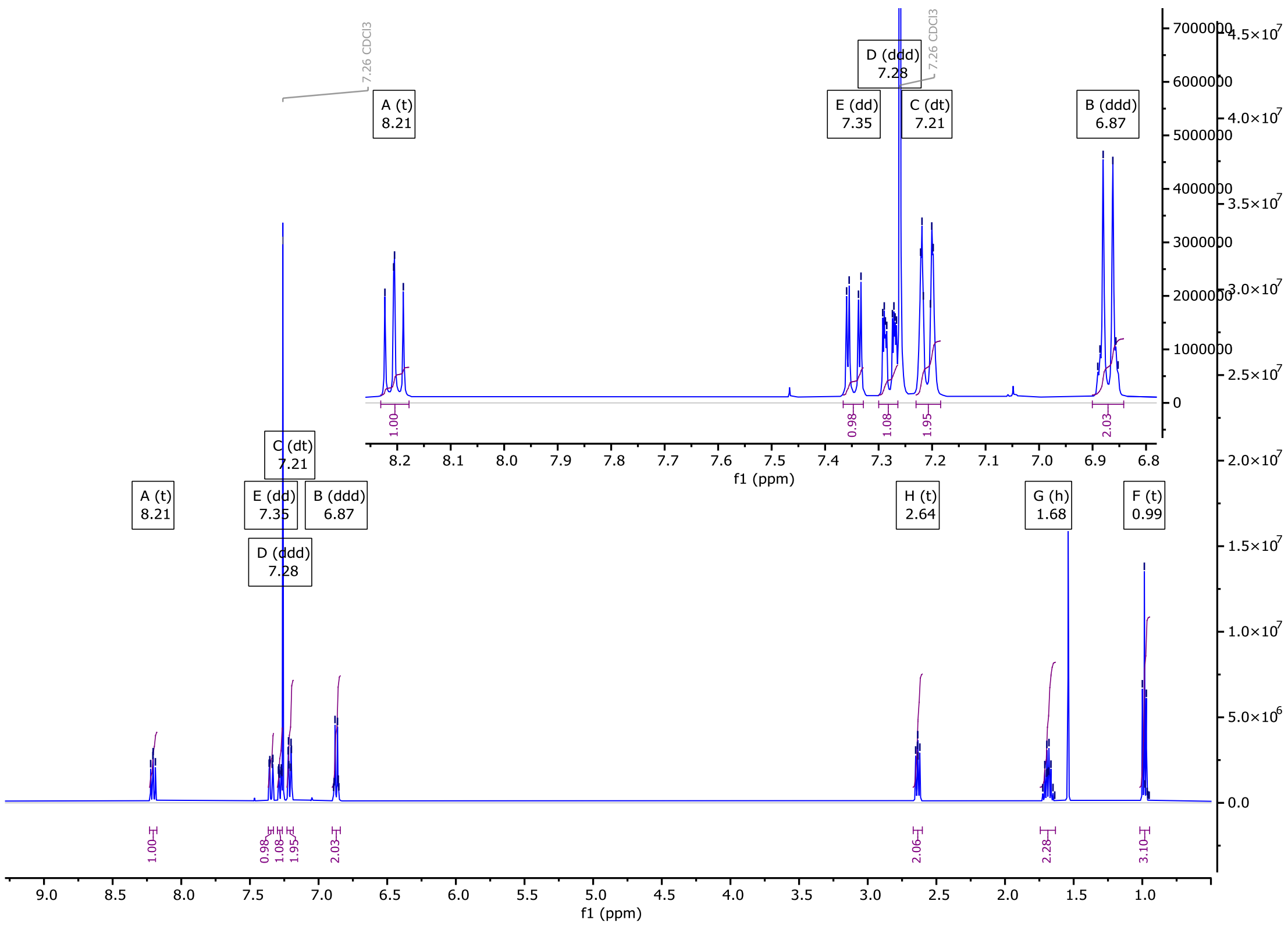

**Fig. S110.** $^{1}H$ NMR of **4e** in $CDCl_3$.

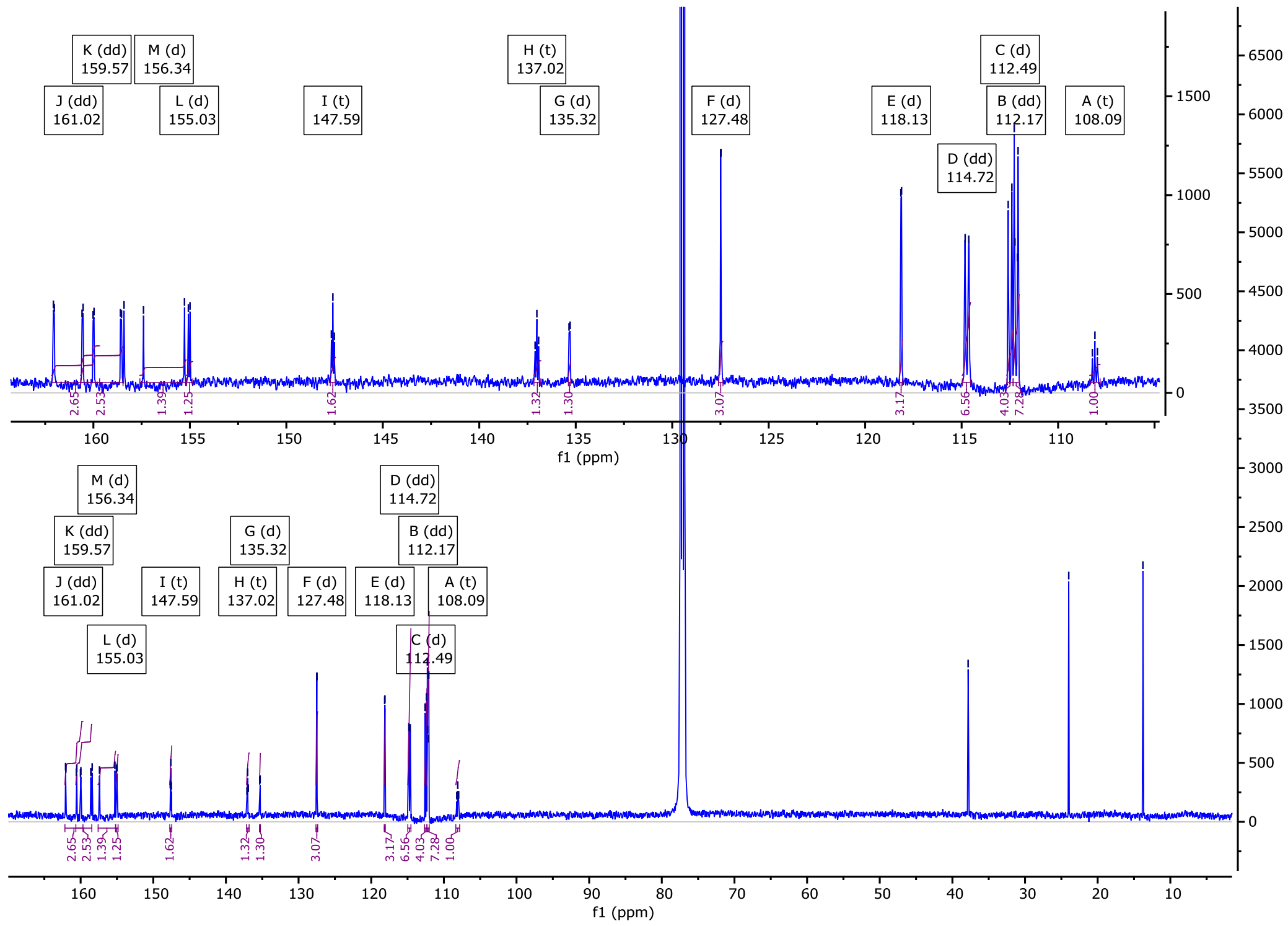

**Fig. S111.** $^{13}C\{^{1}H\}$ NMR of **4e** in $CDCl_3$.

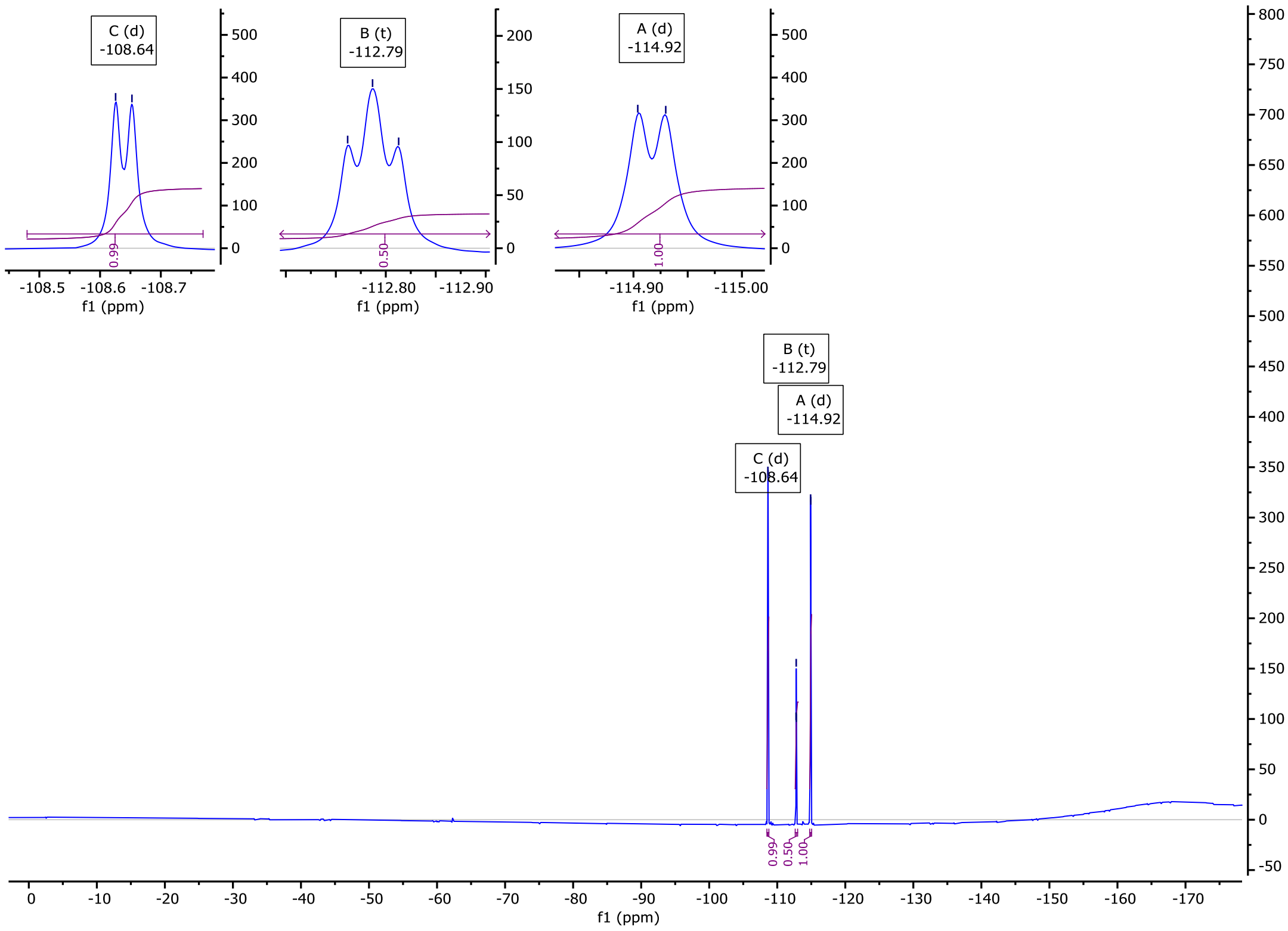

**Fig. S112.** $^{19}$F NMR of **4e** in $CDCl_3$.

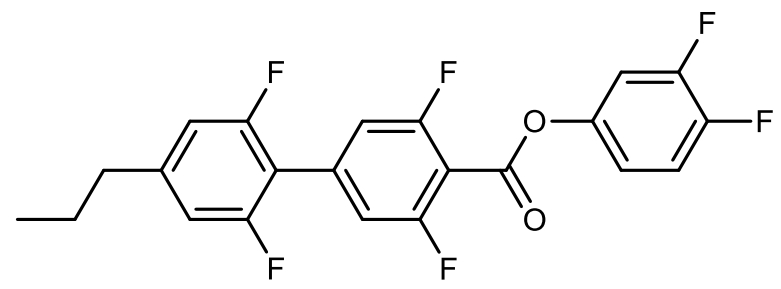

**4f**

3,4-difluorophenyl 2',3,5,6'-tetrafluoro-4'-propyl-[1,1'-biphenyl]-4-carboxylate

Yield: (White Solid) 95 mg, 45 %

$^1$H NMR (501 MHz): 7.23 (dd, *J* = 8.9, 0.9 Hz, 1H, Ar-**H**), 7.21 – 7.15 (m, 3H, Ar-**H**)*, 7.06 – 7.02 (m, 1H, Ar-**H**), 6.86 (ddd, *J* = 9.0, 2.9, 2.2 Hz, 2H, Ar-**H**), 2.63 (t, *J* = 7.5 Hz, 2H, Ar-C**H$_2$**-CH$_2$), 1.68 (h, *J* = 7.3 Hz, 2H, CH$_2$-C**H$_2$**-CH$_3$), 0.98 (t, *J* = 7.3 Hz, 3H, CH$_2$-C**H$_3$**) *Overlapping Signals.

$^{13}$C{$^1$H} NMR (126 MHz): 160.83 (dd, *J* = 257.9, 6.4 Hz), 159.57 (dd, *J* = 251.0, 7.0 Hz), 159.51, 150.13 (dd, *J* = 250.6, 14.2 Hz), 147.78 (dd, *J* = 246.8, 12.4 Hz), 147.35 (t, *J* = 9.5 Hz), 145.94 (dd, *J* = 8.6, 3.4 Hz), 136.23 (t, *J* = 11.3 Hz), 117.82 (dd, *J* = 6.3, 3.9 Hz), 117.60 (d, *J* = 18.9 Hz), 114.60 (dd, *J* = 24.2, 2.5 Hz), 112.51 (t, *J* = 17.2 Hz), 112.12 (dd, *J* = 20.3, 4.8 Hz), 111.90 (d, *J* = 20.2 Hz), 108.96 (t, *J* = 16.9 Hz), 37.82, 23.99, 13.78.

$^{19}$F NMR (376 MHz): -109.39 (d, $J_{F\text{-}H}$ = 10.1 Hz, 2F, Ar-**F**), -114.95 (d, $J_{F\text{-}H}$ = 9.4 Hz, 2F, Ar-**F**), -134.12 – -134.32 (m, 1F, Ar-**F**), -139.95 – -140.22 (m, 1F, Ar-**F**).

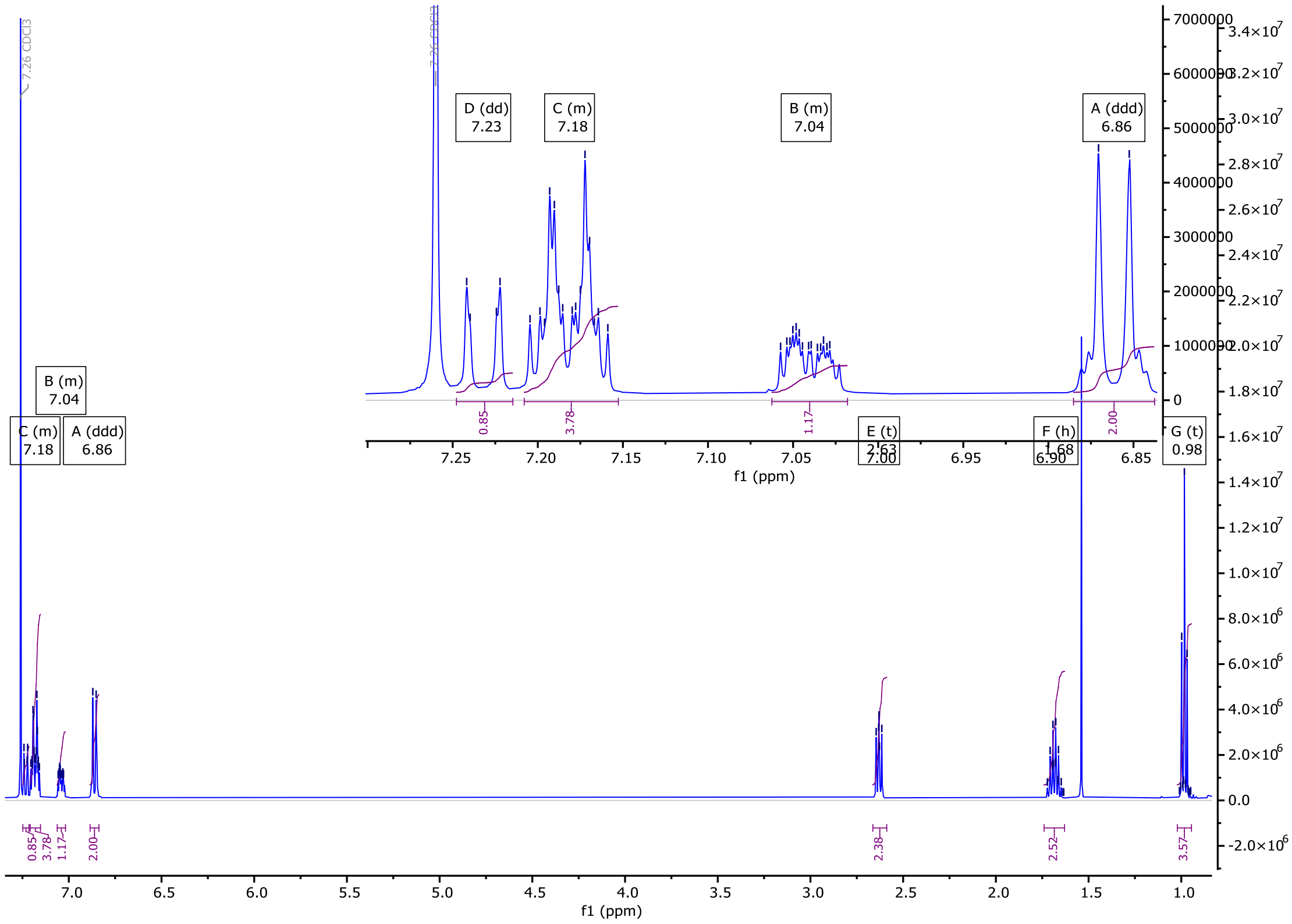

**Fig. S113.** $^{1}H$ NMR of **4f** in $CDCl_3$.

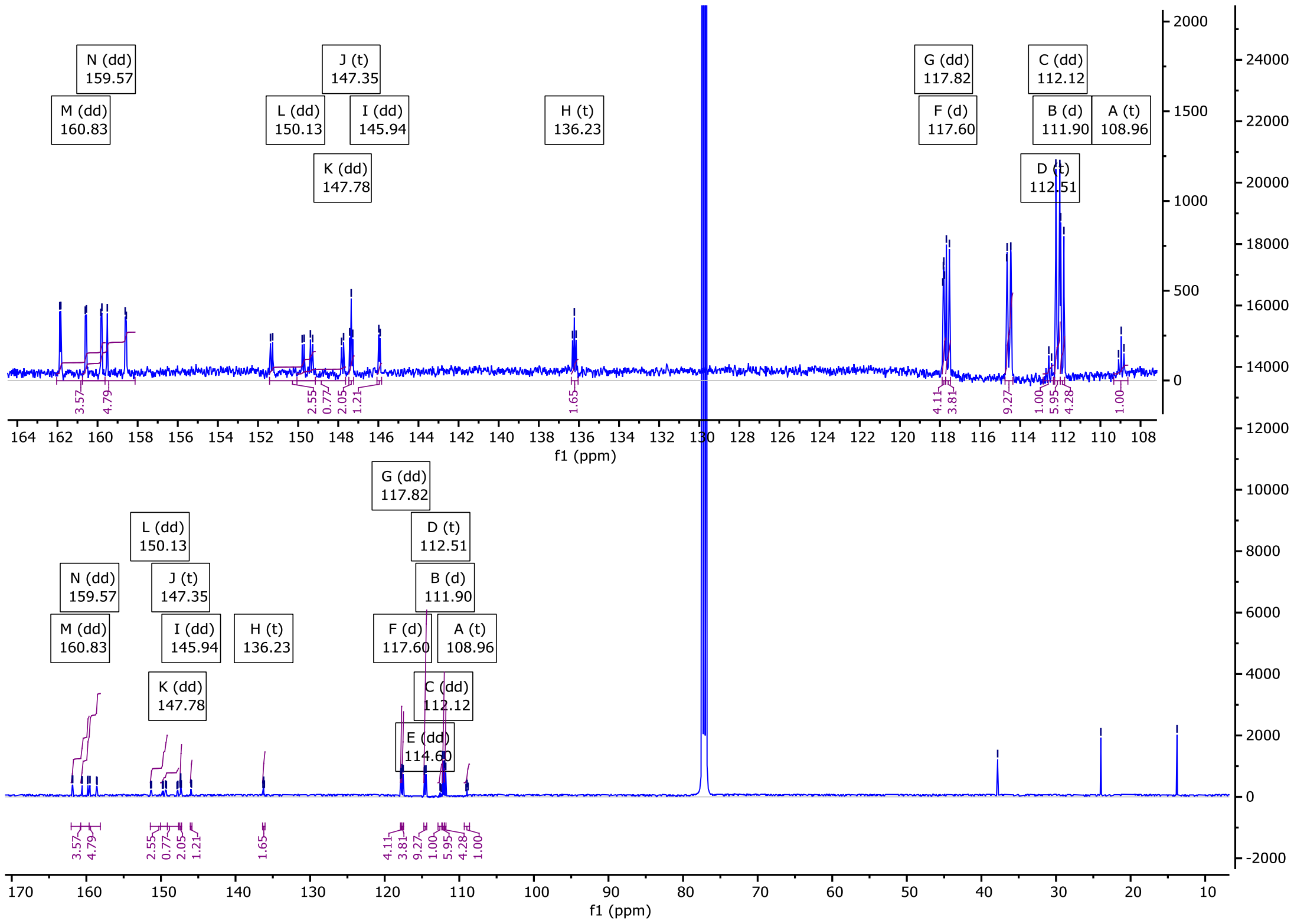

**Fig. S114.** $^{13}C\{^{1}H\}$ NMR of **4f** in $CDCl_3$.

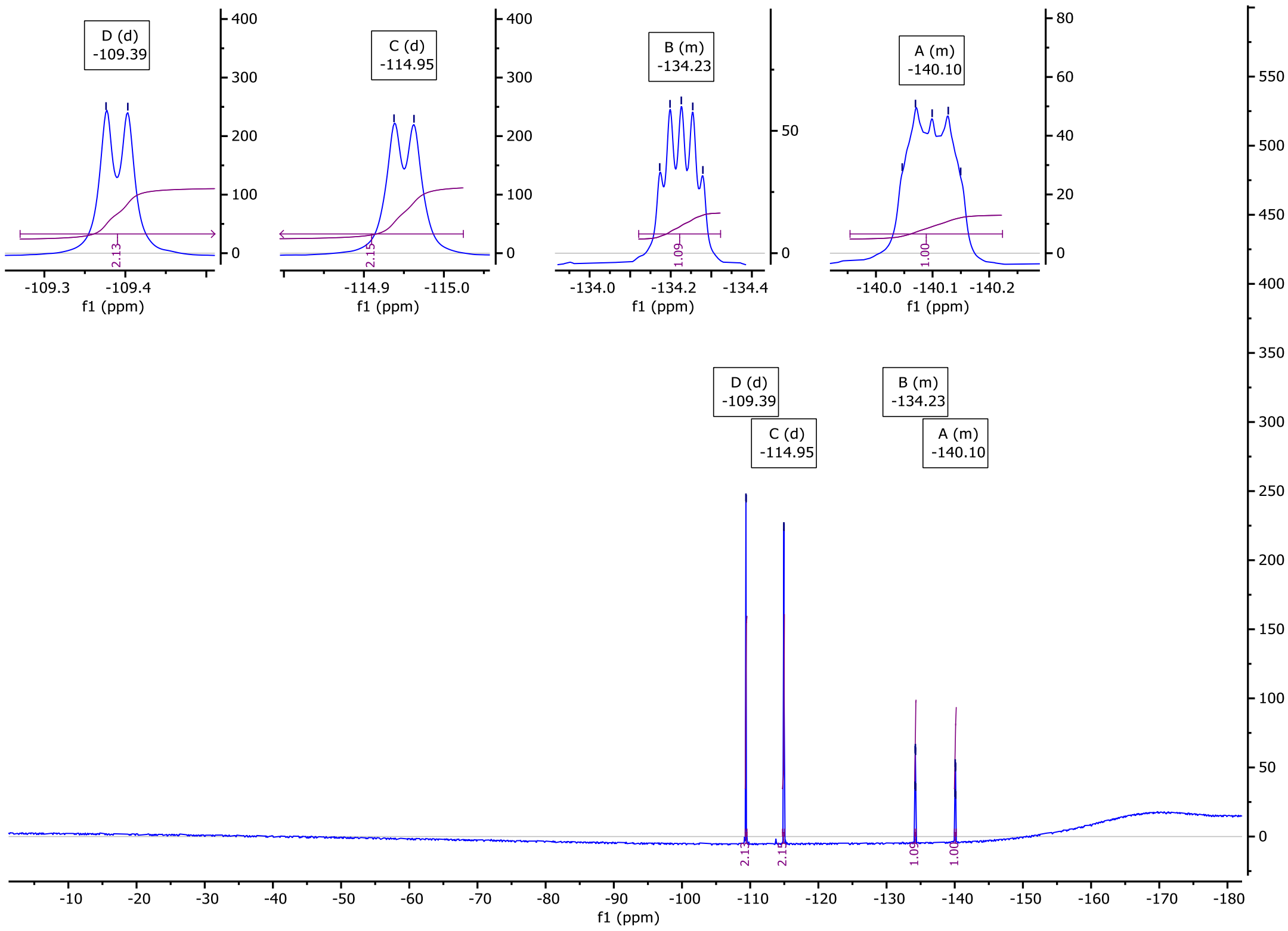

**Fig. S115.** $^{19}F$ NMR of **4f** in $CDCl_3$.

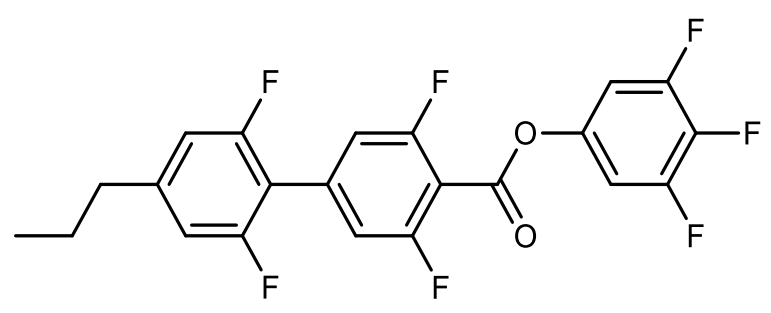

**4g**

*3,4,5-trifluorophenyl 2',3,5,6'-tetrafluoro-4'-propyl-[1,1'-biphenyl]-4-carboxylate*

Yield: (White Solid) 88 mg, 40 %

$^{1}$H NMR (501 MHz): 7.21 – 7.16 (m, 2H, Ar-**H**), 7.04 – 6.97 (m, 2H, Ar-**H**), 6.86 (ddd, *J* = 9.1, 2.8, 2.2 Hz, 2H, Ar-**H**), 2.63 (t, *J* = 7.4 Hz, 2H, Ar-C**H₂**-CH₂), 1.68 (h, *J* = 7.5 Hz, 2H, CH₂-C**H₂**-CH₃), 0.98 (t, *J* = 7.3 Hz, 3H, CH₂-C**H₃**).

$^{13}$C{$^{1}$H} NMR (126 MHz): 160.89 (dd, *J* = 258.3, 6.3 Hz), 159.47 (dd, *J* = 250.8, 7.3 Hz), 158.55, 151.62 (ddd, *J* = 251.6, 11.0, 5.7 Hz), 147.45 (t, *J* = 9.5 Hz), 144.95 (td, *J* = 11.8, 3.9 Hz), 138.60 (dt, *J* = 250.9, 14.9 Hz), 136.58 (t, *J* = 11.3 Hz), 114.59 (dd, *J* = 24.1, 2.4 Hz), 112.49 (t, *J* = 17.6 Hz), 112.14 (dd, *J* = 20.5, 4.9 Hz), 108.44 (t, *J* = 16.7 Hz), 107.28 (dd, *J* = 18.7, 6.1 Hz), 37.83, 23.99, 13.77.

$^{19}$F NMR (376 MHz): -109.10 (d, $J_{F-H}$ = 10.0 Hz, 2F, Ar-**F**), -114.94 (d, $J_{F-H}$ = 9.4 Hz, 2F, Ar-**F**), -132.36 (dd, $J_{F-F}$ = 20.8 Hz, $J_{F-H}$ = 8.3 Hz, 2F, Ar-**F**), -162.67 (tt, $J_{F-F}$ = 20.6 Hz, $J_{F-H}$ = 6.1 Hz, 1F, Ar-**F**).

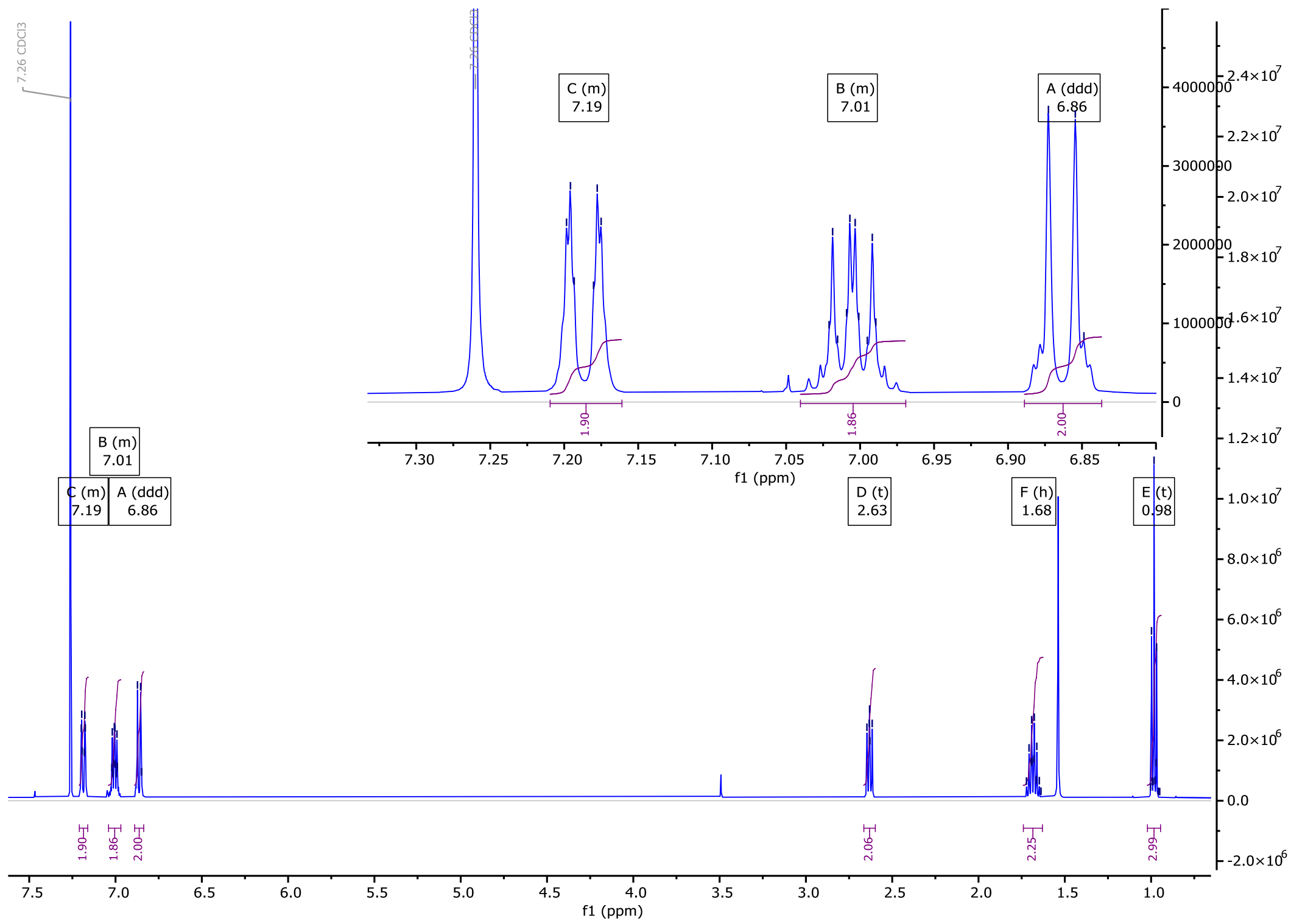

**Fig. S116.** $^{1}$H NMR of **4g** in $CDCl_3$.

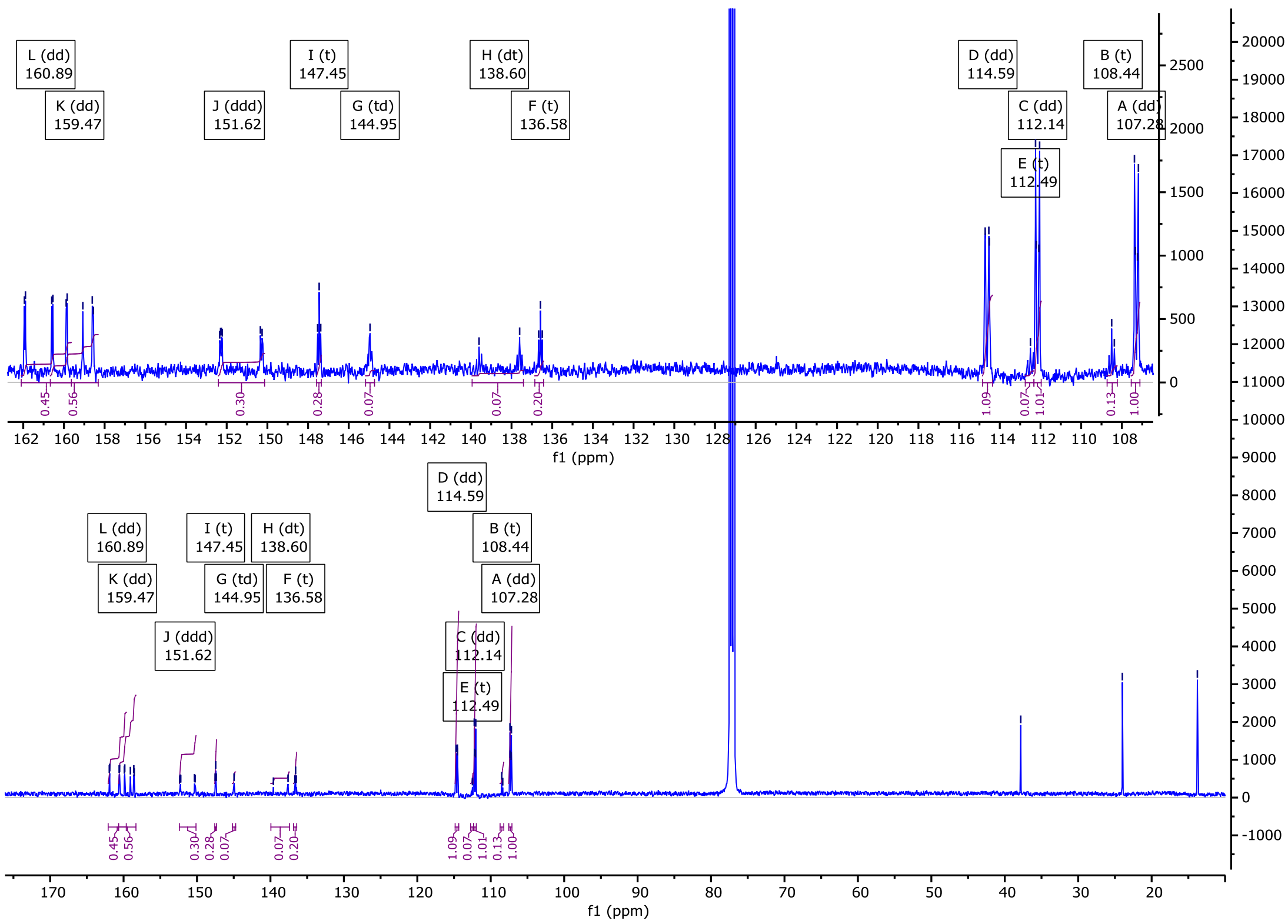

**Fig. S117.** $^{13}C\{^{1}H\}$ NMR of **4g** in $CDCl_3$.

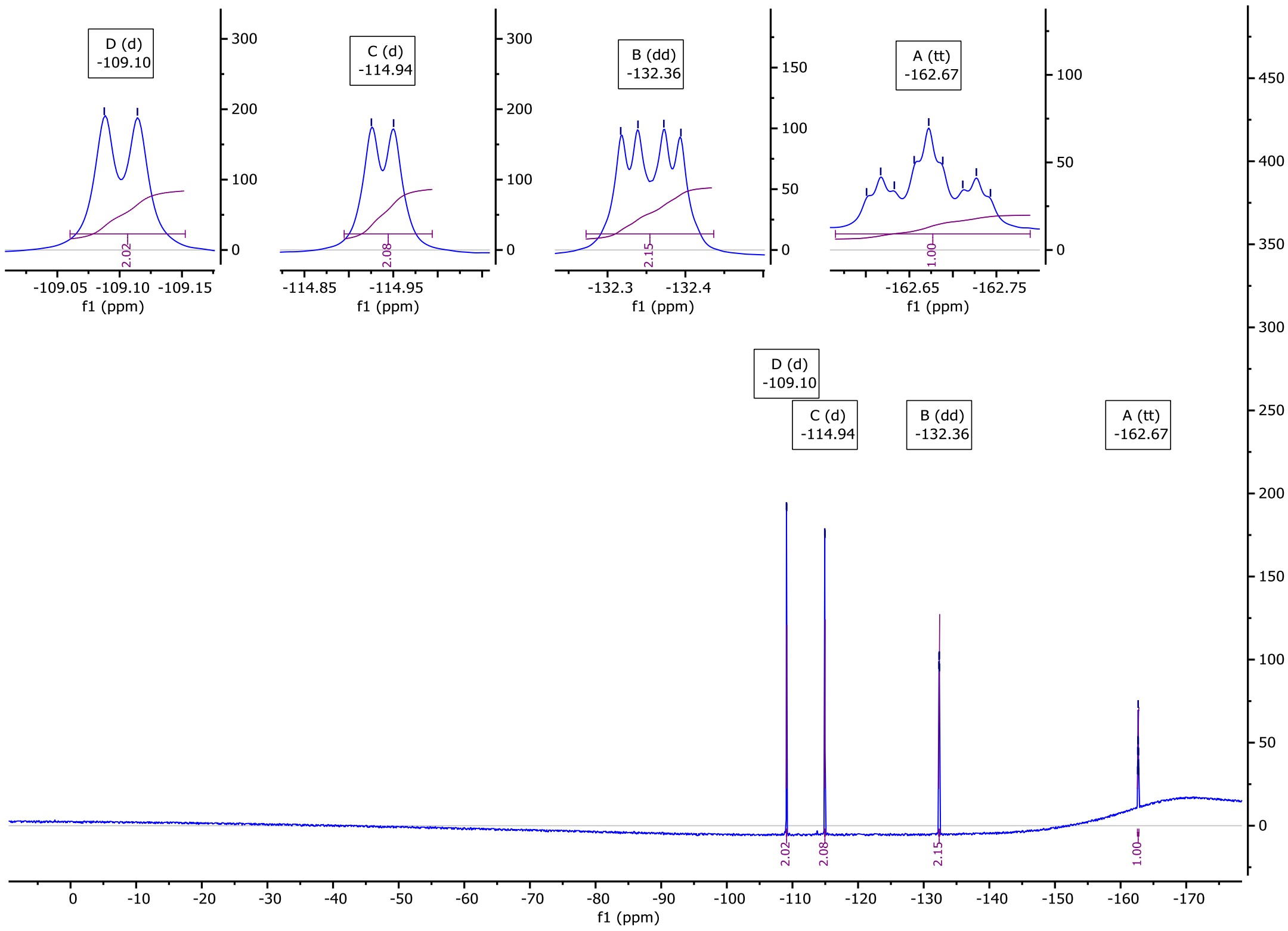

**Fig. S118.** $^{19}F$ NMR of **4g** in $CDCl_3$.

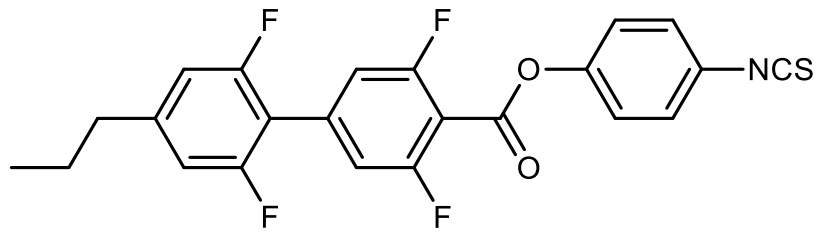

**4i**

*4-isothiocyanatophenyl 2',3,5,6'-tetrafluoro-4'-propyl-[1,1'-biphenyl]-4-carboxylate*

Yield: (Colourless Needles) 51 mg, 23 %

$^1$H NMR (501 MHz): 7.33 – 7.26 (m, 4H, Ar-**H**) *, 7.21 – 7.15 (m, 2H, Ar-**H**), 6.86 (ddd, *J* = 9.0, 3.1, 2.2 Hz, 2H, Ar-**H**), 2.63 (t, *J* = 7.3 Hz, 2H, Ar-C**H₂**-CH₂), 1.68 (h, *J* = 7.6 Hz, 2H, CH₂-C**H₂**-CH₃), 0.98 (t, *J* = 7.3 Hz, 3H, CH₂-C**H₃**). * Overlapping Signals.

$^{13}$C{$^1$H} NMR (126 MHz): 160.82 (dd, *J* = 257.6, 6.4 Hz), 159.57 (dd, *J* = 249.9, 7.4 Hz), 158.56, 149.03, 147.33 (t, *J* = 9.5 Hz), 136.40, 136.15 (t, *J* = 11.3 Hz), 129.51, 127.01, 123.08, 114.57 (dd, *J* = 24.1, 2.3 Hz), 112.58 (t, *J* = 17.6 Hz), 112.11 (dd, *J* = 20.6, 4.8 Hz), 109.10 (t, *J* = 17.0 Hz), 37.81, 23.99, 13.77.

$^{19}$F NMR (376 MHz): -109.40 (d, $J_{F-H}$ = 10.0 Hz, 2F, Ar-**F**), -114.95 (d, $J_{F-H}$ = 9.5 Hz, 2F, Ar-**F**).

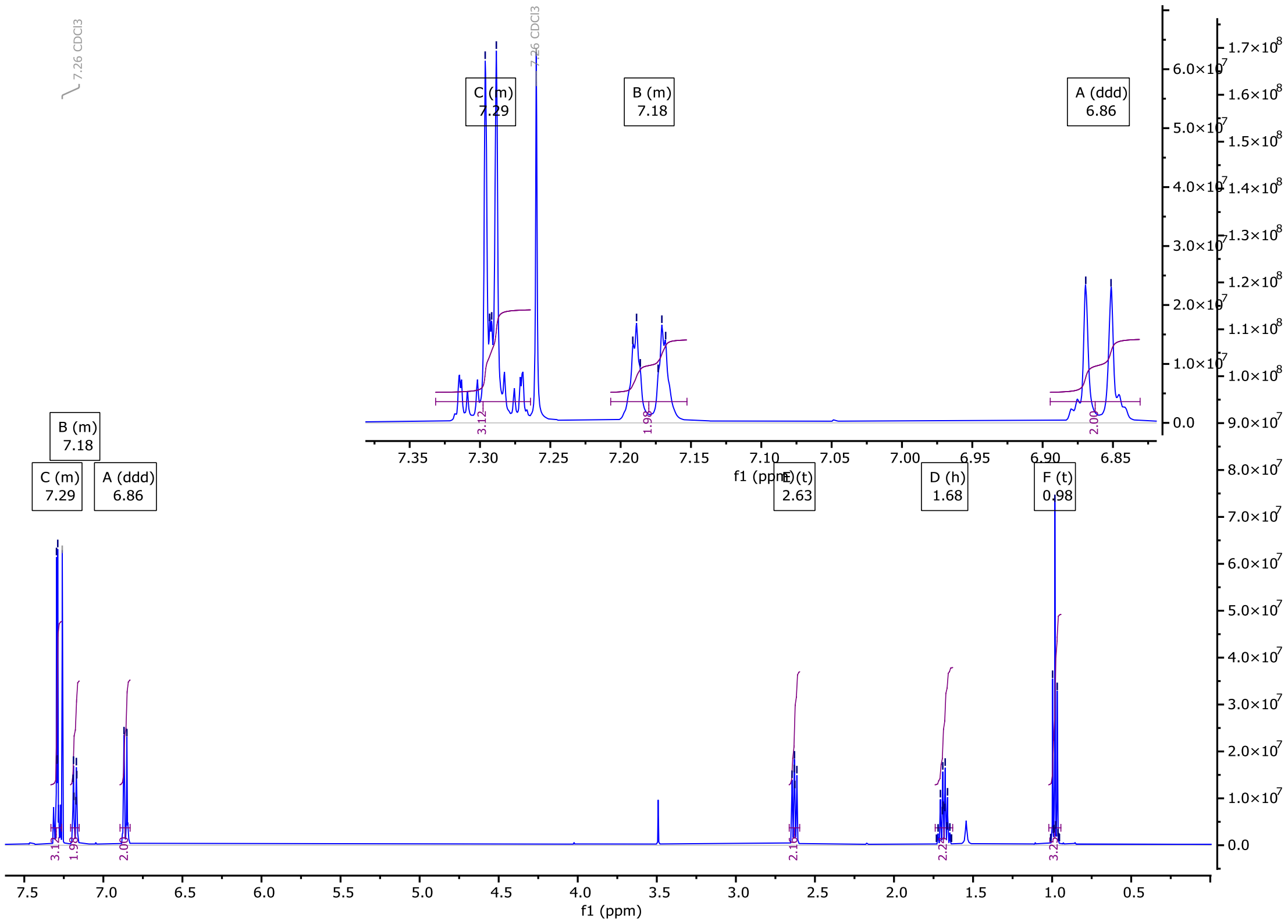

**Fig. S119.** $^{1}H$ NMR of **4i** in $CDCl_3$.

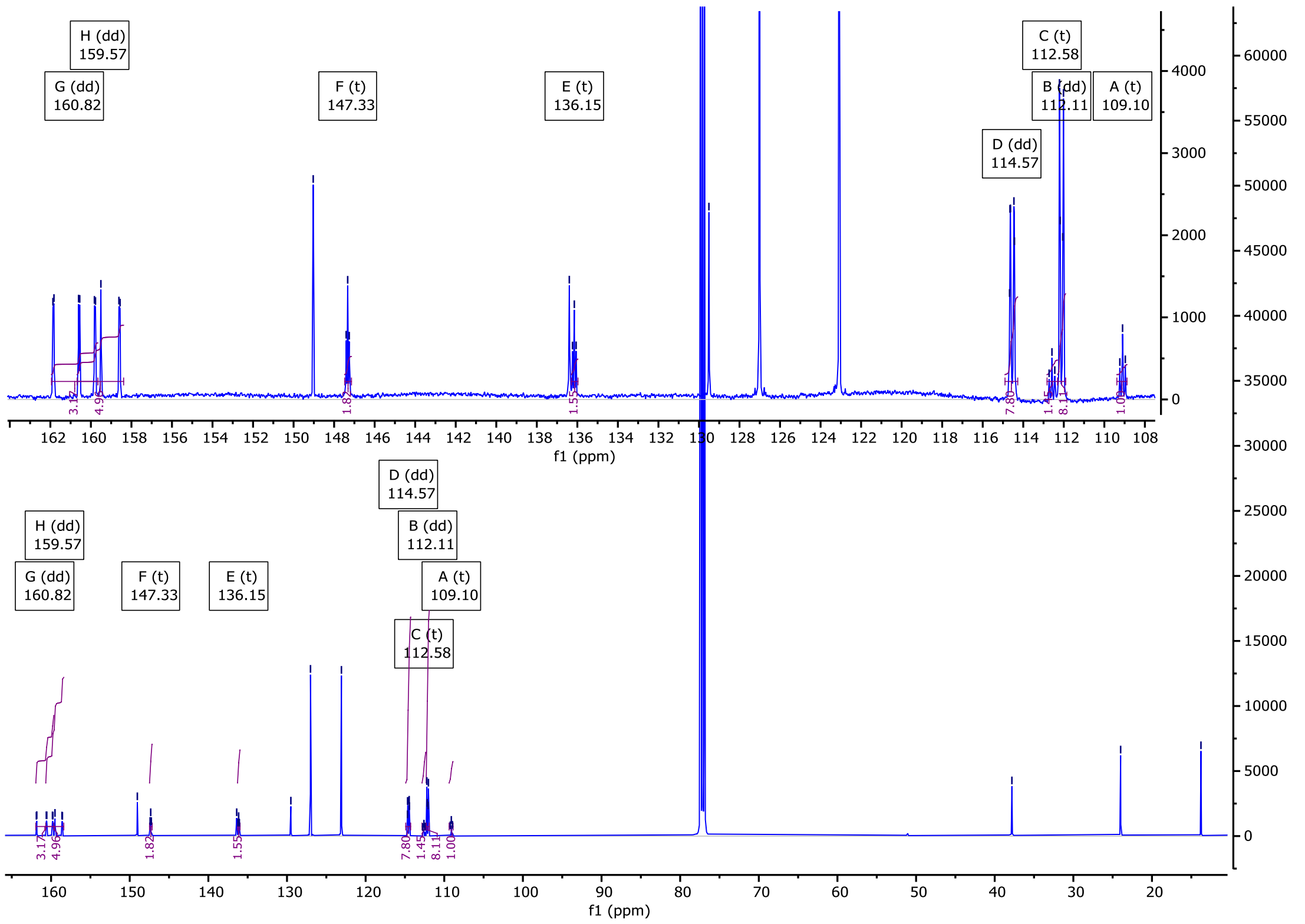

**Fig. S120.** $^{13}C\{^{1}H\}$ NMR of **4i** in $CDCl_3$.

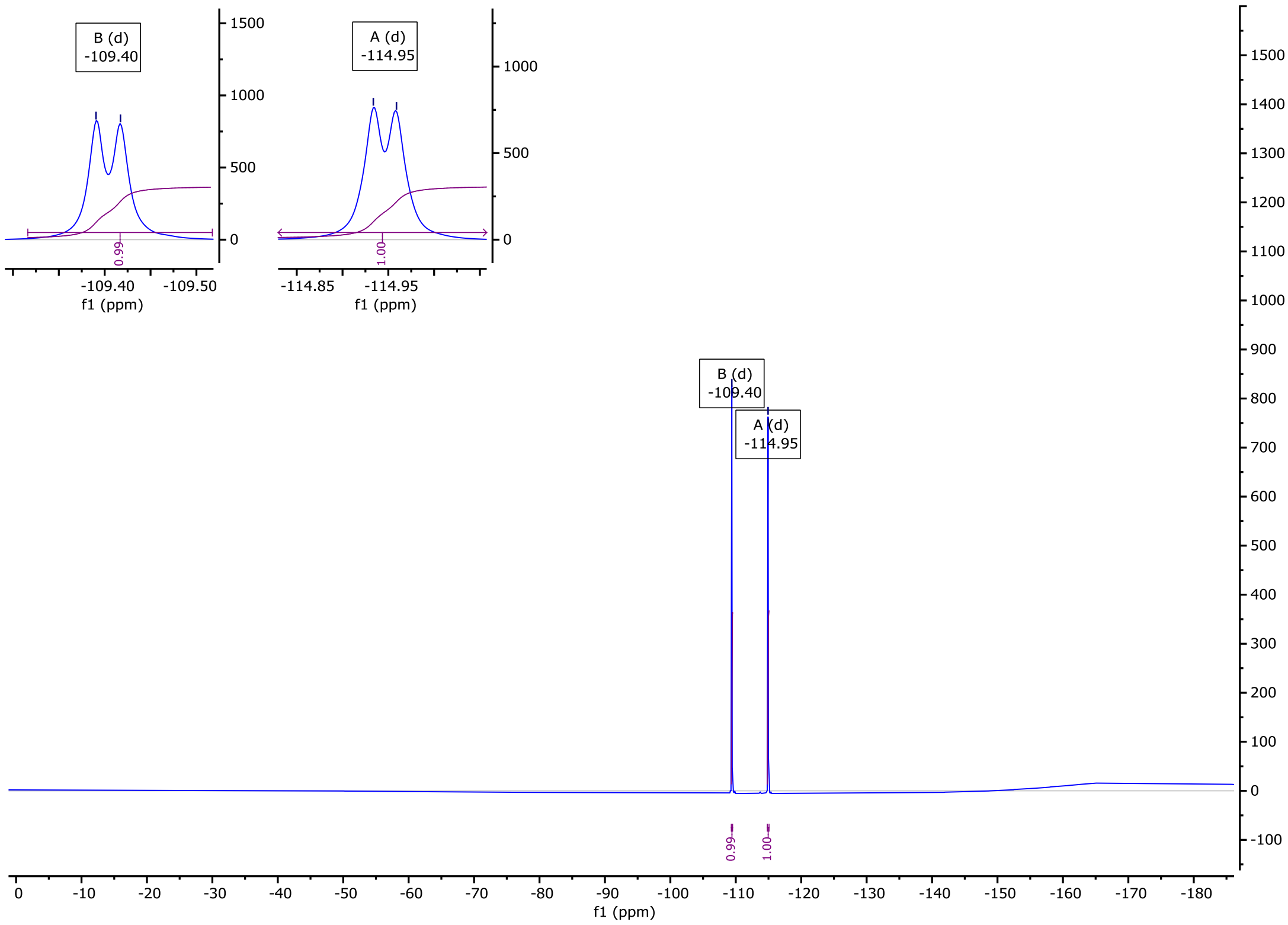

**Fig. S121.** $^{19}F$ NMR of **4i** in $CDCl_3$.

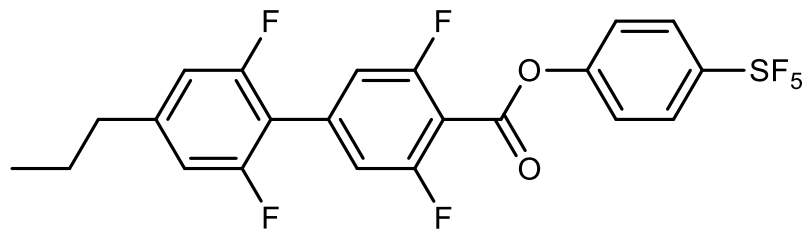

**4j**

*4-(pentafluoro-l6-sulfaneyl)phenyl 2',3,5,6'-tetrafluoro-4'-propyl-[1,1'-biphenyl]-4-carboxylate*

Yield: (White Solid) 50 mg, 20 %

$^1$H NMR (501 MHz): 7.86 (ddd, *J* = 9.2, 3.2, 2.1 Hz, 2H, Ar-**H**), 7.40 (d, *J* = 8.9 Hz, 2H, Ar-**H**), 7.22 – 7.17 (m, 2H, Ar-**H**), 6.87 (ddd, *J* = 9.0, 3.2, 2.1 Hz, 2H, Ar-**H**), 2.63 (t, *J* = 7.4 Hz, 2H, Ar-C$\mathbf{H_2}$-$CH_2$), 1.69 (h, *J* = 7.4 Hz, 2H, $CH_2$-C$\mathbf{H_2}$-$CH_3$), 0.98 (t, *J* = 7.3 Hz, 3H, $CH_2$-C$\mathbf{H_3}$).

$^{13}$C{$^1$H} NMR (126 MHz): 160.90 (dd, *J* = 258.1, 6.3 Hz), 159.50 (dd, *J* = 251.1, 7.1 Hz), 159.17, 158.62, 151.51 (t, *J* = 18.2 Hz), 147.41 (t, *J* = 9.5 Hz), 136.43 (t, *J* = 11.3 Hz), 127.87 (t, *J* = 4.6 Hz), 122.08, 114.63 (dd, *J* = 24.1, 2.5 Hz), 112.54 (t, *J* = 17.6 Hz), 112.14 (dd, *J* = 20.3, 4.7 Hz), 108.81 (t, *J* = 16.7 Hz), 37.82, 23.99, 13.77.

$^{19}$F NMR (376 MHz): 83.93 (p, $J_{F-S}$ = 150.4 Hz, 1F, S-$\mathbf{F_\Lambda}$), 63.46 (d, $J_{F-S}$ = 150.3 Hz, 4F, S-$\mathbf{F_{4\Delta}}$), -109.17 (d, $J_{F-H}$ = 10.0 Hz, 2F, Ar-**F**), -114.94 (d, $J_{F-H}$ = 9.6 Hz, 2F, Ar-**F**).

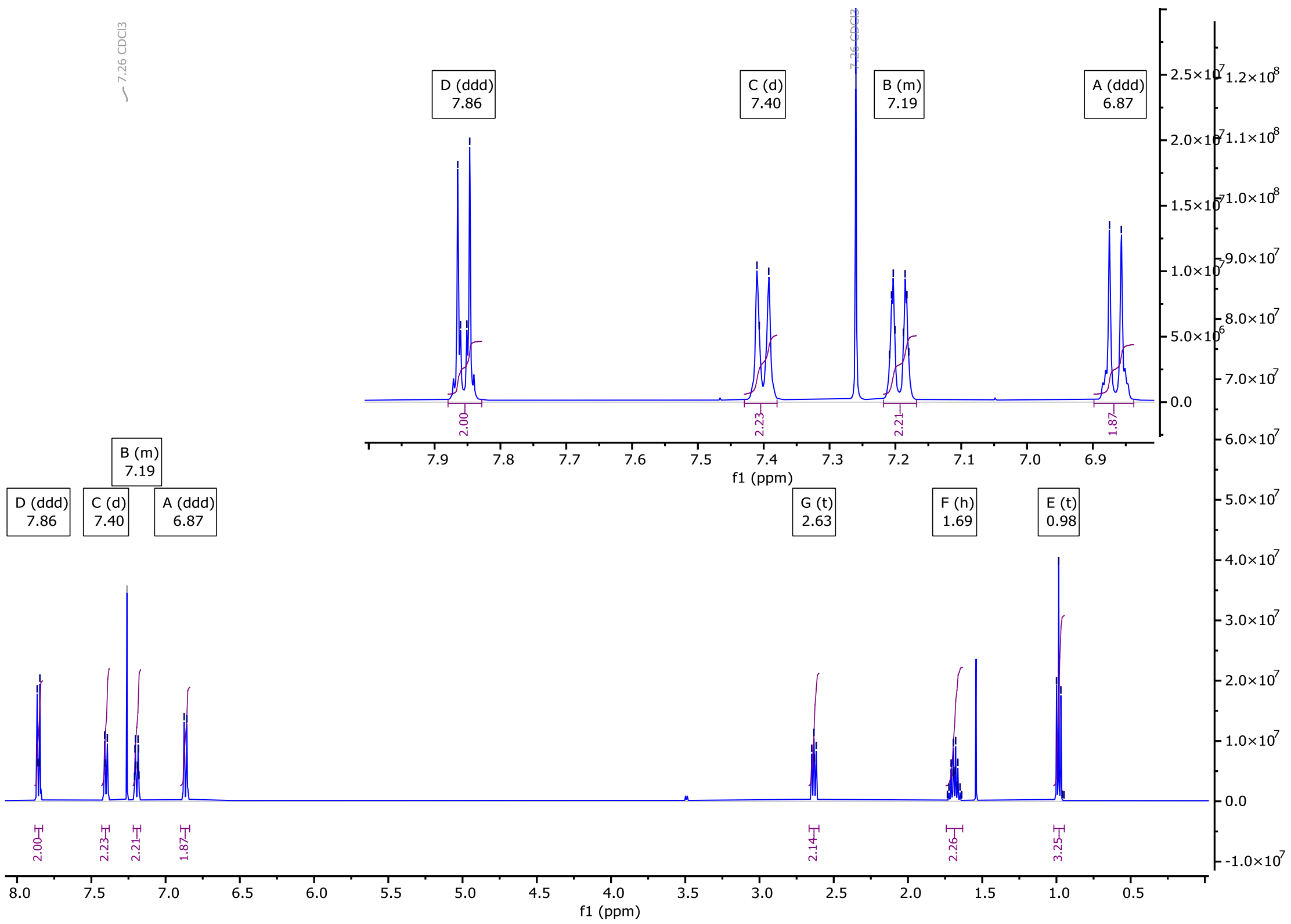

**Fig. S122.** $^{1}H$ NMR of **4j** in $CDCl_3$.

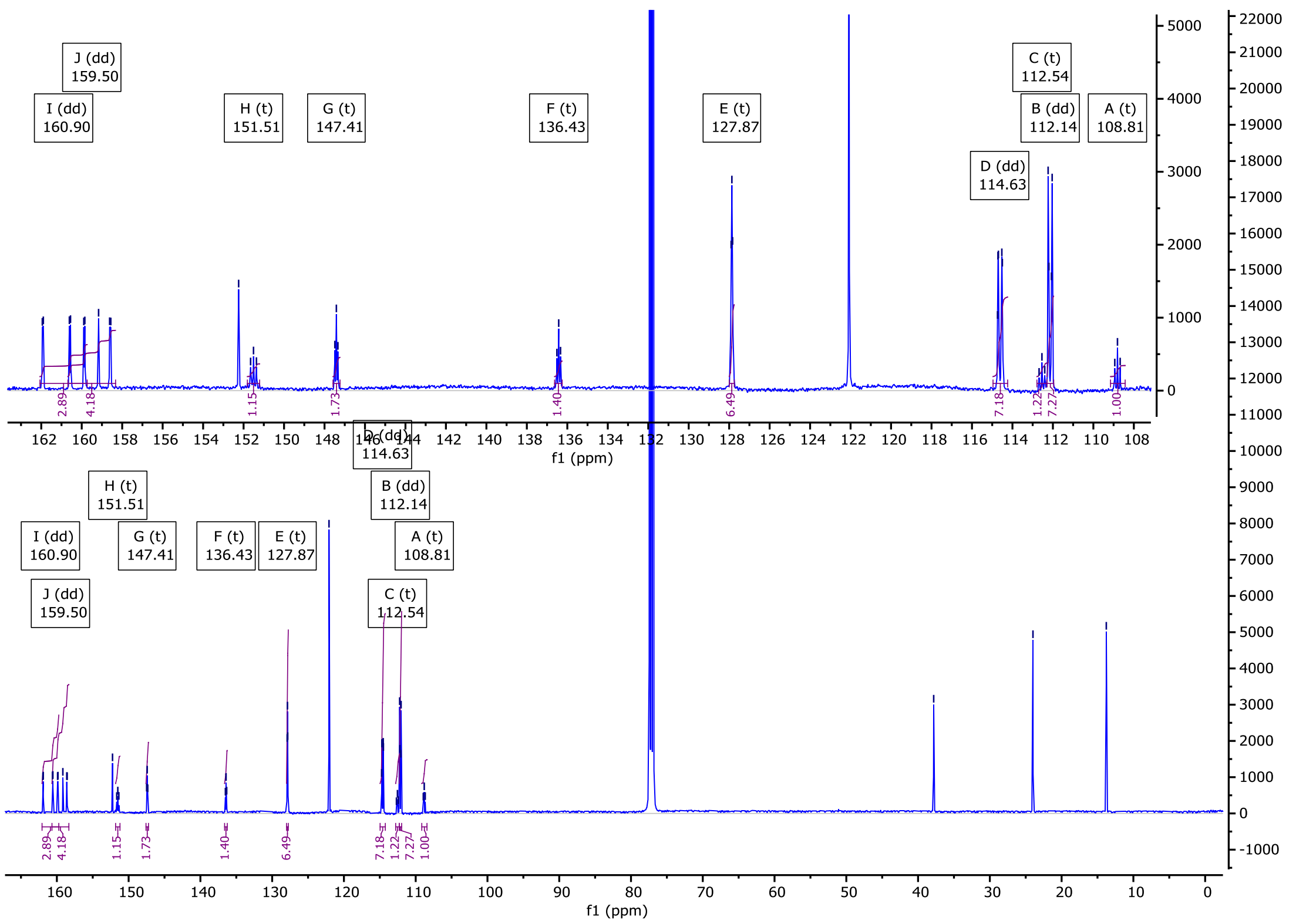

**Fig. S123.** $^{13}C\{^{1}H\}$ NMR of **4j** in $CDCl_3$.

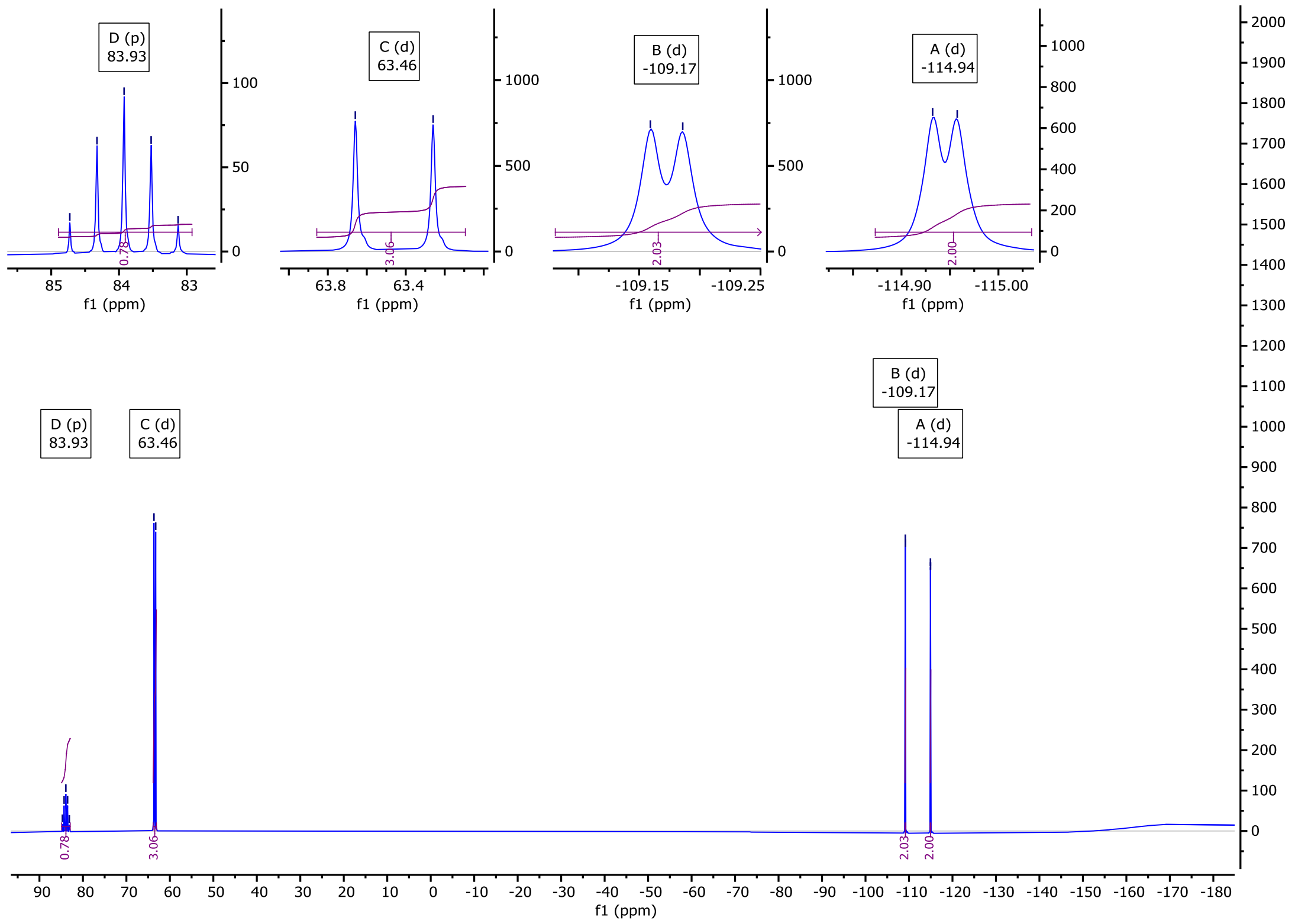

**Fig. S124.** $^{19}F$ NMR of **4j** in $CDCl_3$.

## 4. Supplemental References